\documentclass[10pt,preprint2]{aastex}

\usepackage{amsmath}

\def\p/{\mbox{$^1$}}
\def\pp/{\mbox{$^2$}}
\def\ppp/{\mbox{$^3$}}
\def\pppp/{\mbox{$^4$}}
\def\m/{\mbox{$^{-1}$}}
\def\mm/{\mbox{$^{-2}$}}
\def\mmm/{\mbox{$^{-3}$}}
\def\mmmm/{\mbox{$^{-4}$}}
\def\Ms/{\mbox{M$_\odot$}}

\def\Dm{\mbox{$\Delta m$}}
\def\Teff{\mbox{$T_{\rm eff}$}}
\def\Bap{\mbox{$B_{\rm ap}$}}
\def\Jap{\mbox{$J_{\rm ap}$}}
\newcommand{\HeI}[1]{\mbox{He\,{\sc i}~$\lambda${#1}}}
\newcommand{\HeII}[1]{\mbox{He\,{\sc ii}~$\lambda${#1}}}
\newcommand{\HeIII}[1]{\mbox{He\,{\sc i}~+~He\,{\sc ii}~$\lambda${#1}}}
\newcommand{\CIII}[1]{\mbox{C\,{\sc iii}~$\lambda${#1}}}
\newcommand{\CIV}[1]{\mbox{C\,{\sc iv}~$\lambda${#1}}}
\newcommand{\NII}[1]{\mbox{N\,{\sc ii}~$\lambda${#1}}}
\newcommand{\NIII}[1]{\mbox{N\,{\sc iii}~$\lambda${#1}}}
\newcommand{\NIIId}[1]{\mbox{N\,{\sc iii}~$\lambda\lambda${#1}}}
\newcommand{\NIV}[1]{\mbox{N\,{\sc iv}~$\lambda${#1}}}

\newcommand{\NVd}[1]{\mbox{N\,{\sc v}~$\lambda\lambda${#1}}}
\newcommand{\SiIII}[1]{\mbox{Si\,{\sc iii}~$\lambda${#1}}}

\newcommand{\fracc}[2]{\frac{\raisebox{ .30em}{$#1$}}{\raisebox{-.40em}{$#2$}}}

\shorttitle{The Galactic O-Star Spectroscopic Survey. II}
\shortauthors{Sota et al.}

\slugcomment{Submitted for publication to the Astrophysical Journal Supplement Series}

\begin{document}

\title{The Galactic O-Star Spectroscopic Survey (GOSSS). \linebreak II. Bright Southern Stars\altaffilmark{1}}

\author{A. Sota\altaffilmark{2,3,4}}
\author{J. Ma\'{\i}z Apell\'aniz\altaffilmark{2,3,4,5}}
\affil{Instituto de Astrof\'{\i}sica de Andaluc\'{\i}a-CSIC, Glorieta de la Astronom\'{\i}a s/n, 18008 Granada, Spain}
\author{N. I. Morrell}
\affil{Las Campanas Observatory, Observatories of the Carnegie Institution of Washington, La Serena, Chile}
\author{R. H. Barb\'a\altaffilmark{2}}
\affil{Departamento de F\'{\i}sica, Universidad de La Serena, Av. Cisternas 1200 Norte, La Serena, Chile}
\author{N. R. Walborn}
\affil{Space Telescope Science Institute, 3700 San Martin Drive, Baltimore, MD 21218, USA}
\author{R. C. Gamen}
\affil{Instituto de Astrof\'{\i}sica de La Plata (CCT La Plata-CONICET, Universidad Nacional de La Plata), Paseo del Bosque s/n, 1900 La Plata, Argentina}
\author{J. I. Arias}
\affil{Departamento de F\'{\i}sica, Universidad de La Serena, Av. Cisternas 1200 Norte, La Serena, Chile}
\author{E. J. Alfaro}
\affil{Instituto de Astrof\'{\i}sica de Andaluc\'{\i}a-CSIC, Glorieta de la Astronom\'{\i}a s/n, 18008 Granada, Spain}


\altaffiltext{1}{The GOSSS spectroscopic data in this article were gathered with one primary facility, the 2.5 m du Pont Telescope at 
\facility{Las Campanas Observatory} (LCO), and three auxiliary ones, the 1.5 m Telescope at the 
\facility{Observatorio de Sierra Nevada} (OSN), the 3.5 m Telescope at \facility{Calar Alto Observatory} (CAHA), 
and the 4.2 m William Hershel telescope at \facility{Observatorio del Roque de los Muchachos} (ORM). The OWN spectroscopic data were gathered
at LCO, \facility{La Silla Observatory}, and \facility{CASLEO}. Some of the supporting imaging data 
were obtained by the 2MASS survey and the NASA/ESA \facility{Hubble Space Telescope} (HST). The HST data were obtained at the Space Telescope Science 
Institute, which is operated by the Association of Universities for Research in Astronomy, Inc., under NASA 
contract NAS 5-26555.}

\altaffiltext{2}{Visiting Astronomer, LCO, Chile.}
\altaffiltext{3}{Visiting Astronomer, CAHA, Spain.}
\altaffiltext{4}{Visiting Astronomer, WHT, Spain.}
\altaffiltext{5}{e-mail contact: {\tt jmaiz@iaa.es}.}

\begin{abstract}
We present the second installment of GOSSS, a massive spectroscopic survey of Galactic O stars, based on new homogeneous, high signal-to-noise ratio, $R \sim 2500$ 
digital observations from both hemispheres selected from the Galactic O-Star Catalog (GOSC). In this paper we include bright stars and other objects drawn 
mostly from the first version of GOSC, all of them south of $\delta = -20^{\circ}$, for a total number of 258 O stars. We also revise the northern sample of 
paper I to provide the full list of spectroscopically classified Galactic O stars complete to $B = 8$, bringing the total number of published GOSSS stars to 
448. Extensive sequences of exceptional objects are given, including the early Of/WN, O~Iafpe, Ofc, ON/OC, Onfp, Of?p, and Oe types, as well as 
double/triple-lined spectroscopic binaries. The new spectral subtype O9.2 is also discussed. 
The magnitude and spatial distributions of the observed sample are analyzed.
We also present new results from OWN, a multi-epoch high-resolution
spectroscopic survey coordinated with GOSSS that is assembling the largest sample of Galactic spectroscopic massive binaries ever attained. The OWN data combined
with additional information on spectroscopic and visual binaries from the literature
indicate that only a very small fraction (if any) of the stars with masses above 15-20 M$_\odot$ are born as single systems.
In the future we will publish the rest of the GOSSS survey, which is expected to include over 1000 Galactic O stars.
\end{abstract}


\section{Introduction}
\label{sec:Intr}

The Galactic O-Star Spectroscopic Survey (GOSSS) is a long-term project whose main goals are to obtain homogeneous, high SNR, $R\sim2500$, blue-violet 
spectra of a large number (1000+) of O stars in the Milky Way and to derive accurate and self-consistent spectral types for all of them
\citep{Maizetal11}. In \cite{Sotaetal11a}, from now on paper I, we presented the first installment of the survey, which was comprised of 
the results for 178 northern ($\delta > -20^{\circ}$) O stars and also included an initial grid of spectral classification standards 
from both hemispheres. The GOSSS collaboration has also produced two letters: one on the classification of Of stars with C\,{\sc iii} 
emission lines \citep{Walbetal10a} and another one on rapidly rotating nitrogen-enriched O stars \citep{Walbetal11}. GOSSS data have also 
been used for the discovery of the O star with the strongest magnetic field measured to date (NGC~1624-2, \citealt{Wadeetal12b}) and for the study of a 
shell-like event in the Oe star HD~120\,678 \citep{Gameetal12}.

In parallel to GOSSS, four other surveys (OWN, IACOB, NoMaDS, and CAF\'E-BEANS) are obtaining high-resolution optical spectroscopy of a 
subsample of Galactic O stars. OWN, a high-resolution spectroscopic monitoring survey of southern Galactic O- and WN-type stars,
is obtaining multiple-epoch spectroscopy with the aims of detecting binaries, 
determining their orbits, and deriving their physical parameters \citep{Barbetal10}. The detection of spectroscopic binaries (SBs)
plays an important role in this paper and is described in the next section. IACOB is obtaining spectroscopy of bright northern OB stars 
(minimum of 3 epochs) with the goal of deriving their physical parameters \citep{SimDetal11c,SimDetal11a}. NoMaDS is an extension of
IACOB to fainter stars in the northern hemisphere \citep{Maizetal12,Pelletal12}. CAF\'E-BEANS (Calar Alto Fiber-fed \'Echelle Binary
Evolution Andalusian Northern Survey), the latest addition to the group, is the northern counterpart to the study of southern
SB2 by OWN using the CAF\'E spectrograph at the 2.2~m Calar Alto telescope. Taken together, the five surveys aim to provide the most
complete view to date of Galactic O stars with optical spectroscopy, studying their membership, spectroscopic binarity, intervening ISM,
spatial distribution, and IMF within a few kpc of the Sun. Another complementary survey is obtaining high spatial resolution images of
massive stars to study their visual binarity \citep{Maiz10}. The survey initially used only the AstraLux Norte Lucky Imaging instrument
at the 2.2~m Calar Alto telescope but has been recently extended to the southern hemisphere using its counterpart AstraLux Sur at the 
3.5~m NTT at La Silla. 

This paper is the second installment of the survey and its main goal is to provide a southern ($\delta < -20^{\circ}$) counterpart for
paper I. The basis of the southern star sample is version 1 of the Galactic O-Star Catalog (GOSC, \citealt{Maizetal04b}). To that we
added new stars with $B < 8$, some stars located within a few arcminutes of other O stars, and objects in the Carina region from
version 2 of GOSC \citep{Sotaetal08}. We also eliminated from the sample the O stars in NGC 3603 due to crowding and faintness, leaving a 
total of 258 southern stars, whose spectral types and other information are listed in Table~\ref{spectralclasS}. In addition, we also revised 
the northern sample to [a] include new results from the literature, [b] re-observe SB2 stars near quadrature or visual binaries (VBs) with close 
companions, [c] add new stars with $B < 8$, and [d] uniformize the use of the f and z suffixes. The results for the northern sample are
presented in Table~\ref{spectralclasN}. Those changes bring the total (northern+southern) current GOSSS sample to 448 Galactic O stars.

We will continue GOSSS until we have obtained spectra for more than 1000 O Galactic stars. Of course, that implies observing many stars 
that turn out to be of a different spectral type \citep{Maizetal13b}. Over the upcoming years we plan to publish the spectra not only for
the remaining O stars but also for the other early-type stars we are observing (mostly B stars but also a few Wolf-Rayets or WRs, subdwarfs or sds, 
Planetary Nebula Nuclei or PNNs, Luminous Blue Variables or LBVs, and A 
stars). We will also list the late-type stars that have erroneous classifications in the literature as being of O or early-B type,
obtain a revised grid of classification standards that will extend from O2 to A0, and produce a Magellanic Cloud extension with some
significant O stars there. 

\section{Data and methods}
\label{sec:Data}

\subsection{Blue-violet spectroscopy with $R$~$\sim$~2500}
\label{sec:GOSSS}

The GOSSS data were described in paper I and the reader is referred there for further information. Here we detail the differences with respect
to that previous work.

Most of the spectra presented in paper I were obtained with the Observatorio de Sierra Nevada (OSN) 1.5~m and Calar Alto (CAHA) 3.5~m 
telescopes. This paper concentrates on southern stars, so most of the spectra here were obtained with the 2.5 m du Pont telescope at Las
Campanas (LCO). 

A change in the telescopes used by GOSSS for northern dim stars took place after paper I. We are now using the ISIS spectrograph at the 
4.2~m William Herschel Telescope (WHT) at the Observatorio del Roque de los Muchachos (ORM) in La Palma, Spain instead of the 3.5~m CAHA
telescope. Some of the spectra in this paper were obtained with the WHT or with the OSN and CAHA telescopes. 

The information for the four settings is shown in Table~\ref{settings}. Besides the addition of the WHT information, the values have been 
slightly revised from paper I. The data from each observatory cover slightly different wavelength ranges but the spectrograms shown in 
this paper have been cut to show the same spectral range.

The GOSSS data in this paper were obtained between 2007 and 2013. In some cases, observations were repeated due to 
focus and other instrument issues detected after the fact. For SB2 and SB3 spectroscopic binaries, multiple epochs were
obtained to observe the different orbit phases (see below for OWN data). In cases with known orbits, observations near quadrature were attempted. 

The spectral classifications which are the main content of this paper are presented in Tables~\ref{spectralclasN}~and~\ref{spectralclasS}. As we did in paper I,
we give the GOSSS spectral type for all of the targets and in those cases where an alternative spectral type has been obtained using better spectral or spatial 
resolution than GOSSS we list it in the next column. The classification methodology is presented in subsection~\ref{sec:Clas}, individual stars are discussed in 
section~\ref{sec:Res}, and statistics on the spectral types from papers I and II in subsection~\ref{sec:Spclas}.

\begin{table*}
\caption{Telescopes, instruments, and settings used.}
\centerline{
\begin{tabular}{lccccc}
\\
\hline
\multicolumn{1}{c}{Telescope} & Spectrograph      & Grating & Spectral scale & Spatial scale & Wav. range  \\
                              &                   & (l/mm)  & (\AA/px)       & (\arcsec/px)  & (\AA)       \\
\hline
LCO 2.5 m (du Pont)           & Boller \& Chivens & 1200    & 0.80           & 0.71          & 3900$-$5500 \\
OSN 1.5 m                     & Albireo           & 1800    & 0.62           & 0.83          & 3750$-$5070 \\
CAHA 3.5 m                    & TWIN (blue arm)   & 1200    & 0.55           & 0.58          & 3930$-$5020 \\
ORM 4.2 m (WHT)               & ISIS (blue arm)   &  600    & 0.44           & 0.20          & 3900$-$5600 \\
\hline
\end{tabular}
}
\label{settings}
\end{table*}

\subsection{Spectroscopic binarity}
\label{sec:SB}

GOSSS is limited in its detection capabilities for spectroscopic binaries by its relatively low spectral resolution and the small number of epochs (1-2 in most
cases) used. For those reasons, OWN data \citep{Barbetal10} is more appropriate to detect and study spectroscopic binaries. OWN is using high-resolution
spectrographs at Las Campanas, La Silla, and CASLEO to monitor (as of 2013) 284 O and WN stars visible from those observatories. Of those, 168 [a] are
classified as O stars in GOSSS data, [b] are included in the targets of this paper, [c] have $\delta < -20\arcdeg$, and [d] have at least five OWN epochs 
analyzed. This constitutes the largest uniform, high-quality sample of Galactic O-type spectroscopic binaries ever studied\footnote{Note that most of the archival FEROS data
included in \citet{Chinetal12} were obtained within OWN for the purpose of studying spectroscopic binarity.}. In many cases, OWN have discovered 
new spectroscopic binaries which are presented as such for the first time in section~\ref{sec:Res} of this paper. Here we identify the new binaries and use the
information to discuss the multiplicity of O stars. A forthcoming paper (Barb\'a et al. in preparation) will present the orbits.

The spectroscopic binarity information is summarized in column SB of Tables~\ref{spectralclasN}~and~\ref{spectralclasS}. We have omitted the O (Orbit) qualifier 
for simplicity and because our main interest in this paper is multiplicity statistics, not orbits. The main source used for column SB is OWN, followed by
the references listed at the bottom of the two tables. Most of those references are used for only one or two stars, with the exception of
\citet{Sanaetal08b,Sanaetal11a}. When the 27 O stars with $\delta < -20\arcdeg$ studied by such multi-epoch high-resolution are added to the OWN sample, we obtain a sample
of 194 stars for which binarity has been studied in detail.

We have also used the series of papers by Otero et al. \citep{Oter03,Oter06,Oter07,OterClau04,OterWils05} to include information
on eclipsing binaries. Finally, when no alternative source was available, we resorted to the older compilations of \citet{Masoetal98} and \citet{Pouretal04} to
include additional information. Information about individual stars is presented in section~\ref{sec:Res} and O-star multiplicity is discussed in 
subsection~\ref{sec:Mult}.

\subsection{Visual binarity}
\label{sec:VB}

In paper I we included some of the AstraLux Norte images of \citet{Maiz10}. As previously mentioned, we have extended
the Lucky Imaging survey to the south with AstraLux Sur but the data have not been fully reduced yet, so we cannot use them here.
Nevertheless, for 12 complex fields we present 2MASS images with identifications (Figure~\ref{chart1}), as we already did in paper I, in
order to reduce the possible confusion when comparing different works. Those fields include six multiple systems and their surroundings
(HD~93\,146, HD~93\,632, HD~92\,206, HD~124\,314, HD~150\,136, and HDE~319\,703), four classical stellar clusters (Trumpler~14, NGC~6231,
Havlen-Moffat~1, and Pismis~24), one OB association (IC~2944), and an intermediate cluster/association object (Trumpler~16). 

We give in column VB of Tables~\ref{spectralclasN}~and~\ref{spectralclasS} the number of visual companions detected within 5\arcsec\ and 10\arcsec\ of the target 
separated by a hyphen. When all of the companions are within 5\arcsec, a single number is given. Given the absence of a published large-scale Lucky Imaging or 
Adaptive Optics survey, the information on visual multiplicity is collected in the first place by combining the information from the Washington Double Star 
Catalog (WDS, \citealt{Masoetal01}, more complete for companions within 5\arcsec) and the 2MASS Point Source Catalog (\citealt{Skruetal06}, more complete in some 
cases for companions farther away than 5\arcsec). We have also searched the literature (e.g. \citealt{Nelaetal04}, \citealt{Masoetal09}, Aldoretta et al. in preparation) for 
additional data. For a few selected fields with unsaturated ACS/HRC and WFPC2 exposures from HST GO programs 10\,205 (PI: N.R.W.), 10\,602, 10\,898, and 11\,981 
(P.I.: J.M.A.) we also use those images to supplement the information on multiple visual systems. Information on some individual targets is provided in 
section~\ref{sec:Res}. The quality of the input data and the bound character of the visual companions (as opposed to being foreground objects or 
cluster/association members) is discussed in subsection~\ref{sec:Mult}.

\subsection{Distances and cataloguing}
\label{sec:distcat}

With respect to distances, we follow the policy of paper I of presenting the revised Hipparcos distances of \citet{Maizetal08a} whenever they are significant. 

The spectral types are available through the latest version (currently v3.1) of GOSC, accessible at \url{http://gosc.iaa.es}. Starting in 
version 3, the GOSSS spectral types are the default ones and the basis for the catalog selection, though older classifications and those 
obtained with high-resolution spectra are also kept as possible additional columns. $B$- and $J$-band photometry are also provided in GOSC 
for all stars (see \citealt{Maizetal13b} for details, we use \Bap\ and \Jap, respectively, to refer to the photometry in GOSC, where ``ap'' refers to approximate and is
intended to be significant only to one tenth of a magnitude). The rectified GOSSS spectra can also be obtained from GOSC as FITS tables. 

The GOSSS spectral types were first made available at GOSC in June 2013 as part of the GOSSS Data Release 1.0 (GOSSS-DR1.0, 
\citealt{Sotaetal13}), just in time for the {\it Massive Stars: From $\alpha$ to $\Omega$} meeting that took place in Rhodes, Greece at that time. The spectral 
types in this paper are in some cases slight revisions (mostly suffixes that were omitted or added in error) to those in GOSSS-DR1.0 and constitute 
GOSSS-DR1.1.

\subsection{Spectral classification methodology}
\label{sec:Clas}

The spectral classification methodology was laid out in paper I. We emphasize that spectral classification is an ``art'' that depends on the
effects of spectral, spatial, and temporal resolution as well as S/N. Hence, targets need to be revisited when better data become available
in a process that may seem never ending but which actually converges rapidly in most cases. Since paper I was published, we have filled some 
of the gaps in the standard grid presented there but some still remain. In a future paper of the series we will present an updated grid, 
possibly filling some of the gaps with LMC stars. An important update not related to GOSSS is the building of a standard grid for
Galactic O stars at $R\sim 4000$ (Sana et al. in preparation) which has been used by Walborn et al. (in preparation) to classify the large
sample of O stars in 30 Doradus observed by the VFTS project \citep{Evanetal11a}. We are collaborating with those projects to keep the
standard grids at both resolutions consistent.

Besides the changes introduced by new, better data, three developments have yielded changes in the paper I results. First, we have 
reviewed the suffixes related to the f phenomenon (\NIIId{4634-40-42} emission and \HeII{4686} absorption/emission) according to Table~\ref{fphen}, after 
having found some minor discrepancies in Paper~I. The reader is referred to Table~3 in Paper~I for the meaning of the suffix.
Second, we have recalibrated the z phenomenon\footnote{See \citet{Walb09b} and Sab{\'\i}n-Sanjuli{\'a}n et al. (2014, submitted) for details and
a recent analysis.}, defined as having a $z$ ratio:

\begin{equation}
 z = \frac{{\rm EW}(\HeII{4686})}{{\rm Max}[{\rm EW}(\HeI{4471}),{\rm EW}(\HeII{4542})]}
\end{equation}

\noindent greater than $\sim$1 by measuring the equivalent widths of \HeI{4471}, \HeII{4542}, and \HeII{4686} in our 
main-sequence standard stars. This recalibration will be explained in detail by Arias et al. (in preparation) and has been applied to the 
results in this paper. Third, a new spectral subtype, O9.2, has been introduced as a result of the $R\sim 4000$ work mentioned above, as 
described in the next section.

The spectral classification has been carried out primarily by one of us (J.M.A.) using MGB \citep{Maizetal12}
and later reviewed by another author (N.R.W.).

\begin{table}
\caption{Luminosity classes for the f phenomenon as a function of spectral type.}
\centerline{
\begin{tabular}{lccc}
\\
\hline
Sp. type & ((f)) & (f)    & f      \\
\hline
O2-O5.5  &  V    & III    & I      \\
O6-O6.5  & V-IV  & III-Ib & Iab-Ia \\
O7-O7.5  & V-III & II-Ib  & Iab-Ia \\
O8       & V-II  & Ib     & Iab-Ia \\
O8.5     & V-II  & Ib-Iab & Ia     \\
\hline
\end{tabular}
}
\label{fphen}
\end{table}

\section{Results}
\label{sec:Res}

This section constitutes the main body of the paper and is divided in three parts. First, we present a new spectral subtype, O9.2, 
introduced since paper I.  Second, we revise some of the results on the northern sample of paper I. Third, we present the main block of 
results for this paper, the new southern sample. The information is given in Tables~\ref{spectralclasN}~and~\ref{spectralclasS}, with 
details about each star (sorted by right ascension within each subsection) provided in the text.

\subsection{The O9.2 spectral subtype}
\label{sec:O9.2}

\begin{figure*}
\centerline{\includegraphics*[width=\linewidth]{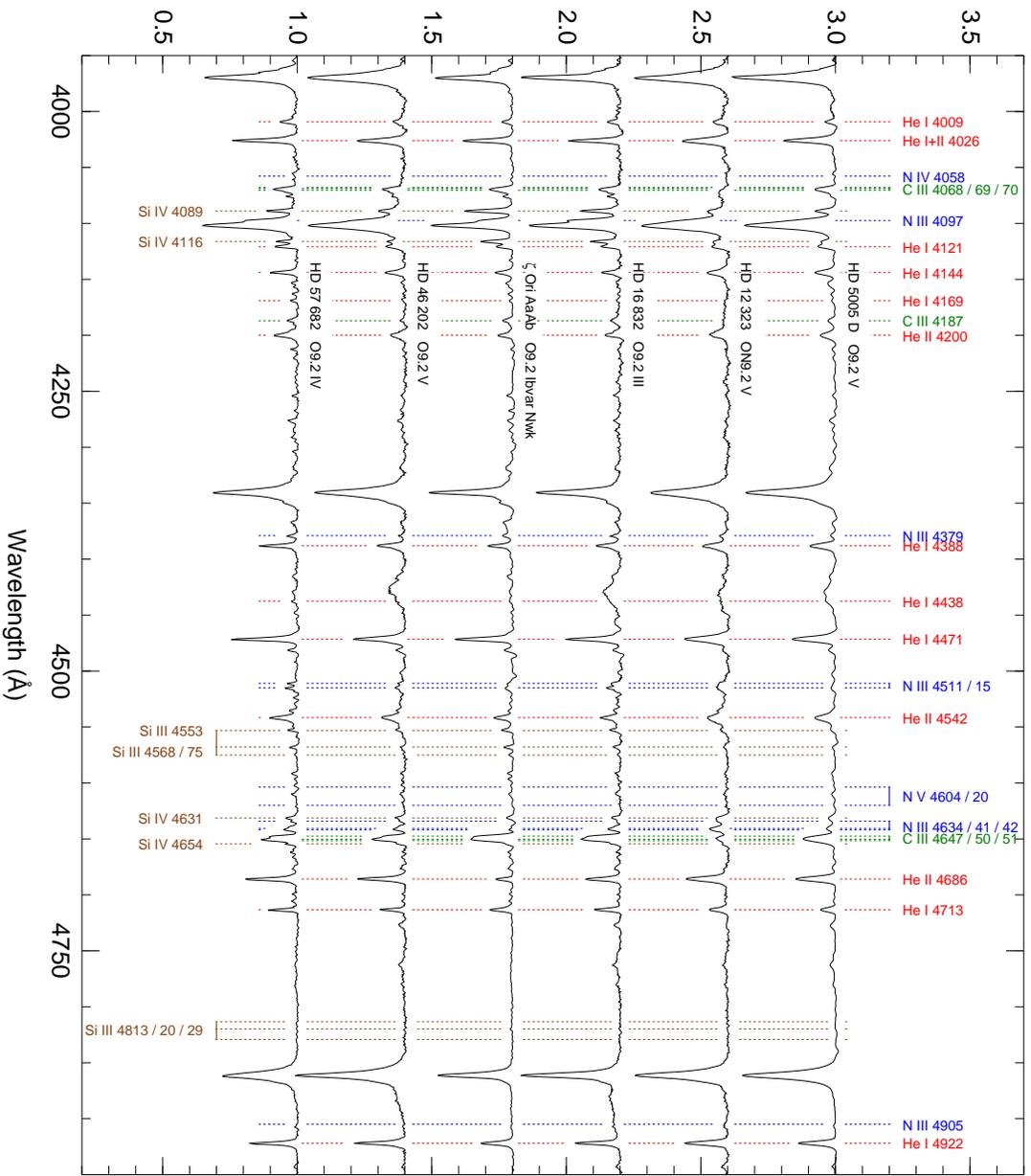}}
\caption{Spectrograms for O9.2 stars. The targets are sorted by right ascension.
[See the electronic version of the journal for a color version of this figure.]}
\label{fig:O9.2}
\end{figure*}	

\addtocounter{figure}{-1}

\begin{figure*}
\centerline{\includegraphics*[width=\linewidth]{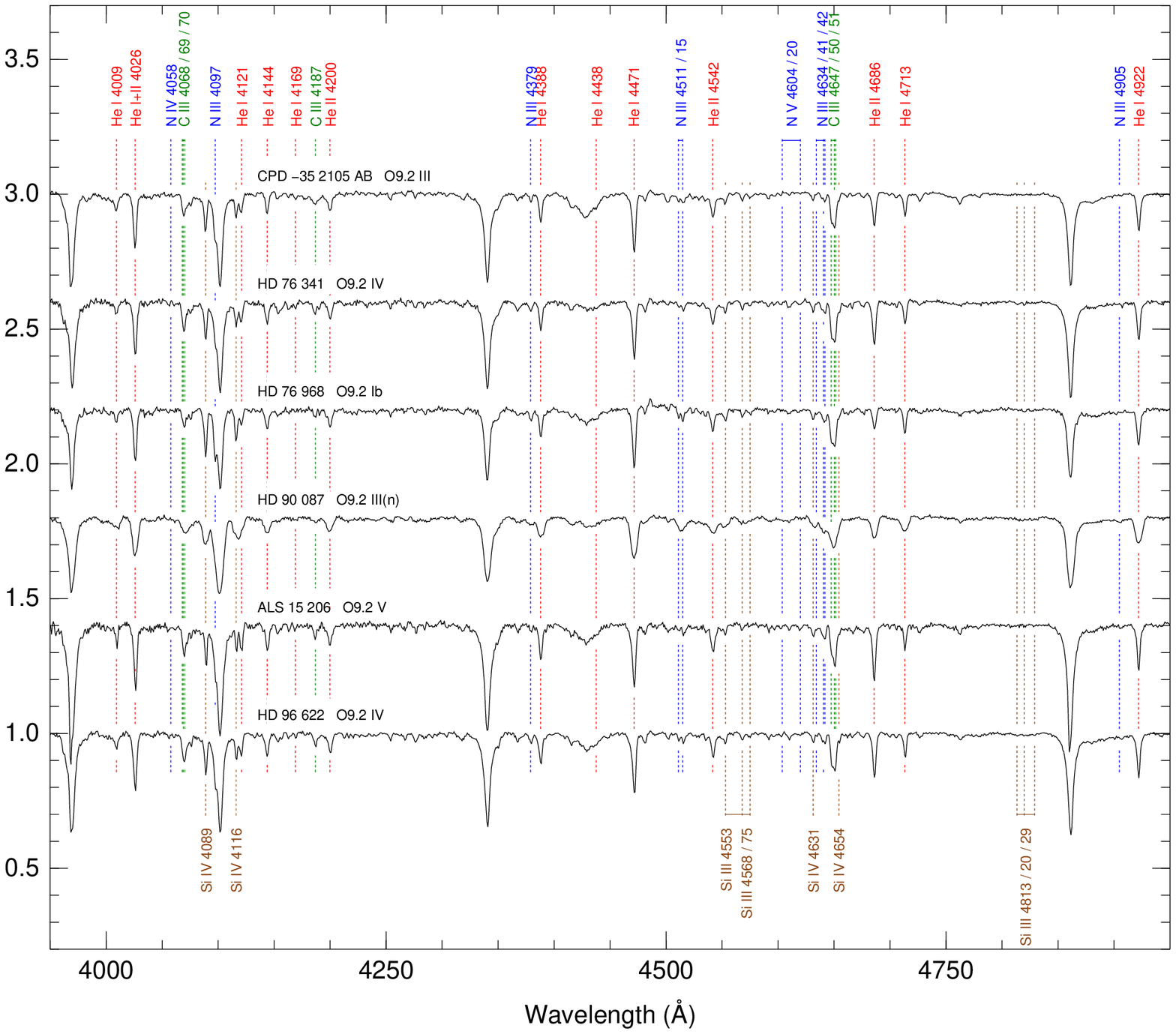}}
\caption{(continued).}
\end{figure*}	

\addtocounter{figure}{-1}

\begin{figure*}
\centerline{\includegraphics*[width=\linewidth]{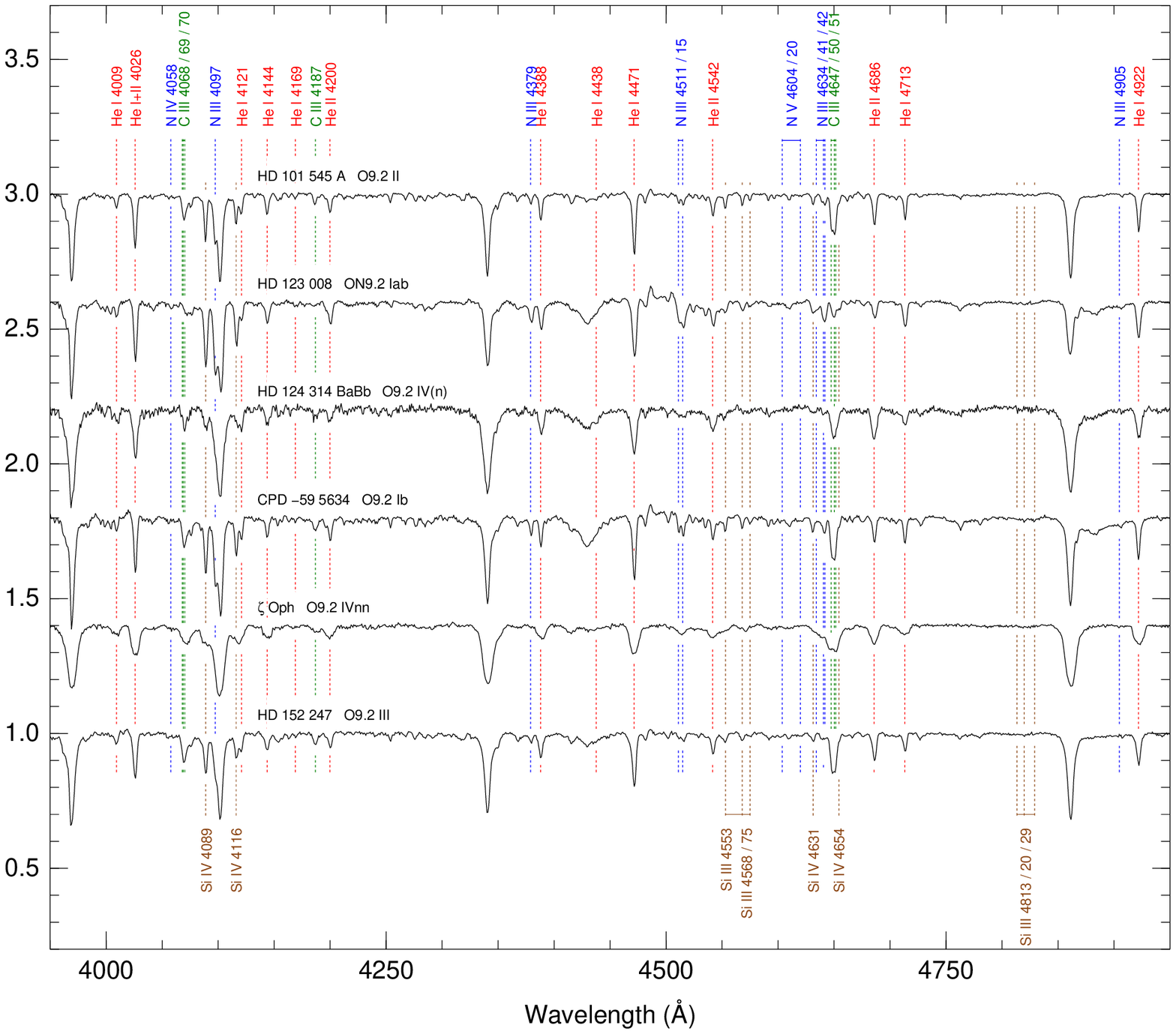}}
\caption{(continued).}
\end{figure*}	

\addtocounter{figure}{-1}

\begin{figure*}
\centerline{\includegraphics*[width=\linewidth]{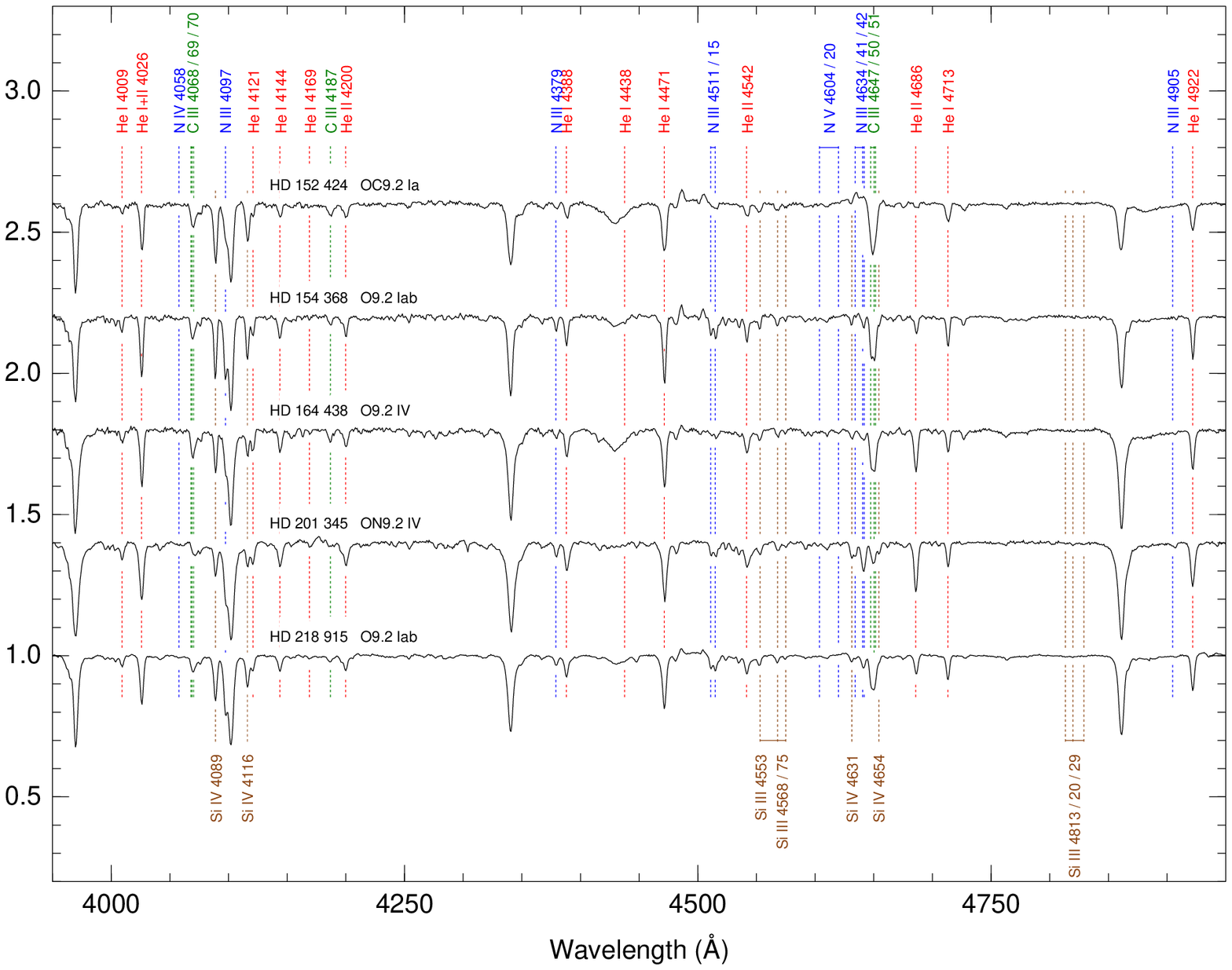}}
\caption{(continued).}
\end{figure*}	

A recent development subsequent to paper I is the definition of subtype O9.2 by Sana et al. and Walborn et al. (both in preparation) based 
on Galactic and 30 Doradus data, respectively, of higher spectral resolution than GOSSS with criteria easily used at $R\sim 2500$. An O9.2 
star has \HeII{4542}/\HeI{4388} and \HeII{4200}/\HeI{4144} ratios just slightly less than unity; thus, this subtype is symmetrical with 
respect to O8.5 on the opposite side of O9. See Table~\ref{O8.5_B0} for the updated criteria for the O8.5-B0 spectral subtype range. 
In this paper we present 23 Galactic O9.2 stars (see Figure~\ref{fig:O9.2}). All but five of them were not classified as such
in GOSSS-DR1.0. Ten of them have $\delta > -20^{\circ}$ and appear in Table~\ref{spectralclasN}. The other thirteen have $\delta < -20^{\circ}$ 
and appear in Table~\ref{spectralclasS}. 

\begin{table}[t]
\caption{Spectral-type criteria at types O8$-$B0 (comparisons between absorption-line pairs based on peak intensities).}
\centerline{
\begin{tabular}{lcc}
\\
\hline
                                  &                                    &                                     \\
                                  & $\fracc{\HeII{4542}}{\HeI{4388}}$  &                                     \\
\multicolumn{1}{c}{Spectral type} & and                                & $\fracc{\SiIII{4552}}{\HeII{4542}}$ \\
                                  & $\fracc{\HeII{4200}}{\HeI{4144}}$  &                                     \\
                                  &                                    &                                     \\
\hline
O8                                & $> 1$                              & N/A                                 \\
O8.5                              & $\geq 1$                           & N/A                                 \\
O9                                & $= 1$                              & $\lll 1$                            \\
O9.2                              & $\leq 1$                           & $\ll 1$                             \\
O9.5                              & $< 1$                              & $< 1$                               \\
O9.7                              & $\ll 1$                            & $\leq 1$ to $\geq 1$                \\ 
B0                                & $\lll 1$                           & $> 1$                               \\
\hline
\end{tabular}
}
\label{O8.5_B0}
\end{table}

\paragraph{HD~5005~D.}
\object[HD 5005]{}

This star was classified as O9.5~V in paper I and is now an O9.2~V.

\paragraph{HD~12\,323.}
\object[HD 12323]{}

This star was classified as ON9.5~V in paper I and is now an ON9.2~V.

\paragraph{HD~16\,832.}
\object[HD 16832]{}

In paper I this star was classified as O9.5~II-III and is now an O9.2~III.

\paragraph{$\zeta$~Ori~AaAb = Alnitak~AaAb = HD~37\,742~AB.}
\object[HD 37742]{}

As in paper I, we were able to extract the individual spectra of A and B (= HD 37\,743), separated by 2\farcs424 and with a $\Delta m$ of 
2.424 magnitudes in the $z$ band. The spectral type has changed from O9.5~Ib~var~Nwk to O9.2~Ib~var~Nwk.

\paragraph{HD~46\,202.}
\object[HD 46202]{}

This star was classified as O9.5~V in paper I and is now an O9.2~V. Aldoretta et al. (in preparation) have recently discovered a companion with a separation of 
86.7~mas and a $\Delta m$ of 2.166 mag.

\paragraph{HD~57\,682.}
\object[HD 57682]{}

\citet{Grunetal12} have recently measured the period, longitudinal magnetic field, and magnetic obliquity of this star. The spectral type has changed
from O9.5~IV in paper I to O9.2~IV here.

\paragraph{CPD~$-$35~2105~AB = CD~$-$35~4384~AB.}
\object[CD-35 4384]{}

This is one of the five O9.2 stars classified as such in GOSSS-DR1.0. This star was not included in \citet{Maizetal04b} but was classified as O9.5~IV
by \citet{Garretal77}. The WDS gives a companion with a separation of 1\farcs1 and a $\Delta m$ of 0.4 magnitudes, which we were unable to spatially resolve
in our long-slit spectra. Also, recently Aldoretta et al. (in preparation) have discovered that the A component is split into Aa and Ab with a separation of 
43~mas and a $\Delta m$ of 1.5 magnitudes. Therefore, the GOSSS spectral type is likely to be a composite.
 
\paragraph{HD~76\,341.}
\object[HD 76341]{}

This star was classified as O9.5~IV in GOSSS-DR1.0 and is now an O9.2~IV. Aldoretta et al. (in preparation) have recently found a faint companion with a separation 
around 0\farcs16. In OWN data the spectra are variable. 
 
\paragraph{HD~76\,968.}
\object[HD 76968]{}

This star was classified as O9.7~Ib by \citet{Walb73a} and here we change the spectral subtype to O9.2 (in Paper I it was O9.5). 
OWN data indicate that this object is an SB1.
 
\paragraph{HD~90\,087.}
\object[HD 90087]{}

The spectral subtype was changed from O9.7 to O9.2 in GOSSS-DR1.1.
 
\paragraph{ALS~15\,206 = CPD~$-$58~2625.}
\object[ALS 15206]{}

This star was not included in \citet{Maizetal04b} but was classified as O9~V by \cite{MassJohn93}. The O9~V classification was kept in GOSSS-DR1.0 but
has been changed to O9.2~V here. See Figure~\ref{chart1} for a chart (Trumpler 14 field).
 
\paragraph{HD~96\,622.}
\object[HD 96622]{}

This star was classified as O9.5~IV in GOSSS-DR1.0 and is now an O9.2~IV. From a preliminary OWN analysis, it is an SB1 with a 98 day period.
 
\paragraph{HD~101\,545~A.}
\object[HD 101545]{}

This system has a B component 2\farcs575 away with a $\Delta m$ of 0.6 magnitudes. We were able to spatially resolve the two stars and determine that the
companion is an early-B star. For the A component the spectral type changed from O9.5~II in GOSSS-DR1.0 to O9.2~II here.
 
\paragraph{HD~123\,008.}
\object[HD 123008]{}

This is one of the five O9.2 stars classified as such in GOSSS-DR1.0. In paper I it appeared as an ON9.5~Iab standard; now it is an 
ON9.2~Iab. OWN data shows it to be variable in \HeII{4686} and H$\alpha$.

\paragraph{HD~124\,314~BaBb.}
\object[HD 124314B]{}

This is one of the five O9.2 stars classified as such in GOSSS-DR1.0. HD~124\,314~A and BaBb are separated by 2\farcs5 and both have O spectral types. 
Ba and Bb cannot be spatially resolved in our data as they are only 0\farcs21 apart.  This star was not included in \citet{Maizetal04b}.
See Figure~\ref{chart1} for a chart (HD~124\,314 field).
 
\paragraph{CPD~$-$59~5634.}
\object[CPD-59 5634]{}

This star was classified as O9.7~Ib in GOSSS-DR1.0 and is now an O9.2~Ib.
 
\paragraph{HD~152\,247.}
\object[HD 152247]{}

This system was found by \citet{Sanaetal08b} to be an SB2 with spectral types of O9 III and O9.7: V. We do not see double lines in the GOSSS data but our 
combined spectral type of O9.2 III (in GOSSS-DR1.0 it was O9.5~III) is in agreement with that result. See Figure~\ref{chart1} for a chart (NGC~6231 field).

\paragraph{HD~152\,424.}
\object[HD 152424]{}

This is one of the five O9.2 stars classified as such in GOSSS-DR1.0. In paper I it appeared as an OC9.7~Ia standard, now it is an 
OC9.2~Ia. From a preliminary OWN analysis, it is an SB1 with a 133 day period.

 
\paragraph{HD~154\,368 = V1074~Sco.}
\object[HD 154368]{}

This is one of the five O9.2 stars classified as such in GOSSS-DR1.0. In paper I it appeared as an O9.5~Iab standard, now it is an 
O9.2~Iab. \citet{Masoetal98} indicate that it is an eclipsing binary.

\paragraph{$\zeta$~Oph = HD~149\,757.}
\object[HD 149757]{}

This is a very fast rotator, a runaway, and the closest O star. In paper I its classification was O9.5~IVnn; it is now O9.2~IVnn

\paragraph{HD~164\,438.}
\object[HD 164438]{}

This object was classified as O9~III in paper I and is now an O9.2~IV. It is an SB1 system according to OWN data.

\paragraph{HD~201\,345.}
\object[HD 201345]{}

This used to be the prototype late-ON dwarf until its luminosity class was changed to IV in paper I. Now its spectral subtype has also changed from
ON9.5 to ON9.2.

\paragraph{HD~218\,915.}
\object[HD 218915]{}

This star was classified as O9.5~Iab in paper I and is now an O9.2~Iab.

\subsection{Revisiting the northern sample from paper I}
\label{sec:PI}

\begin{figure*}
\centerline{\includegraphics*[width=\linewidth]{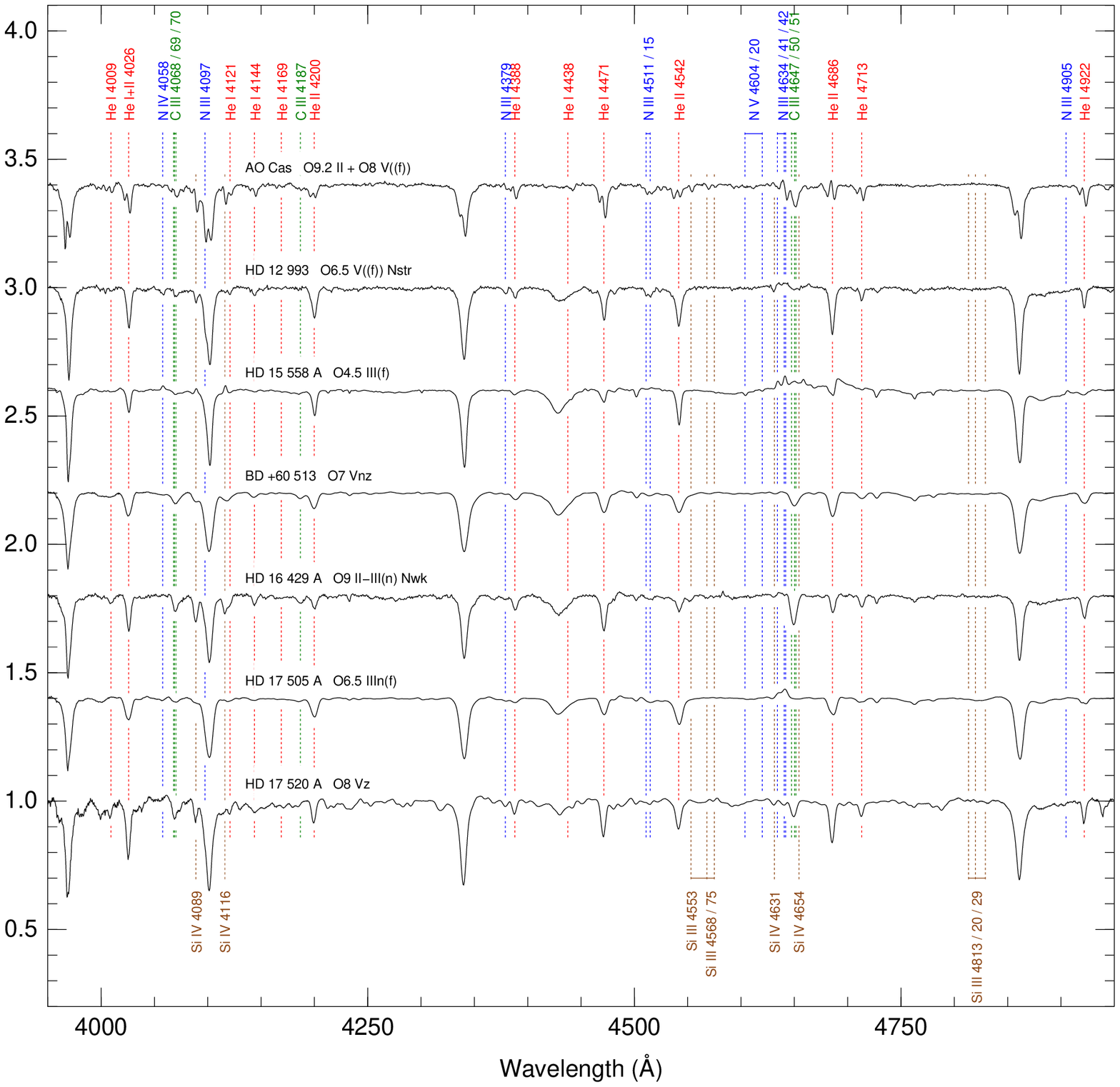}}
\caption{Spectrograms for the northern sample (O9.2 stars not included). The targets are sorted by right ascension with the exception of NGC 1624-2,
whose two spectrograms (for different epochs with dates in YYMMDD) are shown last.
[See the electronic version of the journal for a color version of this figure.]}
\label{fig:PI}
\end{figure*}	

\addtocounter{figure}{-1}

\begin{figure*}
\centerline{\includegraphics*[width=\linewidth]{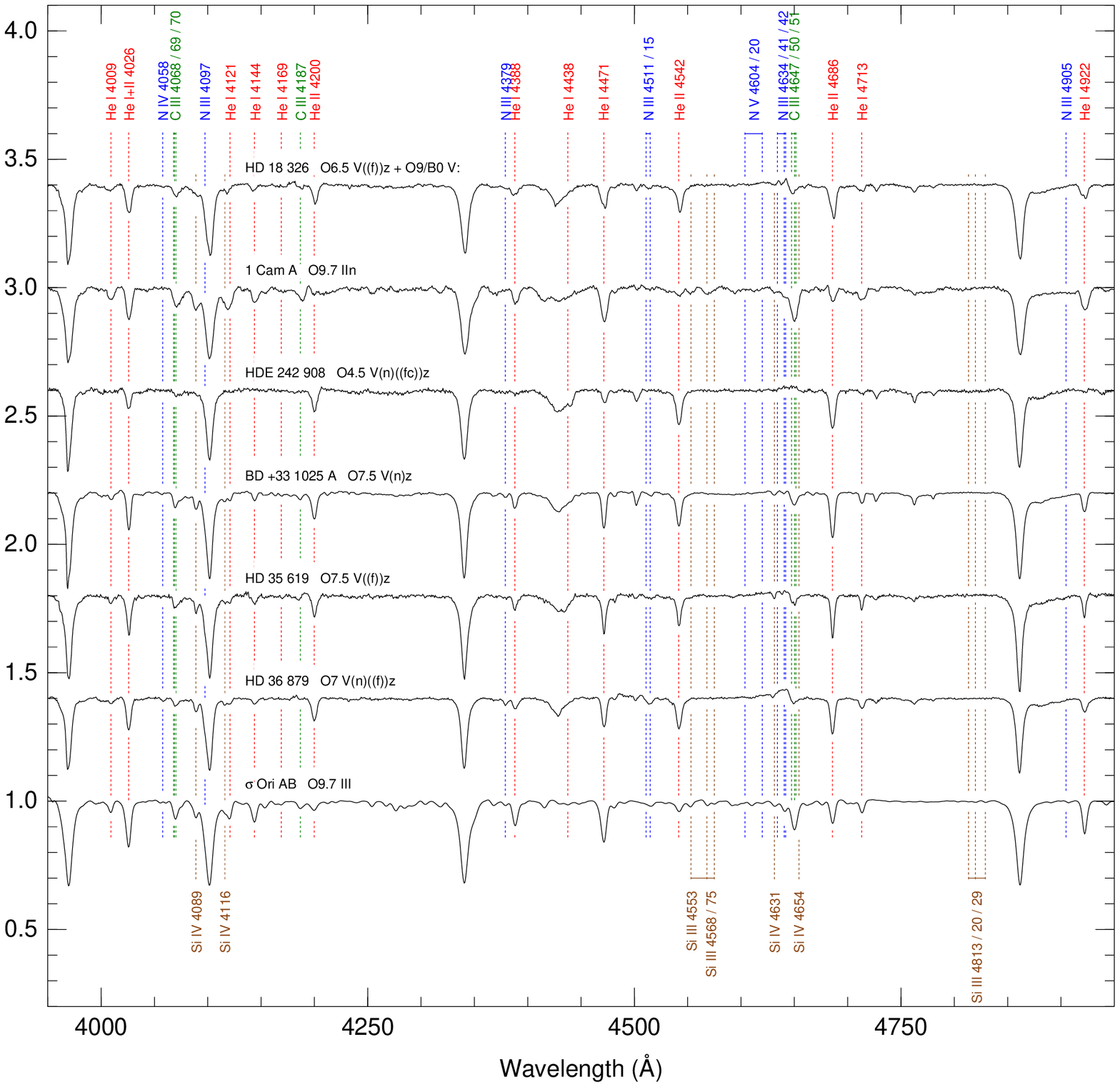}}
\caption{(continued).}
\end{figure*}	

\addtocounter{figure}{-1}

\begin{figure*}
\centerline{\includegraphics*[width=\linewidth]{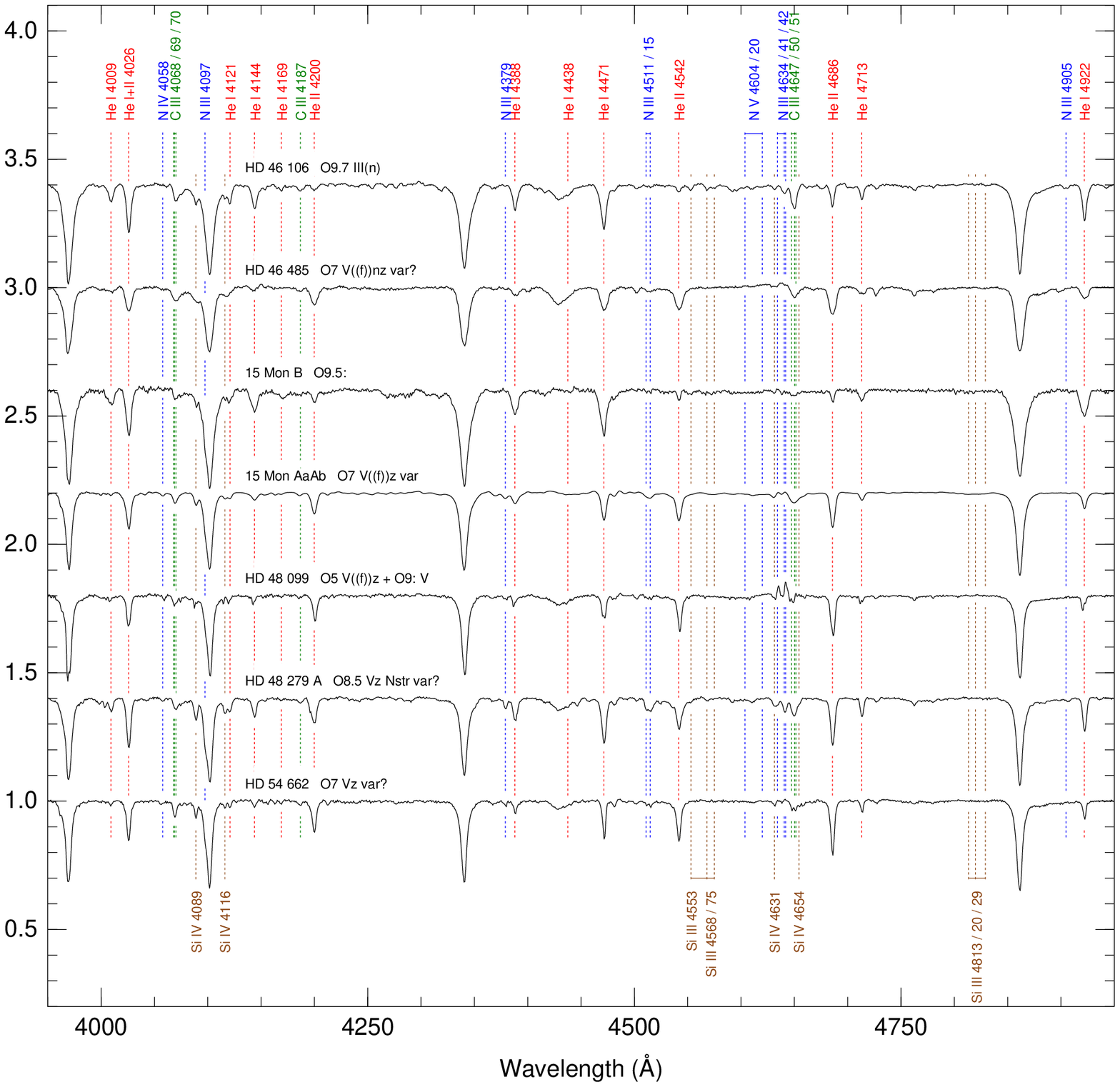}}
\caption{(continued).}
\end{figure*}	

\addtocounter{figure}{-1}

\begin{figure*}
\centerline{\includegraphics*[width=\linewidth]{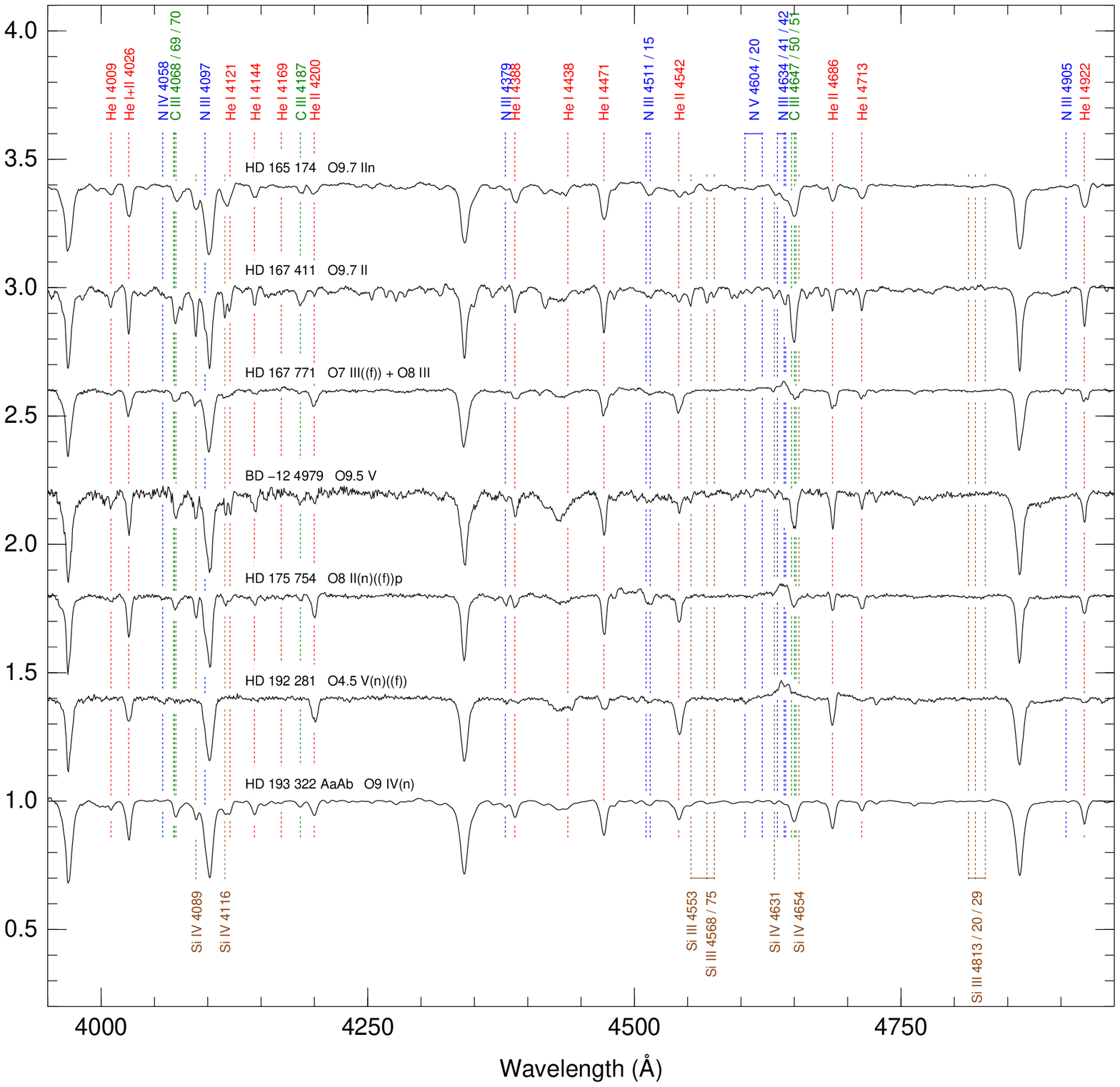}}
\caption{(continued).}
\end{figure*}	

\addtocounter{figure}{-1}

\begin{figure*}
\centerline{\includegraphics*[width=\linewidth]{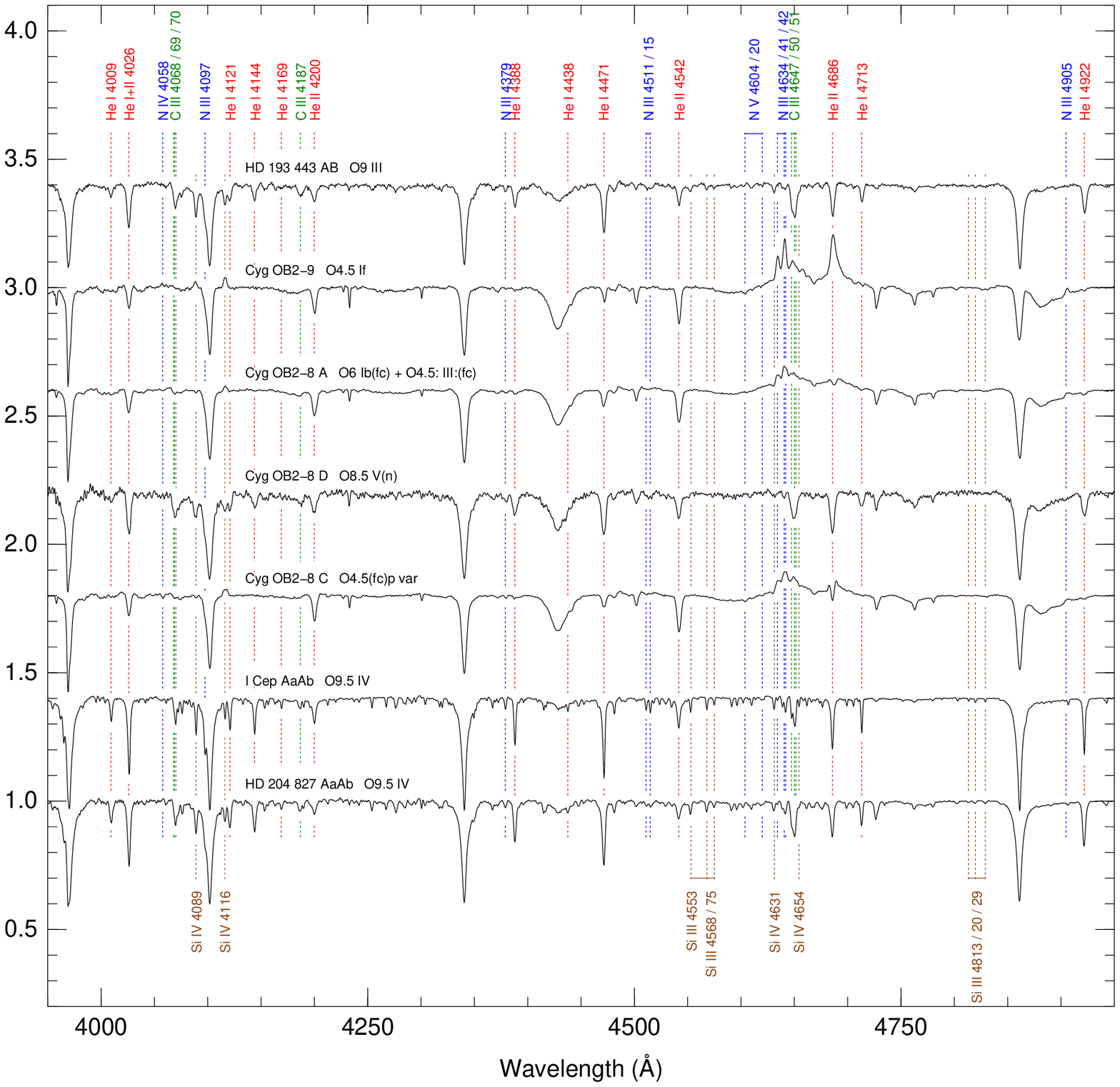}}
\caption{(continued).}
\end{figure*}	

\addtocounter{figure}{-1}

\begin{figure*}
\centerline{\includegraphics*[width=\linewidth]{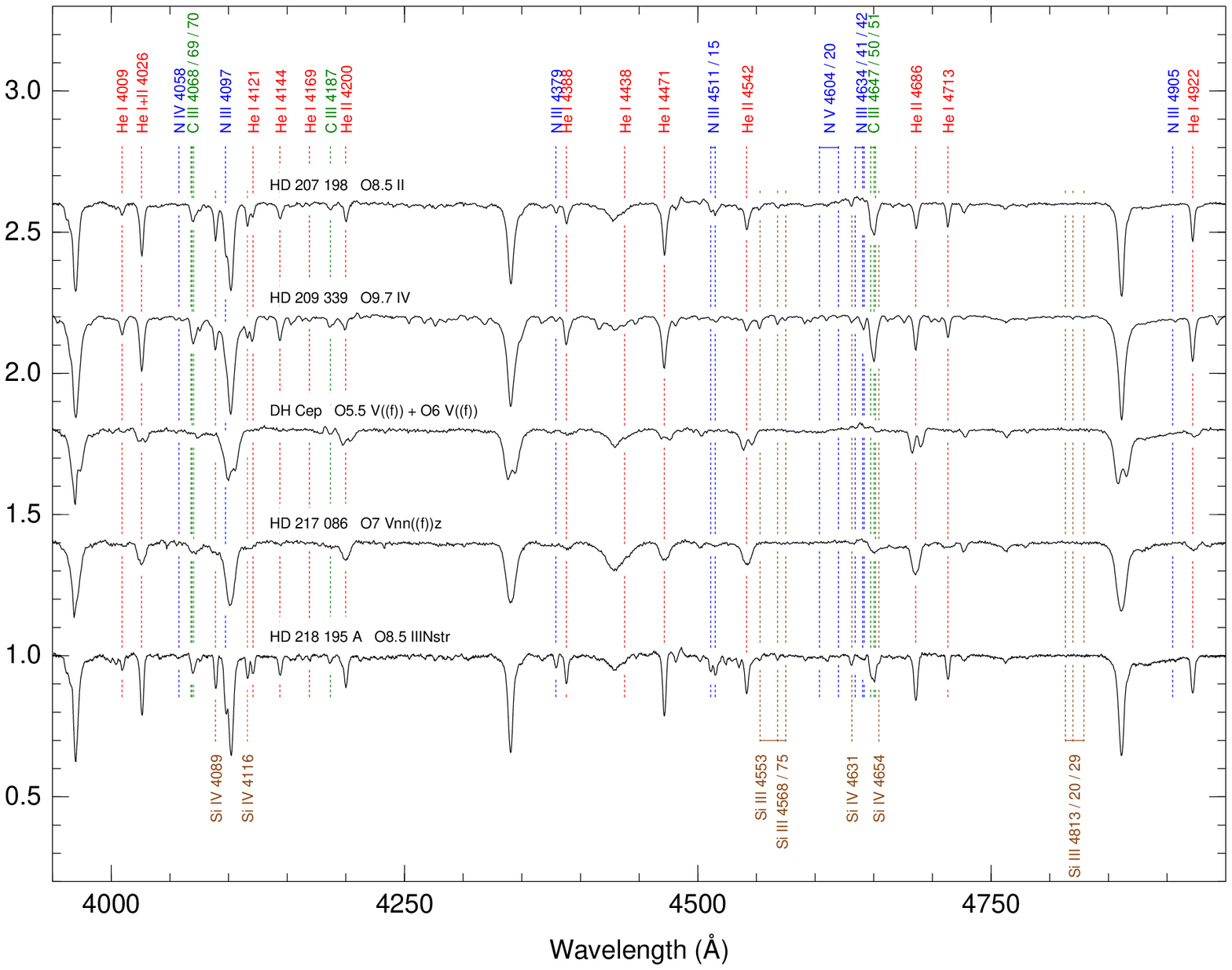}}
\caption{(continued).}
\end{figure*}	

\addtocounter{figure}{-1}

\begin{figure*}
\centerline{\includegraphics*[width=\linewidth]{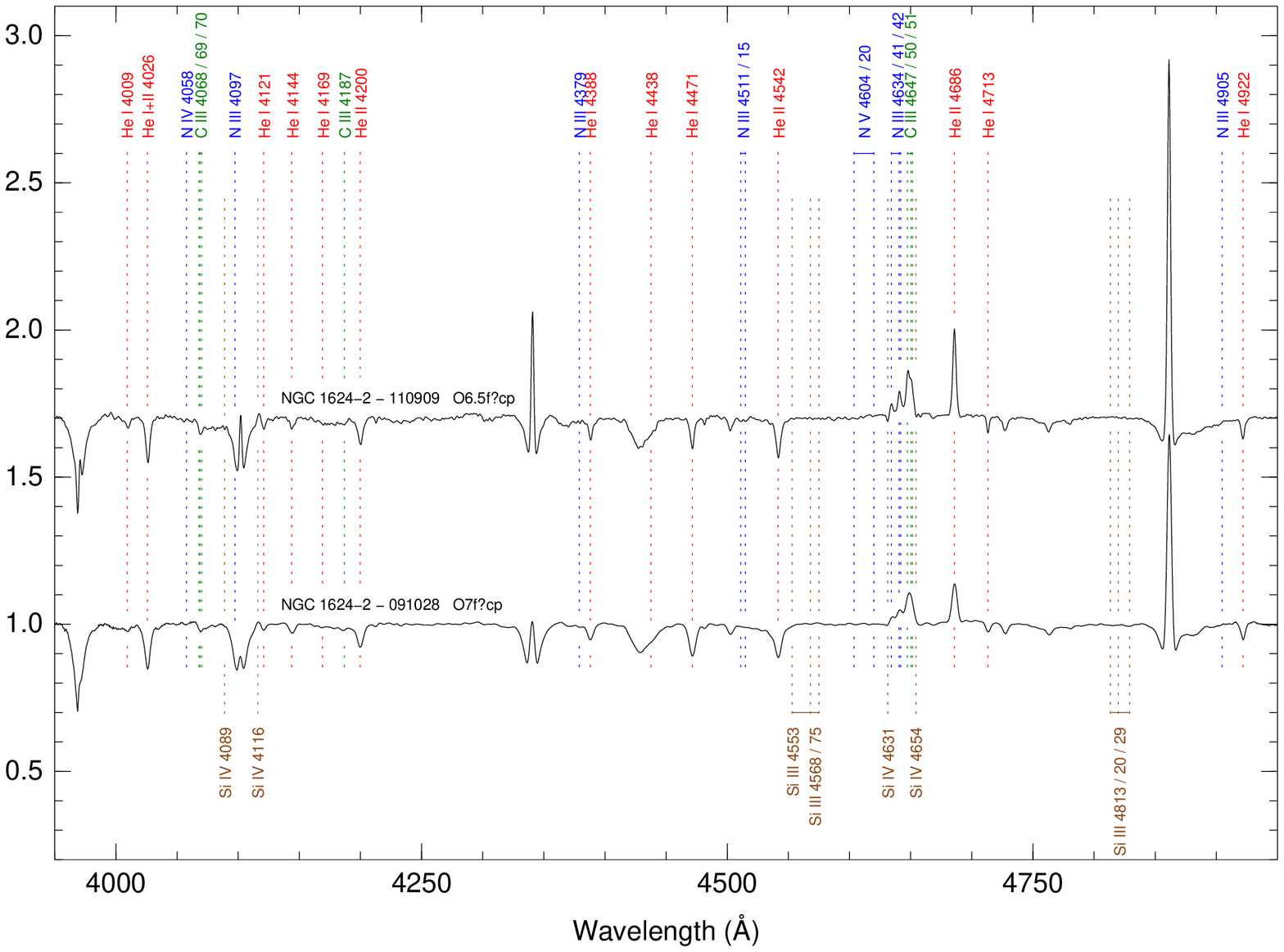}}
\caption{(continued).}
\end{figure*}	

In this subsection we revise the results of paper~I due to four reasons:

\begin{enumerate}
  \item Either we have reobserved an SB2 system with GOSSS or other authors have used high-resolution spectroscopy to obtain resolved (in
        velocity) spectral types for one of such systems. In the second case we simply recall those new results, since a new GOSSS spectral type
        was not obtained.
  \item We have reobserved some of the stars with close companions under better seeing conditions than those in paper I, leading us to
        improved (spatially) resolved spectral types.
  \item As a result of the redefinition of spectral subtype O9.7, we are observing all stars with previous spectral classifications of B0.
        In some cases we have found new cases of bright ($B < 8$) O9.7 stars that are presented in order to achieve completeness for that
        magnitude limit.
  \item We have introduced uniform criteria for the definition of the f and z phenomena (see previous section).
\end{enumerate}

Spectrograms are shown in Figure~\ref{fig:PI}. See also subsection~\ref{sec:O9.2} for additional northern stars.

\paragraph{AO Cas = HD 1337.}
\object[HD 1337]{}

We reobserved this SB2 system and obtained a better velocity resolution for its components. The luminosity class for the secondary has changed 
from GOSSS-DR1.0. Our spectral type for the secondary is the same as that of \citet{BagnGies91} but the one for the primary is slightly different.
For both components small changes were introduced in GOSSS-DR1.1.

\paragraph{HD~12\,993.}
\object[HD 12993]{}

The Nstr suffix was missing in paper I (see Table~3 in paper~I for its definition). Also, its \HeII{4686} is comparable to \HeII{4542}, making the 
star only a marginal z. Both of those changes took place after GOSSS-DR1.0 appeared.

\paragraph{HD~15\,558~A.}
\object[HD 15558A]{}

The target has \CIII{4650} in emission but is not strong enough to warrant a c suffix.

\paragraph{BD~$+$60~513.}
\object[BD +60 513]{}

The z suffix was added in GOSSS-DR1.1. 

\paragraph{HD~16\,429~A.}
\object[HD 16429]{}

The Nwk suffix was added in GOSSS-DR1.1. This complex SB3 system (see paper I) remains unresolved in velocity in GOSSS but see \citet{McSw03} for a
high-resolution study that resolves it.

\paragraph{HD~17\,505 A.}
\object[HD 17505]{}

The ((f)) suffix was changed to (f) in GOSSS-DR1.1, according to the rules in Table~\ref{fphen}. The spectrum of A is spatially resolved from 
that of B but no double lines are detected in our spectra (see paper I and \citealt{Hilletal06}).

\paragraph{HD~17\,520 A.}
\object[HD 17520]{}

The z suffix was added in GOSSS-DR1.1. Note that A and B are spatially separated in GOSSS data (see paper I). 

\paragraph{HD~18\,326.}
\object[HD 18326]{}

The z suffix was added and the (n) suffix dropped in GOSSS-DR1.1 for the primary of this SB2 system. 

\paragraph{1~Cam~A = HD~28\,446~A.}
\object[* 1 Cam]{}

1~Cam~A was classified as B0 V by \citet{Morgetal55} and was not included in \citet{Maizetal04b}. With the new definition of O9.7, we obtain a 
spectral type of O9.7~IIn. According to the WDS, 1~Cam~B is located at a distance of 10\farcs3 with a \Dm\ of 1.0~mag as of 2009. We placed the slit 
along the AB line to obtain a spectrum of B and derive a spectral type\footnote{Note that in this paper we only list O spectral types in the tables.}
of B0 V.

\paragraph{NGC~1624-2 = 2MASS~J04403728+5027410 = MFJ~Sh~2-212~2 = ALS~18~660.}
\object[NGC 1624-2]{}

The Of?p nature of this object was discovered by \citet{Walbetal10a}. \citet{Wadeetal12b} added the c suffix to this star, the O-type with the strongest 
magnetic field measured to date. The change was incorporated in GOSSS-DR1.1. Figure~\ref{fig:PI} shows the spectrum of this object in two epochs that correspond
to the ``high'' and ``low'' states (note that there are is a very slight resolution difference between the two spectra).

\paragraph{HDE~242\,908.}
\object[HDE 242908]{}

\citet{Penn96} flagged this star as a possible SB2. The z suffix was added in GOSSS-DR1.1. 

\paragraph{BD~+33~1025~A.}
\object[BD +33 1025]{}

We observed this star again and obtained a new spectrum with better S/N that made us change its spectral type from O7 to O7.5. We aligned the slit to also measure the B
component of the system, located 13\farcs8 away, and found it to be an early B star. 

\paragraph{HD~35\,619.}
\object[HD 35619]{}

The z suffix was added in GOSSS-DR1.1. 
 
\paragraph{HD~36\,879.}
\object[HD 36879]{}

This star shows peculiar, variable UV Si\,{\sc iv} emission lines \citep{WalbPane84}.
The z suffix was added in GOSSS-DR1.1. 

\paragraph{$\sigma$~Ori~AB = HD~37\,468~AB.}
\object[* sig Ori]{}

As described in paper I, $\sigma$~Ori~AB is a visual binary with a well known orbit and a current separation of 0\farcs26 and a \Dm\ of 1.6~mag
\citep{Turnetal08,Maiz10}.
\citet{SimDetal11b} used multiple-epoch high-resolution spectroscopy to discover that A is a spectroscopic binary and, therefore, that the 
system is an SB3. They derive spectral types of O9.5~V and B0.5~V for the two components of A and a more uncertain of B0/1~V for B. 
$\sigma$~Ori~AB remains unresolved in velocity in GOSSS data. The composite spectrum is O9.7~III with the luminosity class estimated from the 
He lines. However, the Si\,{\sc iv} lines are weak, a sign that the real luminosity class of the components of this system is V. 
 
\paragraph{HD~46\,106.}
\object[HD 46106]{}

The luminosity class changed from II-III to III and an (n) suffix was added in GOSSS-DR1.1. The Si\,{\sc iv} lines are weak and correspond to a luminosity
class of V.

\paragraph{HD~46\,485.}
\object[HD 46485]{}

A weak \NIIId{4634-40-42} emission is detected. Hence, we introduce an ((f)) suffix. Also, $z$ has been measured to be large enough to
warrant the addition of a z suffix. 


\paragraph{15~Mon~AaAb = HD~47\,839~AaAb = S~Mon~AaAb.}
\object[HD 47839]{}

The z suffix was added in GOSSS-DR1.1. See also 15 Mon B. Note that the orbital period of Ab with respect to Aa is of the order of a century 
\citep{Cvetetal10,Maiz10} so we cannot see double lines in our spectra. Indeed, the secondary was resolved in velocity from the primary by \citet{Giesetal93}
using high-resolution spectroscopy thanks to their different rotation speeds. This star was an MK primary standard in the past but GOSSS and other authors have 
shown it in an O7.5 state on occasion, hence the var suffix.

\paragraph{15~Mon~B = HD~47\,839~B = S~Mon~B.}
\object[* S Mon B]{}

In paper I we were able to obtain spatially resolved spectra for 15 Mon AaAb and B (separation of 2\farcs976 and \Dm\ of 3.2~mag, \citealt{Maiz10}) that
yielded an early-B type for the B component. A re-analysis of the data and a new long-slit spectrum obtained under better seeing conditions
show that B has a clear \HeII{4542} absorption line. We estimate a spectral type of O9.5:. We are unable to estimate a luminosity class 
due to the uncertainty in the separation of two nearby spectra with such a large \Dm.

\paragraph{HD~48\,099.}
\object[HD 48099]{}

We obtained a new GOSSS observation and we were able to resolve in velocity the two SB2 components of this system with a 3.078098 day period 
\citep{Stic96}. Our spectral types are reasonably similar to those obtained by \citet{Mahyetal10} with high-resolution spectroscopy.

\paragraph{HD~48\,279~A.}
\object[HD 48279]{}

The z suffix was added in GOSSS-DR1.1. 

\paragraph{HD~54\,662.}
\object[HD 54662]{}

As noted in paper I, \citet{Boyaetal07a} attempted a tomographic reconstruction of the spectra of this system and could only determine it to 
within O6.5~V~+~O7-9.5~V. The OWN analysis indicates a different period for this SB2 (2119$\pm$5 days instead of 557.8 days, Gamen et al. in preparation).
We were unable to detect double lines in our spectra so the spectral classification of O7~Vz~var? refers to the
combined spectrum. In paper I an ((f)) suffix was added but we have now determined the \NIIId{4634-40-42} emission is not detected in our spectra.


\paragraph{HD~165\,174 = V986~Oph.}
\object[HD 165174]{}

\citet{Morgetal55} classified this star as B0.5~III~var and \citet{Lesh68} as B0~IIIn; we are unaware of any classifications as an O star.
Therefore, it was not included in \citet{Maizetal04b} but we observed it in GOSSS as we are doing with all bright stars classified as B0 to check 
whether they are of late-O type or not. This is one case where we indeed obtain a classification of O9.7~IIn with the definition of O9.7 of Table~\ref{O8.5_B0}.
 
\paragraph{HD~167\,411.}
\object[HD 167411]{}

\citet{Morr61} classified this star as B0 IIk and we are unaware of any classifications as an O star. Under the same circumstances as for HD~165\,174,
we observed it and obtained a classification of O9.7 II.

\paragraph{HD~167\,771.}
\object[HD 167771]{}

The (f) suffix of the primary component of this SB2 system with a period of 3.97333 days \citep{Pouretal04} was changed to ((f)) in GOSSS-DR1.1,
according to the rules in Table~\ref{fphen}.

\paragraph{HD 175\,754}

\object[HD 175754]{}
The (f) suffix was changed to ((f)) in GOSSS-DR1.1 to follow the criteria in Table~\ref{fphen}. It has very weak but definite \HeII{4686} emission wings,
making it an Onfp star.

\paragraph{BD~$-$12~4979.}
\object[BD -12 4979]{}

This star was classified as O9.5~III-IV in paper I and is now an O9.5~V.

%

\paragraph{HD~192\,281.}
\object[HD 192281]{}

The suffix was changed to (n)((f)) in GOSSS-DR1.1. We suspect that the broadening is not caused by rotation but instead by binarity, as the position of 
the \NIIId{4634-40-42} emission lines appears shifted.

\paragraph{HD~193\,322~AaAb.}
\object[HD 193322]{}

HD~193\,322 is a highly complex system that includes at least six stars of spectral types O and B. Aa and Ab are separated by 0\farcs05-0\farcs07
and follow a 35 a orbit \citep{tenBetal11,Maiz10}. \citet{tenBetal11} have used multiple-epoch high-resolution spectroscopy to analyze the AaAb
system, determining a spectral type of O9~Vnn for Aa and that Ab is an SB2 system with spectral types of O8.5~III and B2.5:~V:.
HD~193\,322~AaAb remains unresolved in velocity in GOSSS data.
 
\paragraph{HD~193\,443~AB.}
\object[HD 193443]{}

\citet{Mahyetal13} studied this SB2 system with high-resolution spectroscopy and determined a period of 7.467 days, with spectral types of 
O9~III/I and O9.5~V/III, respectively. Note, however that the WDS indicates that this system 
is a visual binary with a separation of 0\farcs1-0\farcs2 and a \Dm\ of 0.3, i.e. there has to be at least a third star in the \citet{Mahyetal13}
aperture with a much longer period (hence, the designation AB). We have unpublished Lucky Imaging data of HD~193\,443 that confirm the 
approximate separation and \Dm\ in the WDS. HD~193\,443~AB remains unresolved in velocity in GOSSS data.
 
%

\paragraph{Cyg~OB2$-$9 = LS~III~$+$41~36 = Schulte~9 = [MT91]~431.}
\object[NAME VI CYG 9]{}

\citet{Nazeetal12c} studied this SB2 system with high-resolution spectroscopy and obtained a period of 2.35 a and spectral types of O5-5.5~I for
the primary and O3-4~III for the secondary. Cyg~OB2$-$9 remains unresolved in velocity in GOSSS data, which is unsurprising given the short amount
of time during the long orbit where velocity differences are large. 
The target has \CIII{4650} in emission but is not strong enough to warrant a c suffix.

\paragraph{Cyg~OB2$-$8~A = BD~$+$40~4227 = Schulte~8~A = [MT91]~465.}
\object[NAME VI CYG 8 A]{}

We have obtained several more spectra of this target and we have been able to separate the two components of this SB2 (see \citealt{DeBeetal04}).
The spectral type for the primary agrees with the previous result but our result for the secondary is somewhat earlier.

\paragraph{Cyg~OB2$-$8~D = Schulte~8~D = [MT91]~473.}
\object[NAME VI CYG 8 D]{}

A new observation with better S/N has led us to revise the spectral type of this system from O9 to O8.5.
 
\paragraph{Cyg~OB2$-$8~C = LS~III~$+$41~38 = Schulte~8~C = [MT91]~483.}
\object[NAME VI CYG 8 C]{}

Using several more spectra, the \HeII{4686} profile has been found to be variable and to display the characteristic profile of an Onfp object on occasion.
By analogy with similar systems, it is possible that there is a companion in this system.

\paragraph{I~Cep~AaAb = HD~202\,214~AaAb.}
\object[* I Cep]{}

This system was classified as B0~V by \citet{Morgetal53b} and by several more authors\footnote{Note that HD~202\,124 is also a late O star 
(see paper I), so it would not be unthinkable that some confusion has existed between the two.}; hence, it was not included in \citet{Maizetal04b}.
Nevertheless, \citet{MannHumb55} dissented and 
indicated that it was an O9. The WDS gives three components (Aa, Ab, and B) within 1\arcsec\ and with small values of \Dm. We initially 
observed this system and attempted to spatially resolve AaAb from B (separation of 1\farcs0, and \Dm\ of 0.68 with respect to the combined AaAb 
system, values confirmed from unpublished Lucky Imaging data) but we were unsuccessful since the observing conditions were not optimal. 
Hence, the spectral type in GOSSS-DR1.0 \citep{Sotaetal13}, O9.5~IV, corresponds to the integrated spectrum from Aa+Ab+B. In a later 
observation with better seeing we were able to spatially resolve AaAb from B (but not Aa from Ab, since they are less than 0\farcs1 apart). AaAb is 
indeed an O9.5~IV while B is an early-B star.

\paragraph{HD~204\,827~AaAb.}
\object[HD 204827]{}

This system was classified as O9.7~III in paper I but the S/N was relatively low compared to the GOSSS typical values. We re-observed the star and
obtained a revised spectral type of O9.5~IV. This change took place after GOSSS-DR1.0.
The WDS gives a separation of 0\farcs1 and a $\Delta m = 1.2$ for the Aa + Ab system, so the spectral type is likely a composite.

\paragraph{HD~207\,198.}
\object[HD 207198]{}

The spectral type changed from O9~II to O8.5~II in GOSSS-DR1.1 as a consequence of the redefinition of the O9-O9.5 spectral subtypes associated with 
the introduction of the O9.2 subtype.

\paragraph{HD~209\,339.}
\object[HD 209339]{}

This star was classified as O9~IV by \citet{Morgetal53b} but was accidentally skipped by \citet{Maizetal04b} even though it is brighter than $B=8$. 
According to the WDS, it has a nearby companion with a separation of 0\farcs8 and a 
\Dm\ of 3.3 (unpublished Lucky Imaging data suggest that the companion may have moved slightly farther away since the last observation recorded
in the WDS), so the companion should not influence the combined spectral type. We obtain O9.7 IV with the new definition of the O9.7 subtype.

\paragraph{DH~Cep = HD~215\,835.}
\object[HD 215835]{}

We obtained a new GOSSS epoch for this SB2 system with a 2.1109 day period \citep{Pouretal04} that allowed for a better separation in velocity of the two 
components. Our spectral types agree with those of \citet{Burketal97} but our luminosity classes are V instead of III.

\paragraph{HD~217\,086.}
\object[HD 217086]{}

The z suffix was added in GOSSS-DR1.1. 

\paragraph{HD~218\,195~A.}
\object[HD 218195]{}

An Nstr suffix was added in GOSSS-DR1.1. Note that the B component is spatially resolved in GOSSS as an early-B star (see paper I).

\subsection{The new southern sample}

In presenting the new southern sample, we follow the same strategy of paper I of introducing first the members of the peculiar categories and then 
those of the normal sample. The reader is referred to paper I for the definitions of most of the peculiar categories. There are two new categories 
in this paper with respect to \citet{Sotaetal11a}, the early Of/WN stars and the O~Iafpe stars (which are usually grouped together under the name of 
``slash'' stars), since none of them were present in the sample of paper I. Note that SB2 stars that belong to another peculiar category are listed 
in the corresponding non-SB2 subsubsection and that O9.2 stars have already been discussed.

\subsubsection{Early Of/WN stars}
\label{sec:earlyslash}

\begin{figure*}
\centerline{\includegraphics*[width=\linewidth]{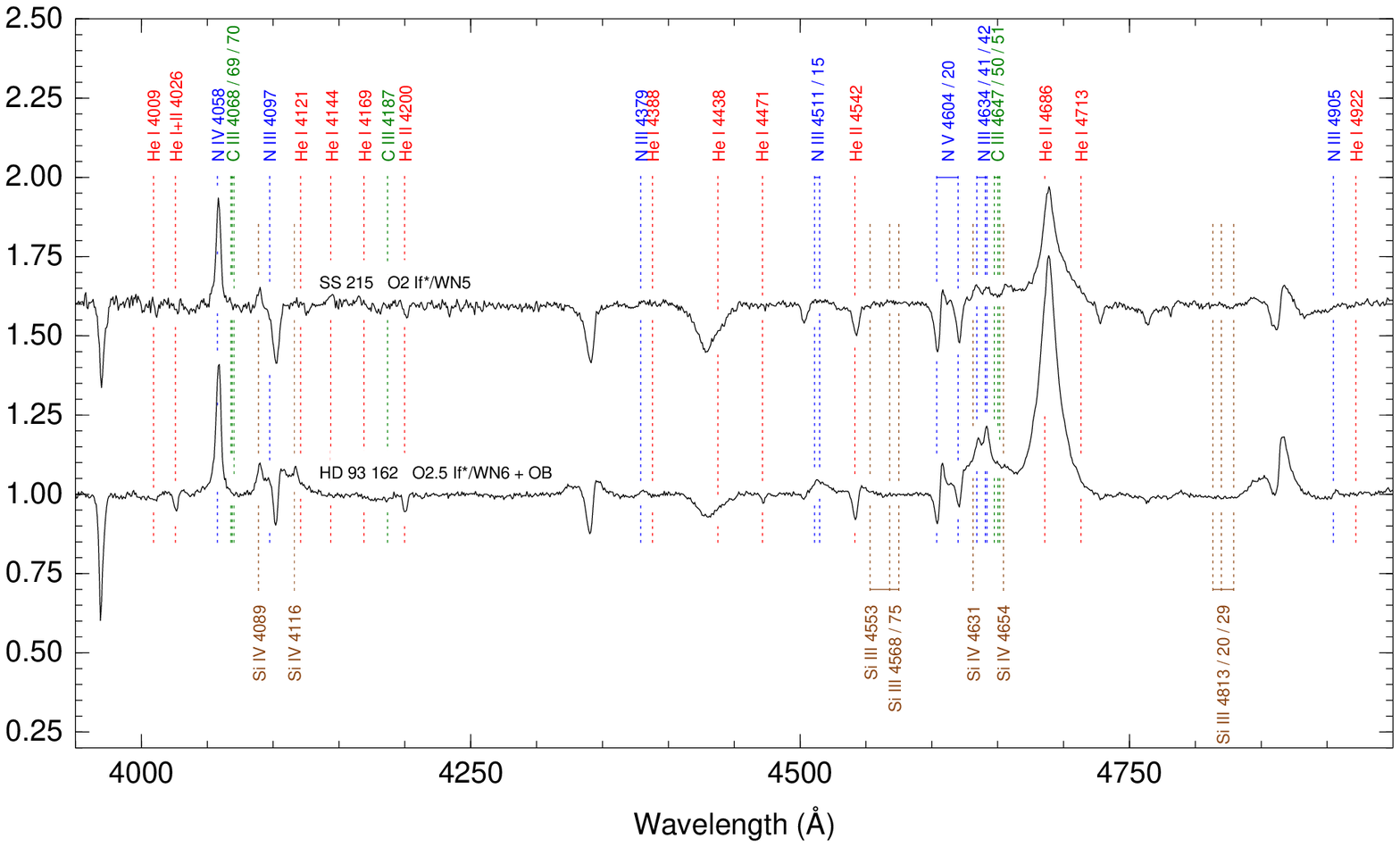}}
\caption{Spectrograms for the early Of/WN stars. The targets are sorted by right ascension.
[See the electronic version of the journal for a color version of this figure.]}
\label{fig:earlyslash}
\end{figure*}	

\citet{Walb71b} was the first to notice a smooth morphological transition between the earliest O supergiants and the H-rich, high-luminosity, narrow-lined WN sequence. 
\citet{Walb82b} introduced the hybrid O3 If*/WN6 classification for Sk~$-67$~22, the first well defined transition object. Such objects are currently called early 
Of/WN stars or ``hot slash'' stars (see below for the ``cool slash'' category) and are thought to be very massive, core hydrogen-burning stars in an intermediate 
evolutionary, or at least wind-development, phase. \citet{CrowWalb11} established the H$\beta$ profile as the morphological criterion to distinguish among the three 
related categories: it is in absorption in O2-3.5 If* stars, it has a P-Cygni profile in O2-3.5 If*/WN5-7 stars (``hot slash'' stars), and it is in emission in 
WN5-7 objects. Note that some early ``slash'' stars have WR numbers for historical reasons though they are no longer considered proper Wolf-Rayet stars from the 
morphological viewpoint. Spectrograms are shown in Figure~\ref{fig:earlyslash}.

\paragraph{SS~215 = WR~20aa.}
\object[SS 215]{}

This O2~If*/WN5 was discovered by \citet{RomLetal11} and could have been ejected from the massive young Galactic cluster Westerlund 2.
This star was not included in \citet{Maizetal04b} due to its dimness.
 
\paragraph{HD~93\,162 = WR~25.}
\object[HD 93162]{}

This system was included in the ``hot slash'' category by \citet{CrowWalb11} on the basis of its P-Cygni H$\beta$ profile. \citet{Gameetal06} presented the first
SB1 orbit and indicated that some of the absorption lines varied in anti-phase, hence indicating they originated in the secondary. The secondary was clearly
detected by \citet{Gameetal08b}. This object was not included in \citet{Maizetal04b} since at that time it was thought to be a Wolf-Rayet star
instead of an early Of/WN. See Figure~\ref{chart1} for a chart (Trumpler 16 field) and Figure~\ref{HRC} for an ACS/HRC image. 
In the existing ACS/HRC images we detected a previously unpublished\footnote{It was presented in JD 13 of the IAU General Assembly in 2009 but did not 
appear in any proceedings.} visual companion. It has a separation of 790~mas, a position angle of 353\arcdeg, and a $\Delta V$ of 5.8 magnitudes. The magnitude 
and colors are consistent with an $\sim$8 M$_\odot$ star.
 
\subsubsection{O~Iafpe stars}
\label{sec:lateslash}

\begin{figure*}
\centerline{\includegraphics*[width=\linewidth]{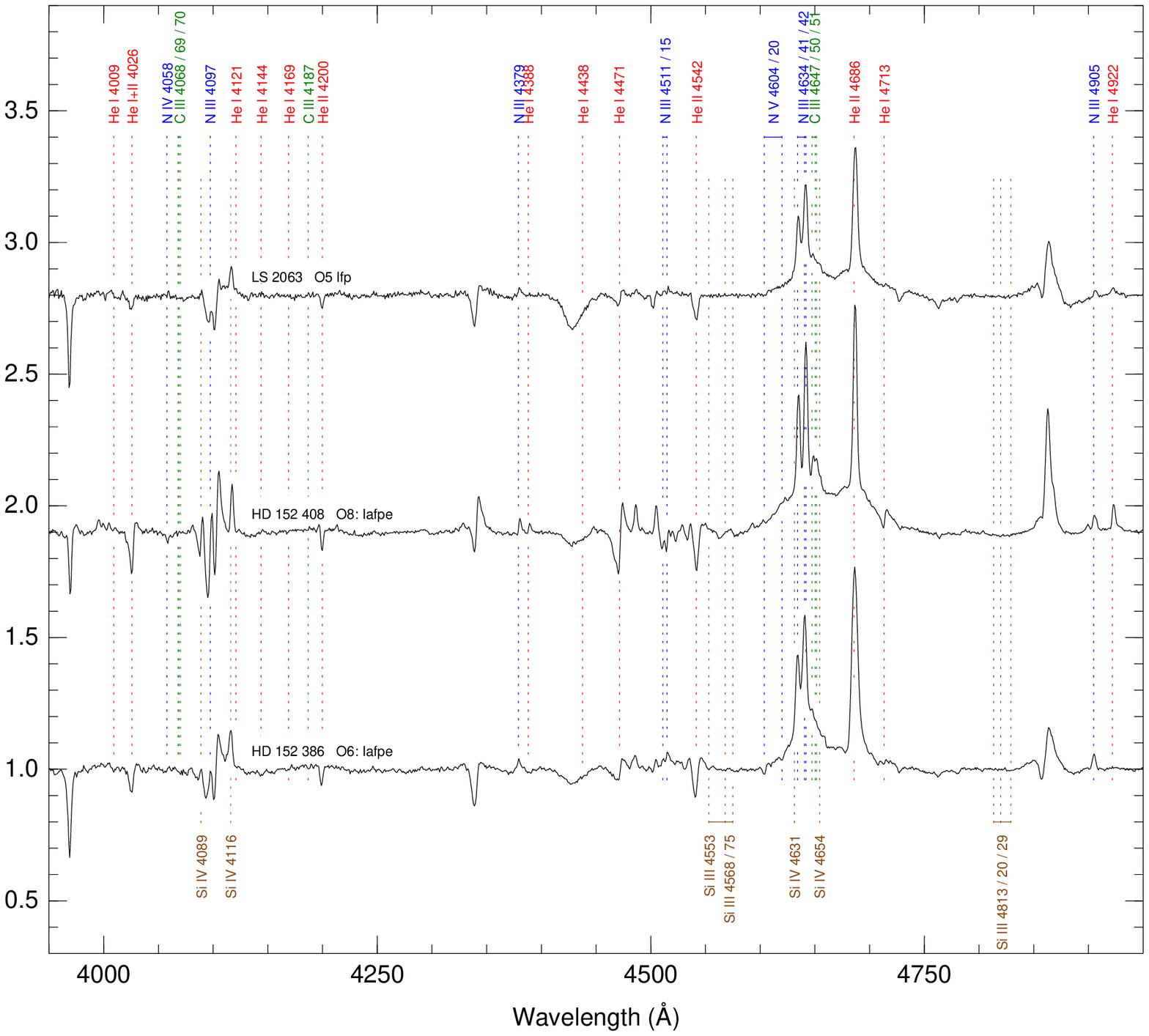}}
\caption{Spectrograms for the O~Iafpe stars. The targets are sorted by right ascension.
[See the electronic version of the journal for a color version of this figure.]}
\label{fig:lateslash}
\end{figure*}	

\addtocounter{figure}{-1}

\begin{figure*}
\centerline{\includegraphics*[width=\linewidth]{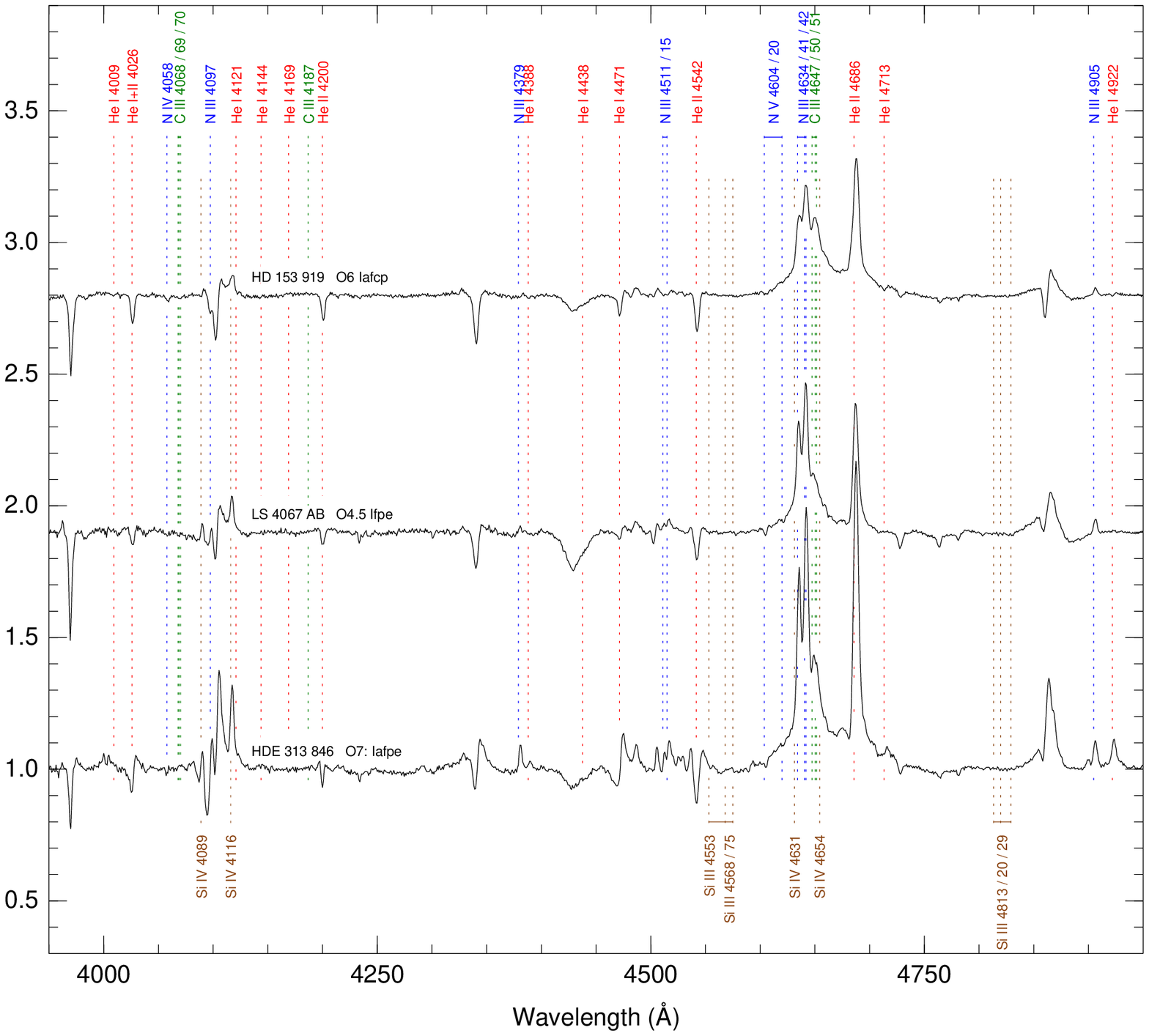}}
\caption{(continued).}
\end{figure*}	

A second, cooler flavor of ``slash'' stars was introduced by \citet{Walb82c} and \cite{BohaWalb89}, the members of which will be referred to here as {\it either} 
O~Iafpe or WNVL (for Very Late) stars, depending upon whether or not distinct absorption lines are present in their spectra, respectively (see also \citealt{WalbFitz00}).
These spectra can be distinguished from their hotter counterparts by the absence of the \NIV{4058} emission line used to classify the hot spectra, while in contrast the 
cool spectra may contain \NII{3995} emission instead. The composite late Of/WN or ``cool slash'' nomenclature will no longer be used in detailed classifications for two 
important reasons\footnote{Nevertheless, the {\it name} ``Ofpe/WN9'' will continue to be useful to refer to that category as originally discovered in the Large Magellanic 
Cloud, although it does not represent a detailed classification of individual spectra. One member of that category, Radcliffe~127, has subsequently undergone a classical 
Luminous Blue Variable outburst (\citealt{Walbetal08} and references therein).}: (1) we include two relatively early-type stars in the O~Iafpe category here; and (2) a 
significant difference from the early Of/WN category is that the cooler objects are {\it not} an intermediate category between Of and WN stars, but rather stars that can 
be described alternatively as one or the other. Indeed, this sometimes implicit distinction between the two ``slash'' categories has caused some confusion in the 
literature. Here we shall follow the criterion of classifying a star as O~Iafpe when \HeI{4471} has a P Cygni profile (which usually introduces some uncertainty in the 
spectral subclass, as indicated by ``:''); if that line is in pure absorption, the spectrum 
is a normal Of type. The blended composite morphology of the H$\delta$ region and generally stronger emission lines also distinguish the O~Iafpe spectra from normal Of.  
Spectrograms are shown in Figure~\ref{fig:lateslash}. There are no WNVL stars in the present sample according to our criteria; however, note that \citet{Crow07}
includes some of our O~Iafpe objects in his WN9-10 classes (see also \citealt{Walb09a}).  

\paragraph{LS~2063.}
\object[LS 2063]{}

This object was described by \citet{WalbFitz00} and is an extreme supergiant, with strong \HeII{4686} and \NIIId{4634-40-42} emission, as well as
H$\beta$ showing a P-Cygni profile. The OWN spectra indicate variability. The suffix was changed in GOSSS-DR1.1.

\paragraph{HD~152\,408 = WR~79a.}
\object[HD 152408]{}

\citet{Walb72} classified this object as O8~Iafpe.
 
\paragraph{HD~152\,386 = WR~79b.}
\object[HD 152386]{}

\citet{Walb73a} classified this object as O6:~Iafpe. The WDS gives a companion with a separation of 0\farcs6 but without a value for $\Delta m$.
 
\paragraph{HD~153\,919 = V884~Sco.}
\object[HD 153919]{}

This system is an SB1 with a period of 3.41~d and is the optical counterpart of the X-ray binary 4U~1700$-$37 \citep{Hutcetal73,Ankaetal01,Claretal02,Hammetal03}.
 
\paragraph{LS~4067~AB = HM~1$-$2~AB = C1715$-$387$-$2~AB.}
\object[LS 4067]{}

This object is the earliest member of the OIafpe category. Its spectral type was changed from O4 to O4.5 in GOSSS-DR1.1. 
\citet{Masoetal09} finds a companion with separation of 1\farcs3 and a $\Delta m$ of 1.2 magnitudes that we were unable to spatially resolve.
See Figure~\ref{chart1} for a chart (Havlen-Moffat 1 field).
  
\paragraph{HDE~313\,846 = WR~108.}
\object[HDE 313846]{}

\citet{Walb82a} classified this object as O7:~Iafpe.

\subsubsection{Ofc stars}
\label{sec:Ofc}

The Ofc category was introduced by \citet{Walbetal10a} and amplified by \citet{Sotaetal11a} in qualitative terms, on the basis of the ratio of peak intensities in \CIII{4650} to \NIII{4634}, 
albeit without an explicit specification of the boundary between definite and marginal cases. Also, that definition is complicated by the occurrence of broad emission pedestals underlying 
the narrow emission lines, especially in the supergiants. To improve this situation, here we have measured the narrow emission-line peaks {\it relative to the pedestals when they exist}
in both Ofc and marginal cases; and a value of the ratio of 1.0 has been adopted as the lower boundary for the Ofc classification. As a result, some prior classifications in terms of this 
parameter have been revised. In the future, EWs will be preferable to peak intensities to define the phenomenon, to avoid resolution effects (\CIII{4650} is a blend barely resolved in 
the GOSSS data), analogously to what has been done for the Vz category here.

\begin{figure*}
\centerline{\includegraphics*[width=\linewidth]{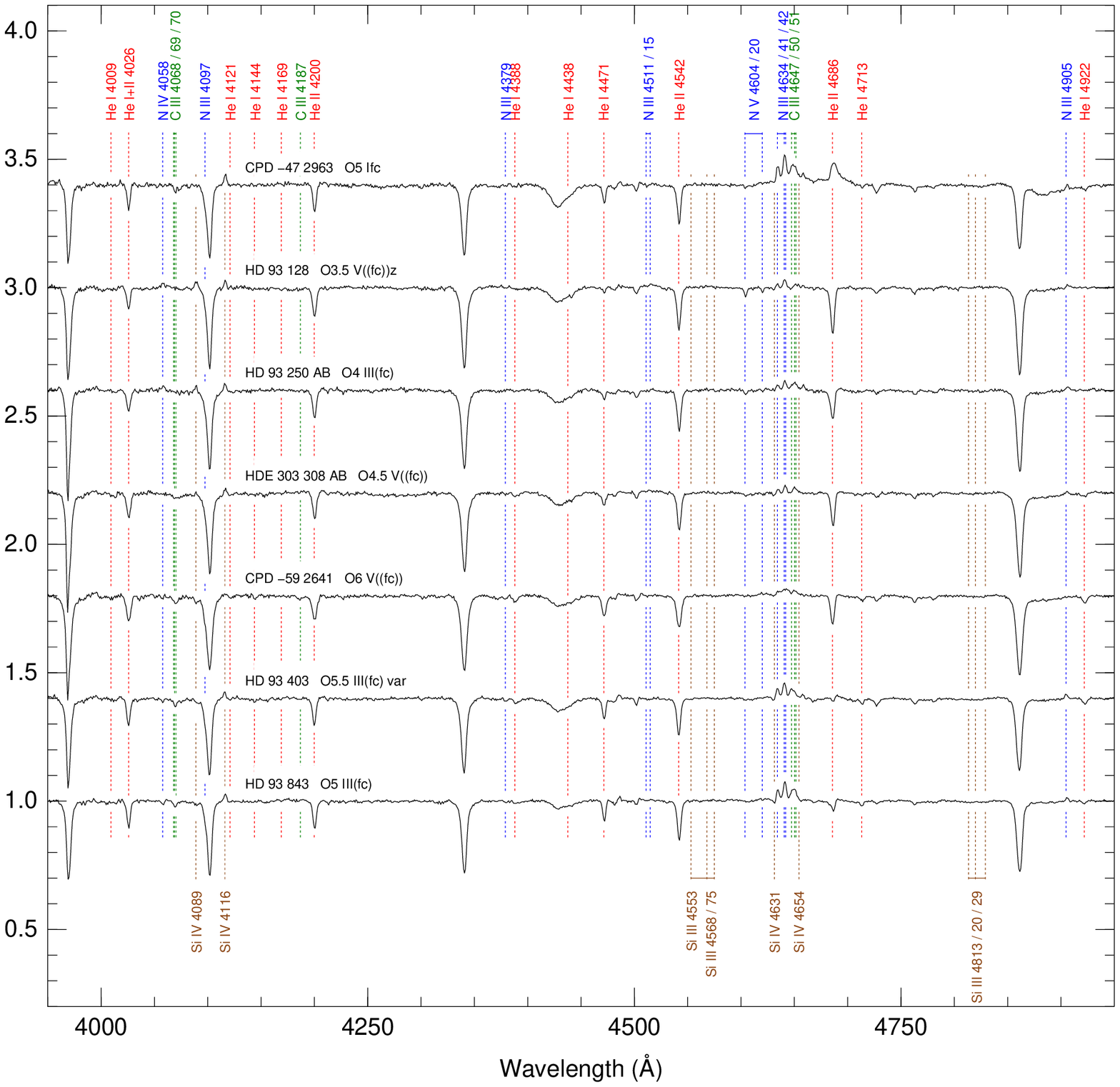}}
\caption{Spectrograms for Ofc stars. The targets are sorted by right ascension.
[See the electronic version of the journal for a color version of this figure.]}
\label{fig:Ofc}
\end{figure*}	

\addtocounter{figure}{-1}

\begin{figure*}
\centerline{\includegraphics*[width=\linewidth]{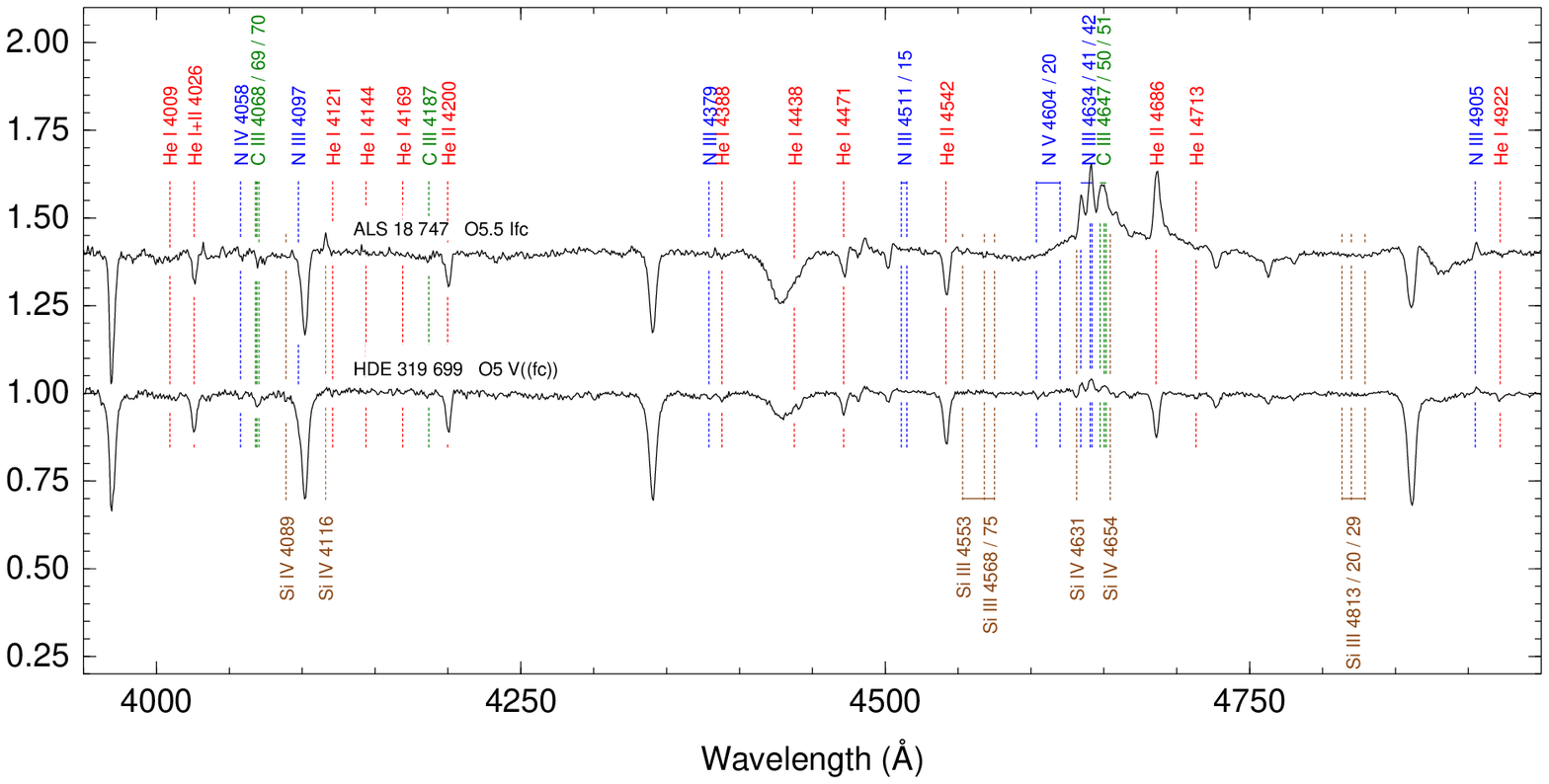}}
\caption{(continued).}
\end{figure*}	

Spectrograms are shown in Figure~\ref{fig:Ofc}. See also section~\ref{sec:PI}.

\paragraph{CPD~$-$47~2963 = ALS~1216 = CD~$-$47~4551.}
\object[CPD-47 2963]{}

\citet{Walb82a} classified this object as O4~III(f) but in GOSSS it is clearly a slightly later supergiant, as already shown in \citet{Walbetal10a}. 
Our classification as O5 Ifc agrees with that of O5~If by \citet{Garretal77}. 
One of us (N.R.W.) checked the original plate of the \citet{Walb82a} 
observation, taken on 25 Nov 1974, and confirmed the presence of weak emission in the 4640-4650~\AA\ region and for \HeII{4686}, indicating that the 
spectrum has not changed in a major way in the last four decades. 
There are no appreciable differences between the two epochs observed with GOSSS but 
in OWN data a radial-velocity variation is apparent with a most likely period of 59 days and a low semi-amplitude $K$ of 9.4 km/s 
(an alternative classification as an O(n)fcp star is also suggested by the variable profile of \HeII{4686} in the OWN data). The variability could be 
caused by a binary companion (colliding wind region) or by the existence of a magnetic field (something not entirely surprising for a star with both 
\CIII{4650} and \HeII{4686} emission, see subsubsection~\ref{sec:Of?p}). The binary hypothesis is supported by the detection of non-thermal radio 
emission from the system \citep{Benaetal01,DeBeRauc13}.
 
\paragraph{HD~93\,250~AB.}
\object[HD 93250]{}

\citet{Rauwetal09} could not find velocity variations in this system but 
\citet{Sanaetal11b} used VLTI to spatially resolve it into two stars separated by 1.5 mas with a $\Delta m$ of 0.8 (such separation is 
unresolvable by GOSSS by more than two orders of magnitude). The (fc) suffix was missing in the GOSSS-DR1.0 spectral classification of 
this O4~III system. No companions are seen in the ACS/HRC images. The system is a colliding-wind binary \citep{DeBeRauc13}.
 
\paragraph{HDE~303\,308~AB.}
\object[HDE 303308]{}

\citet{Nelaetal04} spatially resolved this system with HST/FGS to discover a pair with a separation of 15 mas and a $\Delta m$ of 1.0 magnitudes (obviously
unresolved with GOSSS). See Figure~\ref{chart1} for a chart (Trumpler 16 field).
 
\paragraph{CPD~$-$59~2641 = Trumpler 16-112.}
\object[CPD-59 2641]{}

This object shows an O6 V((fc)) spectrum in the GOSSS data. \citet{Rauwetal09} used high-resolution spectroscopy to derive spectral types for
this SB2 system of O5.5-6~V((f+?p)) and B2~V-III. Their spectral types and luminosity classes are compatible with that of our combined spectrum, 
which does not show double lines. In
order to check whether the primary belongs to the Of?p category we obtained spectra in five different epochs between 2008 and 2013 (once a year except
for 2009). None of the five spectra show \HeII{4686} in emission and \CIII{4650} appears to be constant. Therefore, we conclude that this star is an
Ofc star (note that \citealt{Rauwetal09} appeared before the definition of Ofc in \citealt{Walbetal10a}) and amend the high-resolution spectral types
to O5.5-6~V((fc))~+~B2~V-III. See Figure~\ref{chart1} for a chart (Trumpler 16 field).
 
\paragraph{HD~93\,403.}
\object[HD 93403]{}

\citet{Rauwetal00} classified this SB2 system as O5.5~I~+~O7~V. We observed it three times with GOSSS and we obtained a spectral type of O5.5. 
We do not see double lines but \HeII{4686} is variable, going from a P-Cygni profile to a nearly neutral condition, as a result of the SB2 orbit. 
Therefore, we settled for a III luminosity class, which is consistent with the combination of a dwarf and a supergiant. The spectrum shows 
\CIII{4650} greater than \NIII{4634} in emission; hence, the (fc) suffix. The GOSSS-DR1.1 classification differs from that of GOSSS-DR1.0.
 
\paragraph{HD~93\,843.}
\object[HD 93843]{}

The (fc) suffix was missing in the GOSSS-DR1.0 spectral classification of this O5~III system. This system appears to be variable in OWN data and was already 
reported as such by \citet{Walb73a}; it could be an SB1.
 
\paragraph{ALS~18\,747 = HM~1$-$6 = C1715$-$387$-$6.}
\object[ALS 18747]{}

This star was not included in \citet{Maizetal04b} due to its dimness \citep{HavlMoff77}.
See Figure~\ref{chart1} for a chart (Havlen-Moffat 1 field).
 
\paragraph{HDE~319\,699.}
\object[HDE 319699]{}

The ((fc)) suffix was missing in the GOSSS-DR1.0 spectral classification of this O5~V system. A preliminary analysis of the OWN data indicates that this
system is an SB1 with a period of 12.62 d.
 
\subsubsection{ON/OC stars}
\label{sec:ON/OC}  

\begin{figure*}
\centerline{\includegraphics*[width=\linewidth]{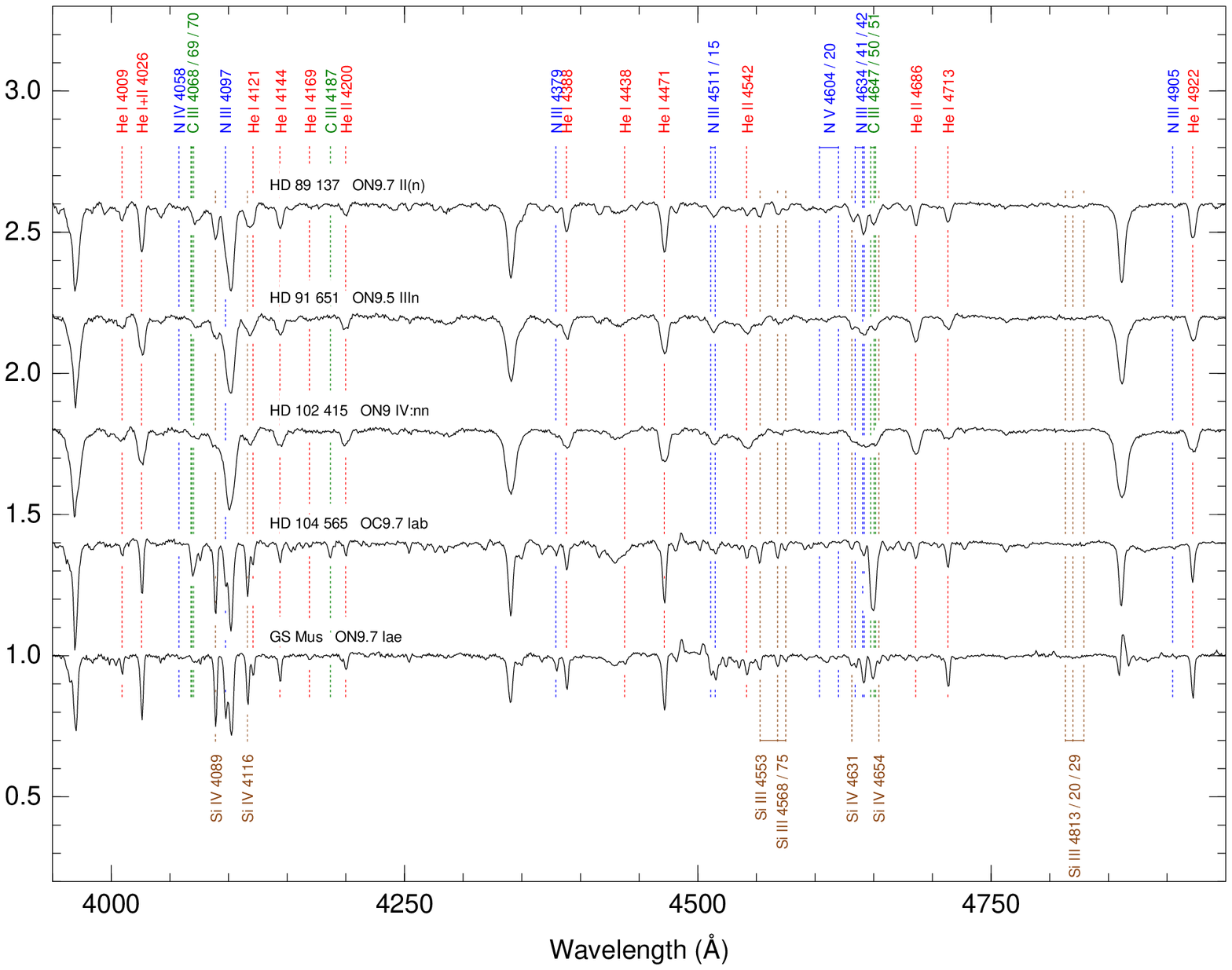}}
\caption{Spectrograms for the ON/OC stars. The targets are sorted by right ascension.
[See the electronic version of the journal for a color version of this figure.]}
\label{fig:ON/OC}
\end{figure*}	

\addtocounter{figure}{-1}

\begin{figure*}
\centerline{\includegraphics*[width=\linewidth]{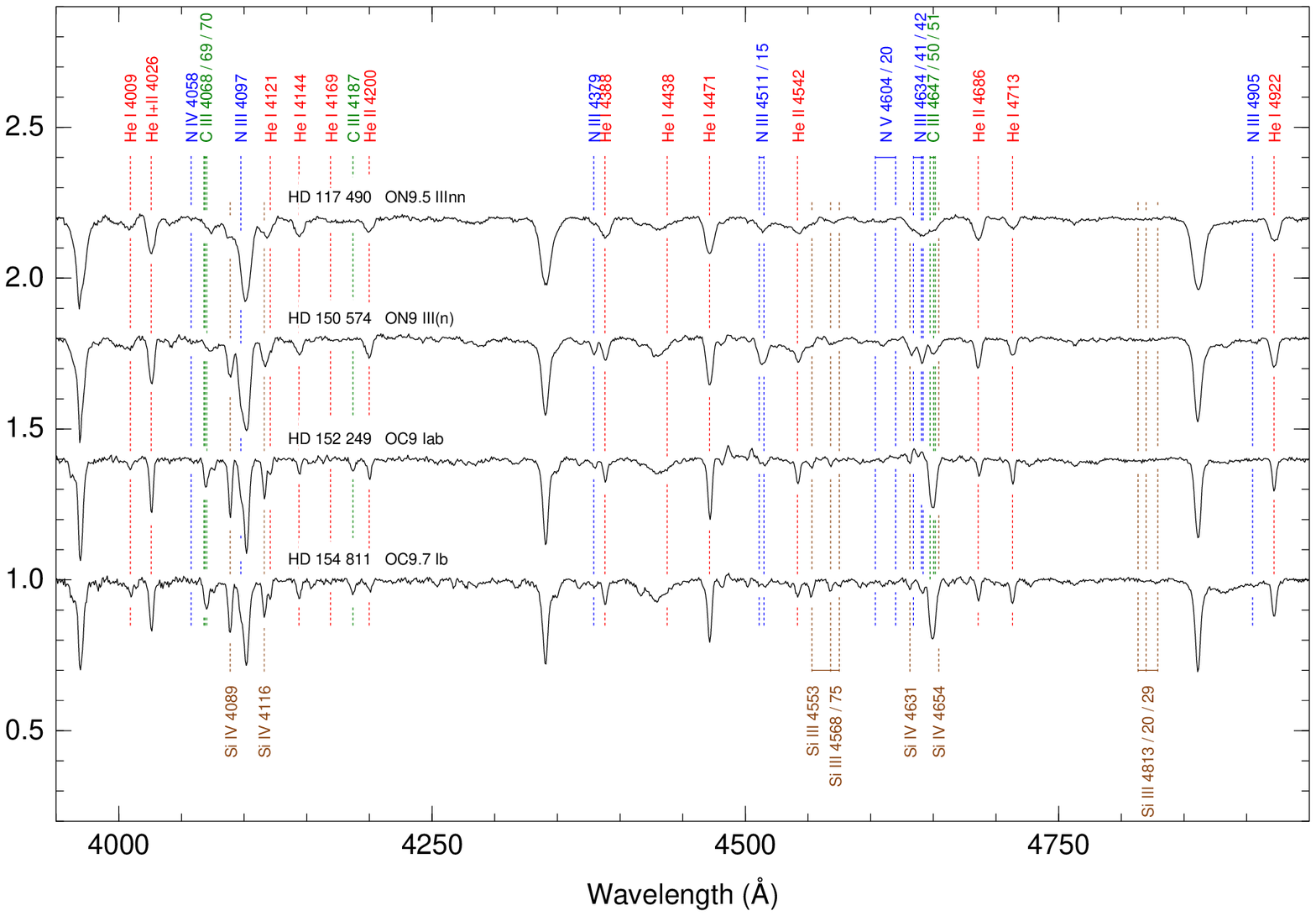}}
\caption{(continued).}
\end{figure*}	

Spectrograms are shown in Figure~\ref{fig:ON/OC}. See sections~\ref{sec:O9.2}~and~\ref{sec:PI} for additional members of this category.

\paragraph{HD~89\,137.}
\object[HD 89137]{}

This object is a nitrogen-rich rapid rotator \citep{Walbetal11}. According to \citet{Levaetal88}, it shows velocity variability that could be caused by a
companion. 
 
\paragraph{HD~91\,651.}
\object[HD 91651]{}

This system was found to be a likely SB2 by \citet{Levaetal88} and is also a nitrogen-rich rapid rotator \citep{Walbetal11}. The OWN data 
clearly show that is an SB2 and possibly an SB3. We do not see double lines in the GOSSS data. The luminosity class was changed in GOSSS-DR1.1.
 
\paragraph{HD~102\,415.}
\object[HD 102415]{}

This star is a nitrogen-rich very rapid rotator \citep{Walbetal11}. The OWN data also indicate variability; it could be an SB2. The luminosity class was changed 
in GOSSS-DR1.1.
 
\paragraph{HD~104\,565.}
\object[HD 104565]{}

This is a nitrogen-deficient star \citep{Walb76}. The luminosity class was changed in GOSSS-DR1.1.
 
\paragraph{GS~Mus = HD~105\,056.}
\object[V* GS Mus]{}

This nitrogen-rich star is variable \citep{Levaetal88} and could be a low-mass (PAGB?) object \citep{Walbetal11}. The OWN data show variability with a
period of 0.71 days that could be due to the system being an SB1, pulsations, or rotating spots.
 
\paragraph{HD~117\,490.}
\object[HD 117490]{}

This star is a nitrogen-rich very rapid rotator \citep{Walbetal11}. It could be a spectroscopic binary according to OWN data.
The n index was changed in GOSSS-DR1.1.
 
\paragraph{HD~150\,574.}
\object[HD 150574]{}

This system is a nitrogen-rich rapid rotator \citep{Walbetal11}. \citet{Levaetal88} suggested that it is an SB2 but that has not been confirmed with OWN
data. We do not see double lines in the GOSSS data.
 
\paragraph{HD~152\,003.}
\object[HD 152003]{}

This is one of the original weakly nitrogen-deficient stars in \citet{Walb76}. The Nwk suffix was omitted by mistake in GOSSS-DR1.0 and the spectral type
there was O9.5 instead of O9.7.
 
\paragraph{HD~152\,147.}
\object[HD 152147]{}

This is one of the original weakly nitrogen-deficient stars in \citet{Walb76}. The Nwk suffix was omitted by mistake in GOSSS-DR1.0. \citet{Willetal13}
indicated that it is an SB1 with a period of 13.8194 days but a preliminary OWN analysis suggests that it is an SB2 with a different orbit.
 
\paragraph{HD~152\,249.}
\object[HD 152249]{}
 
This nitrogen-deficient star \citep{Walb76} shows small velocity variations in OWN data.
See Figure~\ref{chart1} for a chart (NGC~6231 field).

\paragraph{HD~154\,811.}
\object[HD 154811]{}

\citet{Walb73a} classified this star as OC9.7~Ia and here we reclassify it as OC9.7~Ib. The H$\alpha$ line is highly variable in OWN.
  
\subsubsection{Onfp stars}
\label{sec:Onfp}  

\begin{figure*}
\centerline{\includegraphics*[width=\linewidth]{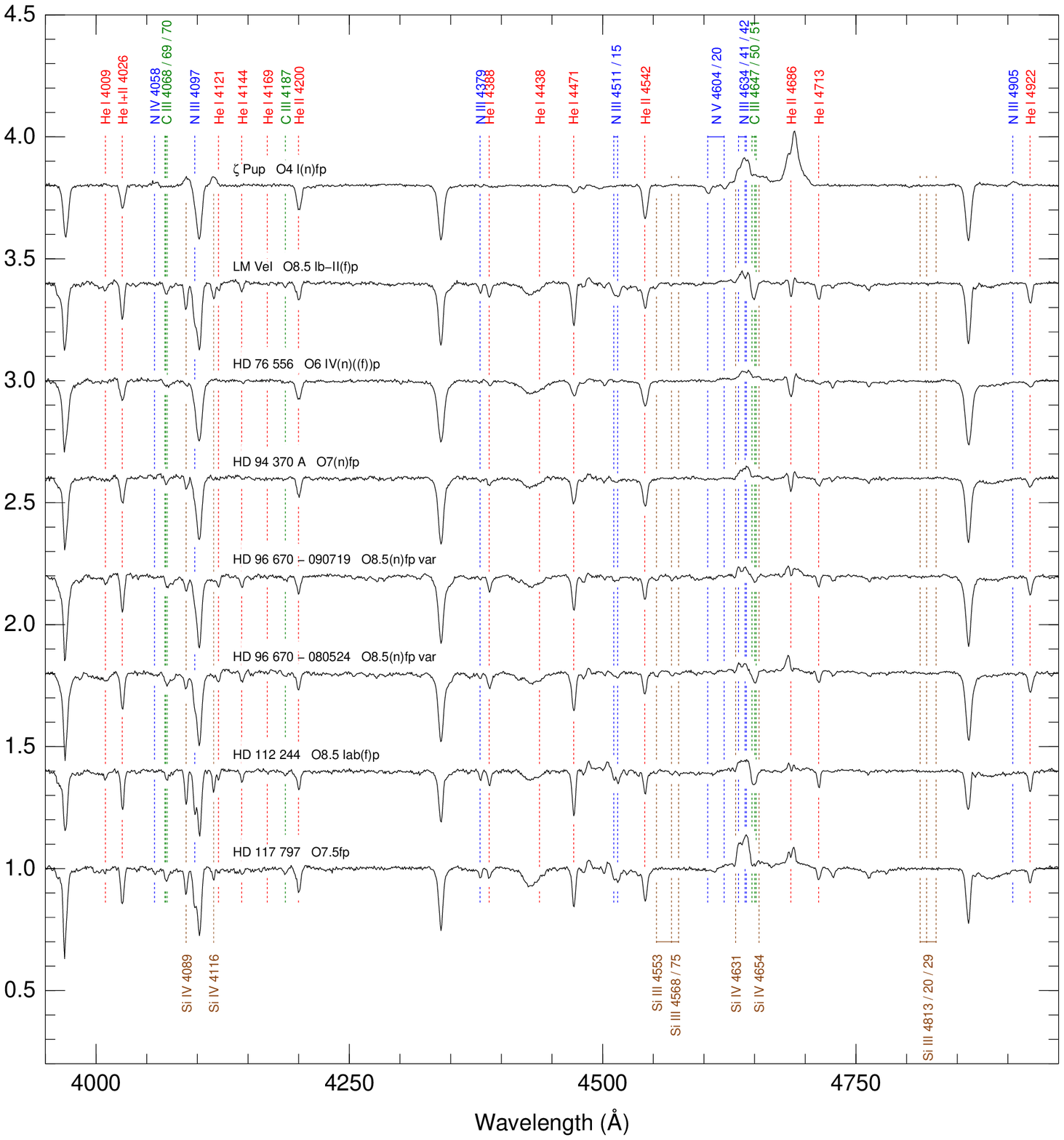}}
\caption{Spectrograms for the Onfp stars. The targets are sorted by right ascension. Two HD~96\,670 epochs are shown (dates in YYMMDD).
[See the electronic version of the journal for a color version of this figure.]}
\label{fig:Onfp}  
\end{figure*}	

See paper I for a description of this class.
Spectrograms are shown in Figure~\ref{fig:Onfp}.

\paragraph{$\zeta$~Pup = HD~66\,811 = Naos.}
\object[* zet Pup]{}

This is one of the two visually brightest O stars in the sky (the other being $\zeta$~Ori~Aa), to the point that it presents a challenge to be observed with a 2-4 m 
telescope at this resolution (the minimum exposure time for our LCO spectra is 1 s) and requires a neutral density filter. The p qualifier was added in GOSSS-DR1.1 
(but see \citealt{ContNiem76} and \citealt{Walbetal10b}). The new Hipparcos calibration gives a revised distance of $335^{+12}_{-11}$ pc \citep{Maizetal08a}.
 
\paragraph{LM~Vel = HD~74\,194.}
\object[V* LM Vel]{}

\citet{Barbetal06} detected variations in radial velocity and in H$\alpha$ that suggest that this system is an SB1 and the possible optical counterpart of the INTEGRAL 
hard X-ray source IGR~J08408-4503. The (f) suffix was omitted by mistake in GOSSS-DR1.0.
 
\paragraph{HD~76\,556.}
\object[HD 76556]{}

\citet{Walb73a} classified this star as O5.5~Vn((f)).  Here we make slight adjustments to the spectral subtype, luminosity class, and rotation index and move it to 
the Onfp category on the basis of the weak \HeII{4686} emission wings revealed by the GOSSS data (some of these changes are also with respect to GOSSS-DR1.0).

\paragraph{HD~94\,370~A.}
\object[HD 94370]{}

This star is a spectroscopic binary with a 2.8 d period according to OWN data. It is possibly an SB2, though the detection of the secondary 
star is marginal in the available data. It has a companion with a separation of 3\farcs6 and a $\Delta m$ of 1.8 magnitudes (according to the WDS) which we were 
able to spatially resolve in our GOSSS spectra: it is an early-B star. The luminosity class was eliminated in GOSSS-DR1.1. 
 
\paragraph{HD~96\,670.}
\object[HD 96670]{}

\citet{Walb72} classified this star as O8p. Here we classify it as O8.5(n)fp~var, with a variable \HeII{4686} profile (this is a change from GOSSS-DR1.0). See 
Figure~\ref{fig:Onfp} for the two GOSSS spectra obtained. According to \citet{SticLloy01} it is an SB1 with a 5.5296~d period.
 
\paragraph{HD~112\,244.}
\object[HD 112244]{}

According to OWN data, this object is an SB2 with a 7.5 day period. The p suffix was added in GOSSS-DR1.1.
We do not see double lines in the GOSSS data. 
 
\paragraph{HD~117\,797.}
\object[HD 117797]{}

This object is a new member of the Onfp category found by GOSSS; it is an SB2 according to OWN data, although we have not seen double lines in the GOSSS data. The 
classification was modified in GOSSS-DR1.1.
 
\subsubsection{Of?p stars}
\label{sec:Of?p}  

\begin{figure*}
\centerline{\includegraphics*[width=\linewidth]{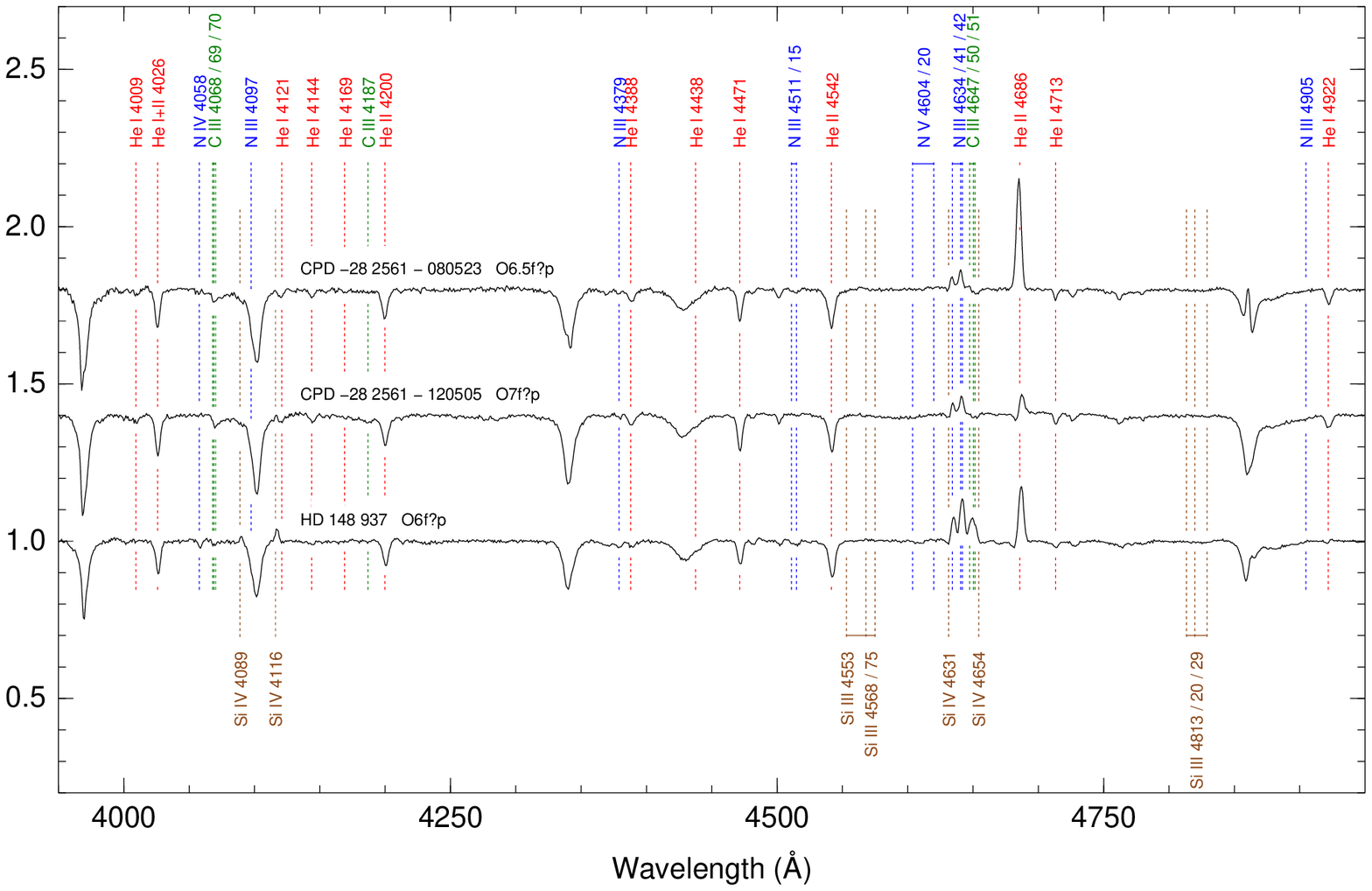}}
\caption{Spectrograms for the Of?p stars. The targets are sorted by right ascension. Two CPD$-$28~2561 epochs are shown (dates in YYMMDD).
[See the electronic version of the journal for a color version of this figure.]}
\label{fig:Of?p}  
\end{figure*}	

See \citet{Walbetal10a} for the characteristics of this class and section~\ref{sec:PI} for another member of this category.
Spectrograms are shown in Figure~\ref{fig:Of?p}.

\paragraph{CPD~$-$28~2561 = CD~$-$28~5104.}
\object[CPD-28 2561]{}

\cite{Walb73a} and \citet{Garretal77} reported this star to be a peculiar Of type of an undetermined nature. One of us (R.H.B.) discovered its Of?p nature, as
reported in \citet{Walbetal10a}. Figure~\ref{fig:Of?p} shows two GOSSS spectra taken in 2008 and 2012 that 
correspond to the ``high'' and ``low'' states, respectively. See Wade et al. for details (in preparation). 
 
\paragraph{HD~148\,937.}
\object[HD 148937]{}

This star was one of the two original Of?p stars \citep{Walb72}.
It differs from the rest in its class in that it only shows small photometric changes on 7 day timescales as opposed to larger variations on longer timescales. 
The period was discovered through the variations in its H$\beta$ profile. A possible explanation is that the object is seen pole-on (\citealt{Nazeetal12a} and 
references therein). It is also the only member of its class to be involved in an ejected, axially symmetric circumstellar nebula (NGC~6164-6165).
 
\subsubsection{Oe stars}
\label{sec:Oe}

\begin{figure*}
\centerline{\includegraphics*[width=\linewidth]{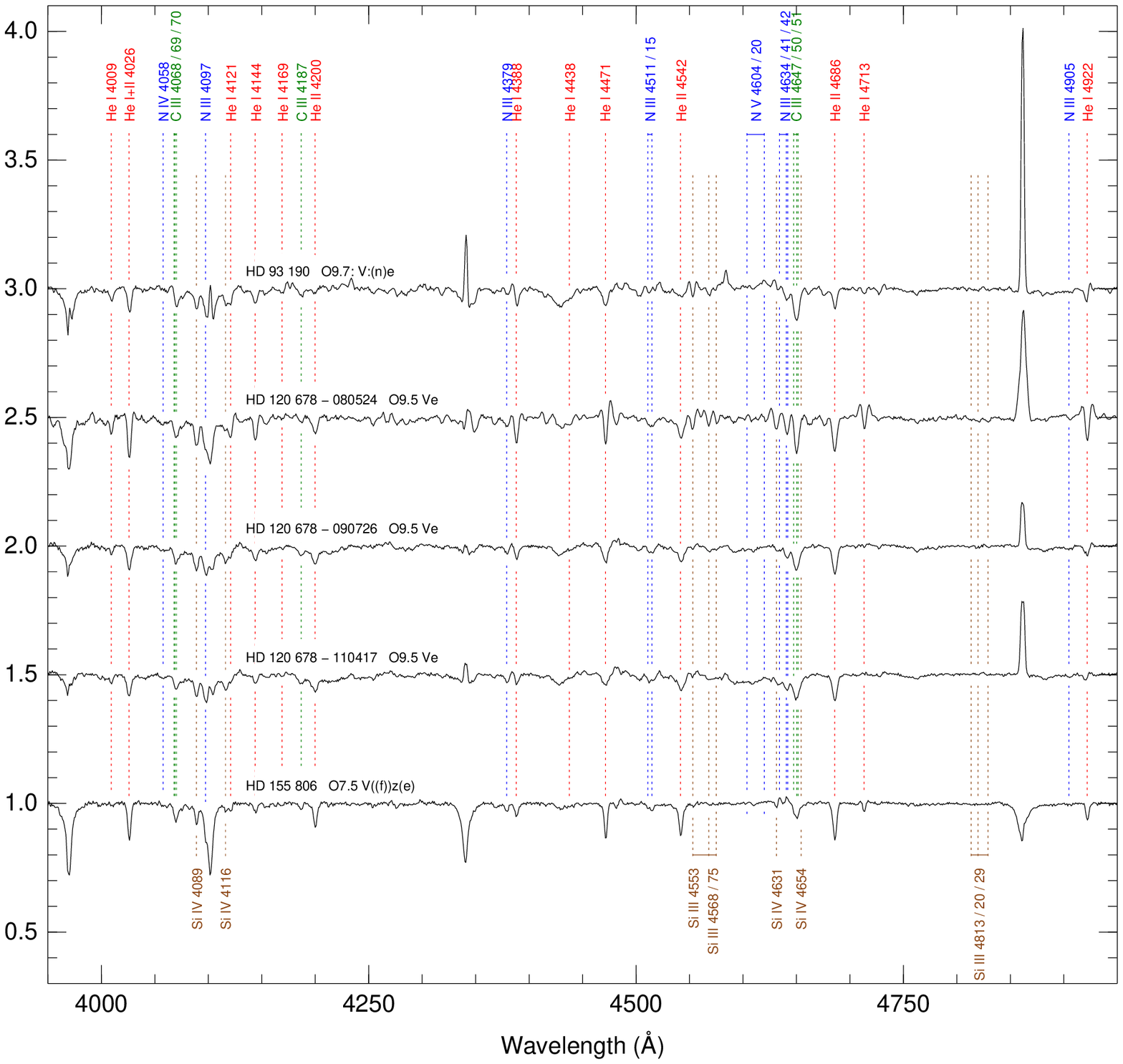}}
\caption{Spectrograms for the Oe stars. The targets are sorted by right ascension. Three HD~120\,678 epochs are shown (dates in YYMMDD).
[See the electronic version of the journal for a color version of this figure.]}
\label{fig:Oe}  
\end{figure*}	

Spectrograms are shown in Figure~\ref{fig:Oe}.

\paragraph{HD~93\,190.}
\object[HD 93190]{}

This star in the Carina Nebula was not included in \citet{Maizetal04b}.
 
\paragraph{HD~120\,678.}
\object[HD 120678]{}

This Oe star experienced a shell-like event in 2008 \citep{Gameetal12} that was detected with OWN and GOSSS data. 
Figure~\ref{fig:Oe} shows three GOSSS spectra taken in May 2008, July 2009, and April 2011 that correspond to the early shell stage, recovery phase, and normal
state, respectively.
 
\paragraph{HD~155\,806 = V1075~Sco.}
\object[HD 155806]{}

This object has been historically considered an Oe star (see \citealt{Neguetal04} for its H$\alpha$ profile). In our blue-violet spectra, H$\beta$ is partially filled
but not enough to show an emission, hence the (e) suffix. The ((f))z suffix was introduced in GOSSS-DR1.1.
 
\subsubsection{Double- and triple-lined spectroscopic binaries}
\label{sec:SB2}

\begin{figure*}
\centerline{\includegraphics*[width=\linewidth]{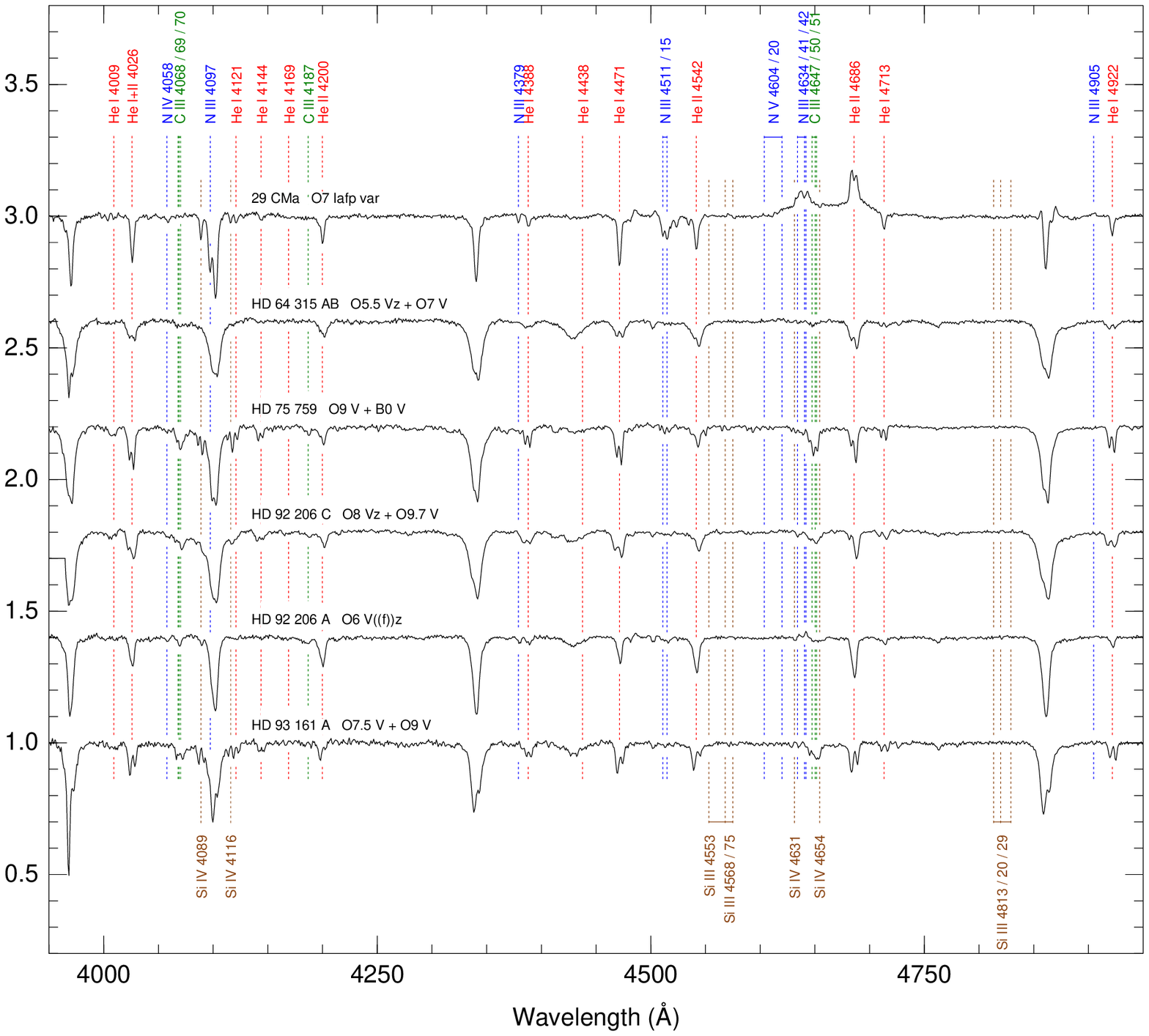}}
\caption{Spectrograms for the SB2 and SB3 stars. The targets are sorted by right ascension.
[See the electronic version of the journal for a color version of this figure.]}
\label{fig:SB2}
\end{figure*}	

\addtocounter{figure}{-1}

\begin{figure*}
\centerline{\includegraphics*[width=\linewidth]{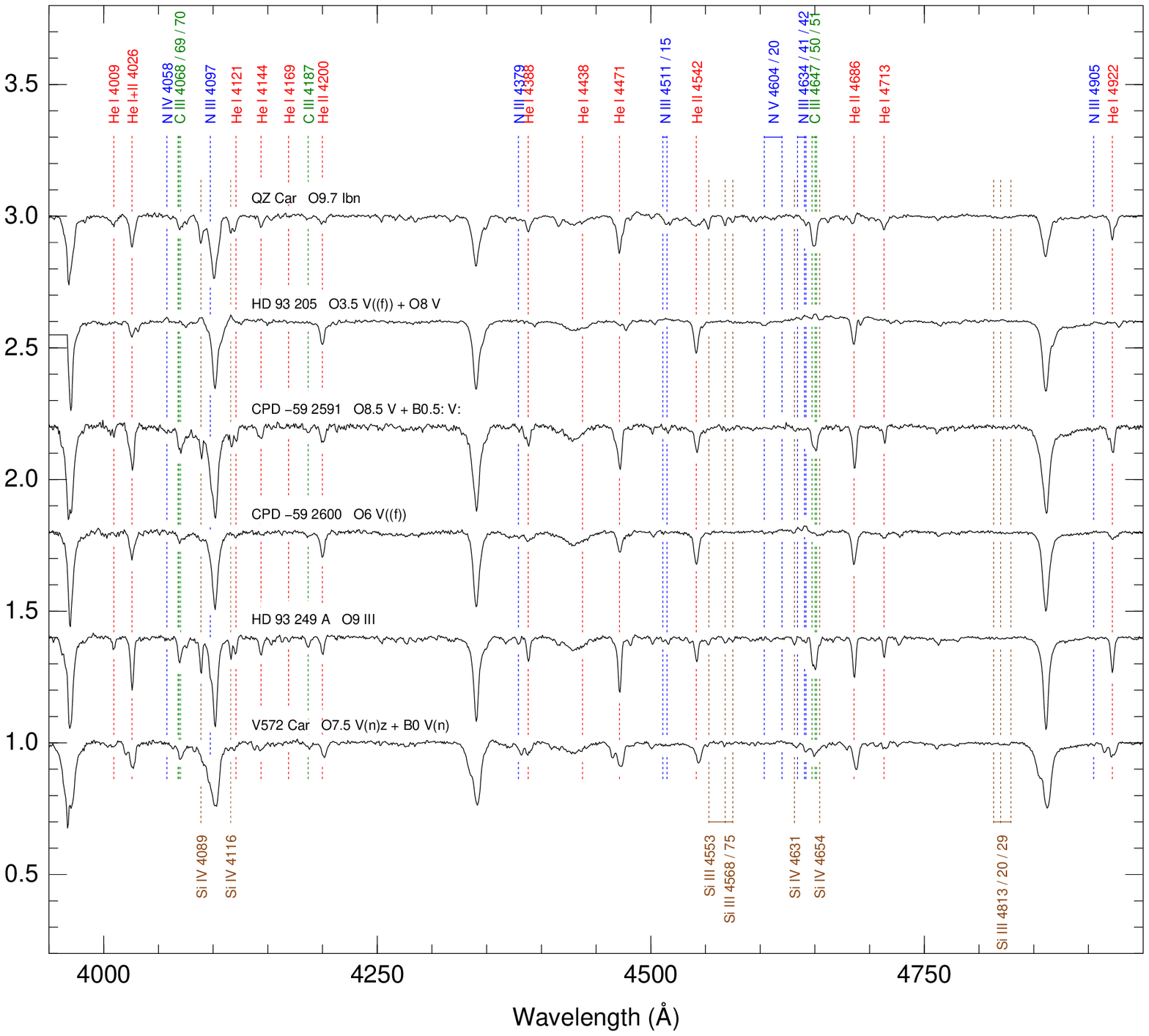}}
\caption{(continued).}
\end{figure*}	

\addtocounter{figure}{-1}

\begin{figure*}
\centerline{\includegraphics*[width=\linewidth]{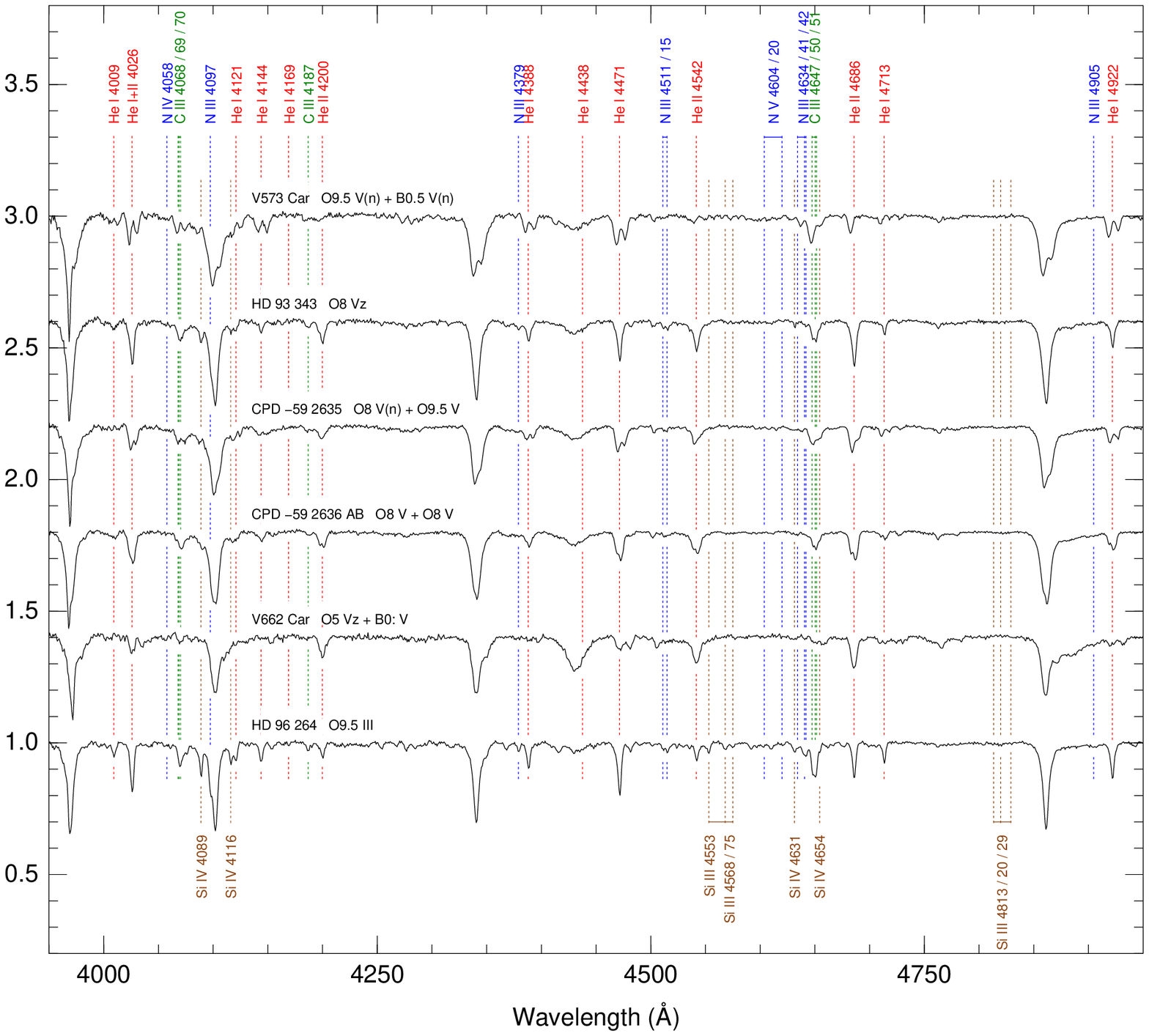}}
\caption{(continued).}
\end{figure*}	

\addtocounter{figure}{-1}

\begin{figure*}
\centerline{\includegraphics*[width=\linewidth]{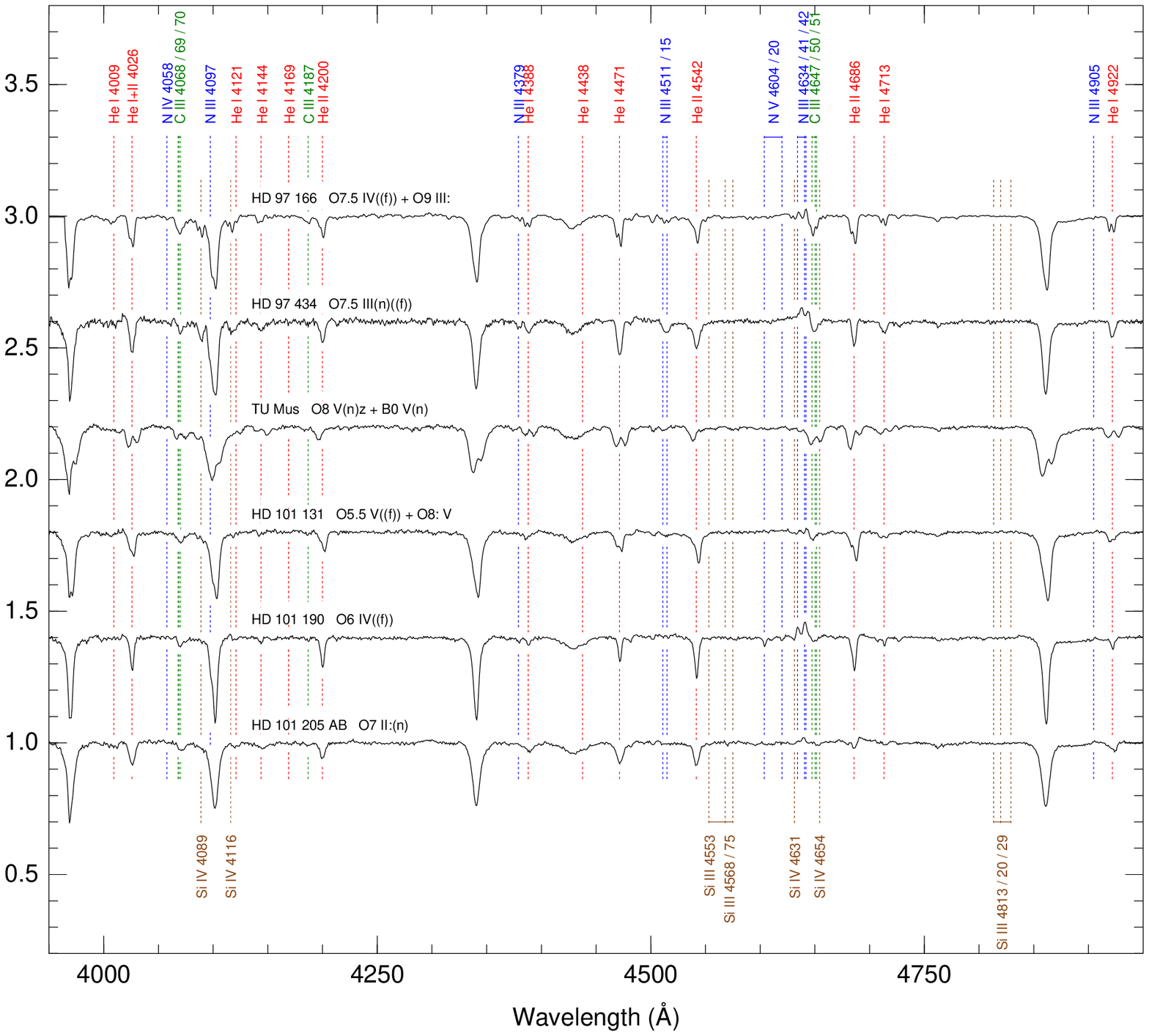}}
\caption{(continued).}
\end{figure*}	

\addtocounter{figure}{-1}

\begin{figure*}
\centerline{\includegraphics*[width=\linewidth]{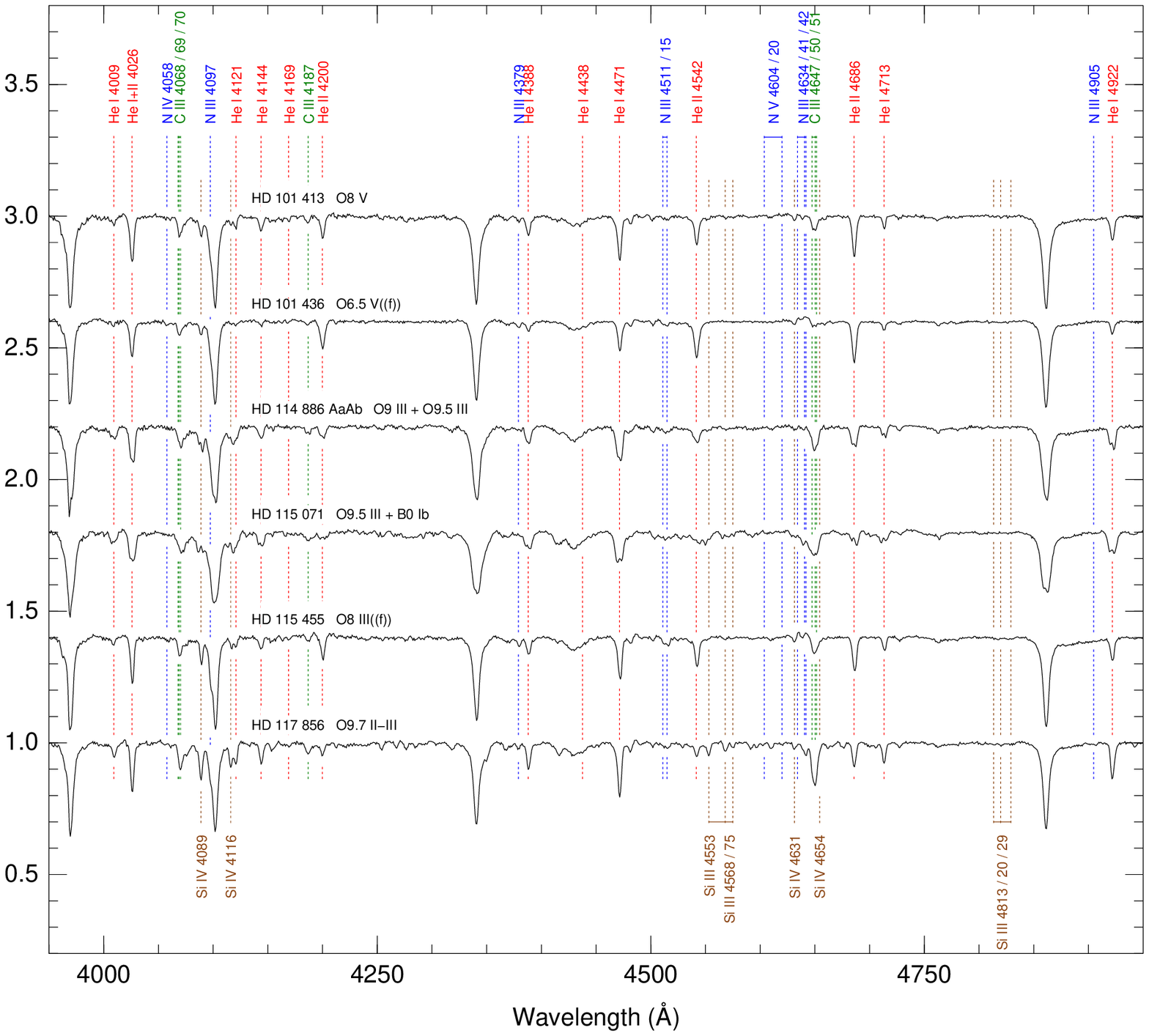}}
\caption{(continued).}
\end{figure*}	

\addtocounter{figure}{-1}

\begin{figure*}
\centerline{\includegraphics*[width=\linewidth]{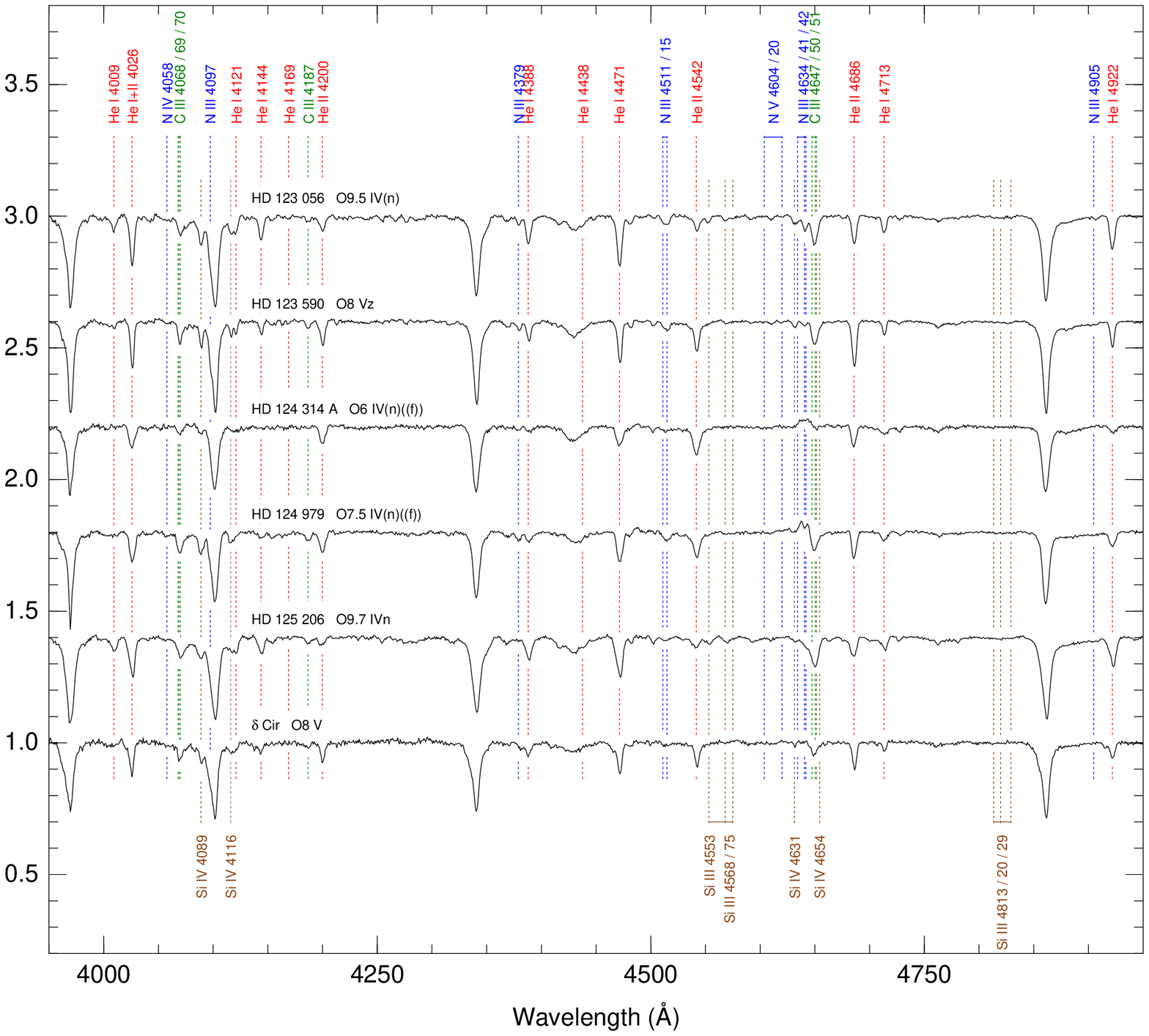}}
\caption{(continued).}
\end{figure*}	

\addtocounter{figure}{-1}

\begin{figure*}
\centerline{\includegraphics*[width=\linewidth]{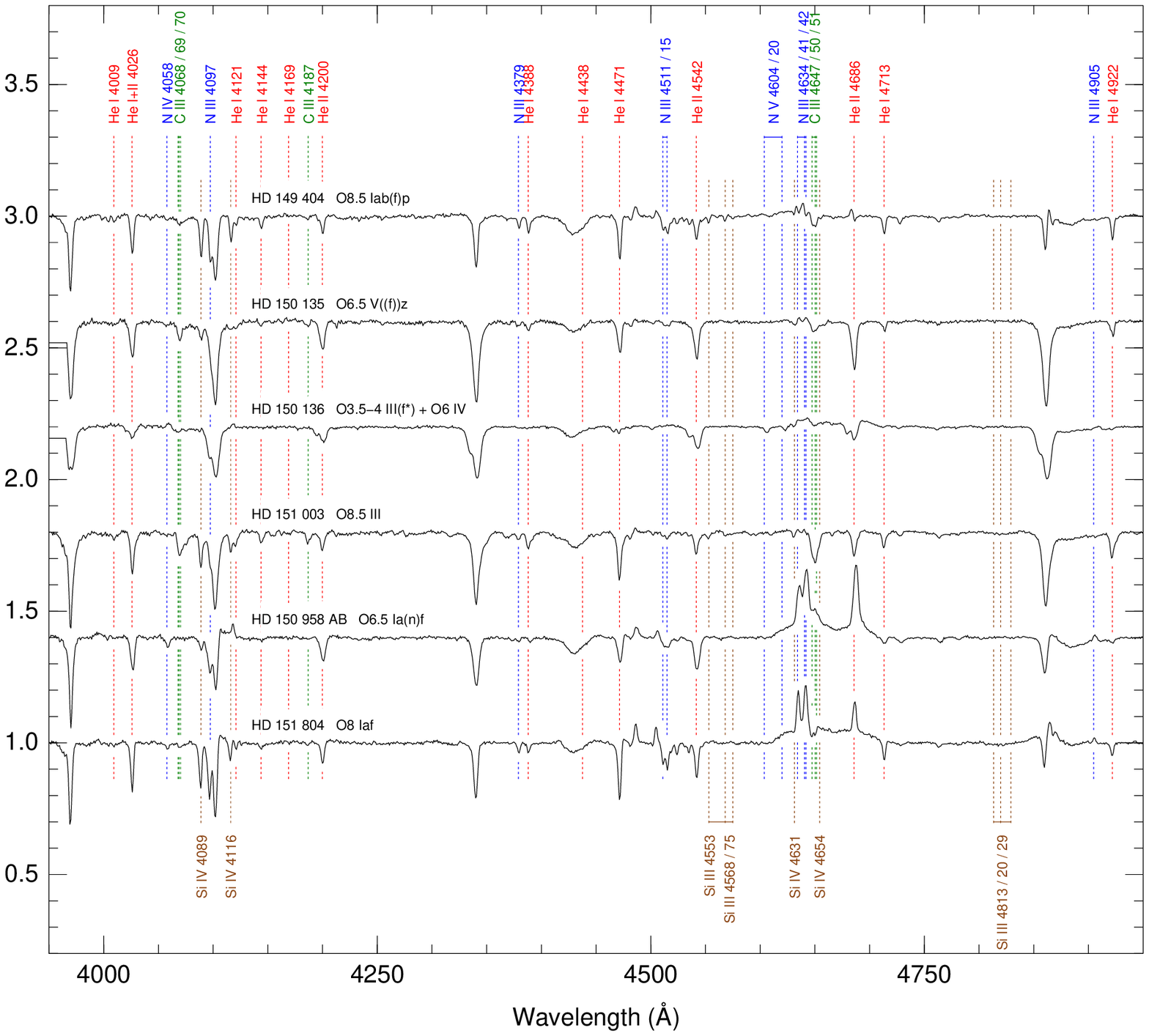}}
\caption{(continued).}
\end{figure*}	

\addtocounter{figure}{-1}

\begin{figure*}
\centerline{\includegraphics*[width=\linewidth]{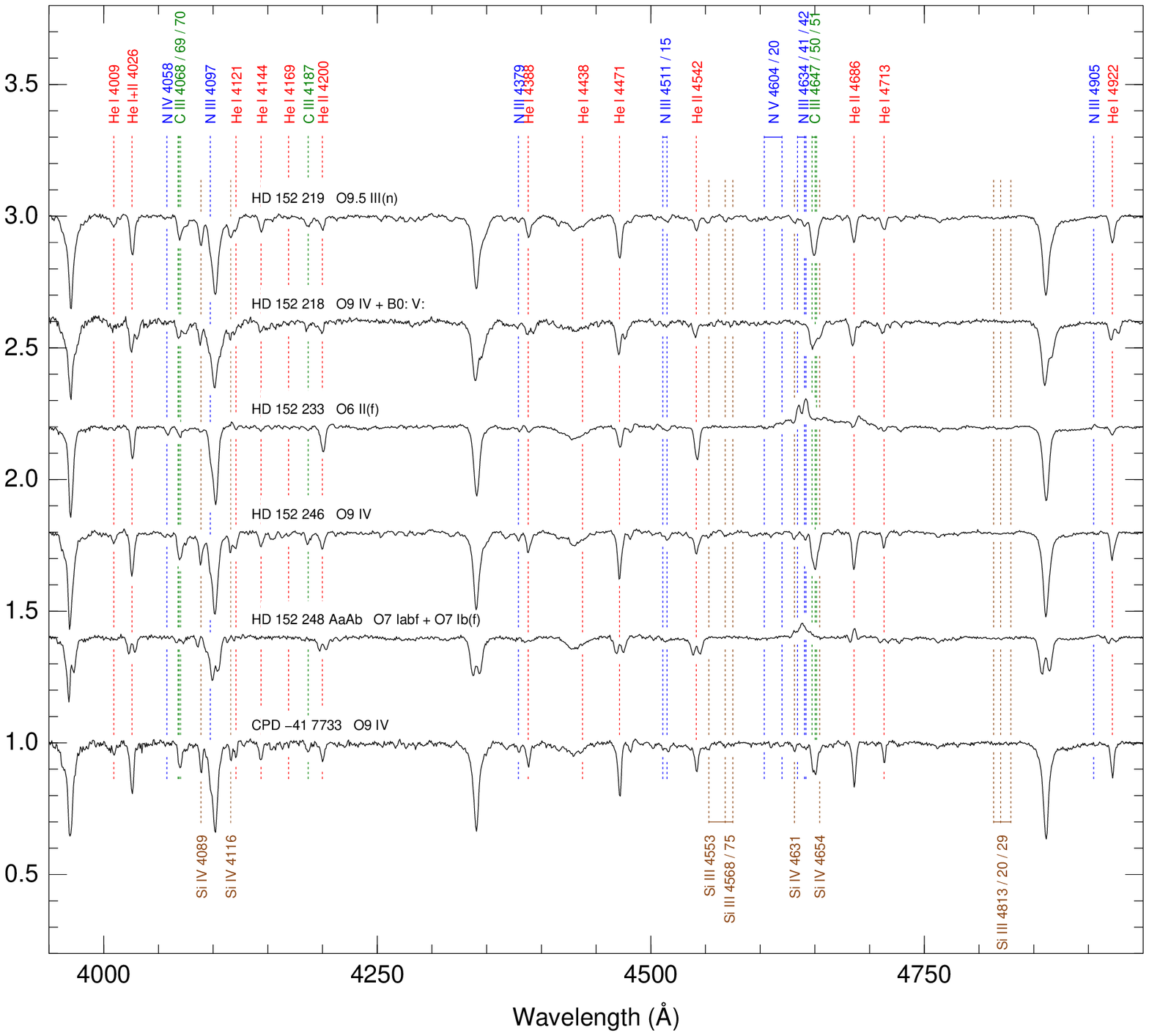}}
\caption{(continued).}
\end{figure*}	

\addtocounter{figure}{-1}

\begin{figure*}
\centerline{\includegraphics*[width=\linewidth]{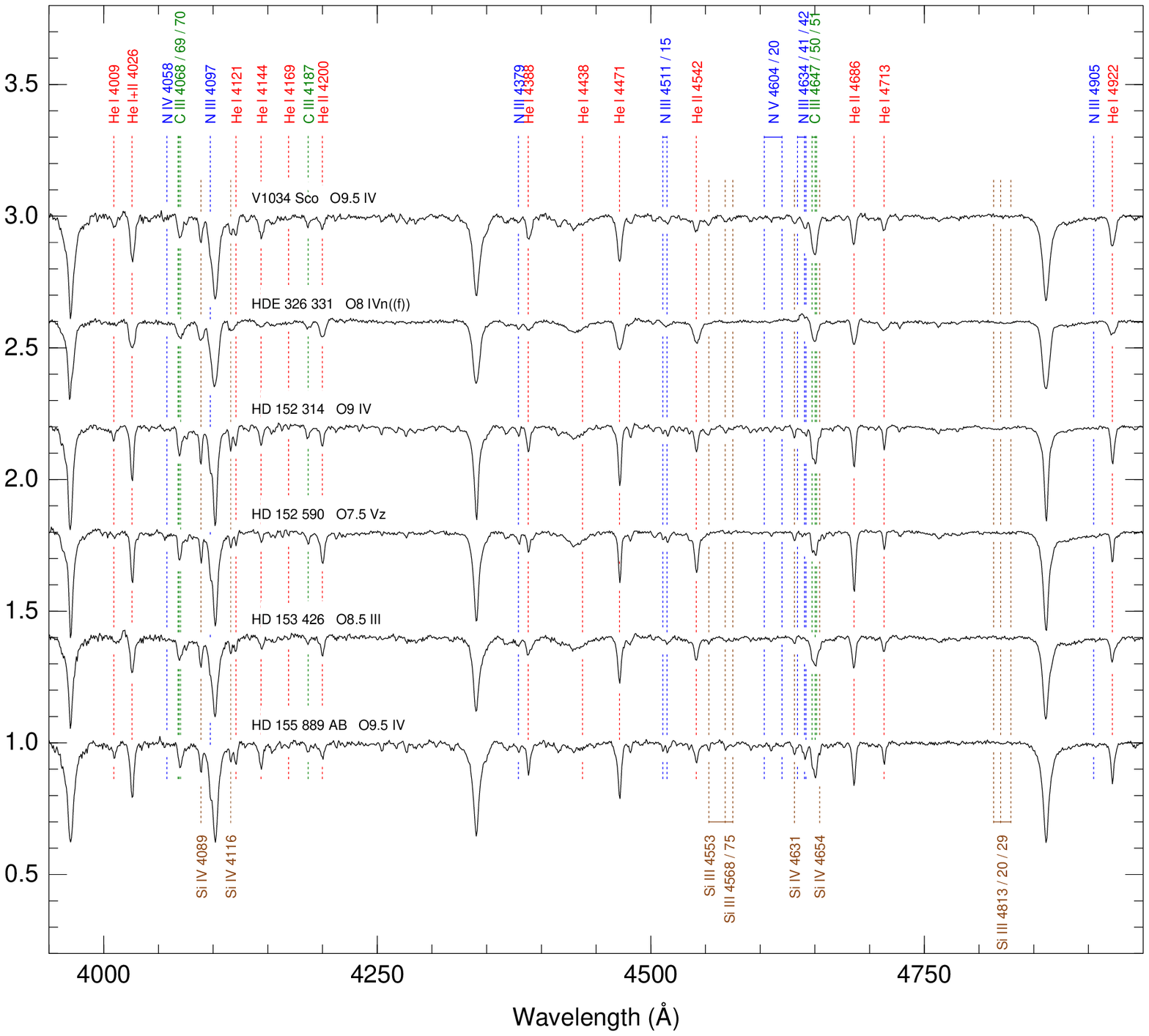}}
\caption{(continued).}
\end{figure*}	

\addtocounter{figure}{-1}

\begin{figure*}
\centerline{\includegraphics*[width=\linewidth]{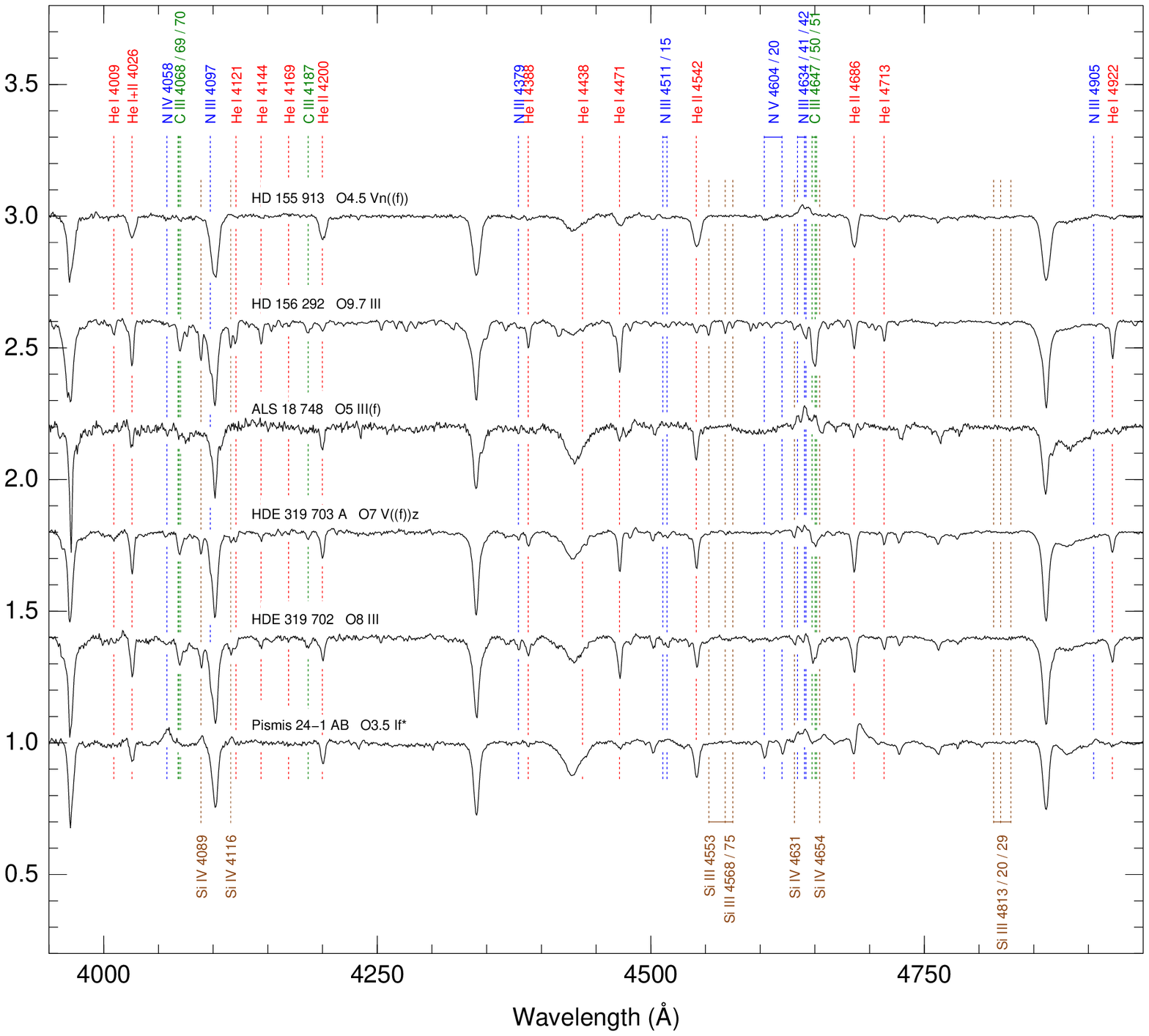}}
\caption{(continued).}
\end{figure*}	

\addtocounter{figure}{-1}

\begin{figure*}
\centerline{\includegraphics*[width=\linewidth]{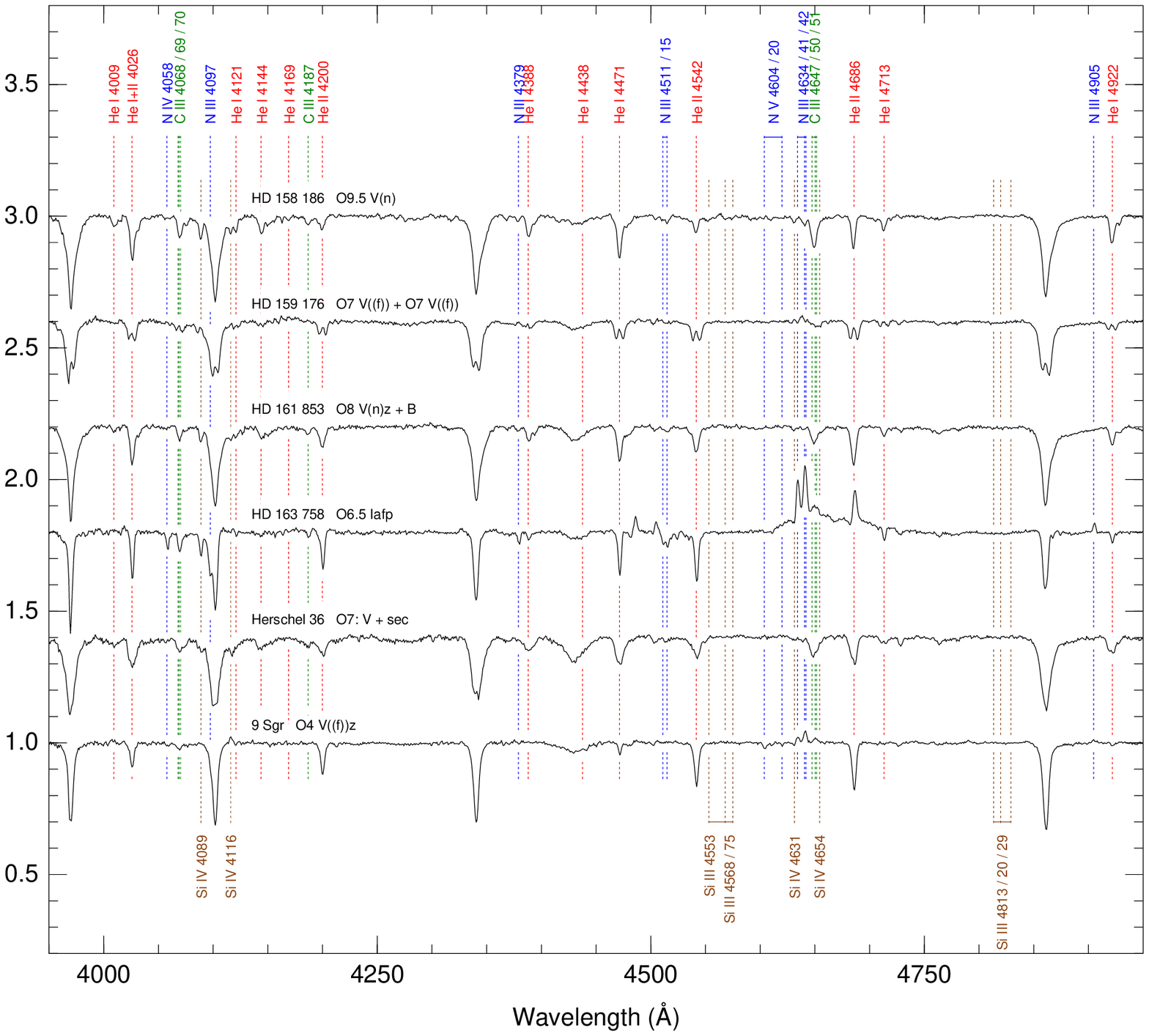}}
\caption{(continued).}
\end{figure*}	

\addtocounter{figure}{-1}

\begin{figure*}
\centerline{\includegraphics*[width=\linewidth]{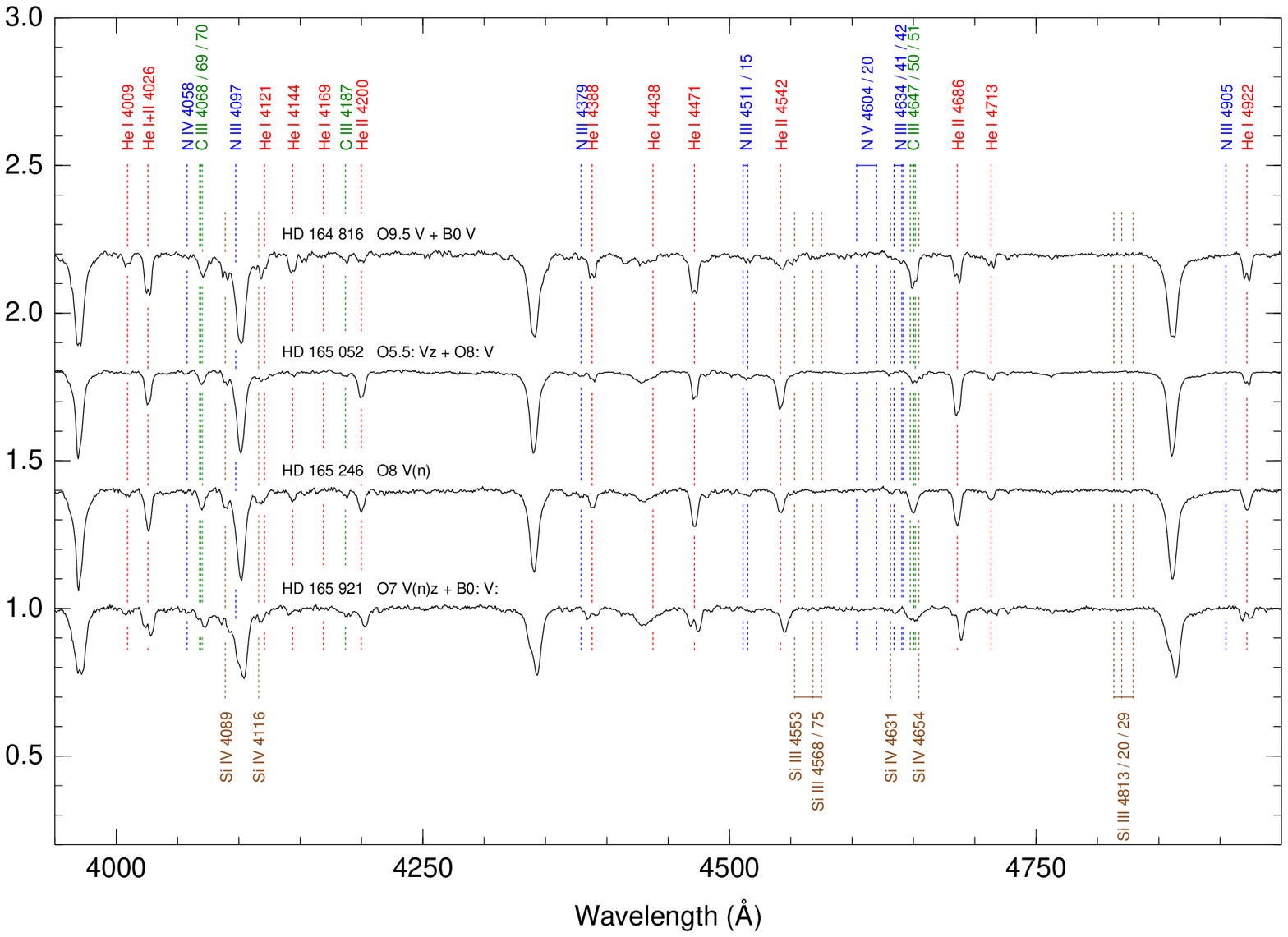}}
\caption{(continued).}
\end{figure*}	

In this part of this subsection we include the double- (SB2) and triple- (SB3) lined spectroscopic 
binaries that do not belong to any of the other peculiar categories. More than for any other peculiar categories,
membership here is determined by spectral resolution and time coverage, given the large ranges of velocity differences 
and periods existent among massive spectroscopic binaries. Therefore, we have included in this category examples that
have been identified as SB2s or SB3s by other authors (in most cases using high-resolution spectroscopy) but that are 
single lined in our spectra. In those cases, we point to the relevant reference. Our classifications 
were obtained with MGB varying seven input parameters: the spectral types, luminosity classes, and velocities of both the
primary and secondary; and the flux fraction of the secondary.
Spectrograms are shown in Figure~\ref{fig:SB2}.

\paragraph{29~CMa = HD~57\,060 = UW~CMa.}
\object[* 29 CMa]{}

We do not see double lines in the GOSSS data of this SB2 system with a 4.39349 day period \citep{Pouretal04}, except for \HeII{4686}.
However, we see clear variabilty in the emission lines among our four epochs, including for the peculiar H$\beta$ profile (in absorption
but with one or two emission peaks at the line edges). The new Hipparcos 
calibration gives a revised distance of $632^{+137}_{-96}$ pc \citep{Maizetal08a}. The analysis of the light curve of this eclipsing
binary by \citet{Antoetal11} suggests a contact configuration.
 
\paragraph{HD~64\,315~AB = V402~Pup~AB.}
\object[HD 64315]{}

\citet{Tokoetal10} spatially resolved this system into three tight components with speckle interferometry although \citet{Masoetal09} and 
Aldoretta et al. (in preparation) only see two of them, separated by 63 mas and with a $\Delta m$ of 0.6 magnitudes. \citet{Loreetal10} discovered 
that HD~64\,315 is a double SB2 system composed of four O stars. We observed this system with GOSSS on four different occasions and 
found the absorption line profiles to be highly varying. For the epoch where there is a better velocity resolution, we see an O5.5~Vz and an 
O7~V component. The spectral types are consistent with the range of effective temperatures measured by \citet{Loreetal10}. The z 
character of the primary component could be caused by its hidden binary nature.
 
\paragraph{HD~75\,759.}
\object[HD 75759]{}

This system is considered by a number of literature sources to have an early-B spectrum but it is an SB2 (period: 33.311 days,
\citealt{Pouretal04}) that contains an O star. The O-type nature was already established by \citet{Hiltetal69} and \citet{Walb73a}; 
the latter also commented on the possible existence of a secondary component. With MGB we obtain spectral types of O9~V and B0~V.
The new Hipparcos calibration gives a revised distance of $947^{+378}_{-203}$~pc \citep{Maizetal08a}.
 
\paragraph{HD~92\,206~C = CPD~$-$57~3580.}
\object[HD 92206 C]{}

This object does not appear in the WDS catalog as HD~92\,206~C (it is located 34\arcsec\ away from the A component), though it is most 
commonly referred to by that name in the literature. \citet{Campetal07} discovered its SB2 nature and assigned it spectral types of O7.5~V and B0~V.
We have observed it on three occasions with GOSSS and in two of them it appears clearly as an SB2, with spectral types of O8~Vz and O9.7~V. It is 
located in NGC 3324 at the NW of the Carina Nebula. See Figure~\ref{chart1} for a chart (HD~92\,206 field).
 
\paragraph{HD~92\,206~A.}
\object[HD 92206 A]{}

HD~92\,206~AB is a closer pair with a separation of 5\farcs3 spatially resolved with ease into two O-type systems with GOSSS. OWN data show 
that it is an SB2. We do not see double lines in the GOSSS data. See Figure~\ref{chart1} for a chart (HD~92\,206 field).
 
\paragraph{HD~93\,161~A.}
\object[HD 93161A]{}

HD~93\,161~A and B are separated by 2\arcsec\ and both are O-type systems, with A itself being an O+O spectroscopic binary 
\citep{Nazeetal05}. When observed within the same aperture, A+B form an SB3 system. We observed A twice and on both occasions we
clearly detected it as an SB2. Our best spectral type for the secondary (O9~V) agrees with that of \citet{Nazeetal05} but that for the
primary (O7.5~V) is half a spectral subtype earlier. See Figure~\ref{chart1} for a chart (Trumpler 14 field).
 
\paragraph{QZ~Car = HD~93\,206.}
\object[V* QZ Car]{}

This complex SB1+SB1 system was studied by \citet{Mayeetal01}, who measured the two periods in the system to be 5.991~d (eclipsing) and 20.73596~d
(non-eclipsing). See also the previous works by \citet{Leunetal79} and \citet{MorrCont80}.
The spectral types of \citet{Parketal11} are consistent with our combined classification (we do not see double lines in the GOSSS data).
 
\paragraph{HD~93\,205 = V560~Car.}
\object[HD 93205]{}

\citet{ContWalb76} classified this massive SB2 system in Carina as O3 V + O8 V. We observed it five times with GOSSS and in two of
them it is clearly in a double-lined state. Our best spectral type for the secondary (O8~V) agrees with that of \citet{Morretal01} but the 
one for the primary (O3.5~V((f))) is half a spectral subtype later (something that was already modified by \citealt{Walbetal02b} when the O3.5 subtype was
created) and adds the ((f)) suffix. 
The 4650~\AA\ region is variable between epochs: it is not clear if what we are seeing in some epochs as emission there
corresponds to \CIII{4650} from the primary (making it an Ofc) or \NIIId{4640-42} from the secondary.
See Figure~\ref{chart1} for a chart (Trumpler 16 field). In the existing ACS/HRC images we detect a previously unpublished visual companion. It has a separation 
of 3\farcs68, a position angle of 272\arcdeg, and a $\Delta V$ of 9.3~magnitudes.
Note that both HD~93\,205 and HD~93\,250 are systems with early-type O stars in them: one should be careful not to confuse them. 
 
\paragraph{CPD~$-$59~2591 = Trumpler~16$-$21.}
\object[CPD-59 2591]{}

This star was not included in \citet{Maizetal04b} but was classified as O8~V by \cite{MassJohn93}. We did not detect that it was an SB2 in GOSSS-DR1.0 
but a later analysis prompted by a spectrum obtained for the Gaia-ESO Survey \citep{Gilmetal12} led us to find the weak secondary B spectrum redshifted by 
$\sim$250~km/s in an existing GOSSS spectrum. See Figure~\ref{chart1} for a chart (Trumpler 16 field).
 
\paragraph{CPD~$-$59~2600.}
\object[CPD-59 2600]{}

OWN data show that this system is an SB1 and, possibly, an SB2 with a 626~d period.
We do not see double lines in the GOSSS data. See Figure~\ref{chart1} for a chart (Trumpler 16 field).
 
\paragraph{HD~93\,249~A.}
\object[HD 93249]{}

OWN data show that this system is an SB2 with a 2.9797~d period.
We do not see double lines in the GOSSS data. This is the brightest star in Trumpler~15.
 
\paragraph{V572~Car = CPD~$-$59~2603 = Trumpler~16-104.}
\object[V* V572 Car]{}

\citet{Rauwetal01a} classified this system as a hierarchical triple consisting of an inner eclipsing O7~V + O9.5~V binary and an outer
B0.2~IV component. In one of our two GOSSS epochs there are two components visible, an O7.5~V(n)z and a B0~Vn, the latter likely
being a combination of the two later spectral types detected by \citet{Rauwetal01a} in their high-resolution spectra.
See Figure~\ref{chart1} for a chart (Trumpler 16 field).
 
\paragraph{V573~Car = CPD~$-$59~2628.}
\object[V* V573 Car]{}

This SB2 system has a period of 1.47 days and was classified by \citet{Freyetal01} as O9.5 V + B0.2 V. We observed it twice with GOSSS 
and in both occasions we clearly detect double lines. Our spectral types (O9.5 V and B0.5 V) are similar to those of 
\citet{Freyetal01} though in both cases qualified with (n) suffixes. This star was not included in \citet{Maizetal04b}.
See Figure~\ref{chart1} for a chart (Trumpler 16 field).
 
\paragraph{HD~93\,343.}
\object[HD 93343]{}

\citet{Rauwetal09} classified this SB2 system as O7-8.5 + O8. We do not see double lines in the GOSSS data but our combined spectral
type, O8 Vz, is consistent with the \citet{Rauwetal09} results. 
See Figure~\ref{chart1} for a chart (Trumpler 16 field).
 
\paragraph{CPD~$-$59~2635 = Trumpler 16-34.}
\object[CPD-59 2635]{}

\citet{Albaetal01} classified this SB2 system as O8~V + O9.5~V and \citet{Oter06} discovered its eclipses (period of 2.29995 d). We obtain the same spectral 
types with GOSSS data with the only difference of an (n) suffix for the primary. See Figure~\ref{chart1} for a chart (Trumpler 16 field).
 
\paragraph{CPD~$-$59~2636~AB = Trumpler 16-110 AB.}
\object[CPD -59 2636]{}

This SB3 system was classified as O7~V + O8~V + O9~V by \citet{Albaetal02}. According to the WDS, the B component (called C in \citealt{Albaetal02}) 
is a visual component 0\farcs3 away from the SB2 system and is an SB1 system by itself, making CPD~$-$59~2636 a quadruple system with one undetected
component. We see two of those components in one of the six GOSSS epochs and we obtain spectral types of O8~V + O8~V. One of the set of lines is
deeper than the other, likely indicating that is a composite of the O7 V and O9 V components.  See Figure~\ref{chart1} for a chart (Trumpler 16 
field).
 
\paragraph{V662~Car = FO~15.}
\object[V* V662 Car]{}

\citet{Niemetal06} classified this SB2 system as O5.5~Vz + O9.5~V. It is an eclipsing binary with a period of 1.41355 d \citep{Oter06,FerLNiem06b}.
We detect double lines in \HeI{4471} for three of the five GOSSS epochs, yielding a classification of O5~Vz + B0:~V, which is reasonably consistent with the 
previous result and confirms the z nature of the primary. This star was not included in \citet{Maizetal04b}.
 
\paragraph{HD~96\,264.}
\object[HD 96264]{}

This object is an SB2 according to OWN data. Here we do not see double lines and we classify it as O9.5~III.
 
\paragraph{HD~97\,166.}
\object[HD 97166]{}

According to OWN data, this system is an SB2 and possibly an SB3.
We do not see double lines in the GOSSS data. The ((f)) suffix was added in GOSSS-DR1.1.
 
\paragraph{HD~97\,434.}
\object[HD 97434]{}

According to OWN data, this system is an SB2.
We do not see double lines in the GOSSS data. 
 
\paragraph{TU~Mus = HD~100\,213.}
\object[V* TU Mus]{}

\citet{Lindetal07} classified this SB2 system as O7.5~V + O9.5~V. We obtained two epochs with GOSSS and double lines were clearly seen in this very
fast system (1.3873~days period). Our spectral types are O8~V(n)z + B0~V(n), which are not too different from those in \citet{Lindetal07}. In
particular, an (n) or n index is expected in a contact binary like TU Mus.
 
\paragraph{HD~101\,131 = V1051~Cen.}
\object[HD 101131]{}

\citet{Giesetal02} obtained spectral types of O6.5~V((f)) + O8.5~V. From the GOSSS spectra we derived slightly earlier types of O5.5~V((f)) + O8:~V.
In GOSSS-DR1.1 we added the missing ((f)) suffix to the primary. See Figure~\ref{chart1} for a chart (IC~2944 field).
 
\paragraph{HD~101\,190.}
\object[HD 101190]{}

According to \citet{Sanaetal11a}, this system is an SB2 with spectral types O4~V((f)) + O7~V, but they caution that the system could include a third component
based on the lack of motion of the N\,{\sc iii} and N\,{\sc v} lines. We do not detect double lines in the GOSSS data and assign this system a classification of
O6~IV((f)), intermediate between that of the primary and the secondary of \citet{Sanaetal11a}. The WDS lists two dim companions within 15\arcsec. 
The ((f)) suffix was added in GOS-DR1.1. See Figure~\ref{chart1} for a chart (IC~2944 field).
 
\paragraph{HD~101\,205~AB = V871~Cen.}
\object[HD 101205]{}

This system is an SB2 according to OWN data. The WDS lists a B companion with a separation of 0\farcs4 and a $\Delta m$ of 0.3 magnitudes 
that we were unable to spatially resolve in the GOSSS data (we do not see double lines, either). Either A or B is an eclipsing binary with a period of 2.090704~d 
\citep{Oter07}, which is different from the SB2 period. The luminosity class of II: (changed in GOSSS-DR1.1) is a compromise between the behavior of \NIIId{4634-40-42} 
and of \HeII{4686}, possibly a consequence of the composite nature of the spectrum. See Figure~\ref{chart1} for a chart (IC~2944 field).
 
\paragraph{HD~101\,413.}
\object[HD 101413]{}

\citet{Sanaetal11a} obtained spectral types of O8~V + B3:~V for this SB2 system. We do not see double lines in the GOSSS data, which is unsurprising given the
large expected $\Delta m$ (our combined spectral type is the same that \citet{Sanaetal11a} obtain for the primary). See Figure~\ref{chart1} for a chart (IC~2944 field).
 
\paragraph{HD~101\,436.}
\object[HD 101436]{}

\citet{Sanaetal11a} obtained spectral types of O6.5~V + O7~V for this SB2 system. We do not see double lines in the GOSSS data and obtain a combined spectral
type of O6.5~V, consistent with the \citet{Sanaetal11a} result.
In GOSSS-DR1.1 we eliminated the z suffix of the primary. See Figure~\ref{chart1} for a chart (IC~2944 field).
 
\paragraph{HD~114\,886~AaAb.}
\object[HD 114886]{}

According to the WDS (supplemented with additional information), this system is composed of three close components, Aa, Ab, and B with $B$ magnitudes 
of 7.5, 8.6, and 9.2, respectively. Aa and Ab have a separation of 0\farcs2 and cannot be spatially resolved with GOSSS while B is 1\farcs7 from the other two
and is indeed spatially resolved in our spectra. We obtain spectral types of O9~III + O9.5~III for AaAb, consistent with the O9~II-III classification of
\citet{Walb73a}. From OWN data, a period of 13.559~d is derived. The B component is an early B star.
 
\paragraph{HD~115\,071 = V961~Cen.}
\object[HD 115071]{}

\citet{Pennetal02} classified this SB2 system as O9.5~V + B0.2~III. With GOSSS we obtain O9.5~III + B0~Ib, which are similar in spectral subtypes but
correspond to higher luminosity classes. 
 
\paragraph{HD~115\,455.}
\object[HD 115455]{}

This SB2 system has a period of 15.08~d according to OWN data.
We do not see double lines in the GOSSS data. The ((f)) suffix was missing in GOSSS-DR1.0.
 
\paragraph{HD~117\,856.}
\object[HD 117856]{}

This SB2 system has a period of 27.6~d according to OWN data.
We do not see double lines in the GOSSS data. The classification was changed from O9.7~II to O9.7~II-III in GOSSS-DR1.1.
 
\paragraph{HD~123\,056.}
\object[HD 123056]{}

This system is at least an SB2 in the OWN data. We do not see double lines in the GOSSS data. 
 
\paragraph{HD~123\,590.}
\object[HD 123590]{}

This SB2 system has a period of 58.9~d according to OWN data.
We do not see double lines in the GOSSS data. The z suffix was added in GOSSS-DR1.1.
 
\paragraph{HD~124\,314~A.}
\object[HD 124314]{}

HD~124\,314~A and BaBb are separated by 2\farcs5 and both have O spectral types. 
A is a likely SB2 system according to preliminary OWN results but is not seen with double lines in the existing GOSSS spectrum.
The luminosity class was changed from III to IV in GOSSS-DR1.1.
The system is a colliding-wind binary \citep{DeBeRauc13}.
See Figure~\ref{chart1} for a chart (HD~124\,314 field).
 
\paragraph{HD~124\,979.}
\object[HD 124979]{}

This object is an SB2 system according to OWN data. We do not see double lines in the GOSSS data. The (n) suffix was added in GOSSS-DR1.1.
 
\paragraph{HD~125\,206.}
\object[HD 125206]{}

This object is an SB2 system according to OWN data. We do not see double lines in the GOSSS data.
 
\paragraph{$\delta$~Cir = HD~135\,240.}
\object[V* del Cir]{}

\citet{Pennetal01} classified this SB3 system as O7~III-V + O9.5~V + B0.5~V. We do not see double lines in the GOSSS data and we obtain a combined
classification of O8~V (changed from O7.5~V in GOSSS-DR1.1), which is consistent with the \citet{Pennetal01} result.
 
\paragraph{HD~149\,404 = V918~Sco.}
\object[HD 149404]{}

\citet{Rauwetal01b} classified this SB2 system\footnote{Note that the (f) suffix cannot be used for an O7.5 supergiant according to
Table~\ref{fphen}.} as O7.5~I(f) + ON9.7~I (see also \citet{Thaletal01}). 
We do not see double lines in our three GOSSS epochs, though both \HeII{4686} and H$\beta$ have variable
and anomalous profiles due to the binarity of the system. The combined spectral type is O8.5~Iab(f), consistent with the \citet{Rauwetal01b}
classification. Note that the two unidentified emission lines at 4486 \AA\ and 4504 \AA\ discussed by those authors (see also \citealt{Walb01}) were 
soon afterwards discovered to originate in S\,{\sc iv} by \citet{WernRauc01}. The new Hipparcos calibration gives a revised distance of 
$458^{+96}_{-67}$ pc \citep{Maizetal08a}.
 
\paragraph{HD~150\,135.}
\object[HD 150135]{}

This SB2 system has a period of 181~d according to OWN data.
We do not see double lines in the GOSSS data. 
See Figure~\ref{chart1} for a chart (HD~150\,136 field).
 
\paragraph{HD~150\,136.}
\object[HD 150136]{}

\citet{NiemGame05} classified this object as O3, making it the closest O2/3.5 system, and discovered it was an SB3. \citet{Mahyetal12} classified it 
as O3-3.5~V((f)) + O5.5-6~V((f)) + O6.5-7~V((f)). In the 
GOSSS data we see it only as SB2, with spectral types of O3.5-4~III(f*) and O6~IV, which is roughly consistent with the three individual spectral types of \citet{Mahyetal12} 
merged into two. Recently, \citet{Sanaetal13a} and \citet{SanBetal13} have independently spatially resolved one of the three components with VLTI with a 
separation of 7-9~mas and an 8 a orbit around the inner pair of O stars (which are expected to be separated by only $\sim$0.1 mas according to the 2.67454~d 
period of their spectroscopic orbit). Note that in addition to those three stars (all part of the A component), a fainter B component is listed by WDS with a
separation of 1\farcs6. See Figure~\ref{chart1} for a chart (HD~150\,136 field).
The visual multiplicity for this target was checked in ACS/HRC images (the VLTI-detected companion is obviously not seen there).
The system is a colliding-wind binary \citep{DeBeRauc13}.
 
\paragraph{HD~151\,003.}
\object[HD 151003]{}

This SB2 system has a period of 199~d according to OWN data.
We do not see double lines in the GOSSS data. The spectral classification was changed from O9~III to O8.5~III in GOSSS-DR1.1 due to the revision of
criteria around spectral type O9.
 
\paragraph{HD~150\,958~AB.}
\object[HD 150958]{}

This object is an SB2 system according to OWN data. We do not see double lines in the GOSSS data.
The WDS lists a B component with a separation of 0\farcs3 and a $\Delta m$ of 1.7 magnitudes that we cannot spatially resolve.
 
\paragraph{HD~151\,804 = V973~Sco.}
\object[HD 151804]{}

This object is an SB2 system according to OWN data. We do not see double lines in the GOSSS data.
The system is a colliding-wind binary \citep{DeBeRauc13}.
 
\paragraph{HD~152\,219 = V1292~Sco.}
\object[HD 152219]{}

\citet{Sanaetal08b} classified this system as O9~III + B1-2~V/III. We do not see double lines in the GOSSS data and we get a
combined spectrum of O9.5 III(n), which is consistent with the previous one. See Figure~\ref{chart1} for a chart (NGC~6231 field).
 
\paragraph{HD~152\,218 = V1294~Sco.}
\object[HD 152218]{}

\citet{Sanaetal08b} classified this system as O9~IV + O9.7~V. We also resolve in velocity the two components in the GOSSS data and we obtain the very similar
classification of O9~IV + B0:~V:. See Figure~\ref{chart1} for a chart (NGC~6231 field).
 
\paragraph{HD~152\,233.}
\object[HD 152233]{}

\citet{Sanaetal08b} classified this system as O5.5 + O7.5. We do not see double lines in the GOSSS data and we get a
combined spectrum of O6 II(f), whose spectral subtype is compatible with the previous measurement. The GOSSS luminosity class was adjusted from Ib to
II in GOSSS-DR1.1. See Figure~\ref{chart1} for a chart (NGC~6231 field).
 
\paragraph{HD~152\,246.}
\object[HD 152246]{}

This object is an SB2 system according to OWN data. We do not see double lines in the GOSSS data. The luminosity class was changed from V to IV in GOSSS-DR1.1.
 
\paragraph{HD~152\,248~AaAb = V1007~Sco~AaAb.}
\object[HD 152248]{}

\citet{Sanaetal08b} classified this system as O7~III(f) + O7.5~III(f). We also resolve in velocity the two components in the GOSSS data and we obtain a
classification of O7~Iabf + O7~Ib(f) i.e. similar spectral subtypes but somewhat brighter luminosity classes (note that \citealt{Sanaetal08b} showed convincingly 
that both components have \HeII{4686} in absorption and all the emission in that line arises in the colliding wind region, so their luminosity classes are likely 
correct). In GOSSS-DR1.0 the luminosity class of
the primary was Ib and its suffix (f). According to \citet{Masoetal09} there is a visual Ab companion at a distance of 0\farcs05 with a $\Delta m$ of 2.0 mag, 
which we were unable to spatially resolve in our data. See Figure~\ref{chart1} for a chart (NGC~6231 field).

 
\paragraph{CPD~$-$41~7733.}
\object[CPD-41 7733]{}

\citet{Sanaetal08b} classified this system as O8.5~V + B3. We do not see double lines in the GOSSS data and we get a combined
classification of O9 IV, which is consistent with the previous result. See Figure~\ref{chart1} for a chart (NGC~6231 field).
 
\paragraph{V1034~Sco = CPD~$-$41~7742.}
\object[V* V1034 Sco]{}

\citet{Sanaetal08b} classified this system as O9.5~V + B1.5~V. We do not see double lines in the GOSSS data and we get a combined
classification of O9.5 IV, which is not far from what one would expect from the previous result. In GOSSS-DR1.0 the luminosity class for this system
was III. See Figure~\ref{chart1} for a chart (NGC~6231 field).
 
\paragraph{HDE~326\,331.}
\object[HDE 326331]{}

This object is an SB2 system according to OWN data. We do not see double lines in the GOSSS data.
The ((f)) suffix was added in GOSSS-DR1.1. See Figure~\ref{chart1} for a chart (NGC~6231 field).
 
\paragraph{HD~152\,314.}
\object[HD 152314]{}

This system was found by \citet{Sanaetal08b} to be an SB2 with spectral types of O8.5 III and B1-3 V. We do not see double lines in the GOSSS data but our 
combined spectral type of O9 IV (in GOSSS-DR1.0 it was O9.5 IV) is in agreement with that result. See Figure~\ref{chart1} for a chart (NGC~6231 field).
  
\paragraph{HD~152\,590 = V1297~Sco.}
\object[HD 152590]{}

This eclipsing system has a period of 4.487~d \citep{OterClau04} and is an SB2 according to OWN data.
We do not see double lines in the GOSSS data. 
 
\paragraph{HD~153\,426.}
\object[HD 153426]{}

This object is an SB2 system with a period of 22.4~d according to OWN data.
We do not see double lines in the GOSSS data. The classification was changed from O9~II-III to O8.5~III in GOSSS-DR1.1.
 
\paragraph{HD~155\,889~AB.}
\object[HD 155889]{}

This object is an SB2 system (and possibly an SB3) according to OWN data. We do not see double lines in the GOSSS data.
The WDS lists a companion with a separation of 0\farcs2 and a $\Delta m$ of 0.6 mag than cannot be spatially resolved in our long-slit spectra.
 
\paragraph{HD~155\,913.}
\object[HD 155913]{}

This object is an SB2 system according to OWN data. We do not see double lines in the GOSSS data.
Aldoretta et al. (in preparation) have found a new dim companion at a small distance from the primary.
 
\paragraph{HD~156\,292.}
\object[HD 156292]{}

This object is an SB2 system with a period of 4.94~d according to OWN data.
We do not see double lines in the GOSSS data. 
 
\paragraph{ALS~18\,748 = HM~1$-$8 = C1715$-$387$-$8.}
\object[ALS 18748]{}

This star was not included in \citet{Maizetal04b} due to its dimness \citep{HavlMoff77}. The OWN collaboration has recently discovered that it is an SB2 with a
5.8783 d period. We do not see double lines in our data. 
The target has \CIII{4650} in emission but the \CIII{4650}/\NIII{4634} intensity ratio of 0.9 places this spectrum just below the fc boundary. It is interesting 
that it is associated with the Ofc star ALS~18\,747 in the Havlen-Moffat 1 cluster shown in Figure~\ref{chart1}, which may be significant in the context of the 
\citet{Walbetal10a} finding that this phenomenon occurs in some clusters but not others.
 
\paragraph{HDE~319\,703~A.}
\object[HDE 319703]{}

A chart of the complex HDE~319\,703 system in NGC 6334 can be seen in Figure~\ref{chart1} (HDE~319\,703 field, see also Figure~1 in \citealt{Walb82a}). 
The A component is clearly spatially resolved from the 
others (see \ref{sec:Nor} subsubsection for two additional O stars) and we do not see double lines in its spectrum. This object is an SB2 system with a 16.4~d
period according to OWN data. The ((f))z suffix was added in GOSSS-DR1.1, where the star was also moved from O7.5 to O7.
 
\paragraph{HDE~319\,702.}
\object[HDE 319702]{}

This object in NGC 6334 was found to be an eclipsing binary with a period of 2.0~d by \citet{BarDetal13}. OWN data reveal that it is actually an SB3
system. We do not see double lines in the GOSSS data.
 
\paragraph{Pismis~24$-$1~AB = HDE~319\,718~AaAb.}
\object[Cl Pismis 24 1]{}

\citet{Maizetal07} were able to spatially resolve this close pair (separation of 0\farcs364 and $\Delta m$ of 0.1 mag) with IMACS at the Baade telescope
at LCO to obtain spatially resolved spectral types of O3.5~If* and O4~III(f) for A and B, respectively\footnote{Note that the f+ qualifier became obsolete in
paper I.}. With the GOSSS data we were unable to spatially resolve the pair, so we can only produce a combined spectral type of O3.5~If*, which is
consistent with the expected merged spectrum. Note however, that the merged spectrum shows a P-Cygni-like profile in \HeII{4686}.
\citet{Maizetal07} also cited a private communication by Phil Massey about the photometric variability 
of the system that indicated the eclipsing binary nature of one of its components, making Pismis 24-1 a system with three very massive O stars\footnote{Note that
the alternative classification in Table~\ref{spectralclasS} refers to the two visual components, not to a classification based on a velocity resolution of the
spectroscopic binary.}. The eclipsing binary nature and the period of 2.36 days has been recently confirmed by \citet{BarDetal13}. OWN data 
also reveal the period in radial velocity data. See Figure~\ref{chart1} for a chart (Pismis~24 field).
The visual multiplicity for this target was measured in ACS/HRC images.
 
\paragraph{HD~158\,186 = V1081~Sco.}
\object[HD 158186]{}

This object is an SB3 system according to OWN data.
We do not see double lines in the GOSSS data. The (n) suffix was added in GOSSS-DR1.1.
Aldoretta et al. (in preparation) have recently detected a companion with a $\Delta m$ of 2.3 magnitudes and a separation larger than 34 mas.
 
\paragraph{HD~159\,176 = V1036~Sco.}
\object[HD 159176]{}

\cite{Lindetal07} classified this system as O7~V((f)) + O7~V((f)), see also \citet{Contetal75}. 
We obtain exactly the same classification with GOSSS (after adding the ((f))
suffixes missing in GOSSS-DR1.0). According to the WDS, the HD~159\,176 system is rather complex. The B and C components are rather faint and well 
separated from A. A itself is composed of Aa, Ab, and Ac, all less than 1\arcsec\ from each other and spatially unresolved in the GOSSS data. Ac is considerably
fainter than Aa but Ab has no magnitude listed in the WDS. We have unpublished Lucky Imaging obtained with AstraLux Sur at the NTT where Ab does not
appear. That indicates that it either does not exist or is too faint. In either case, we do not need to add any component to the star name.
 
\paragraph{HD~161\,853.}
\object[HD 161853]{}

This object is an SB2 (and possibly an SB3) system according to OWN data. In one of the GOSSS observations we see a weak secondary component in the He~{\sc i} 
lines that indicate a B-type secondary. This object has a history of confusion with being a post AGB star. However, \cite{Szczetal07} 
disqualify it from that category and \citet{Urquetal09} identify it as the main ionizing source of RCW 134.
 
\paragraph{HD~163\,758.}
\object[HD 163758]{}

This object is an SB2 system according to OWN data. We do not see double lines in the GOSSS data. 
We observe both enhanced C and N absorption lines. That prompted the introduction of the p suffix in GOSSS-DR1.1. \citet{Leep78} had already suggested that this
was a carbon-rich star.
 
\paragraph{Herschel~36 = HD~164\,740.}
\object[NAME Herschel 36]{}

\citet{Ariaetal10} discovered that Herschel~36 is an SB3 system with spectral types O7.5~V + O9~V + B0.5~V. We see hints of the secondary in two of
the five GOSSS epochs but not full-fledged double lines. The combined spectral type is O7:~V + sec. A number of companions can be detected in
the IR in this region with heavy and patchy extinction \citep{Ariaetal06}, including one with a separation of just 0\farcs25 \citep{Gotoetal06}.
The visual multiplicity for this target was measured in WFPC2 images.
 
\paragraph{9~Sgr = HD~164\,794.}
\object[* 9 Sgr]{}

\citet{Rauwetal12} classified this system as O3.5~V((f*))~+~O5-5.5~V((f)). We do not see double lines in our spectrum, which is unsurprising
given the long period (8.6 a) of the orbit. Our combined spectral subtype (O4) and luminosity class (V) are consistent with those expected from the
\citet{Rauwetal12} results. The suffixes are different for two reasons: there is no * at O3.5/O4~V
and there is a z suffix due to the depth of the \HeII{4686}
absorption with respect to that of the \HeII{4542} absorption (this could be a case where the z phenomenon appears due to the existence of a binary).
The target has \CIII{4650} in emission but is not strong enough to warrant a c suffix.
The system is a colliding-wind binary \citep{DeBeRauc13}.
Note that both 9~Sgr and 9~Sge are O stars: one should be careful not to confuse them.
 
\paragraph{HD~164\,816.}
\object[HD 164816]{}

This object is an SB2 system according to OWN data.
Double lines are clearly seen in two of our five GOSSS epochs. We derive spectral types of O9.5~V + B0~V for the two components. 
 
\paragraph{HD~165\,052.}
\object[HD 165052]{}

\citet{Ferretal13} used multiple-epoch high-resolution spectroscopy to determine the apsidal motion rate of this SB2 system, calculate the masses
of the two components, and assign to them spectral types of O7 Vz and O7.5Vz, respectively (see also the previous work by \citealt{Ariaetal02}). 
We obtained GOSSS spectra at six different epochs and
in one of them we barely resolve in velocity the two SB2 components when they had a separation of $\sim$200 km/s. That is close to our velocity
resolution limit, so the GOSSS spectral types are rather uncertain. 
 
\paragraph{HD~165\,246.}
\object[HD 165246]{}

\citet{Oter07} discovered that this is an eclipsing binary and \citet{Mayeetal13} derived spectral types for this SB2 of O8~V + B7~V. The GOSSS classification 
is O8~V(n). Other than the difference in the rotation index, the GOSSS spectral type is what one would expect, since detecting a companion with such a large $\Delta m$ 
as that between O8 and B7 dwarfs should not be possible with our data.
 
\paragraph{HD~165\,921 = V3903~Sgr.}
\object[HD 165921]{}

\citet{NiemMorr88} obtained spectral types of O7~V + O9~V for this SB2 system. We detect double lines in one of our two GOSSS epochs and derive
spectral types of O7~V(n)z + B0:~V:, which agree (given the uncertainty in the secondary) reasonably well with the previous classification.
The z suffix of the primary was added in GOSSS-DR1.1.
 
\subsubsection{Normal sample}
\label{sec:Nor}

\begin{figure*}
\centerline{\includegraphics*[width=\linewidth]{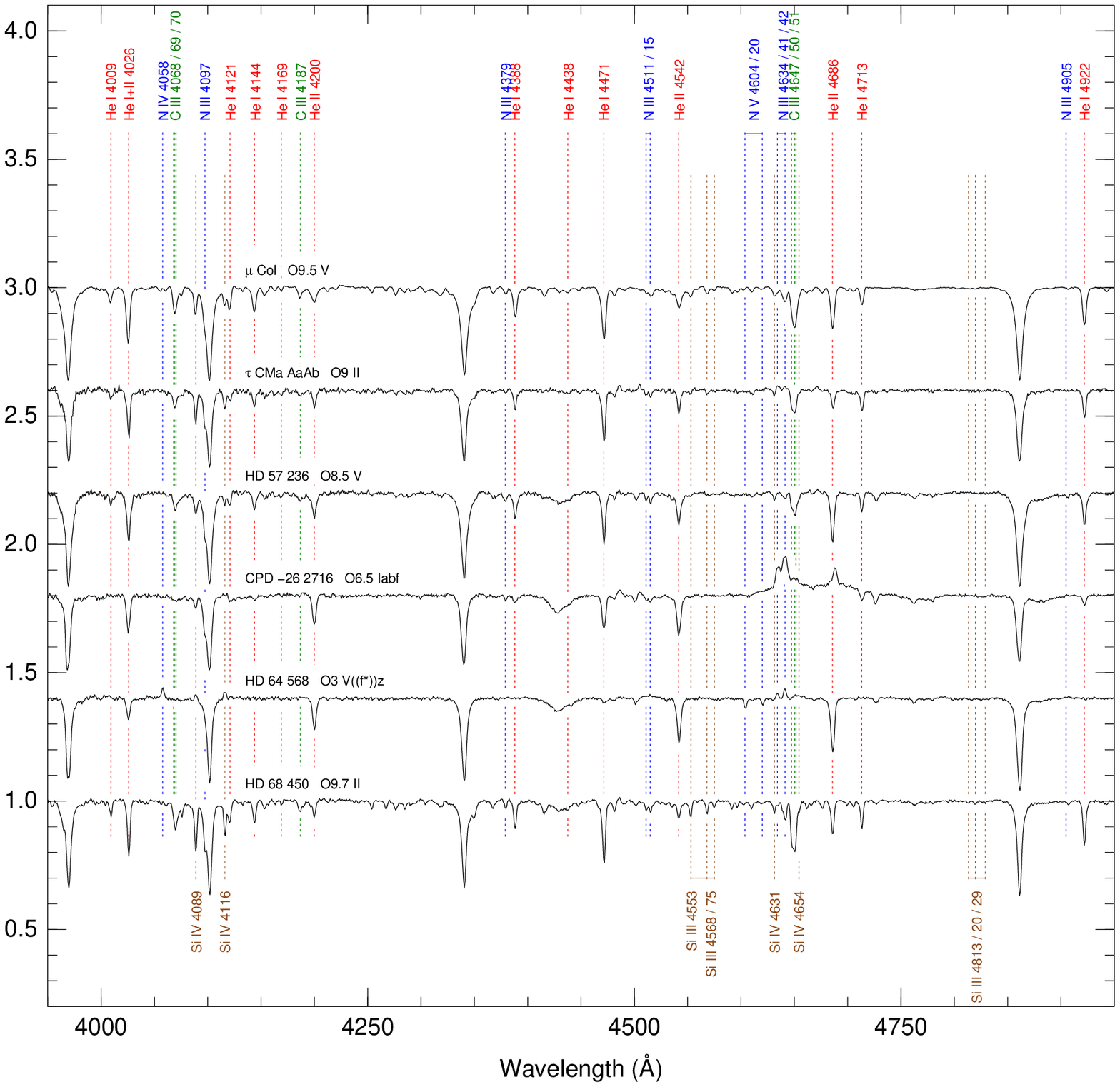}}
\caption{Spectrograms for the stars in the normal sample. The targets are sorted by right ascension.
[See the electronic version of the journal for a color version of this figure.]}
\label{fig:Nor}
\end{figure*}	

\addtocounter{figure}{-1}

\begin{figure*}
\centerline{\includegraphics*[width=\linewidth]{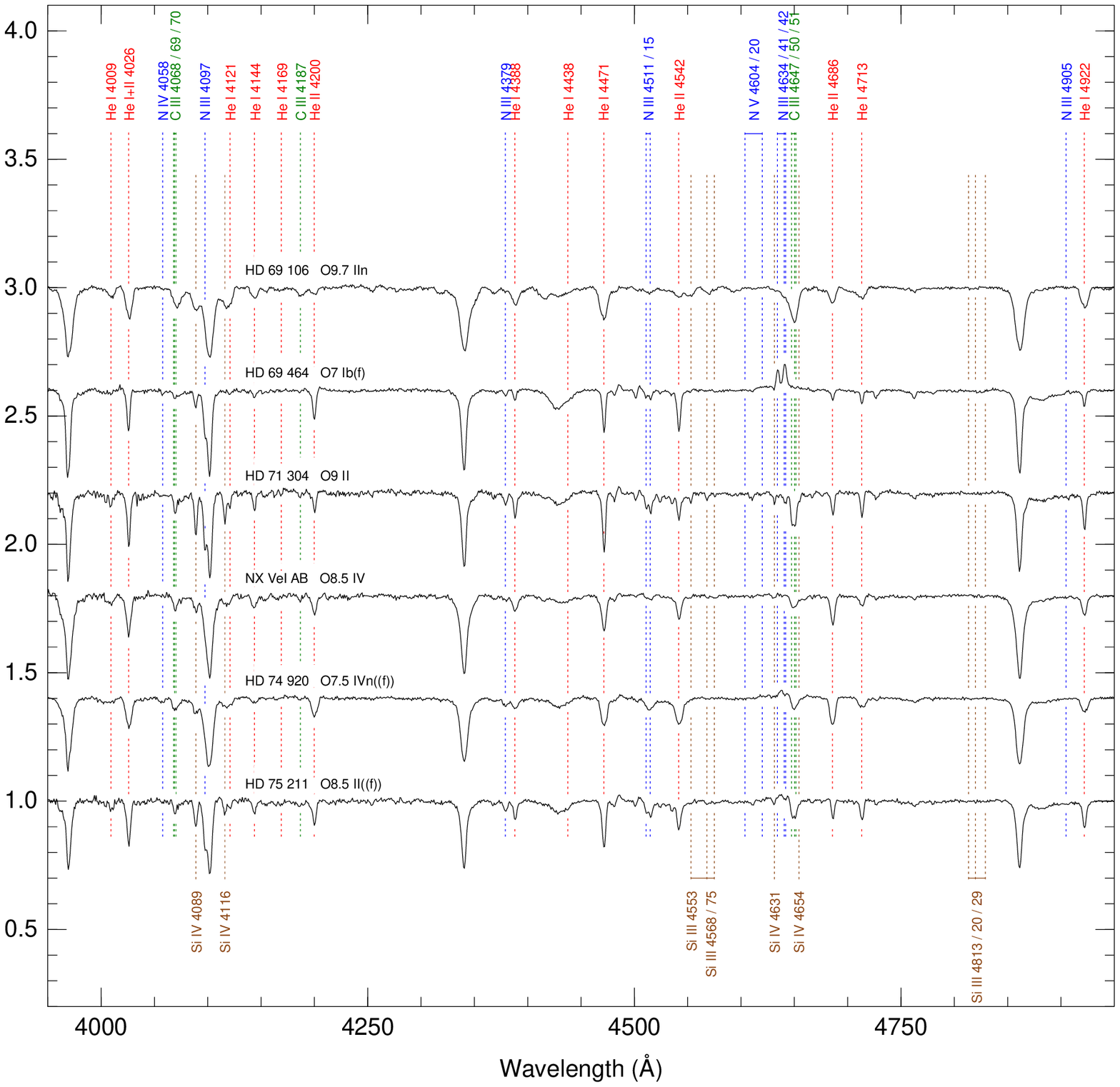}}
\caption{(continued).}
\end{figure*}	

\addtocounter{figure}{-1}

\begin{figure*}
\centerline{\includegraphics*[width=\linewidth]{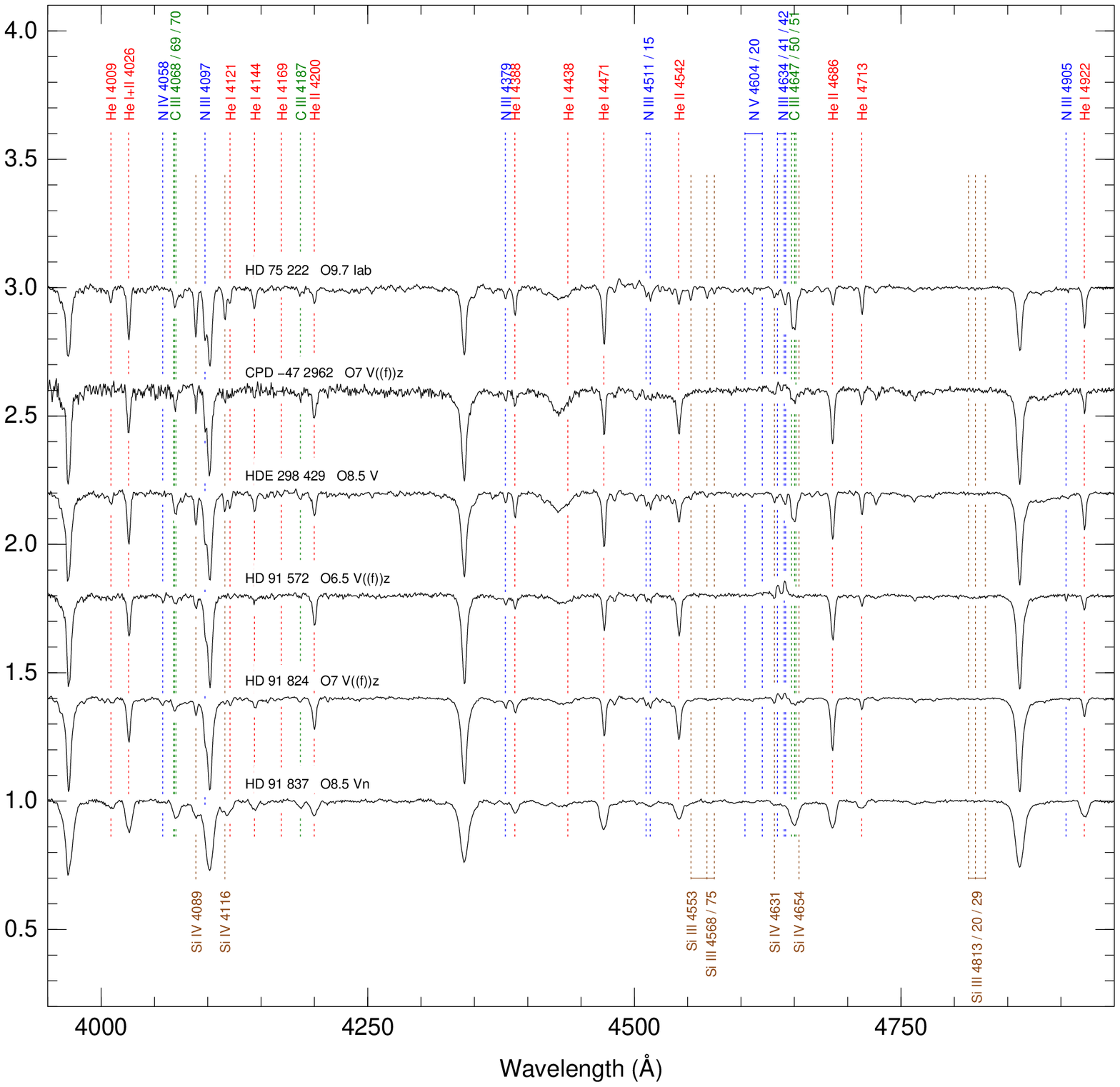}}
\caption{(continued).}
\end{figure*}	

\addtocounter{figure}{-1}

\begin{figure*}
\centerline{\includegraphics*[width=\linewidth]{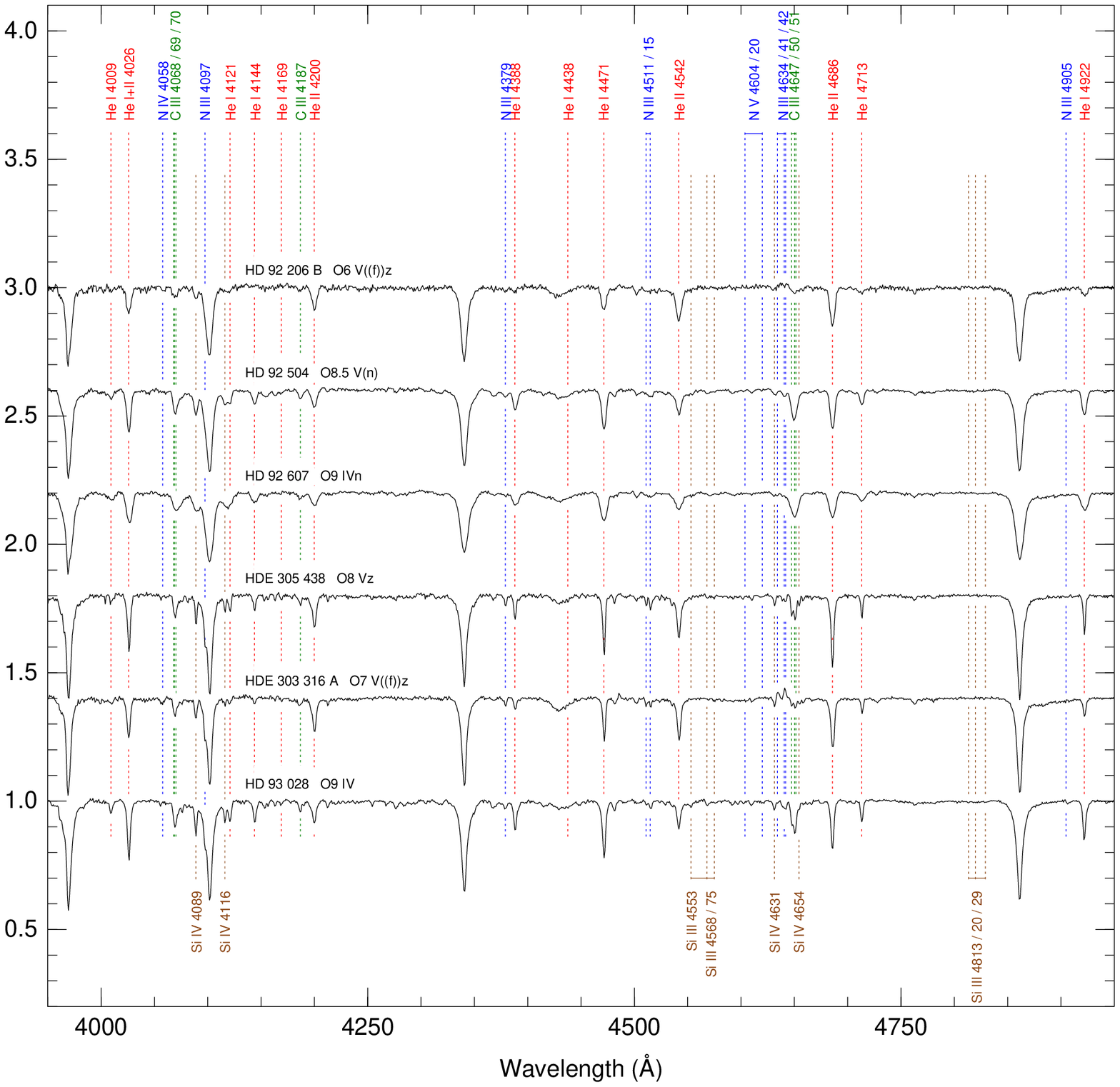}}
\caption{(continued).}
\end{figure*}	

\addtocounter{figure}{-1}

\begin{figure*}
\centerline{\includegraphics*[width=\linewidth]{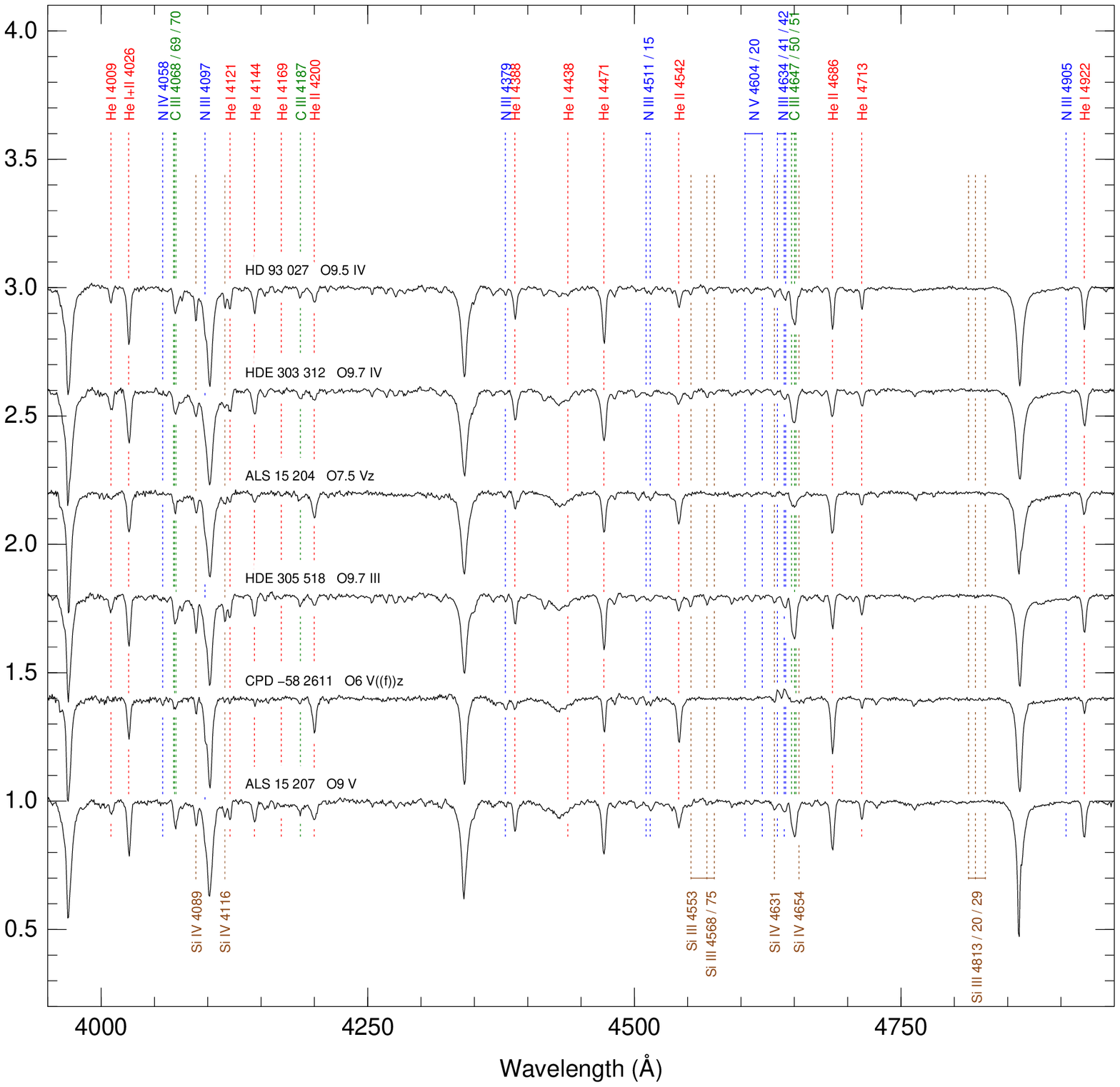}}
\caption{(continued).}
\end{figure*}	

\addtocounter{figure}{-1}

\begin{figure*}
\centerline{\includegraphics*[width=\linewidth]{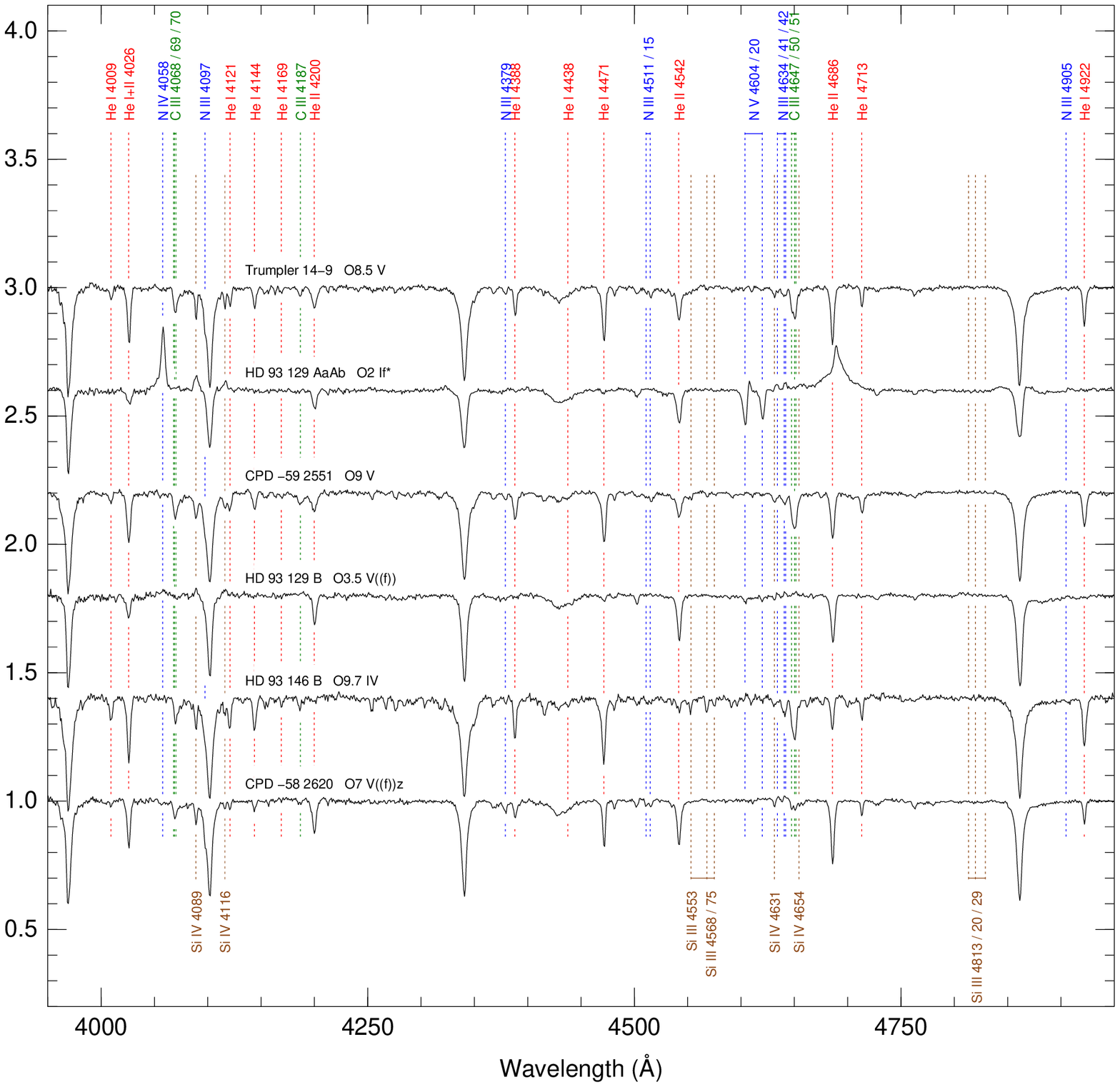}}
\caption{(continued).}
\end{figure*}	

\addtocounter{figure}{-1}

\begin{figure*}
\centerline{\includegraphics*[width=\linewidth]{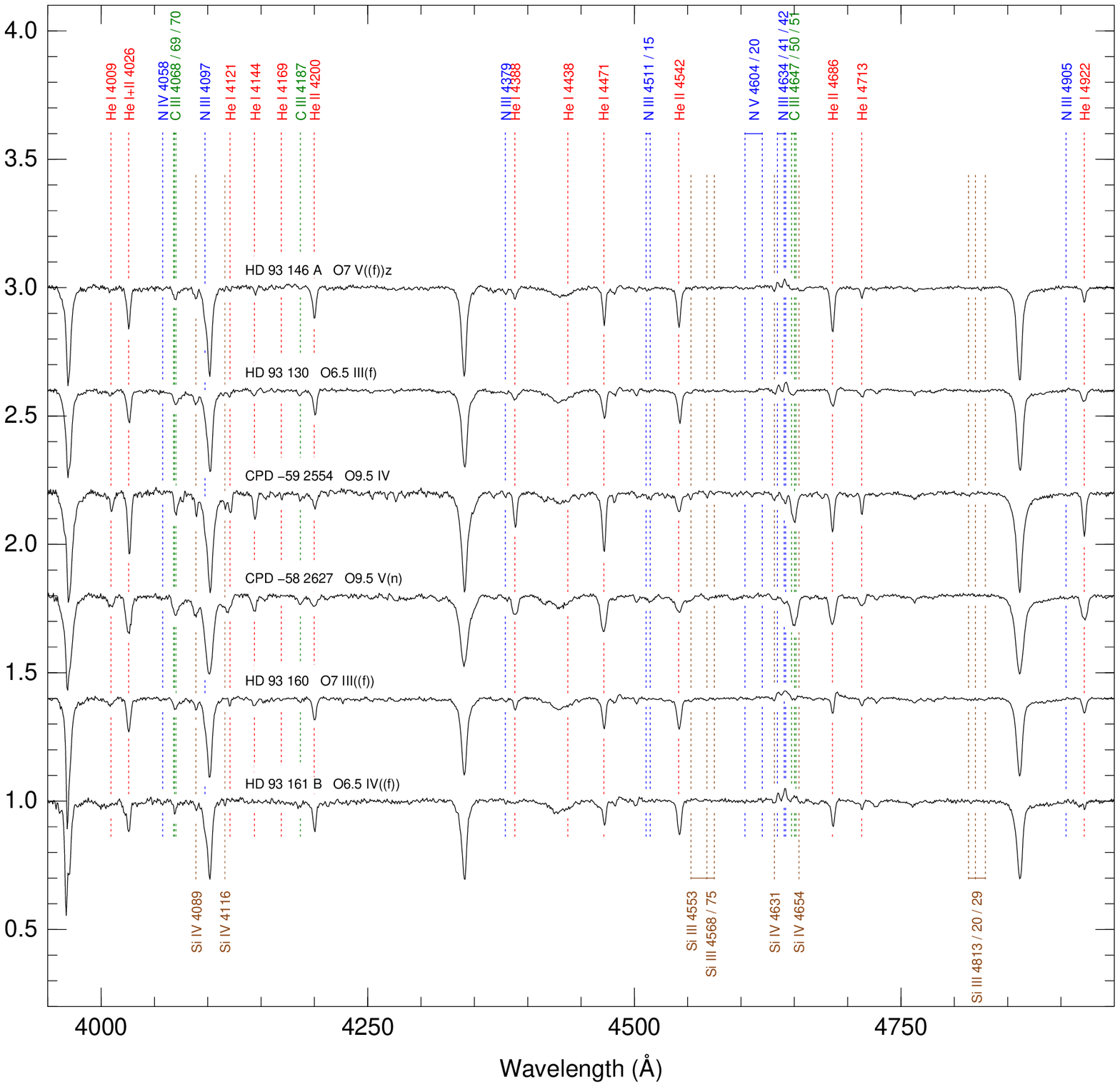}}
\caption{(continued).}
\end{figure*}	

\addtocounter{figure}{-1}

\begin{figure*}
\centerline{\includegraphics*[width=\linewidth]{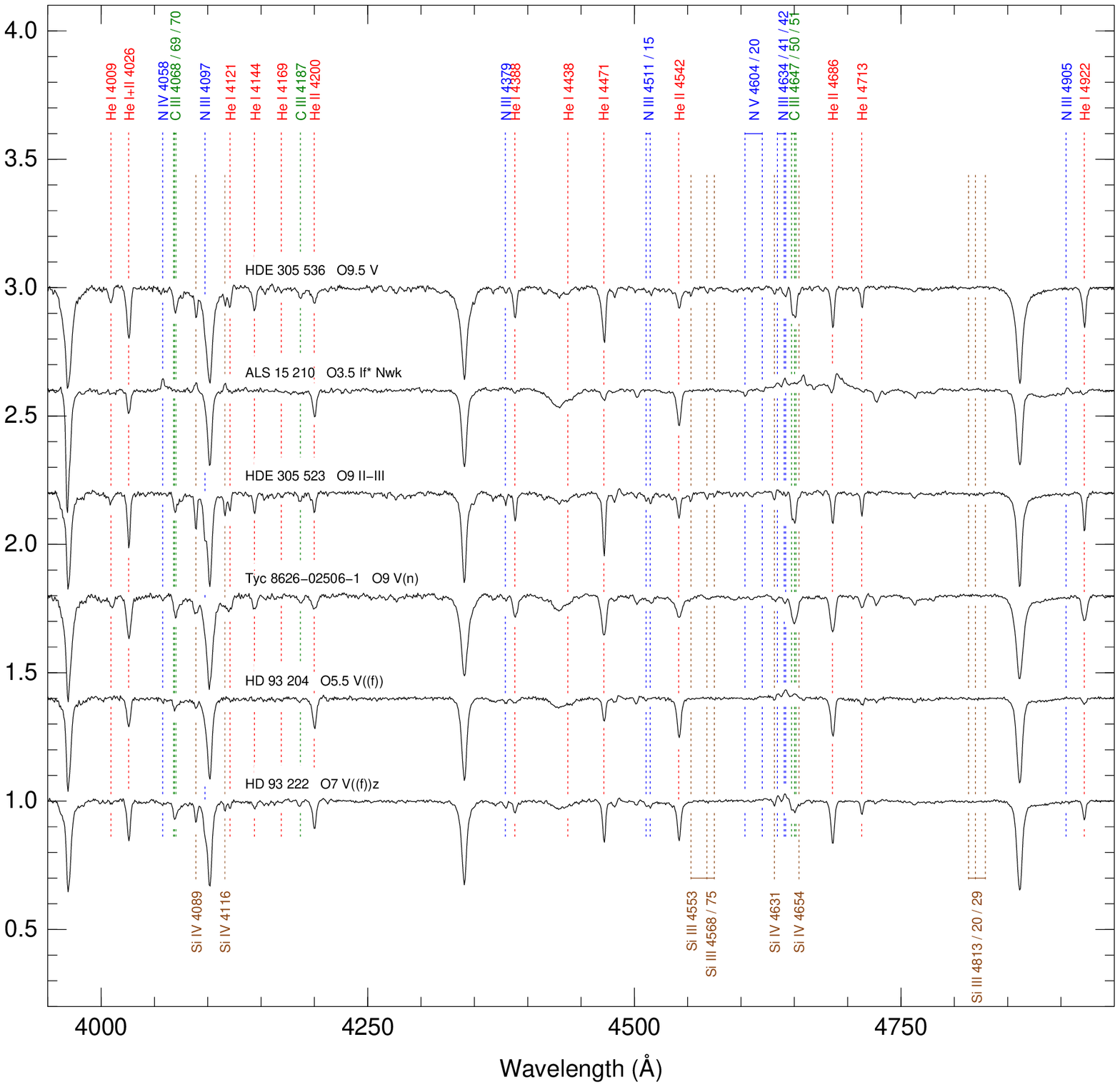}}
\caption{(continued).}
\end{figure*}	

\addtocounter{figure}{-1}

\begin{figure*}
\centerline{\includegraphics*[width=\linewidth]{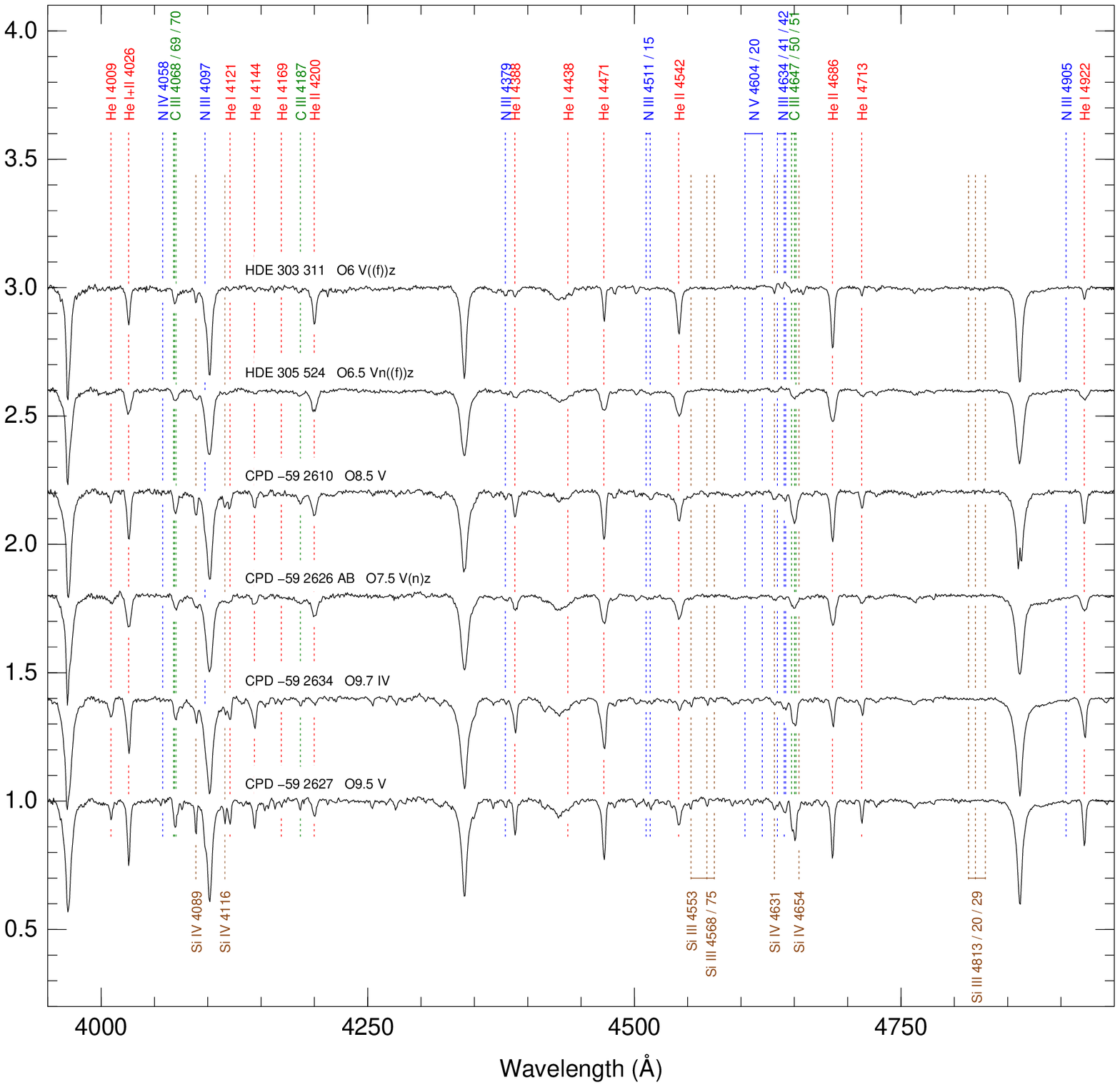}}
\caption{(continued).}
\end{figure*}	

\addtocounter{figure}{-1}

\begin{figure*}
\centerline{\includegraphics*[width=\linewidth]{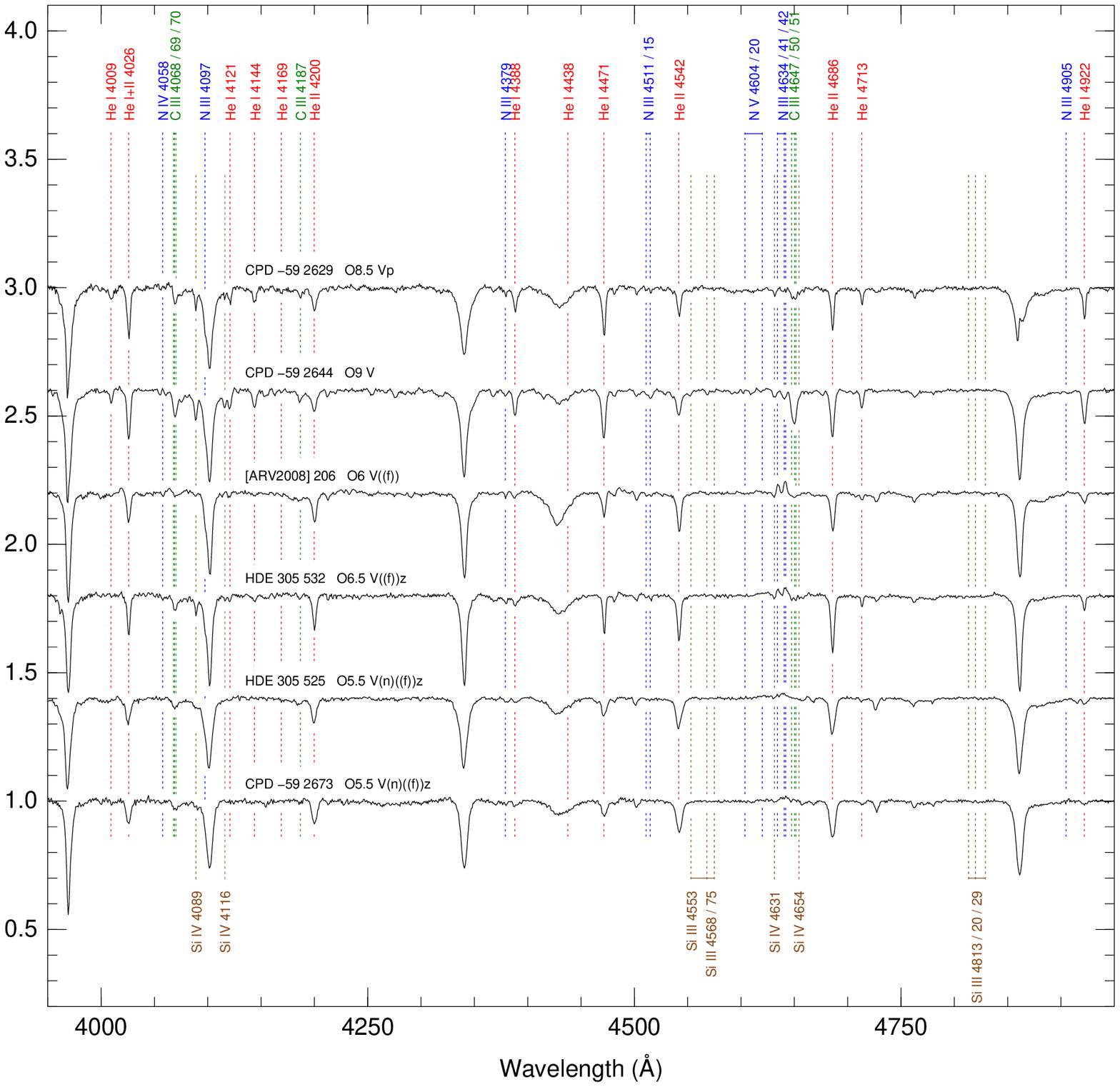}}
\caption{(continued).}
\end{figure*}	

In this subsubsection we briefly describe the O stars in our sample that do not belong to any of the categories of the previous subsubsections.
Spectrograms are shown in Figure~\ref{fig:Nor}.
 
\paragraph{$\mu$~Col = HD~38\,666.}
\object[* mu. Col]{}

$\mu$~Col is a runaway star, likely ejected from the Trapezium 2.5 Ma ago \citep{Hoogetal01}.
The new Hipparcos calibration gives a revised distance of $412^{+38}_{-32}$ pc \citep{Maizetal08a}.
	
\paragraph{$\tau$~CMa~AaAb = HD~57\,061~AaAb.}
\object[V* tau CMa]{}

This is a complex system, with three visual components within 1\arcsec\ and a number more outside of it ($\tau$~CMa is by far the brightest system in 
the open cluster NGC 2362 and it is unclear where the multiple system ends and the cluster begins). The GOSSS spectrum corresponds to the combined light 
of Aa+Ab+E. E is too dim to modify the spectral type\footnote{Hence, it is not included in the GOSSS name. Note that Aldoretta et al. (in preparation) have recently 
discovered that the PA of E with respect to Aa is 268\arcdeg, an offset of 180\arcdeg\ with respect to the previous value in the WDS, an effect that 
we have confirmed in unpublished AstraLux data.} but the $\Delta m$ between Aa and Ab is only 0.8 mag and the separation is 0\farcs122; hence, 
the O9~II classification is a composite. The system is further complicated by being [a] a single-lined spectroscopic binary with a period of 
154.9~days \citep{Sticetal98} and [b] an eclipsing binary with a 1.282~day period \citep{vanLvanG97}. Therefore, the system includes at least five 
bodies within 1\arcsec.
 
\paragraph{HD~57\,236.}
\object[HD 57236]{}

This star was classified as O8~V((f)) by \citet{Walb82a}. Based on the new criteria around spectral subtype O9 we reclassify it as O8.5~V.

\paragraph{CPD~$-$26~2716 = CD~$-$26~5136.}
\object[CPD-26 2716]{}

\HeII{4686} is clearly in emission in this star, prompting us to change the luminosity class from Ib \citep{Walb82a} to Iab. This object is variable in OWN
spectra on time scales of 1 day.
 
\paragraph{HD~64\,568.}
\object[HD 64568]{}

This was one of the ten O3 stars in \citet{Walb82b}. \citet{SoliNiem86} indicated that is a possible SB1.
The ((f*))z suffix was omitted by mistake in GOSSS-DR1.0.
The visual multiplicity for this target was measured in ACS/HRC images.
 
\paragraph{HD~68\,450.}
\object[HD 68450]{}

The WDS lists two weak companions far away from the primary.
 
\paragraph{HD~69\,106.}
\object[HD 69106]{}

This bright star was not included in \citet{Maizetal04b} due to the lack of previous classifications as O type we were aware of. The likely reason for that is
the broadening of \HeII{4542} in this fast O9.7 rotator. 
 
\paragraph{HD~69\,464.}
\object[HD 69464]{}

Aldoretta et al. (in preparation) have recently discovered a weak secondary 0\farcs92 away from the primary. The (f) suffix was omitted by mistake in GOSSS-DR1.0.
 
\paragraph{HD~71\,304.}
\object[HD 71304]{}

This star was classified as O9.5~Ib by \citet{Walb73a} and here we reclassify it as O9~II according to the revised criteria around O9.
 
\paragraph{NX~Vel~AB = HD~73\,882~AB.}
\object[V* NX Vel]{}

We cannot spatially resolve the B component indicated by the WDS with a separation of 0\farcs7 and a $\Delta m$ of 1.3 magnitudes. 
\citet{Oter03} discovered that this system is an eclipsing binary with a period of 2.9199 days. OWN data show an SB1 orbit with a preliminary period of
20.6 days, raising the total number of bodies in the system to four.

\paragraph{HD~74\,920.}
\object[HD 74920]{}

This O star was inadvertently excluded from \citet{Maizetal04b} even though it was classified as O8: by \citet{ThacAndr74}. 
The ((f)) suffix was omitted by mistake in GOSSS-DR1.0.
 
\paragraph{HD~75\,211.}
\object[HD 75211]{}

The ((f)) suffix was omitted by mistake in GOSSS-DR1.0. OWN data indicate that it is an SB1 with a 20.4~d period.
 
\paragraph{HD~75\,222.}
\object[HD 75222]{}

This star was classified as O9.7~Iab by \citet{Walb73a}, which is maintained here.
 
\paragraph{CPD~$-$47~2962 = ALS~1215 = CD~$-$47~4550.}
\object[CPD-47 2962]{}

This star had no previous classifications as being of O type and was not included in \citet{Maizetal04b}. We placed it on the same slit as
CPD~$-$47~2963 and discovered it was an O star.
 
\paragraph{HDE~298\,429.}
\object[HDE 298429]{}

This star was classified as O8~III((f)) by \citet{Walb82a} and here we reclassify it as O8.5~V.
 
\addtocounter{figure}{-1}

\begin{figure*}
\centerline{\includegraphics*[width=\linewidth]{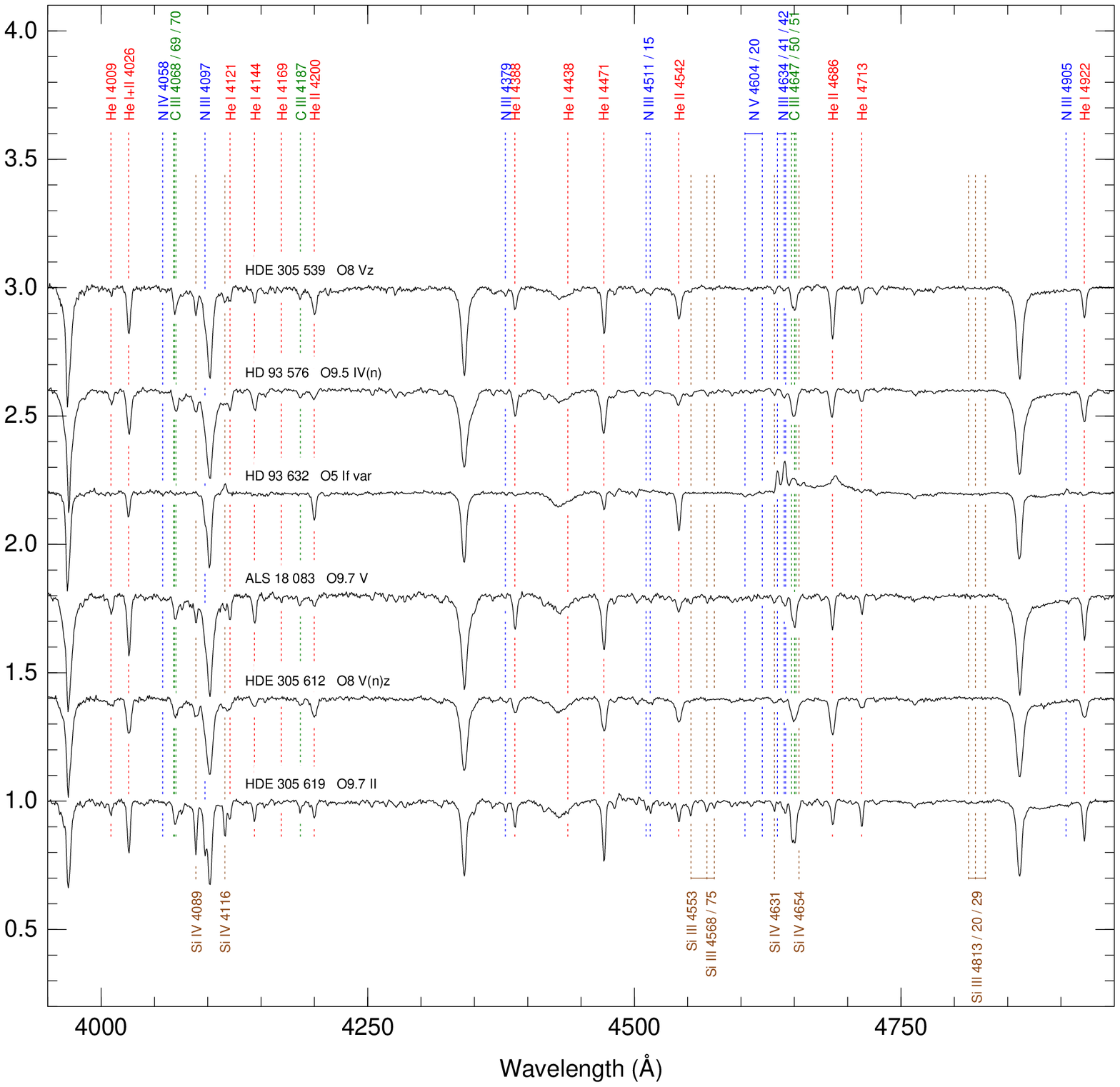}}
\caption{(continued).}
\end{figure*}	

\addtocounter{figure}{-1}

\begin{figure*}
\centerline{\includegraphics*[width=\linewidth]{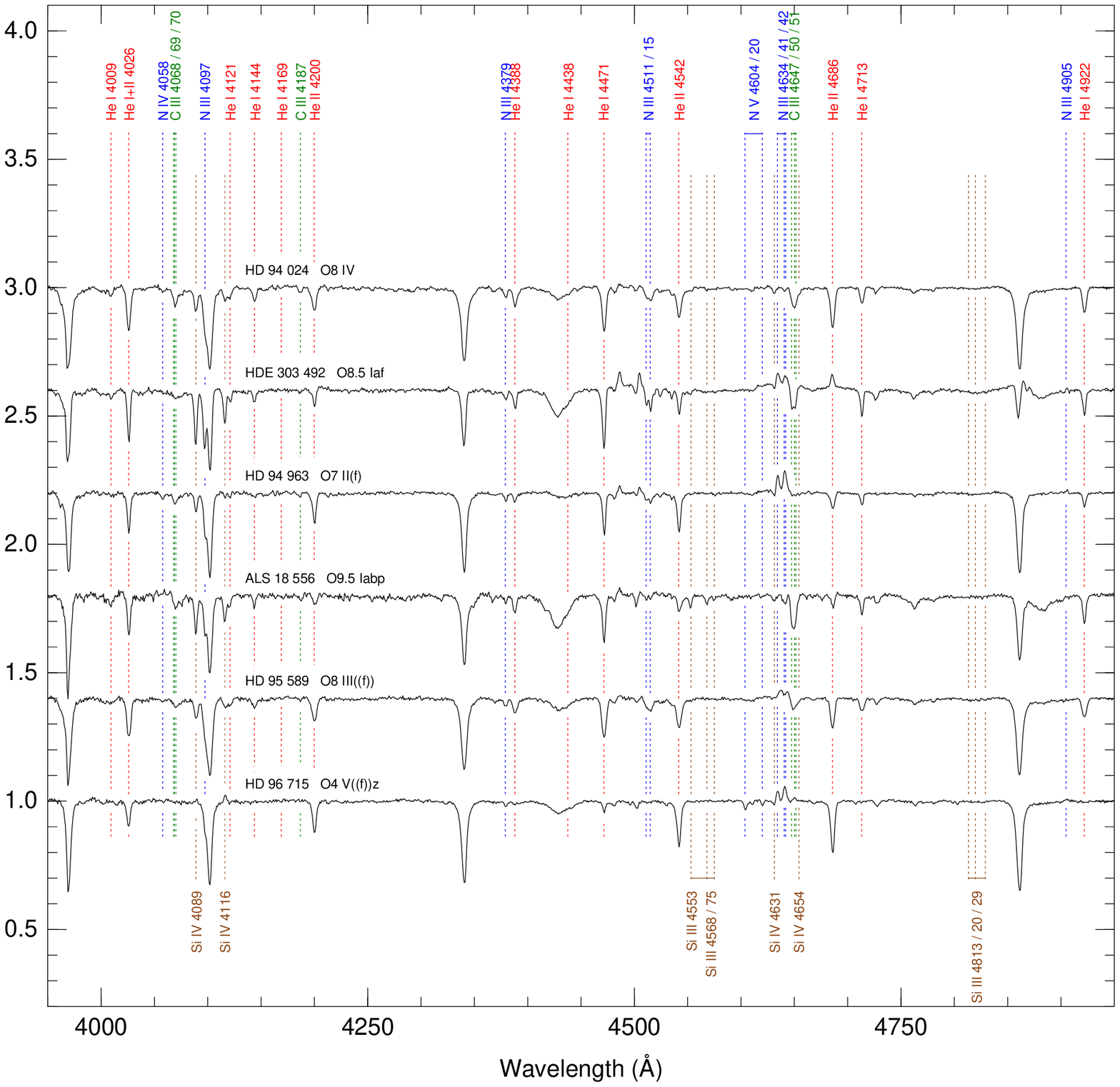}}
\caption{(continued).}
\end{figure*}	

\addtocounter{figure}{-1}

\begin{figure*}
\centerline{\includegraphics*[width=\linewidth]{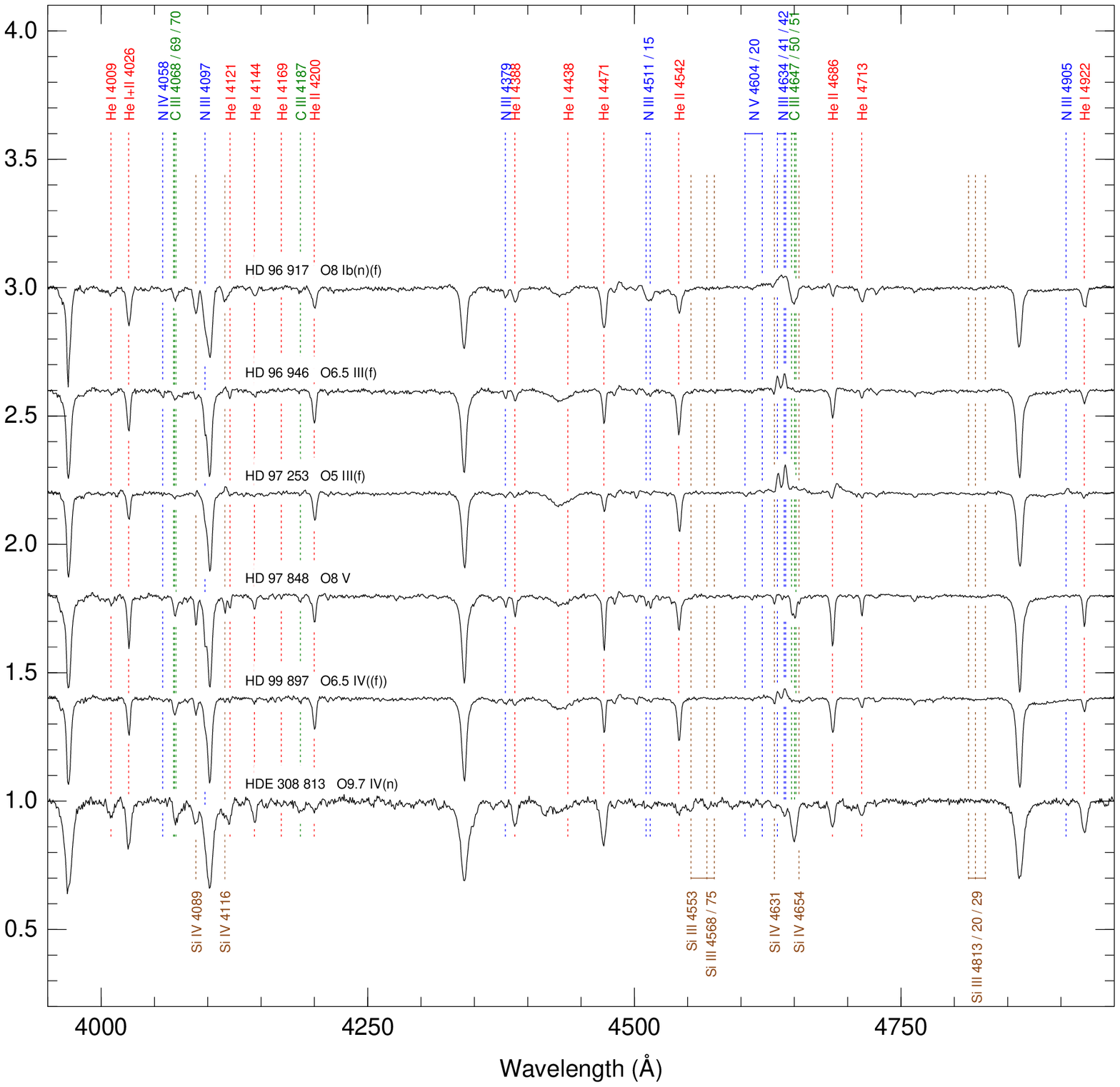}}
\caption{(continued).}
\end{figure*}	

\addtocounter{figure}{-1}

\begin{figure*}
\centerline{\includegraphics*[width=\linewidth]{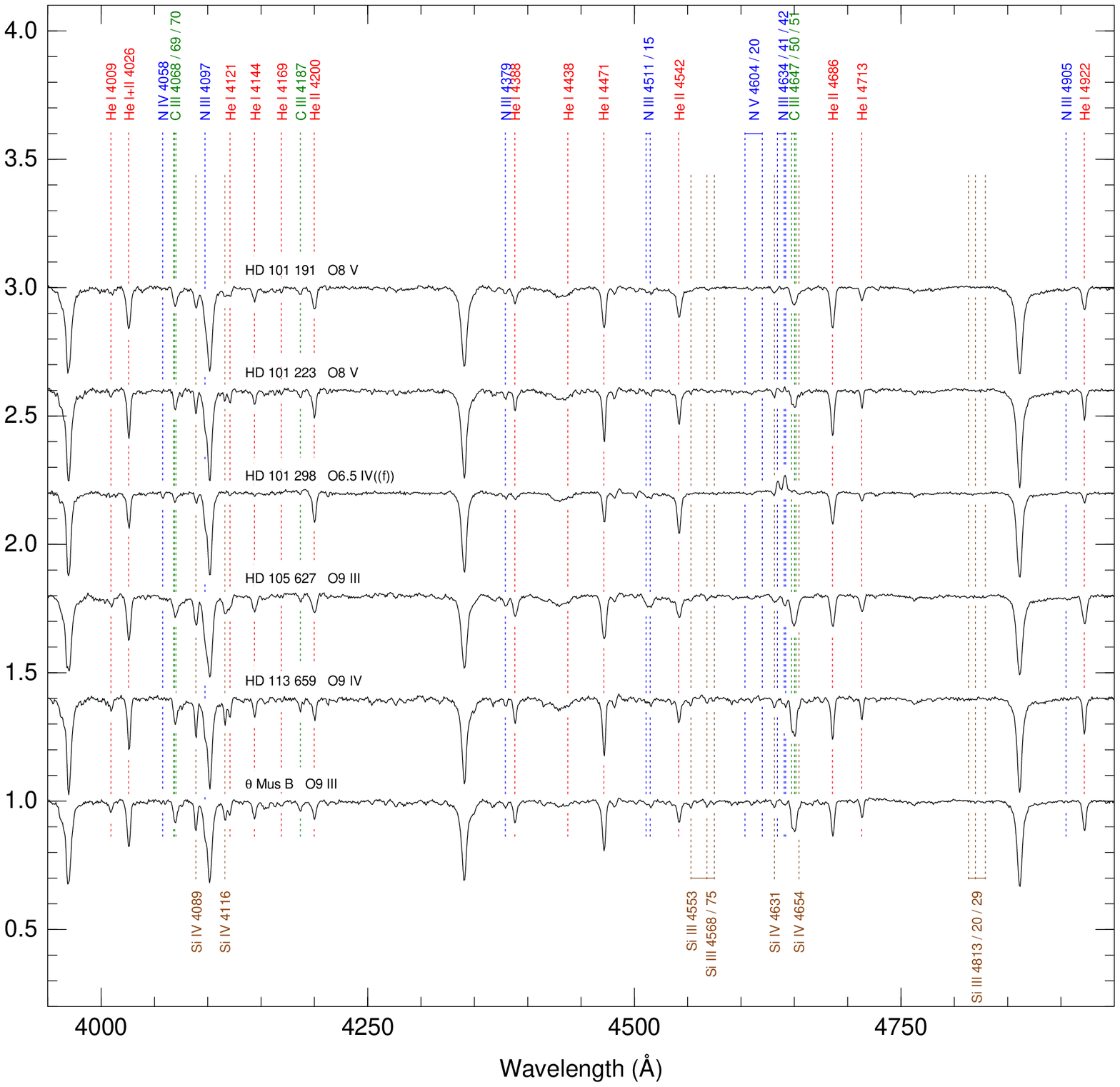}}
\caption{(continued).}
\end{figure*}	

\addtocounter{figure}{-1}

\begin{figure*}
\centerline{\includegraphics*[width=\linewidth]{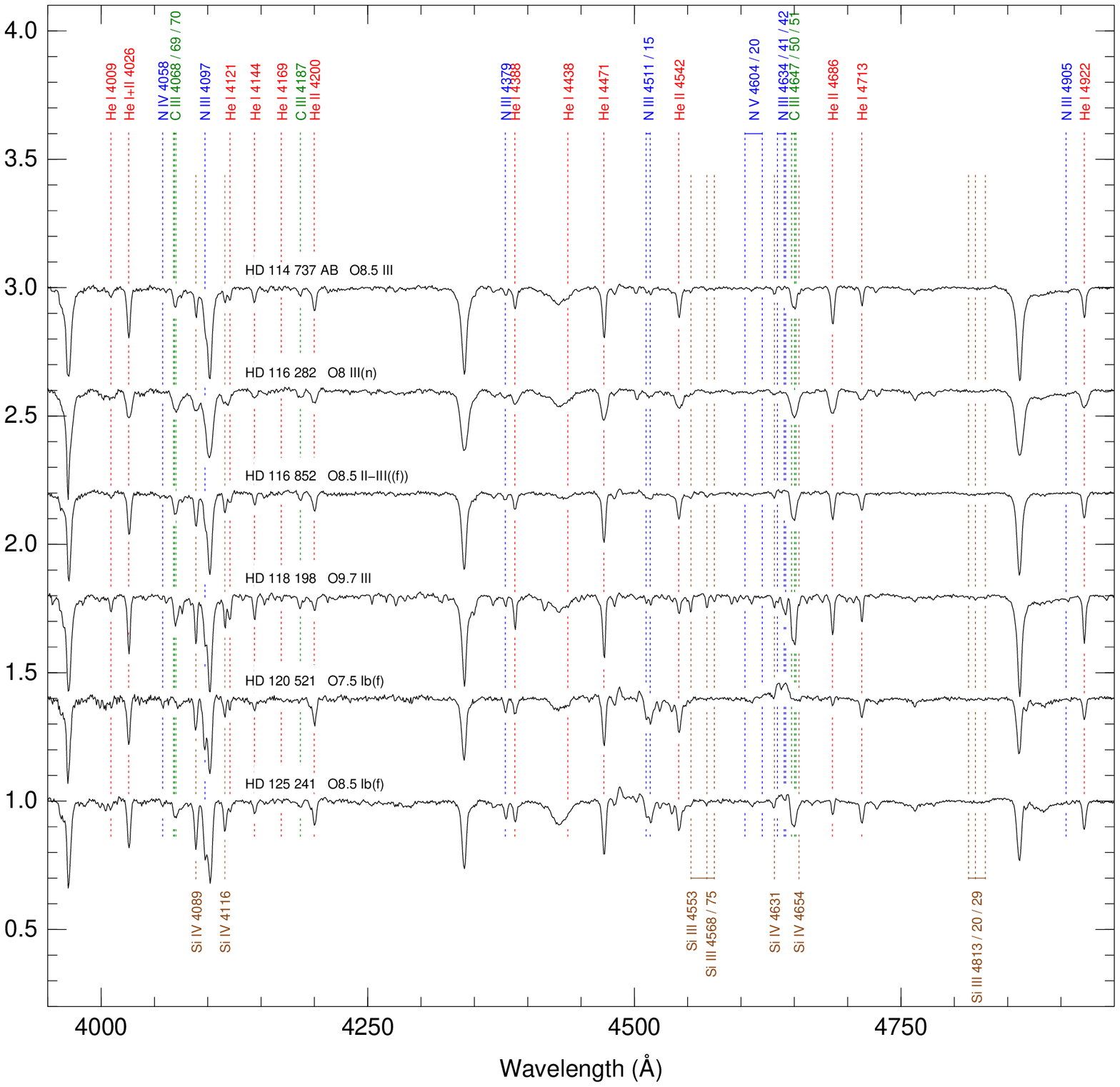}}
\caption{(continued).}
\end{figure*}	

\addtocounter{figure}{-1}

\begin{figure*}
\centerline{\includegraphics*[width=\linewidth]{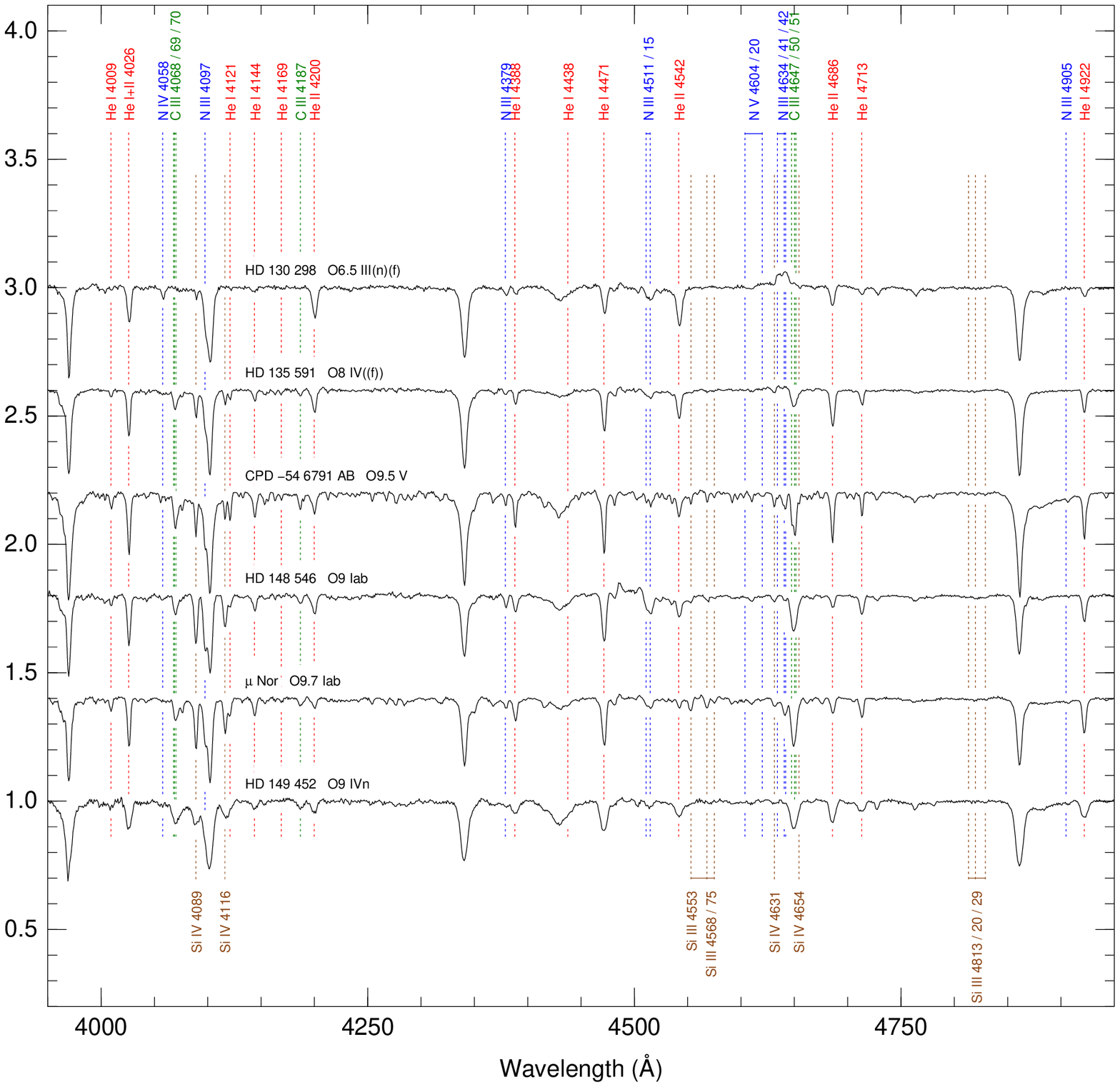}}
\caption{(continued).}
\end{figure*}	

\addtocounter{figure}{-1}

\begin{figure*}
\centerline{\includegraphics*[width=\linewidth]{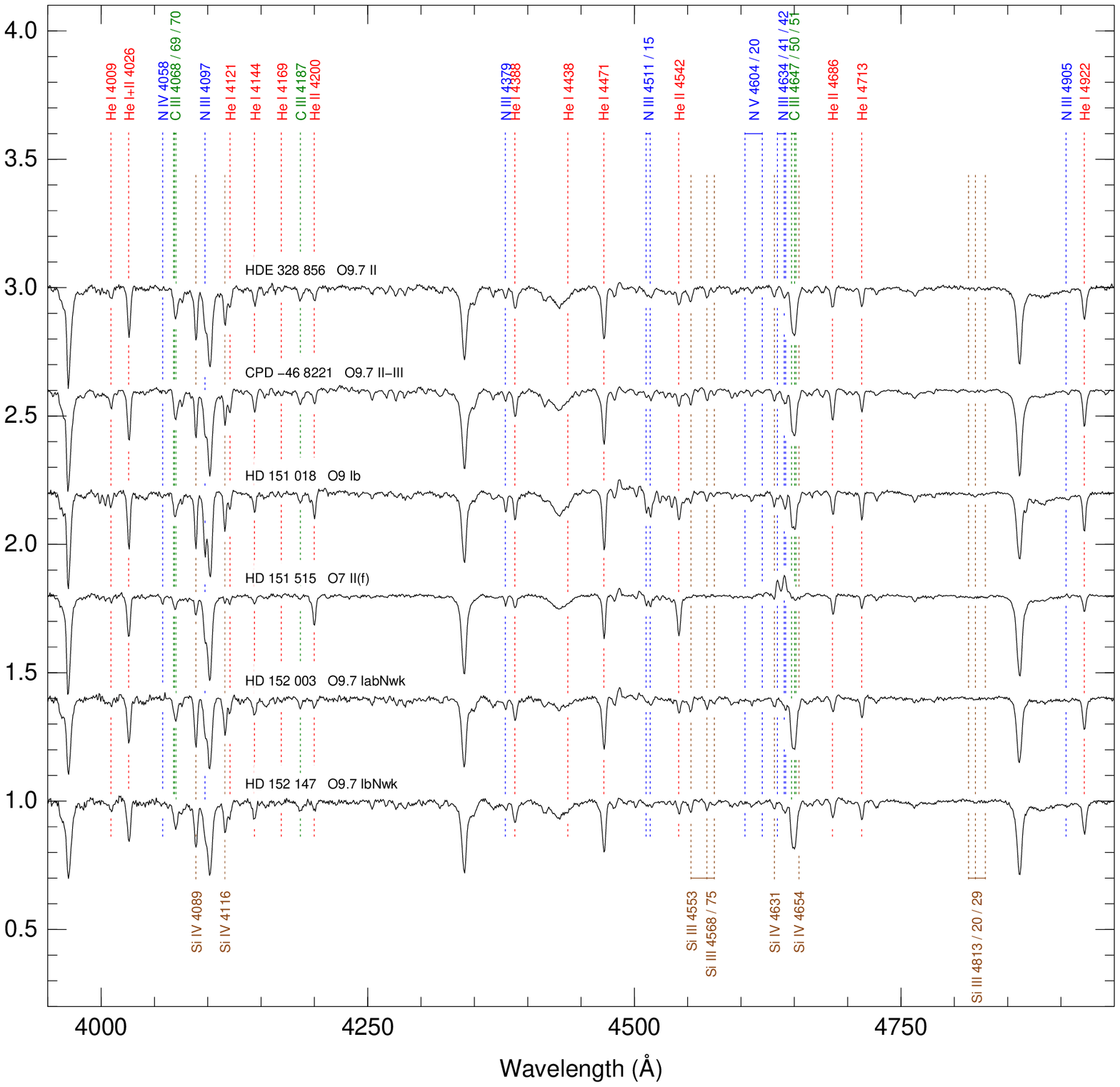}}
\caption{(continued).}
\end{figure*}	

\addtocounter{figure}{-1}

\begin{figure*}
\centerline{\includegraphics*[width=\linewidth]{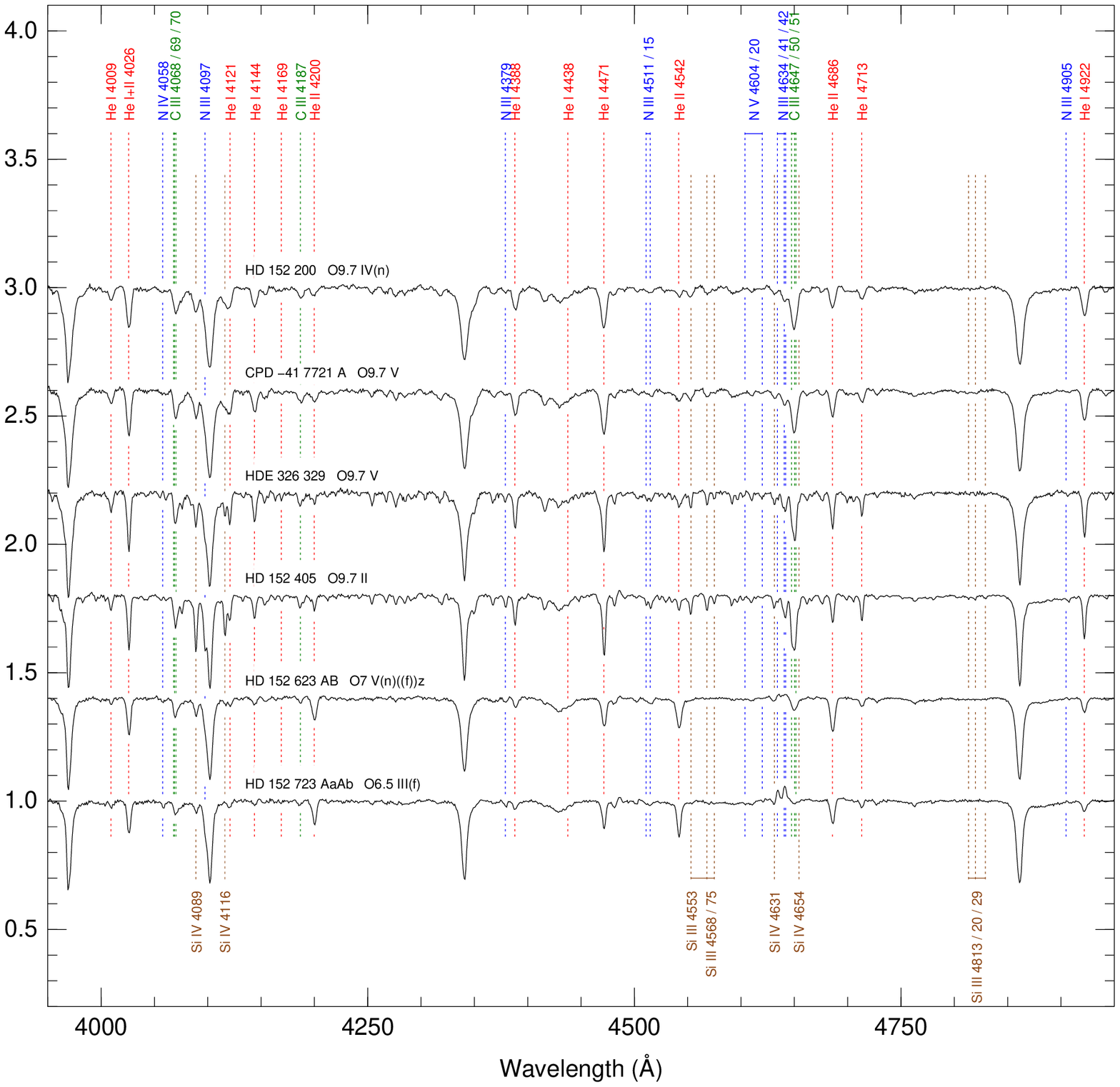}}
\caption{(continued).}
\end{figure*}	

\addtocounter{figure}{-1}

\begin{figure*}
\centerline{\includegraphics*[width=\linewidth]{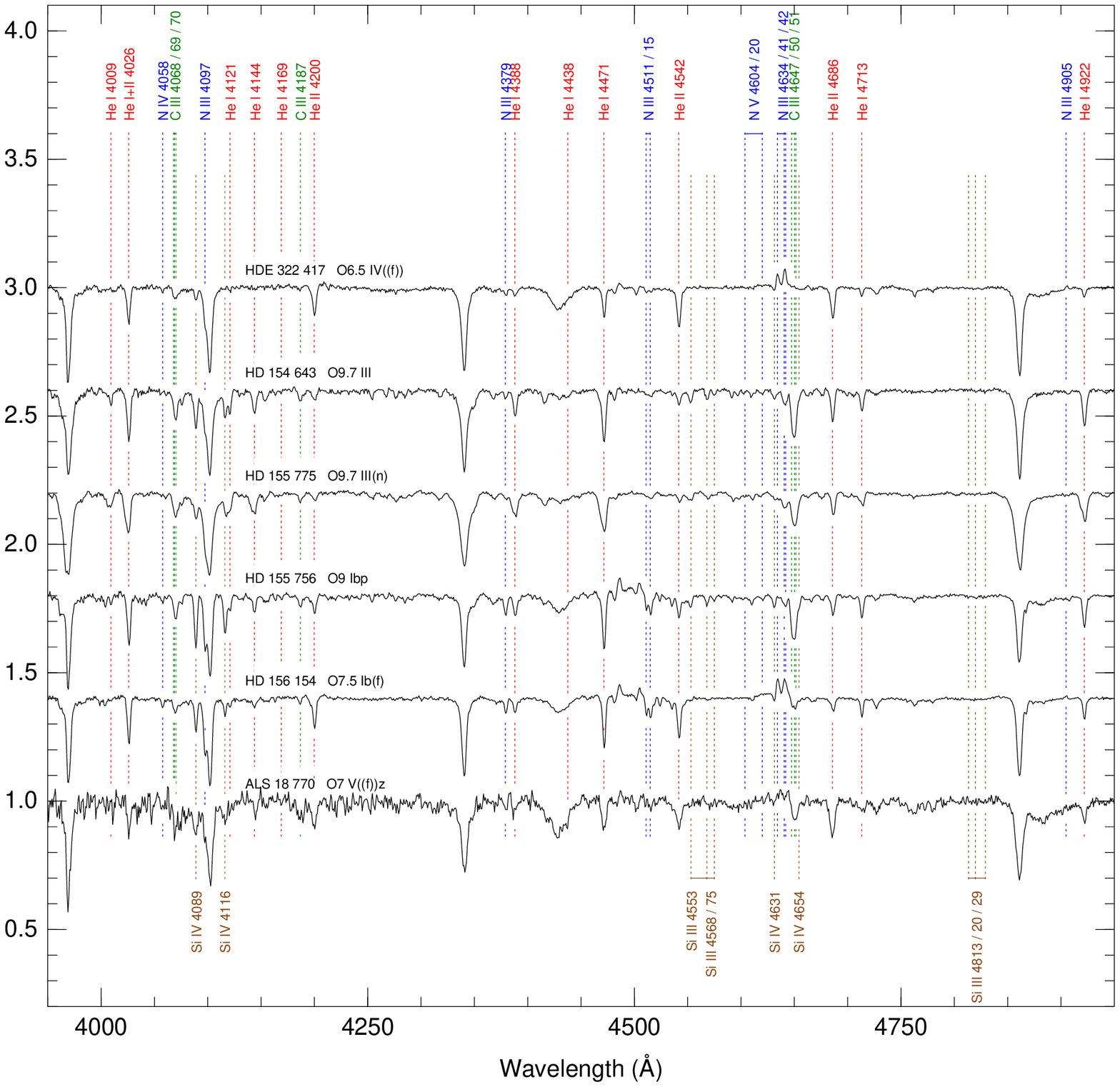}}
\caption{(continued).}
\end{figure*}	

\addtocounter{figure}{-1}

\begin{figure*}
\centerline{\includegraphics*[width=\linewidth]{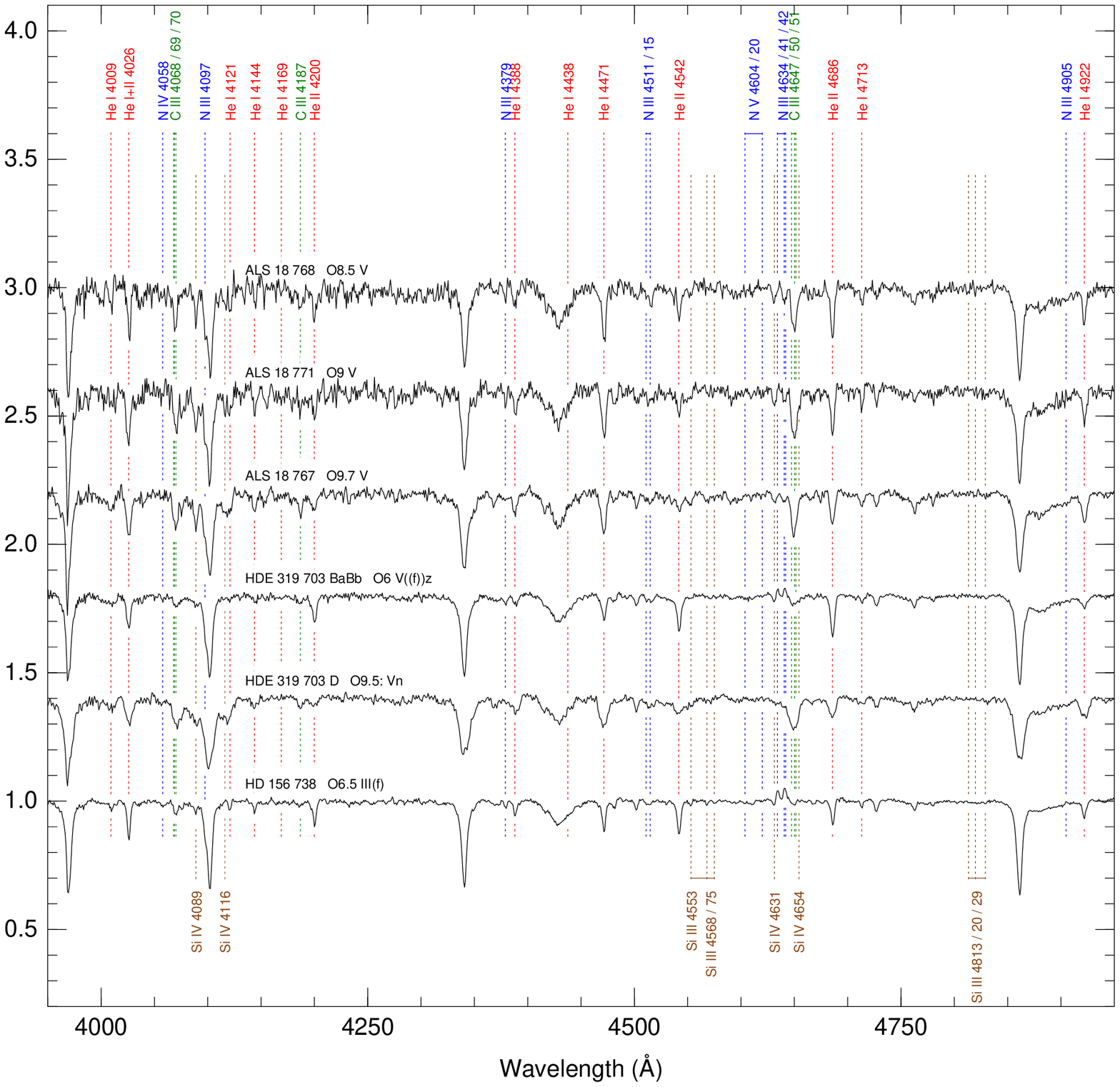}}
\caption{(continued).}
\end{figure*}	

\addtocounter{figure}{-1}

\begin{figure*}
\centerline{\includegraphics*[width=\linewidth]{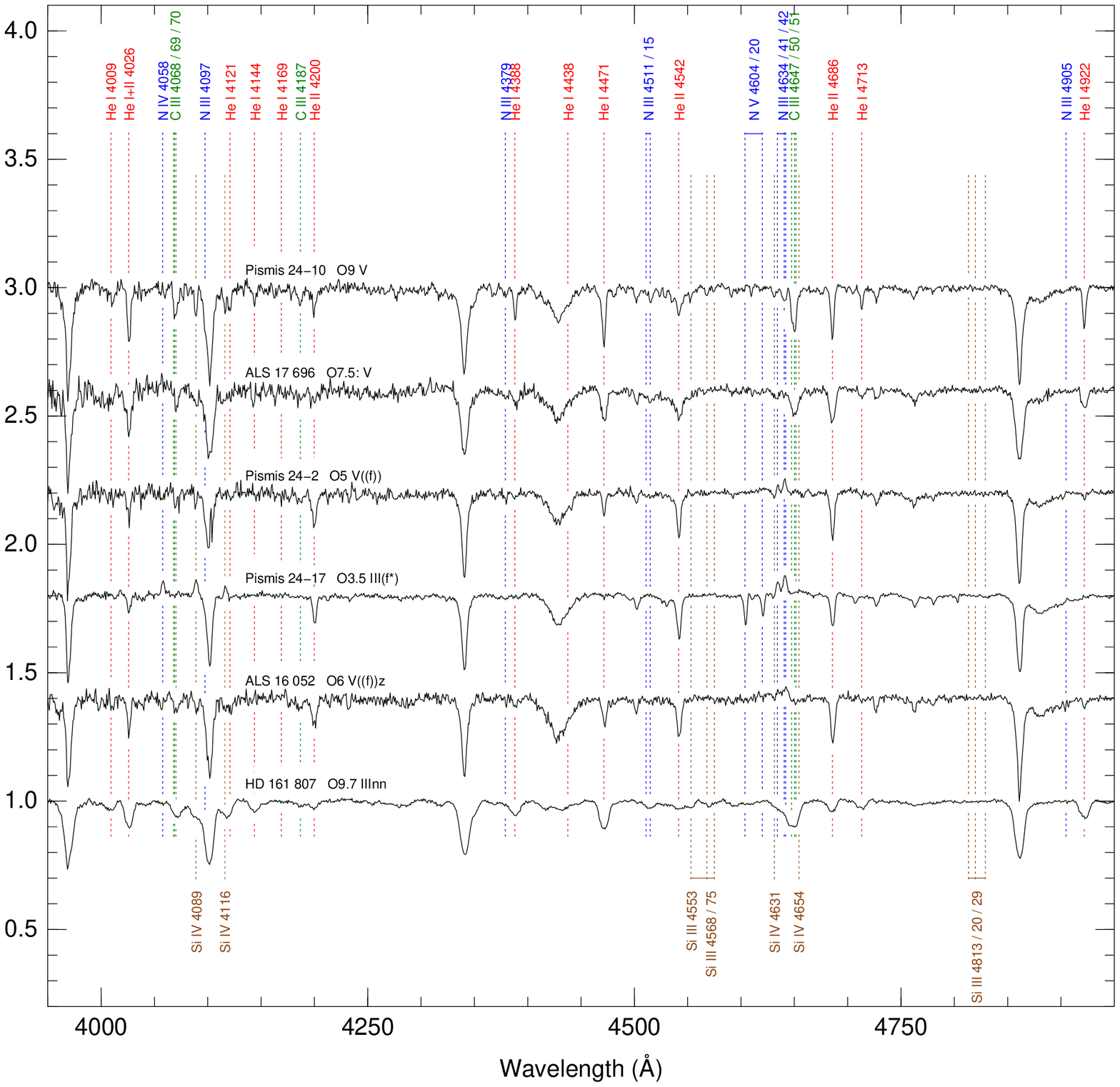}}
\caption{(continued).}
\end{figure*}	

\addtocounter{figure}{-1}

\begin{figure*}
\centerline{\includegraphics*[width=\linewidth]{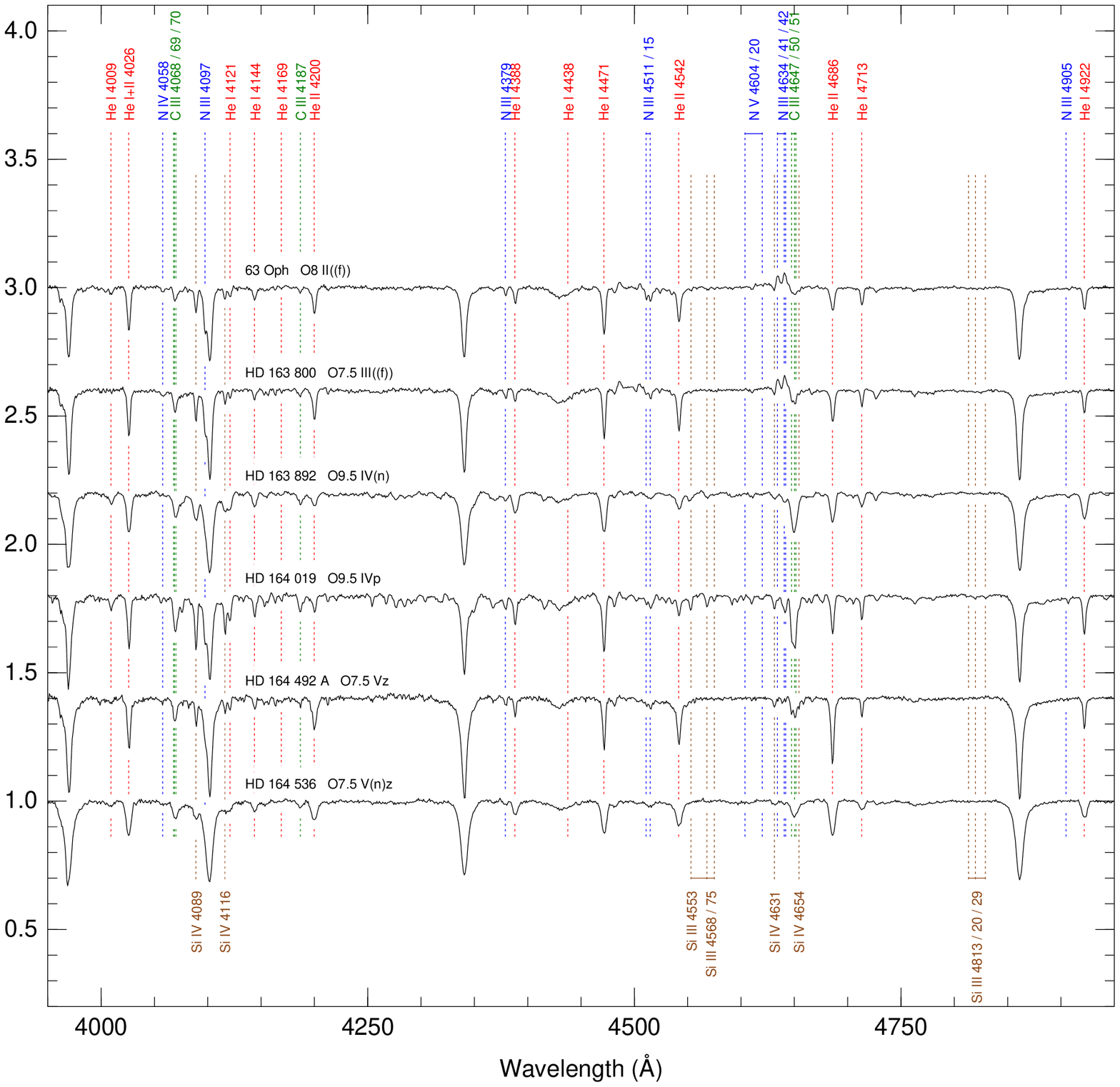}}
\caption{(continued).}
\end{figure*}	

\addtocounter{figure}{-1}

\begin{figure*}
\centerline{\includegraphics*[width=\linewidth]{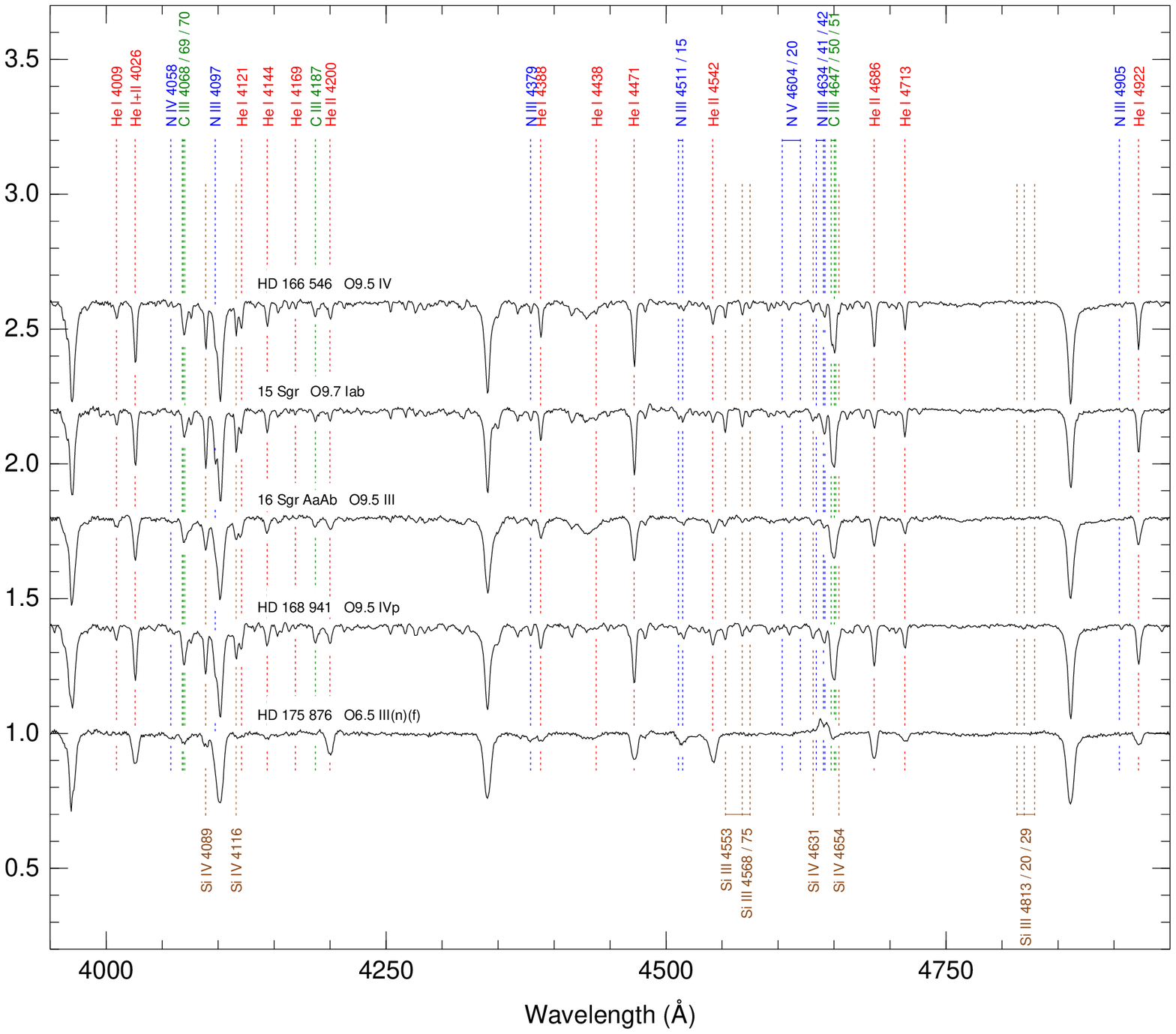}}
\caption{(continued).}
\end{figure*}	

\paragraph{HD~91\,572.}
\object[HD 91572]{}

This target is an SB1 according to OWN data.
The z suffix was added in GOSSS-DR1.1.
 
\paragraph{HD~91\,824.}
\object[HD 91824]{}

The ((f))z suffix was added in GOSSS-DR1.1. OWN data indicate that it is an SB1
with a 112~d period.
 
\paragraph{HD~91\,837.}
\object[HD 91837]{}

This star was not included in \citet{Maizetal04b} and, to our knowledge, had been never classified as an O star.
 
\paragraph{HD~92\,206~B.}
\object[HD 92206 B]{}

This star in NGC 3324 was not included in \citet{Maizetal04b}, probably because its spectrum had been previously analyzed only together with that of A (see e.g.
\citealt{Walb82a}). Note that in this system, each of the A, B, and C components contain at least one O star (see previous comments on A and C).
See Figure~\ref{chart1} for a chart (HD~92\,206 field).
 
\paragraph{HD~92\,504.}
\object[HD 92504]{}

The WDS lists a companion with a separation of 8\farcs7 and a small $\Delta m$. A look at images available in Aladin and the 2MASS Point Source Catalog reveals
that the separation is correct but the magnitude difference is significantly larger (over 2 magnitudes).
 
\paragraph{HD~92\,607.}
\object[HD 92607]{}

This star in the Carina Nebula was not included in \citet{Maizetal04b} but was previously classified as an O star by \citet{FortOrsa81}.
 
\paragraph{HDE~305\,438.}
\object[HDE 305438]{}

This star in the Carina Nebula was not included in \citet{Maizetal04b} but was previously classified as an O star by \citet{ThacAndr74}. 
 
\paragraph{HDE~303\,316~A.}
\object[HDE 303316]{}

This star in the Carina Nebula was not included in \citet{Maizetal04b} but was previously classified as an O star by \citet{FortOrsa81}. 
A companion with a separation of 10\farcs9 and a $\Delta m$ of 1.9 magnitudes was spatially resolved with GOSSS; it is an early-B star.
 
\paragraph{HD~93\,028.}
\object[HD 93028]{}

According to \citet{Levaetal90}, this object is an SB1 with a 51.554~d period. OWN data indicate that the period is close to four times that
value. This star was classified as O9~V by \citet{Walb72} and that was revised to O9~IV by \citet{Sotaetal11a}.
 
\paragraph{HD~93\,027.}
\object[HD 93027]{}

This star was classified as O9.5~V by \citet{Walb73c} and that was revised to O9.5~IV by \citet{Sotaetal11a}.
 
\paragraph{HDE~303\,312 = CPD~$-$58~2604 = V725~Car = CD~$-$58~3523.}
\object[HDE 303312]{}

This star was not included in \citet{Maizetal04b} but was previously classified as an O star by \citet{FortOrsa81}. \citet{Oter06} finds it to be an eccentric
eclipsing binary with a 9.4109 d period.
 
\paragraph{ALS~15\,204 = Trumpler~14~MJ~92.}
\object[ALS 15204]{}

This object is a serendipitous discovery (hence, it was not included in \citealt{Maizetal04b}): it was placed on the slit because another O star was observed
nearby. A weak redward component may indicate the presence of a dim spectroscopic B companion. The z suffix was added in GOSSS-DR1.1.
See Figure~\ref{chart1} for a chart (Trumpler 14 field).
 
\paragraph{HDE~305\,518.}
\object[HDE 305518]{}

This star was not included in \citet{Maizetal04b} but was previously classified as an O star by \citet{LevaMala82}.
 
\paragraph{CPD~$-$58~2611 = Trumpler~14$-$20.}
\object[CPD-58 2611]{}

\citet{Walb82a} classified this star as O6~V((f)). We have now added the z suffix but kept the rest unchanged. This star was flagged as a possible SB1 by
\citet{Garcetal98}. See Figure~\ref{chart1} for a chart (Trumpler 14 field). The visual multiplicity for this target was measured in ACS/WFC images.
 
\paragraph{ALS~15\,207 = Trumpler~14$-$21.}
\object[ALS 15207]{}

This star was not included in \citet{Maizetal04b} but was previously classified as an O star by \citet{Morretal88}.
See Figure~\ref{chart1} for a chart (Trumpler 14 field).
 
\paragraph{HD~93\,128 = 2MASS~J10435441$-$5932574.}
\object[HD 93128]{}

This object in the core of Trumpler 14 was one of the O3 stars in \citet{Walb82b}. Its spectral subtype was revised to O3.5 by \citet{Walbetal02b}.
The ((fc))z suffix was inadvertently omitted in GOSSS-DR1.0. Note that there is an error in Table 3 of paper I: The range of validity of the ((f*)) suffix is O2-O3,
not O2-O3.5 (the latter is applicable only to (f*) and f*). \citet{Levaetal91} indicate that it is an SB1 but its radial velocity appears to be constant in OWN data.
See Figure~\ref{chart1} for a chart (Trumpler 14 field).
The visual multiplicity for this target was measured in ACS/HRC images.
 
%
 
\paragraph{Trumpler~14$-$9 = Trumpler~14~MJ~165 = 2MASS~J10435540$-$5932493.}
\object[Cl Trumpler 14 9]{}

This star was not included in \citet{Maizetal04b} but was previously classified as an O star by \citet{Morretal88}.
The z suffix was removed in GOSSS-DR1.1.
See Figure~\ref{chart1} for a chart (Trumpler 14 field) and Figure~\ref{HRC} for an ACS/HRC image.
The visual multiplicity for this target was measured in ACS/HRC images.
 
\paragraph{HD~93\,129~AaAb.}
\object[HD 93129 A]{}

This object was considered an O3 star in \citet{Walb82b} and later became the prototype O2~If* \citep{Walbetal02b}. It is the most massive system at the core of 
Trumpler 14 in the Carina Nebula. It was split by \citet{Nelaetal04} into two components (Aa and Ab) using HST/FGS. Later observations
with HST/HRC and HST/FGS by \citet{Maizetal05b,Maizetal08b} detected a proper motion along the radius vector (i.e. near-constant position angle)
between Aa and Ab with an average value (updated with unpublished data) of 2.14$\pm$0.21~mas/a and a separation of $\approx$40 mas in early 2009. The $\Delta m$ of
0.9 indicates that both Aa and Ab must be very early-type O stars. 
See Figure~\ref{chart1} for a chart (Trumpler 14 field), Figure~\ref{HRC} for an ACS/HRC image, and below for the B component.
The visual multiplicity for this target was measured in ACS/HRC images. The system is a colliding-wind binary \citep{Benaetal10,DeBeRauc13}.
 
\paragraph{CPD~$-$59~2551 = Collinder~228$-$21.}
\object[CPD-59 2551]{}

This star was not included in \citet{Maizetal04b} but was previously classified as an O star by \citet{LevaMala81}.
See Figure~\ref{chart1} for a chart (HD~93\,146 field).
 
\paragraph{HD~93\,129~B.}
\object[HD 93129B]{}

This object was the third O3 star in \citet{Walb82b} (besides HD~93\,129~A - now AaAb and an O2 - and HD~93\,128) located in the core of Trumpler 14. It is
separated by 2\farcs5 from AaAb and has magnitudes similar to its siblings HD~93\,129~Ab and HD~93\,128. 
The target has \CIII{4650} in emission but is not strong enough to warrant a c suffix.
See Figure~\ref{chart1} for a chart (Trumpler 14 field) and Figure~\ref{HRC} for an ACS/HRC image.
The visual multiplicity for this target was measured in ACS/HRC images.
 
\paragraph{HD~93\,146~B.}
\object[HD 93 146 B]{}

This star was not included in \citet{Maizetal04b} but was previously classified as an O star by \citet{LevaMala81}. According to the WDS, the A (see below) and 
B components of HD~93\,146 have a $\Delta m$ of 1.5 magnitudes and a separation of 6\farcs5. Both are O stars.
See Figure~\ref{chart1} for a chart (HD~93\,146 field).
 
\paragraph{CPD~$-$58~2620 = Trumpler~14$-$8.}
\object[CPD-58 2620]{}

The ((f)) suffix was omitted by mistake in GOSSS-DR1.0.
See Figure~\ref{chart1} for a chart (Trumpler 14 field).
 
\paragraph{HD~93\,146~A.}
\object[HD 93146]{}

The ((f))z suffix was added in GOSSS-DR1.1. See Figure~\ref{chart1} for a chart (HD~93\,146 field) and above for a brief description of HD~93\,146~B. OWN data
indicate that it is an SB1 with a 1130~d period.
 
\paragraph{HD~93\,130 = V661~Car.}
\object[HD 93130]{}

This object is an eccentric eclipsing binary with a 23.944 d period \citep{Oter06}. OWN data also show that it is an SB1.
 
\paragraph{CPD~$-$59~2554 = Collinder~228$-$67.}
\object[CPD-59 2554]{}

This star was not included in \citet{Maizetal04b} but was classified as an O star by \citet{Massetal01}.
See Figure~\ref{chart1} for a chart (HD~93\,146 field).
 
\paragraph{CPD~$-$58~2627.}
\object[CPD-58 2627]{}

This star was not included in \citet{Maizetal04b} but was classified as an O star by \citet{FortOrsa81}.
 
\paragraph{HD~93\,160.}
\object[HD 93160]{}

The (f) suffix was changed to ((f)) in GOSSS-DR1.1 according to the rules in Table~\ref{fphen}. HD~93\,160 is 12\farcs6 away from HD~93\,161~A.
See Figure~\ref{chart1} for a chart (Trumpler 14 field).
 
\paragraph{HD~93\,161~B.}
\object[HD 93161B]{}

HD~93\,161~A and B are separated by 2\arcsec\ and both are O-type systems. This star was not included in \citet{Maizetal04b} because \citet{Walb72} produced a
spectral type for the combined AB system (but see \citealt{Nazeetal05}). The luminosity class was changed from V to IV in GOSSS-DR1.1. 
The target has \CIII{4650} in emission but is not strong enough to warrant a c suffix. One of us (N.R.W.) disagrees about the spectral subtype of this star and
thinks that it should be O6 instead of O6.5 based on the ratio of \HeIII{4026}/\HeII{4200}. However, any of the combinations among the strong He lines excluding 
\HeIII{4026} (\HeII{4200}, \HeI{4471}, \HeII{4542}, and \HeII{4686}) yields ratios essentially identical to those of the O6.5~IV standard used in paper I and in
this work, hence the O6.5 classification. See Figure~\ref{chart1} for a chart (Trumpler 14 field).
 
\paragraph{HDE~305\,536.}
\object[HDE 305536]{}

This star was not included in \citet{Maizetal04b} but \citet{LevaMala81} classified it as an O type. \citet{Levaetal90} discovered that it is an SB1 with a
2.018~d period. See Figure~\ref{chart1} for a chart (HD~93\,146 field). 
 
\paragraph{ALS~15\,210 = Trumpler~16$-$244.}
\object[ALS 15210]{}

This star was not included in \citet{Maizetal04b} but \citet{MassJohn93} classified it as O3/4~If. We obtain O3.5~If*~Nwk, making it the most extincted
O2-O3.5 star in the Carina Nebula known to date. \HeII{4686} shows a P-Cygni profile which, by analogy with Pismis~24-1~A, could imply the existence of a
companion. The Nwk suffix was added in GOSSS-DR1.1 based on weak \NVd{4604-4620} and strong \CIV{4658}.
\citet{Smitetal04} considered that this star and/or the adjacent HD~93\,162 are the main sources responsible
for creating Carina's defiant finger\footnote{See \url{http://www.spacetelescope.org/images/heic0822b/}.}. 
See Figure~\ref{chart1} for a chart (Trumpler 16 field) and Figure~\ref{HRC} for an ACS/HRC image.
In the existing ACS/HRC images we detect a previously unpublished\footnote{It was presented in the JD 13 of the IAU General Assembly in 2009 but did not 
appear in any proceedings.} visual companion. It has a separation of 1\farcs09, a position angle of 314\arcdeg, and a $\Delta V$ of 8.9~magnitudes. The magnitude 
and colors are consistent with those of a $\sim$1 M$_\odot$ star. An even fainter companion (separation of 1\farcs49, position angle of 102\arcdeg, and 
$\Delta V$ of 10.7~magnitudes) is seen in the saturated ACS/WFC images.
 
\paragraph{HDE~305\,523.}
\object[HDE 305523]{}

\citet{Walb73a} classified this star as O9~II. Here we change its luminosity class to II-III.
 
\paragraph{Tyc~8626$-$02506$-$1.}
\object[Tyc 8626-02506-1]{}

This star was not included in \citet{Maizetal04b} and, to our knowledge, this is the first time it receives a clear classification as an O type.
 
\paragraph{HD~93\,204.}
\object[HD 93204]{}

See Figure~\ref{chart1} for a chart (Trumpler 16 field). In the existing ACS/HRC images we detect three previously unpublished visual companions. They have 
separations of 1\farcs40, 1\farcs67, and 4\farcs81; position angles of 42\arcdeg, 209\arcdeg, and 0\arcdeg; and $\Delta V$ of 6.7, 8.9, and 8.7 mag,
respectively. The \CIII{4650}/\NIII{4634} intensity ratio (in emission) of 0.9 places this spectrum just below the boundary of the fc category.
 
\paragraph{HD~93\,222.}
\object[HD 93222]{}

The ((f))z suffix was added in GOSSS-DR1.1. 
 
\paragraph{HDE~303\,311.}
\object[HDE 303311]{}

The WDS lists a companion with a separation of 2\farcs3 and a $\Delta m$ of 2.9 magnitudes i.e. not bright enough to influence the spectral type.
The ((f))z suffix was added in GOSSS-DR1.1.
 
\paragraph{HDE~305\,524.}
\object[HDE 305524]{}

This star was not included in \citet{Maizetal04b} but was classified as O type by \cite{LevaMala81}.
 
\paragraph{CPD~$-$59~2610 = Collinder~228$-$39.}
\object[CPD-59 2610]{}

This star was not included in \citet{Maizetal04b} but was classified as O type by \cite{LevaMala81}.
 
\paragraph{CPD~$-$59~2626~AB = Trumpler~16$-$23~AB.}
\object[CPD-59 2626]{}

This star was not included in \citet{Maizetal04b} but was classified as O type by \cite{LevaMala82}. A spatially unresolved B component has a separation of 
16~mas and a $\Delta m$ of 1.6 magnitudes \citep{Nelaetal04,Nelaetal10}. The z suffix was added in GOSSS-DR1.1. See Figure~\ref{chart1} for a chart (Trumpler 16 field).
 
\paragraph{CPD~$-$59~2634 = Trumpler~16$-$9.}
\object[CPD-59 2634]{}

This star was not included in \citet{Maizetal04b} but was classified as O type by \citet{MassJohn93}. A spatially unresolved B component has a separation of 
77~mas and a $\Delta m$ of 2.4 magnitudes \citep{Nelaetal04}. The luminosity class was changed to IV in GOSSS-DR1.1. See Figure~\ref{chart1} for a chart (Trumpler 16 field).
 
\paragraph{CPD~$-$59~2627 = Trumpler~16$-$3.}
\object[CPD-59 2627]{}

This star was not included in \citet{Maizetal04b} but was classified as O type by \citet{MassJohn93}.
See Figure~\ref{chart1} for a chart (Trumpler 16 field).
 
\paragraph{CPD~$-$59~2629 = Trumpler~16$-$22.}
\object[CPD-59 2629]{}

This star was not included in \citet{Maizetal04b} but was classified as O8.5 V by \citet{MassJohn93}.
\citet{Nazeetal12d} discovered a magnetic field in this star with a strong H$\alpha$ emission. 
We observe a weak emission in H$\beta$ in the 
middle of the stellar absorption, something that is also detected in other magnetic O stars. 
See Figure~\ref{chart1} for a chart (Trumpler 16 field).
 
\paragraph{CPD~$-$59~2644.}
\object[CPD-59 2644]{}

This star was not included in \citet{Maizetal04b} but was classified as O type by \cite{LevaMala82}. The spectral subtype was changed to O9 in GOSSS-DR1.1.
See Figure~\ref{chart1} for a chart (Trumpler 16 field).
 
\paragraph{[ARV2008]~206 = Trumpler~16~MJ~568.}
\object[Cl* Trumpler 16 MJ 568]{}

This star was not included in \citet{Maizetal04b} and, to our knowledge, had never been optically classified as an O star.
 
\paragraph{HDE~305\,532.}
\object[HDE 305532]{}

This star was classified as O6~V((f)) by \citet{Walb82a}. Here we reclassify it as O6.5~V((f))z. \citet{Levaetal90} found variable velocities, possibly 
indicating a spectroscopic binary. 
 
\paragraph{HDE~305\,525.}
\object[HDE 305525]{}

This star was not included in \citet{Maizetal04b} but was classified as O type by \cite{LevaMala81}. The WDS lists a companion with a separation of 4\farcs7 and
a $\Delta m$ of 3.7 magnitudes.
 
\paragraph{CPD~$-$59~2673 = Feinstein~97 = Collinder~228$-$97.}
\object[CPD-59 2673]{}

The ((f)) suffix was added in GOSSS-DR1.1.
 
\paragraph{HDE~305\,539.}
\object[HDE 305539]{}

This star was classified as O7p by \citet{Walb82a}. Here it appears as a relatively normal O8~Vz, with the strength of its metallic lines typical for that
spectral type. See Figure~\ref{chart1} for a chart (HD~93\,632 field).
 
\paragraph{HD~93\,576.}
\object[HD 93576]{}

This star was not included in \citet{Maizetal04b} but was classified as O type by \cite{LevaMala81}. According to \citet{Levaetal90}, it is an SB1 with a 2.02~d
period. See Figure~\ref{chart1} for a chart (HD~93\,632 field).
 
\paragraph{HD~93\,632.}
\object[HD 93632]{}

\citet{Walb73a} classified HD~93\,632 as O5~III(f)~var but in our GOSSS spectra it clearly has a 
luminosity class of I, based on \HeII{4686} strongly in emission. Recent OWN spectra 
show a weaker \HeII{4686}, indicating unusual variability, possibly related to the presence of a magnetic field. 
The target has \CIII{4650} in emission but is not strong enough to warrant a c suffix.
See Figure~\ref{chart1} for a chart (HD~93\,632 field).
 
\paragraph{ALS~18\,083 = Bochum~11$-$5.}
\object[ALS 18083]{}

This star was not included in \citet{Maizetal04b} but was classified as O type by \cite{FitzMeht87}.
See Figure~\ref{chart1} for a chart (HD~93\,632 field).
 
\paragraph{HDE~305\,612.}
\object[HDE 305612]{}

This star was not included in \citet{Maizetal04b} but was classified as O type by \cite{FitzMeht87}.
See Figure~\ref{chart1} for a chart (HD~93\,632 field).
 
\paragraph{HDE~305\,619.}
\object[HDE 305619]{}

The luminosity class of this star was changed from Ib to II in GOSSS-DR1.1.
 
\paragraph{HD~94\,024.}
\object[HD 94024]{}

\citet{Walb73a} classified this star as O8~V((n)). Here we find it to be O8~IV. OWN data indicate that this system is an SB1.
 
\paragraph{HDE~303\,492.}
\object[HDE 303492]{}

\citet{Walb82a} classified this star as O9~Ia. In Paper I it was adopted as the O8.5 Iaf standard in accordance with the revised criteria around O9 discussed there.
 
\paragraph{HD~94\,963.}
\object[HD 94963]{}

The (f) suffix was added in GOSSS-DR1.1.
 
\paragraph{ALS~18\,556.}
\object[ALS 18556]{}

This star was not included in \citet{Maizetal04b}. Simbad gives a spectral type of O5 but we have been unable to find any reference that has previously obtained
an optical spectrum for this target. Apparently, the source of the ``spectral type'' in Simbad is a purely photometric estimate by \citet{Wram76} assuming that
the star is a main-sequence object. Actually, its spectral type is O9.5~Iabp, with the p arising from discrepant ratios between the He lines used to determine the
precise spectral type for late-O stars.
 
\paragraph{HD~95\,589.}
\object[HD 95589]{}

The ((f)) suffix was added in GOSSS-DR1.1.
 
\paragraph{HD~96\,715.}
\object[HD 96715]{}

The ((f))z suffix was added in GOSSS-DR1.1.
 
\paragraph{HD~96\,917.}
\object[HD 96917]{}

\citet{Walb73a} classified this star as O8.5~Ib(f) and here we change the classification to O8~Ib(n)(f), which is slightly different from the one in GOSSS-DR1.0. 
According to OWN data, it is an SB1 with a likely period of 4.03~d.
 
\paragraph{HD~96\,946.}
\object[HD 96946]{}

According to OWN data, this system is an SB1 with a likely period of 4.34~d. The (f) suffix was added in GOSSS-DR1.1.
 
\paragraph{HD~97\,253.}
\object[HD 97253]{}

\citet{Walb73a} classified this star as O5.5~III(f). We obtain the same luminosity class and suffix but as an O5.
 
\paragraph{HD~97\,848.}
\object[HD 97848]{}

\citet{Walb82a} classified this star as O8~V and that same spectral type is also found here.
This star is a marginal z.
 
\paragraph{HD~99\,897.}
\object[HD 99897]{}

\citet{Walb73a} classified this star as O6~V(f). Here we obtain O6.5~IV((f)).
 
\paragraph{HDE~308\,813.}
\object[HDE 308813]{}

This star was not included in \citet{Maizetal04b} but \citet{Schi70} classified it as an O star.
According to \citet{Willetal13}, it is an SB1 with a 6.340 d period. The spectral type was changed from O9.5 to O9.7 in GOSSS-DR1.1.
See Figure~\ref{chart1} for a chart (IC~2944 field).
 
\paragraph{HD~101\,191.}
\object[HD 101191]{}

\citet{Walb73a} classified this star as O8~V((n)). We obtain O8~V. \citet{Sanaetal11a} indicate that it is a long-period SB1 system. OWN data,
however, point towards a 25.8 d period. See Figure~\ref{chart1} for a chart (IC~2944 field).
 
\paragraph{HD~101\,223.}
\object[HD 101223]{}

\citet{Walb73a} classified this star as O8~V((f)). We obtain the same spectral type without the suffix.
See Figure~\ref{chart1} for a chart (IC~2944 field).
 
\paragraph{HD~101\,298.}
\object[HD 101298]{}

The classification was changed from O6~V((f)) to O6.5~IV((f)) in GOSSS-DR1.1.
See Figure~\ref{chart1} for a chart (IC~2944 field).
 
\paragraph{HD~105\,627.}
\object[HD 105627]{}

According to OWN data, this system is an SB1 with a likely period of 2.692~d.
 
\paragraph{HD~113\,659.}
\object[HD 113659]{}

This object was not included in \citet{Maizetal04b} even though it has $B=8.0$. The spectral type changed from O9.5~V in GOSSS-DR1.0 to O9~IV here.
It is an eclipsing binary with a period of 3.4273 days \citep{Oter07}.

\paragraph{$\theta$~Mus~B = HD~113\,904~B.}
\object[* tet Mus B]{}

This O star is located 5\farcs5 to the south of the brighter A component, a WR~+~O binary (WR~48) which itself has a companion separated by 47 mas 
\citep{Hartetal99}. We easily spatially resolve the spectra of A and B. $\theta$~Mus~B is an SB1 according to OWN data.
 
\paragraph{HD~114\,737~AB.}
\object[HD 114737]{}

The WDS lists a B component with a separation of 0\farcs2 and a $\Delta m$ of 1.5 magnitudes. According to OWN data, this system is an SB1 with a 12.38~d period.
 
\paragraph{HD~116\,282.}
\object[HD 116282]{}

\cite{Walb82a} classified this star as O8~III(n)((f)). We obtain the same classification but only with (n) as suffix.
 
\paragraph{HD~116\,852.}
\object[HD 116852]{}

The ((f)) suffix was added in GOSSS-DR1.1.
 
\paragraph{HD~118\,198.}
\object[HD 118198]{}

\citet{Walb73a} classified\footnote{Note that a luminosity class of II-III means ``intermediate between II and III'' while one of III-II means ``uncertain between II and III''.} this star as O9.5~II-III. 
Here we obtain O9.7~III (in GOSSS-DR1.0 the luminosity class was IV).
 
\paragraph{HD~120\,521.}
\object[HD 120521]{}

\citet{Walb73a} classified this star as O8~Ib(f). Here we obtain O7.5~Ib(f) (in GOSSS-DR1.0 the luminosity class was Ib-II). OWN data indicate it is variable.
 
\paragraph{HD~125\,241.}
\object[HD 125241]{}

The (f) suffix was added in GOSSS-DR1.1. OWN data indicate it is variable.
 
\paragraph{HD~130\,298.}
\object[HD 130298]{}

This object is an SB1 with a 14.63~d period according to OWN data.
The (n)((f)) suffix was changed to (n)(f) in GOSSS-DR1.1.
 
\paragraph{HD~135\,591.}
\object[HD 135591]{}

The ((f)) suffix was added in GOSSS-DR1.1.
 
\paragraph{CPD~$-$54~6791~AB = Muzzio~I$-$116.}
\object[CPD-54 6791]{}

The WDS lists a B component with a separation of 1\farcs1 and a $\Delta m$ of 0.4 mag that we are unable to spatially resolve in the GOSSS data.
OWN data indicate that a long-period SB1 may be present in the system but not necessarily involving the brightest component.
 
\paragraph{HD~148\,546.}
\object[HD 148546]{}

\citet{Walb73a} classified this object as O9~Ia and here the luminosity class is changed to Iab.
OWN data indicate it is variable.
 
\paragraph{$\mu$~Nor = HD~149\,038.}
\object[V* mu. Nor]{}

\citet{Walb72} classified this star as O9.7~Iab and that spectral type is also the one derived from GOSSS data.
 
\paragraph{HD~149\,452.}
\object[HD 149452]{}

\citet{Walb72} classified this star as O8~Vn((f)) and with GOSSS data we obtain O9~IVn.
 
\paragraph{HDE~328\,856.}
\object[HDE 328856]{}

This star was not included in \citet{Maizetal04b} but was classified as O type by \citet{Whit63}. 
The luminosity class was changed to II in GOSSS-DR1.1.
 
\paragraph{CPD~$-$46~8221.}
\object[CPD-46 8221]{}

This star was not included in \citet{Maizetal04b} but was classified as O type by \citet{VijaDrill93}.
The luminosity class was changed to II-III in GOSSS-DR1.1.
 
\paragraph{HD~151\,018.}
\object[HD 151018]{}

\citet{Garretal77} classified this star as O9~Ia. We derive the same spectral subtype but a Ib luminosity class.
 
\paragraph{HD~151\,515.}
\object[HD 151515]{}

The (f) suffix was omitted by mistake in GOSSS-DR1.1.
 
\paragraph{HD~152\,200.}
\object[HD 152200]{}

This star was not included in \citet{Maizetal04b} but was classified as O type by \citet{Schietal69}.
See Figure~\ref{chart1} for a chart (NGC~6231 field).
 
\paragraph{CPD~$-$41~7721~A = CD~$-$41~11027~p.}
\object[CPD-41 7721]{}

This star was not included in \citet{Maizetal04b} but was classified as O type by \citet{Sanaetal08b}. The WDS lists a B companion with a separation of 5\farcs7
and a $\Delta m$ of 1.0 magnitude that was not included in the GOSSS spectrum. See Figure~\ref{chart1} for a chart (NGC~6231 field).
 
\paragraph{HDE~326\,329 = CPD~$-$41~7735.}
\object[HDE 326329]{}

This star was not included in \citet{Maizetal04b} but was classified as O type by \citet{Morgetal53a}. The WDS lists a dim companion 7\farcs7 away.
See Figure~\ref{chart1} for a chart (NGC~6231 field).
 
\paragraph{HD~152\,405.}
\object[HD 152405]{}

This object is an SB1 system with a period of 25.5~d according to OWN data.
 
\paragraph{HD~152\,623~AB.}
\object[HD 152623]{}

This object is an SB1 system according to OWN data. The WDS lists a companion with a separation of 0\farcs3 and a $\Delta m$ of 1.3 magnitudes
that we were unable to spatially resolve in the GOSSS data.
The system is a colliding-wind binary \citep{DeBeRauc13}.
 
\paragraph{HD~152\,723~AaAb.}
\object[HD 152723]{}

This object is an SB1 system with a period of 18.9~d according to OWN data. The WDS lists three (B, C, and D) well resolved companions within
15\arcsec\ that are not included in the GOSSS data. Ab, however, has a separation of 0\farcs098 and a $\Delta m$ of 1.7 magnitudes \citep{Masoetal09}, so it
likely contributes to the spectral classification.
 
\paragraph{HDE~322\,417 = Trumpler~24$-$430.}
\object[HDE 322417]{}

This object is an SB1 system with a period of 223~d according to OWN data.
The ((f)) suffix was omitted by mistake in GOSSS-DR1.0
 
\paragraph{HD~154\,643.}
\object[HD 154643]{}

This object is an SB1 system with a period of 28.6~d according to OWN data.
 
\paragraph{HD~155\,775 = V1012~Sco.}
\object[HD 155775]{}

This star was not included in \citet{Maizetal04b} but was classified as O type by \citet{Goy73}. It is an eclipsing binary according to \citet{Malketal06}.

\paragraph{HD~155\,756.}
\object[HD 155756]{}

This star was classified as O9.5~Iab in GOSSS-DR1.0 and is now an O9~Ibp. The peculiarity arises from the discrepancies between luminosity criteria and
the presence of both strong N and C lines (except for \NIIId{4634-40-42}).
 
\paragraph{HD~156\,154.}
\object[HD 156154]{}

This object is an SB1 system according to OWN data. The (f) suffix was added in GOSSS-DR1.1.
 
\paragraph{ALS~18\,770 = HM~1$-$18 = C1715$-$387$-$18.}
\object[ALS 18770]{}

This star was not included in \citet{Maizetal04b} and, to our knowledge, had never been classified as O type before.
The z suffix was changed to ((f))z in GOSSS-DR1.1. See Figure~\ref{chart1} for a chart (Havlen-Moffat 1 field).
 
\paragraph{ALS~18\,768 = HM~1$-$10 = C1715$-$387$-$10.}
\object[ALS 18768]{}

This star was not included in \citet{Maizetal04b} and, to our knowledge, had never been classified as O type before.
Our data are noisy but there is a hint of the C and Si absorption lines being stronger than expected while N lines are not weak. 
That could be a sign of high metallicity. See Figure~\ref{chart1} for a chart (Havlen-Moffat 1 field).
 
\paragraph{ALS~18\,771 = HM~1$-$19 = C1715$-$387$-$19.}
\object[ALS 18771]{}

This star was not included in \citet{Maizetal04b} and, to our knowledge, had never been classified as O type before.
The spectral type was changed from O8.5 to O9 in GOSSS-DR1.1.
See Figure~\ref{chart1} for a chart (Havlen-Moffat 1 field).
 
\paragraph{ALS~18\,767 = HM~1$-$9 = C1715$-$387$-$9.}
\object[ALS 18767]{}

This star was not included in \citet{Maizetal04b} and, to our knowledge, had never been classified as O type before.
See Figure~\ref{chart1} for a chart (Havlen-Moffat 1 field).
 
\paragraph{HDE~319\,703~BaBb.}
\object[HDE 319703 B]{}

The B component in HDE~319\,703 is well separated (14\arcsec) from A but is closer to C (5\farcs4). According to the WDS, B itself comprises Ba
and Bb, with a $\Delta m$ of 1.5 magnitudes and a separation of 0\farcs2. Our spectrum corresponds to the composite BaBb. C is clearly spatially resolved in the
GOSSS data: it is an early-B star. See Figure~\ref{chart1} for a chart (HDE~319\,703 field), where C is to the north of BaBb.
 
\paragraph{HDE~319\,703~D = 2MASS~J17194894$-$3606029.}
\object[HDE 319703D]{}

This star was not included in \citet{Maizetal04b}. Indeed, to our knowledge, no spectra had ever been obtained. We placed it on the slit with
HDE~319\,703~A because of its proximity and we discovered it was also an O star. See Figure~\ref{chart1} for a chart (HDE~319\,703 field).
 
\paragraph{HD~156\,738.}
\object[HD 156738]{}

The (f) suffix was added in GOSSS-DR1.1. The O6.5~III(f) classification is the same as that of \citet{Walb82a}.
 
\paragraph{Pismis~24$-$10.}
\object[Cl Pismis 24 10]{}

This star was not included in \citet{Maizetal04b} but was classified as O type by \citet{Massetal01}. 
The spectral subtype was changed from O9.5 to O9 in GOSSS-DR1.1. See Figure~\ref{chart1} for a chart (Pismis~24 field).
 
\paragraph{ALS~17\,696 = Pismis~24$-$3.}
\object[ALS 17696]{}

This star was not included in \citet{Maizetal04b} but was classified as O type by \citet{Massetal01}. 
The luminosity class was changed to V in GOSSS-DR1.1.
See Figure~\ref{chart1} for a chart (Pismis~24 field).
 
\paragraph{Pismis~24$-$2 = ALS~17\,695.}
\object[Cl Pismis 24 2]{}

This star was not included in \citet{Maizetal04b} but was classified as O type by \citet{Massetal01}. 
See Figure~\ref{chart1} for a chart (Pismis~24 field).
 
\paragraph{Pismis~24$-$17 = HDE~319\,718~B = ALS~18\,752.}
\object[Cl Pismis 24 17]{}

Due to an error in GOSSS-DR1.0, the (f*) suffix was omitted there. The spectral type we obtain in GOSSS is the same as that in \citet{Maizetal07}.
See Figure~\ref{chart1} for a chart (Pismis~24 field).
The visual multiplicity for this target was measured in ACS/HRC images.
 
\paragraph{ALS~16\,052 = Pismis~24$-$13.}
\object[ALS 16052]{}

This star was not included in \citet{Maizetal04b} but was classified as O type by \citet{Massetal01}. 
See Figure~\ref{chart1} for a chart (Pismis~24 field). This is the star that has evacuated the cave-like hole at the bottom of the iconic HST image
\url{http://www.spacetelescope.org/images/heic0619a/} and is likely to be one of the youngest stars in our sample (which is consistent with the interpretation
of the z suffix indicating a young age). The pillar above it has been likely produced by Pismis~24-1~AB and Pismis~24-17 (the two brightest point
sources in the image), since those are the main sources of ionizing photons in the region. 
 
\paragraph{HD~161\,807.}
\object[HD 161807]{}

This star was not included in \citet{Maizetal04b} and, to our knowledge, it had never been classified as O type. We observed it because it was a bright star
classified as B0 \citep{Garretal77}. \citet{Garretal83} note that it is an eclipsing binary.
 
\paragraph{63~Oph = HD~162\,978.}
\object[* 63 Oph]{}

The spectral subtype was changed from O7.5 to O8 in GOSSS-DR1.1.
 
\paragraph{HD~163\,800.}
\object[HD 163800]{}

The ((f)) suffix was added in GOSSS-DR1.1.
 
\paragraph{HD~163\,892.}
\object[HD 163892]{}

This object is an SB1 system with a period of 7.83~d according to OWN data.
 
\paragraph{HD~164\,019.}
\object[HD 164019]{}

\citet{Walb82a} classified this star as O9.5~III and here we revise the luminosity class to IV. The metallic lines are particularly strong, hence the p suffix.
 
\paragraph{HD~164\,492~A.}
\object[HD 164492A]{}

This star is the main ionizing source of the Trifid Nebula and has three companions within 20\arcsec\ according to the WDS. Of those, the brightest is C, with a 
separation of 10\farcs9 and a $\Delta m$ of 1.1 magnitudes. We placed C on the slit and determined it is an early-B star.
 
\paragraph{HD~164\,536.}
\object[HD 164536]{}

This star was not included in \citet{Maizetal04b} but \citet{MacCBide76} classified it as O type.
This system is an SB1 with a 13.4~d period \citep{Willetal13}. The WDS lists a faint companion 1\farcs7 away.
The (n) suffix was changed to (n)z in GOSSS-DR1.1.
 
\paragraph{HD~166\,546.}
\object[HD 166546]{}

\citet{Walb73a} classified this star as O9.5~II-III and here we revise the luminosity class to IV.
 
\paragraph{15~Sgr = HD~167\,264.}
\object[* 15 Sgr]{}

This system is an SB1 with a 668~d period according to OWN data. The WDS lists a faint companion with a separation of 1\farcs3.
 
\paragraph{16~Sgr~AaAb = HD~167\,263~AaAb.}
\object[* 16 Sgr]{}

This system is an SB1 with a 14.75825~d period \citep{SticLloy01}.
\citet{Masoetal09} measure a $\Delta m$ of 2.0 magnitudes and a separation of 69 mas between Aa and Ab, obviously spatially unresolved in GOSSS. 
Aldoretta et al. (in preparation) give a similar separation but with a lower $\Delta m$ (as well as a position angle that differs by nearly 180\arcdeg).
The luminosity class was changed from II-III to III in GOSSS-DR1.1.
 
\paragraph{HD~168\,941.}
\object[HD 168941]{}

\citet{Walb82a} classified this star as O9.5~II-III and here we revise the luminosity class to IV. The metallic lines are particularly strong, hence the p suffix.
 
\paragraph{HD~175\,876.}
\object[HD 175876]{}

\citet{Walb73a} classified this star as O6.5~III(n)(f) and here we confirm that classification.

\begin{figure*}
\centerline{\includegraphics*[width=\linewidth, bb=50 225 545 725]{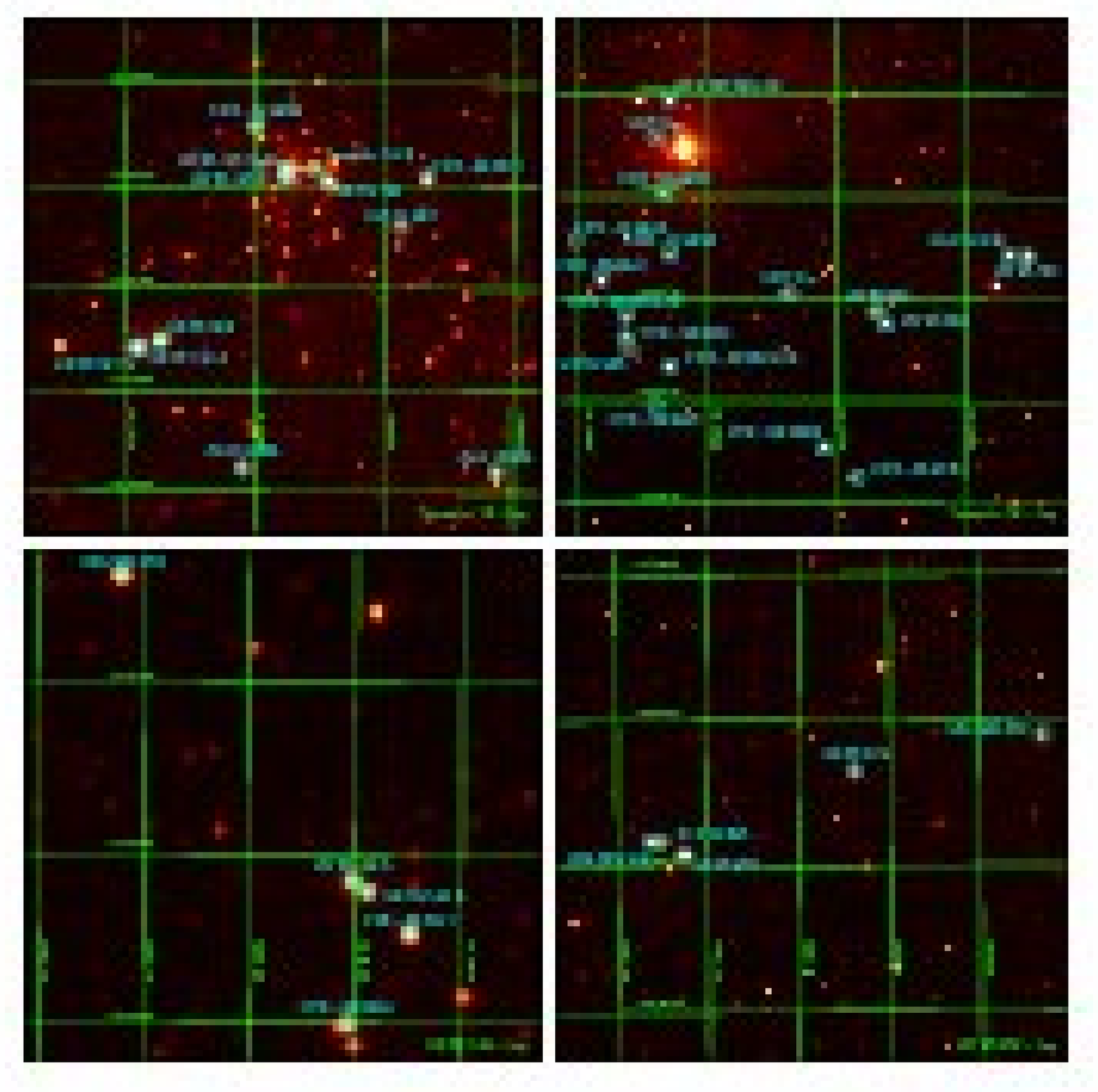}}
\caption{Twelve fields that include O stars for which we have obtained spectral types (marked in cyan or black). 
The background image is the 2MASS J band and is shown with a logarithmic stretch. The four panels displayed here (from left to right
and from top to bottom) have linear sizes of 5\arcmin, 10\arcmin, 3\arcmin, and 7\arcmin, respectively, and correspond to the Trumpler~14,
Trumpler~16, HD~93\,146, and HD~93\,632 fields. The low quality of the figure is caused by arXiv file size limitations.
[See the electronic version of the journal for a color version of this figure.]}
\label{chart1}
\end{figure*}	

\addtocounter{figure}{-1}

\begin{figure*}
\centerline{\includegraphics*[width=\linewidth, bb=50 225 545 725]{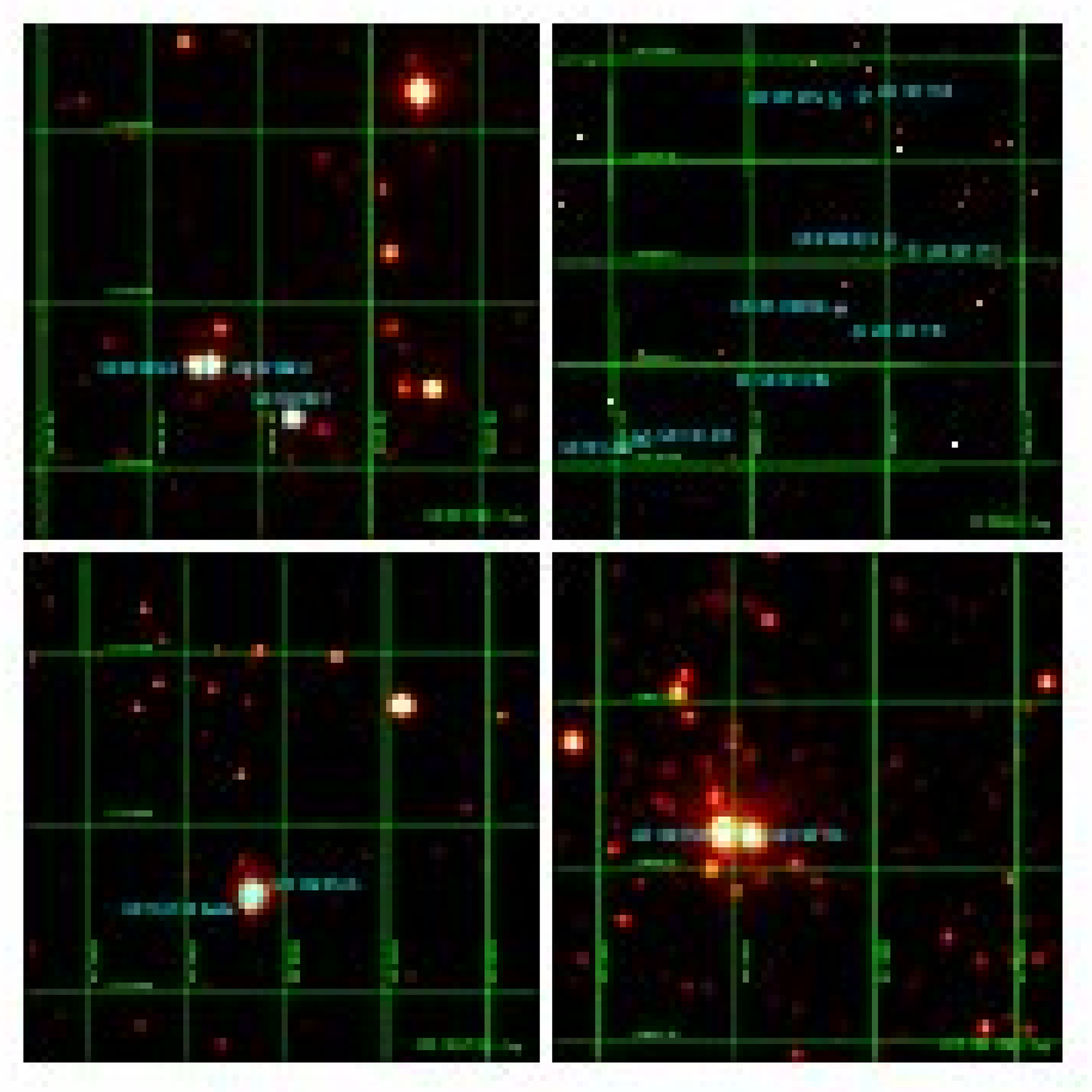}}
\caption{(continued).The four panels displayed here (from left to right and from top to bottom) have linear sizes of 3\arcmin, 25\arcmin, 
3\arcmin, and 3\arcmin, respectively, and correspond to the HD~92\,206, IC~2944, HD~124\,314, and HD~150\,136 fields. The low quality of the figure is caused by arXiv file size limitations.}
\end{figure*}	

\addtocounter{figure}{-1}

\begin{figure*}
\centerline{\includegraphics*[width=\linewidth, bb=50 225 545 725]{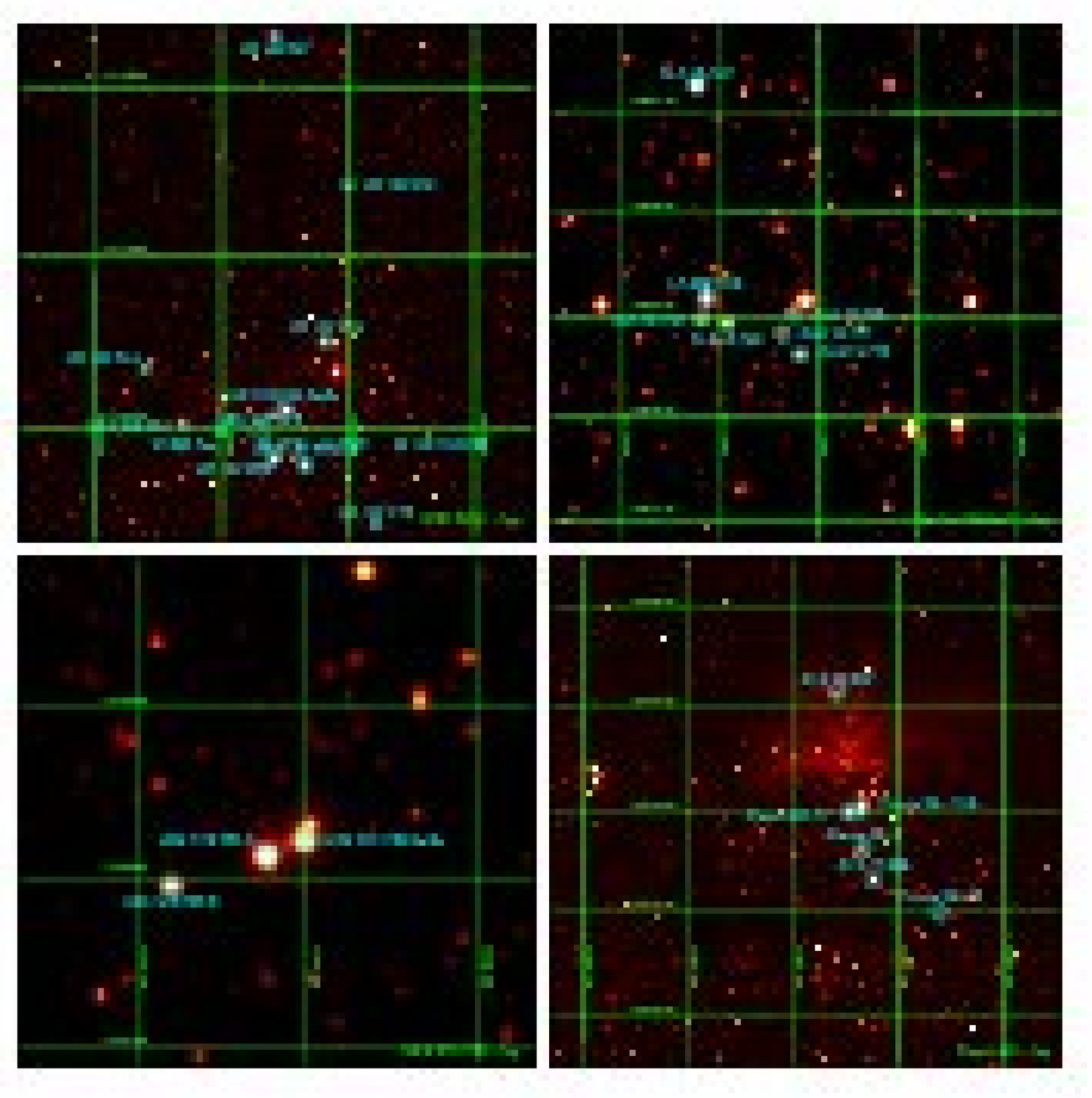}}
\caption{(continued).The four panels displayed here (from left to right and from top to bottom) have linear sizes of 15\arcmin, 5\arcmin, 
3\arcmin, and 10\arcmin, respectively, and correspond to the NGC~6231, Havlen-Moffat~1, HDE~319\,703, and Pismis~24 fields. The low quality of the figure is caused by arXiv file size limitations.}
\end{figure*}	

\section{Analysis}   
\label{sec:Anal}

\subsection{Sample selection and completeness}
\label{sec:Sample}

The definition of an O star is a purely spectroscopic one: it is a star that shows \HeII{4542} in absorption and, if \SiIII{4552} in absorption is present
(as it happens for late-O stars), it has to be equal to or weaker than \HeII{4542} (see Table~\ref{O8.5_B0}). Some low-mass stars can show those characteristics at the end 
of their lives (as sdO, PAGB, or PNN) but are excluded specifically from our sample since our interest lies in the study of massive stars (see \citealt{Driletal13} for the
spectral classification of sdO objects). Objects with 
masses above 20 M$_\odot$ spend most of their lives as O stars\footnote{With the exception of the most massive ones, that likely show a Wolf-Rayet spectrum 
during most of their existence.}, so using the O-type selection criterion provides the most straightforward way of choosing a large and uniform sample of massive 
stars. The main objections that can be made to this strict criterion to study massive stars are that [a] we are excluding the later stages of stellar evolution, 
[b] the limit between O and non-O stars is not a fixed boundary in \Teff, and [c] multiplicity and rotation may complicate the otherwise 
straightforward definition of what an O star is. 

\renewcommand{\labelenumi}{\bf [\alph{enumi}]}

\begin{enumerate}
 \item The first objection is indeed valid and one should always keep it in mind not to identify massive stars exclusively with O stars. For a complete picture, one should also 
       study the relatively short-lived high-luminosity B stars, supergiants of any type (including LBVs), and Wolf-Rayet stars. If we want to extend the lower mass limit to 
       stars that explode as core-collapse supernovae, we should also include the stars that are of spectral types B0-B2.5 during their main sequence lives. 
 \item The \Teff\ limit between the O and B spectral types depends not only on luminosity class (for the same subtype, a supergiant is cooler than a dwarf) but also on
       metallicity. The \SiIII{4552} line becomes weaker when metallicity decreases, so an O9.7 star at low metallicity will have a lower \Teff\ than one at solar
       metallicity. Therefore, when considering how much of its lifetime does a massive star consume as an O star using evolutionary tracks, those effects have to be 
       considered. 
 \item Multiplicity is a ubiquitous issue with massive stars, as the results of this paper emphasize. Probably nothing else complicates more the determination
       of the spectral type of an O star, forcing the use of multiple epoch spectroscopy and high-resolution imaging to establish how many objects one is looking at and
       whether a straightforward spectral type derived from a single observation can be trusted. Indeed, it is possible that some stars classified as B0 are in reality
       e.g. O9.7 + B0.2 binaries yet undiscovered. Fast rotation also complicates the identification of O stars as such, as some of the examples in GOSSS demonstrate
       (e.g. HD~161\,807). This occurs because at the O-B boundary, \HeII{4542} and \SiIII{4552} are relatively weak and close together, so in a fast rotator they may
       merge or become so diluted as to be hard to identify. In this respect, a tool such as MGB becomes quite useful since it artificially broadens the standard spectra
       for a more adequate comparison. Finally, multiplicity can also lead to interactions, which may alter the observed spectra and even change the fate of the star.
\end{enumerate}

GOSSS aims to build a collection of spectra of optically observable O stars as complete as possible. That implies observing many candidates that turn out to be non-O 
stars a posteriori as well as a small fraction of massive stars that are known a priori to be non-O but that are needed to establish an adequate knowledge of the rest of
the spectra. More specifically, an object is included in the GOSSS sample if it is within the accessible magnitude range and:

\begin{enumerate}
 \item It has previously received a classification as O type by some author (the primary criterion).
 \item It has previously received a classification as B0 by some author (the spectral proximity criterion).
 \item It is located within 3\arcmin\ of another star in the sample and has a $\Delta B$ small enough to allow the two stars to be observed within the slit 
       (the spatial proximity or opportunity criterion).
 \item It is a known or suspected early-type star of an interesting small-sized category (the ``zoo'' criterion).
 \item Additional information such as cluster membership, photometry, or X-ray data suggests that it can be an O star (the ``other'' criterion).
 \item It is a known or potential standard for early-type spectral classification (the standard criterion).
\end{enumerate}

The goal is to attain completeness in the first two categories down to a certain $B$ magnitude. Most objects ($\sim 90\%$) observed in the survey so far belong to the 
first three categories. As previously mentioned, in the first two papers we only present objects that turn out to be O stars.

Given the selection criteria and the ongoing status of the project, the GOSSS-DR1.1 sample is somewhat heterogeneous and covers a wide range in magnitude
(Fig.~\ref{BapJap}). Nevertheless, we have concentrated on observing all of the targets with $\Bap < 8$ and 175 (henceforth, the bright sample) out of the total 448
(henceforth, the full sample) are within that range. We believe 
that the O-star sample in GOSSS-DR1.1 is quite complete up to that magnitude, given that most bright stars have been classified before and that the tendency is to have 
more non-O stars classified as O than the other way around and that most missed O stars had been previously classified as B0 \citep{Maizetal13b}, so they are included in
the GOSSS sample\footnote{Note, however, that the GOSSS sample specifically excludes unresolved O+WR systems due to the difficulty of deducing information about the O
star from the combined spectrum.}. We will 
continue evaluating the completeness of the sample by searching for additional sources in the literature and, in the long term, by using photometric methods to select 
O-star candidates \citep{MaizSota08,Maiz13b}. That solution will become more necessary as we reach into dimmer magnitudes, where most stars have never been classified 
before. A word of caution: there are 175 systems with $\Bap < 8$ containing an O star but that is not the total number of individual O stars within that 
magnitude range because some of those contain two or more unresolved O stars. An entry in GOSSS may have two or more individual O stars due to multiplicity. 

\subsection{Magnitude and spatial distributions}
\label{sec:Distr}

\begin{figure*}
\centerline{\includegraphics*[width=0.465\linewidth, bb=28 10 566 566]{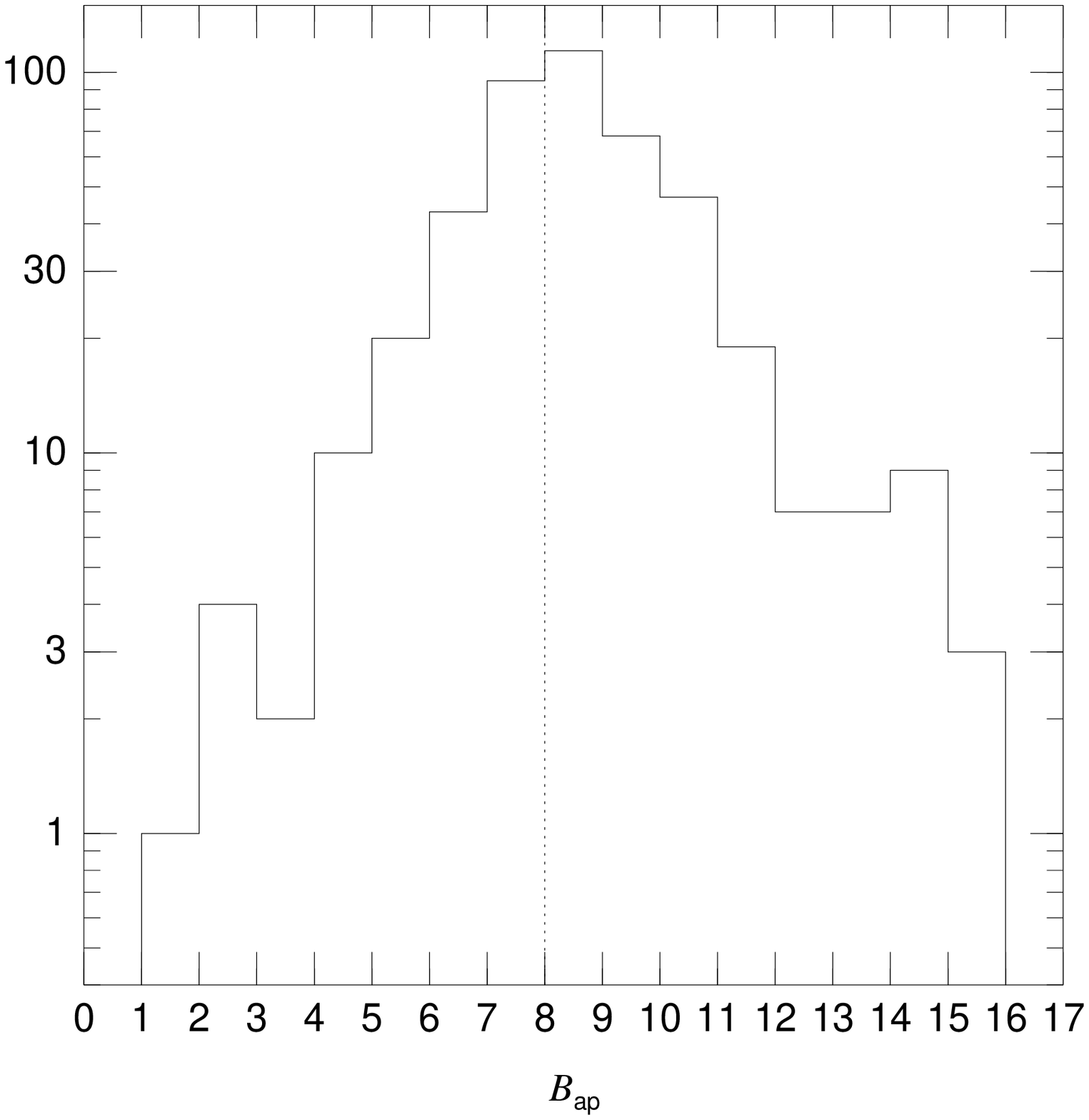} \
            \includegraphics*[width=0.520\linewidth                  ]{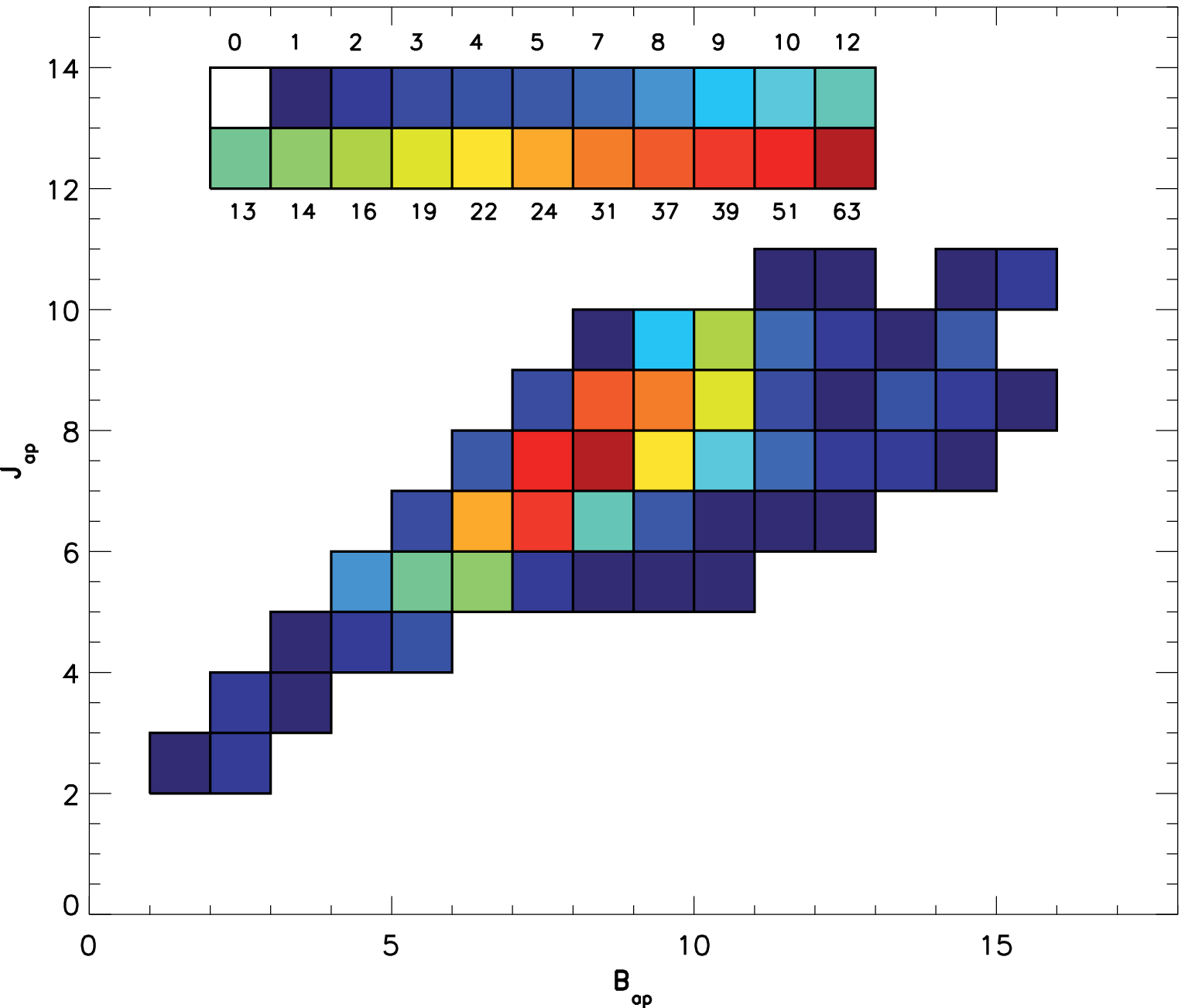}}
\caption{[left] \Bap\ histogram and [right] \Bap-\Jap\ density diagram for the stars in GOSSS-DR1.1. The dotted line in the left panel marks the expected completeness
limit in GOSSS-DR1.1. [See the electronic version of the journal for a color version of this figure.]}
\label{BapJap}
\end{figure*}	

\begin{figure*}
\centerline{\includegraphics*[width=0.55\linewidth, angle=90, bb=140 20 550 822]{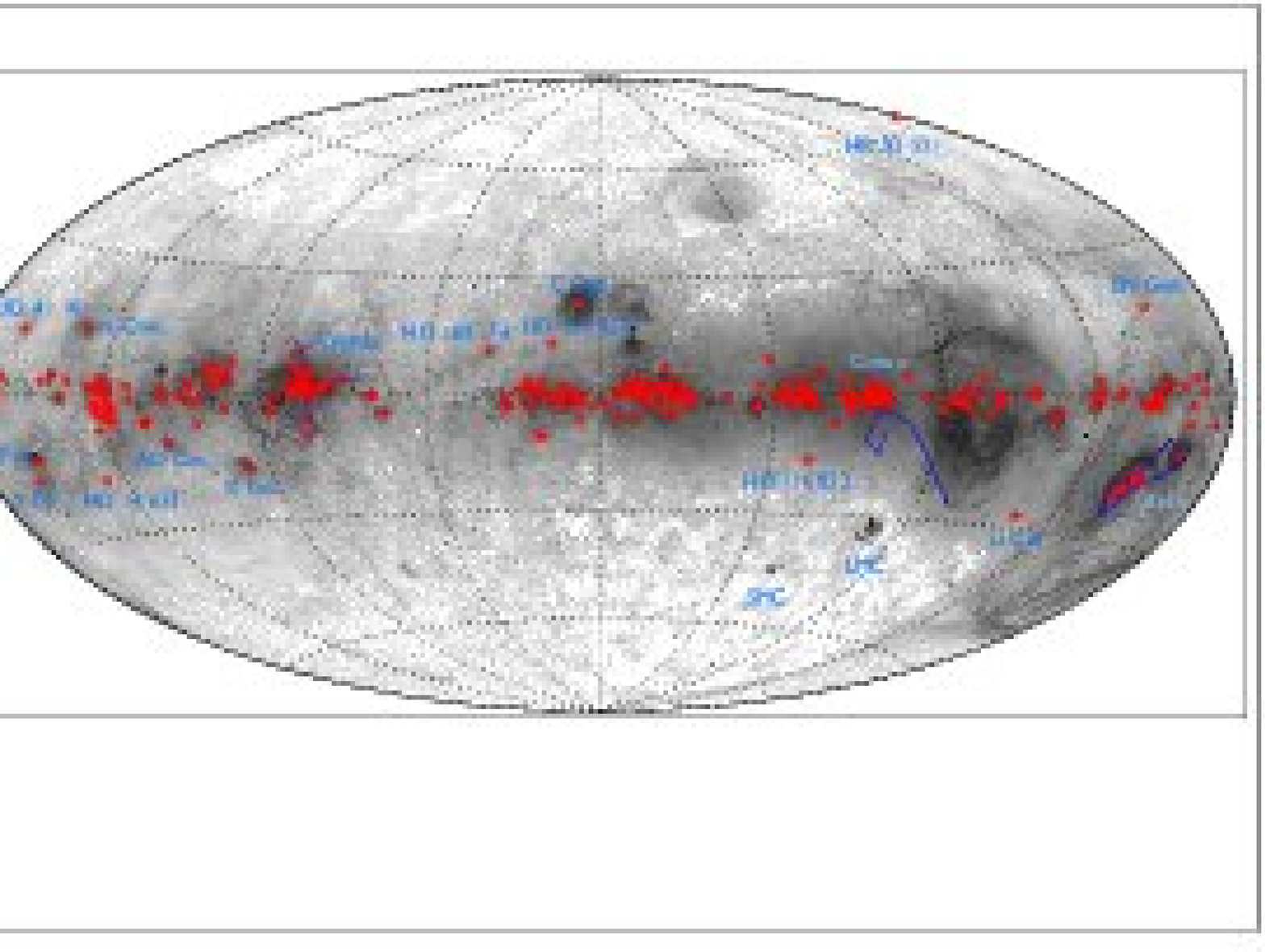}}
\caption{Spatial distribution of the GOSSS-DR1.1 sample in Galactic coordinates. The background image is the H$\alpha$ full-sky map of \citet{Fink03}. Some regions of the
sky and the stars located far from the Galactic Plane are labeled. The low quality of the figure is caused by arXiv file size limitations.
[See the electronic version of the journal for a color version of this figure.]}
\label{allsky}
\end{figure*}	

\begin{figure*}
\centerline{\includegraphics*[width=0.47\linewidth]{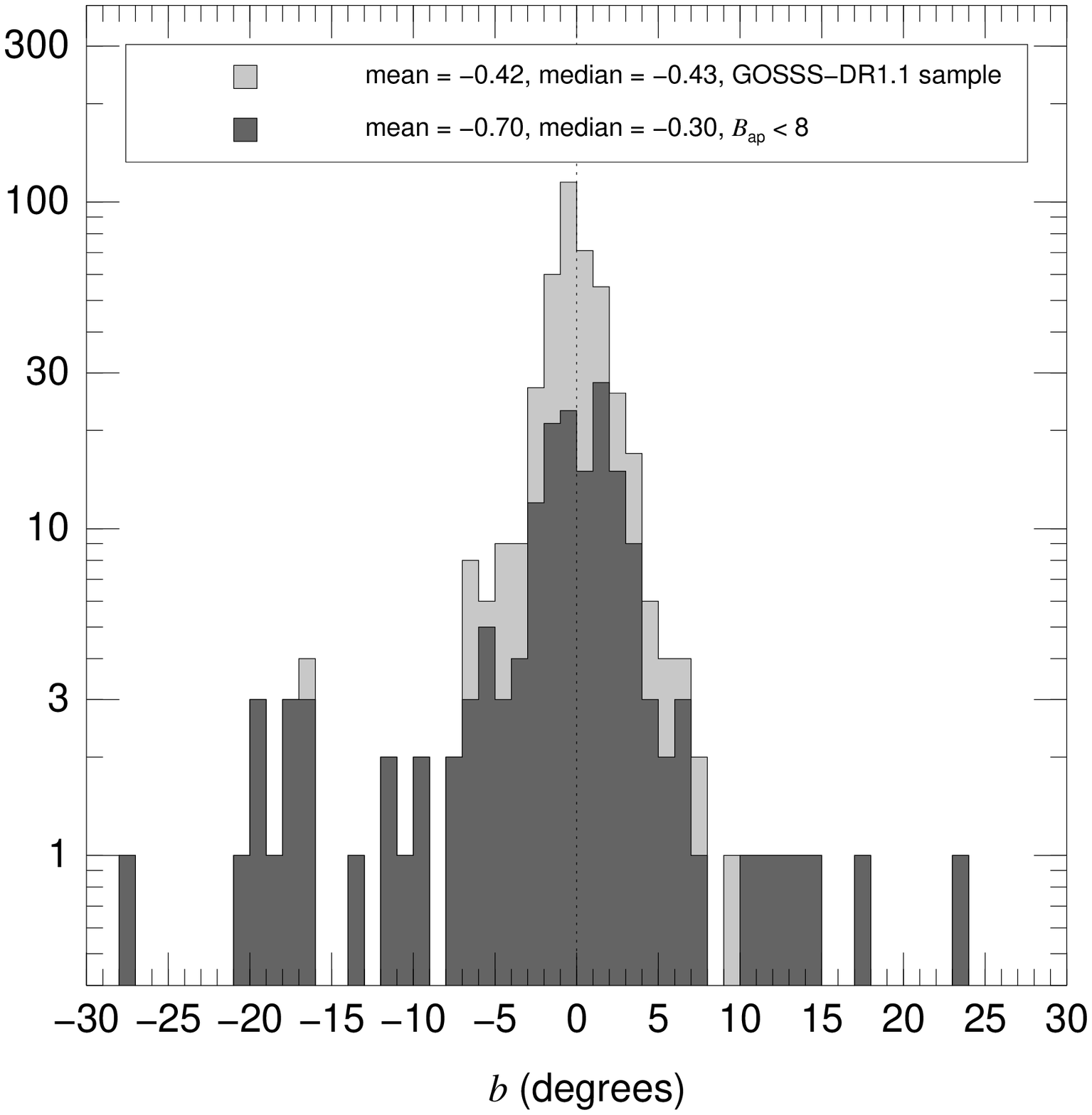} \
            \includegraphics*[width=0.47\linewidth]{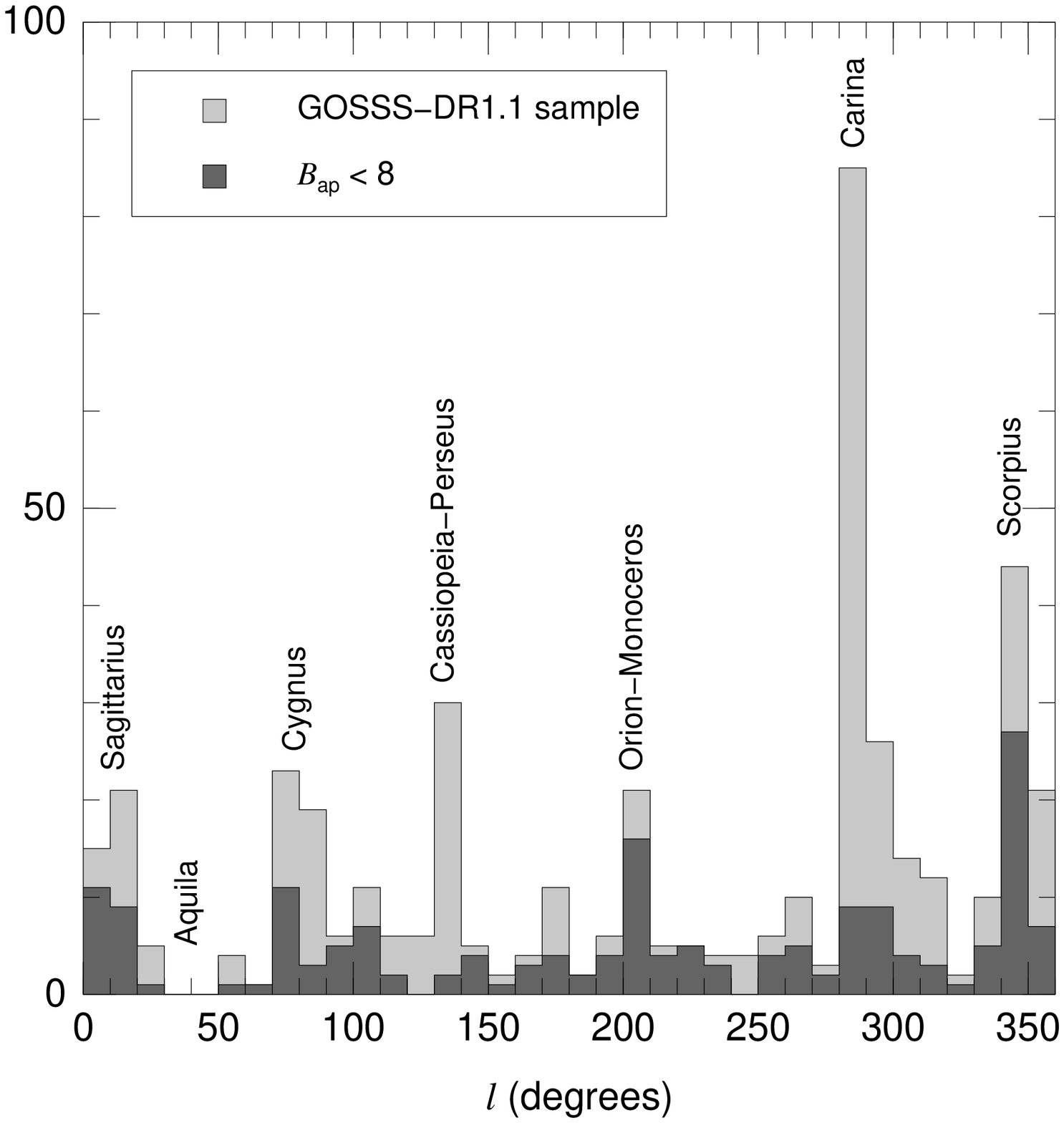}}
\caption{Galactic latitude [left] and longitude [right] histograms for the full and bright samples. HD~93\,521 is outside the range in the left panel
(see Fig.~\ref{allsky}). Some regions are labeled in the right panel.}
\label{latlong}
\end{figure*}	

The \Bap\ histogram in Figure~\ref{BapJap} shows that the number density per unit magnitude as a function of \Bap\ for $\Bap < 8$ grows approximately as 
$10^{\alpha B_{\rm ap}}$, with $\alpha$ slightly under 0.4, the value expected for an unextincted uniform spatial disk distribution. Beyond $\Bap = 8$, extinction and 
the decrease in the spatial density (especially towards the outer Galaxy) are expected to bend the distribution below the extrapolated behavior from bright magnitudes 
\citep{Maizetal13b}, but even taking into account that effect it is obvious that the GOSSS-DR1.1 sample is heavily incomplete already in the $\Bap = 9-10$ bin.

The magnitude density diagram in Figure~\ref{BapJap} also reflects the effect of extinction. Nearby stars have $\Bap-\Jap$ values close to the unreddened ones 
(between $-1.0$ and $-0.7$), clustering close to the $\Bap = \Jap$ line. As more distant stars are included, they become dimmer in \Jap\ but even more so in \Bap\ due to 
reddening. Hence, in the density diagram they concentrate around $\Jap \sim 8$. If dimmer stars were included, we would expect the locus in \Bap\ as a function of \Jap\ 
to become broader and more distant from the $\Bap = \Jap$ line.

\citet{Maizetal13b} have evaluated the number of systems with O stars in the GOSSS-DR1.1 sample and included additional systems with $\Bap \ge 8$ already observed by 
GOSSS, using a simple model to evaluate the number of O-type systems in the Galaxy. They conclude that, at a minimum, 14\,000-18\,000 such systems exist in the Milky Way. 
Allowing for effects such as the radial density gradient in the Galaxy, spectroscopic binaries (which are counted just once), WR+O systems, and patchy extinction may raise this 
number to 30\,000-50\,000 individual O stars. Those numbers can be compared with the 6500 predicted Galactic WR stars \citep{vadH01,Sharetal09} and are roughly
consistent in terms of the expected evolutionary paths and relative lifetimes of each phase.

Figure~\ref{allsky} shows the distribution in the sky of the GOSSS-DR1.1 sample and Figure~\ref{latlong} shows the latitude and longitude histograms. The latitude
distribution is significantly more irregular for the bright sample than for the full one. This effect is caused by the off-the-plane stars, which are concentrated in
the bright sample and the southern hemisphere (hence, the relatively large difference between the mean and median latitudes for the bright sample). In particular, the
classical Gould's belt structure is clearly seen in the anticenter direction but not so towards the Galactic Center (where only three O stars, $\zeta$ Oph, HD~157\,857 and
HD~165\,174 are clearly above the plane). This is the same asymmetry detected in the cluster population by \citet{Eliaetal09}. Both samples have negative values for their
mean and median latitudes, a sign of the location of the Sun above the Galactic Plane, detected for early-type stars with Hipparcos measurements 
\citep{Maiz01a,Maizetal08a}.

In the longitude histogram the largest OB associations (or groups of them) are easily identifiable. Also, the Aquila Rift and its surrounding area are notorious in that
histogram, as there are no GOSSS-DR1.1 stars between longitudes 30 and 55 degrees. The differences between the two samples are remarkable. Carina represents the largest 
peak in the full sample but is difficult to identify in the bright sample, as expected for a moderately distant and extincted region. Similar contrasts are seen for 
Cygnus and Cassiopeia-Perseus. In the other extreme, Orion-Monoceros dominates the two outer Galactic quadrants in the bright sample due to its proximity and low 
extinction.

\subsection{Spectral classification statistics}
\label{sec:Spclas}

\begin{table*}
\caption{Distribution by spectral subtypes and luminosity classes in GOSSS-DR1.1. The other category includes the Of?p class as well as stars without 
accurate luminosity classiﬁcations.}
\label{sptydist}
\begin{tabular}{lrrrrrrrr}
\\
\hline
 & \multicolumn{1}{c}{O2-3.5}  & \multicolumn{1}{c}{O4-5.5}  & \multicolumn{1}{c}{O6-7.5} & \multicolumn{1}{c}{O8-8.5} & \multicolumn{1}{c}{O9-9.2} & \multicolumn{1}{c}{O9.5-9.7} & \multicolumn{1}{c}{Total} \\
\hline
Ia     & \nodata & \nodata &      8 &      4 &      2 &        3 &    17 \\
I/Iab  &       7 &      13 &      7 &      4 &      8 &       10 &    49 \\
Ib     & \nodata & \nodata &      7 &      3 &      7 &        7 &    24 \\
II     & \nodata & \nodata &      9 &      4 &      3 &       14 &    30 \\
III    &       1 &       8 &     17 &     11 &     12 &       22 &    71 \\
IV     & \nodata & \nodata &     10 &      6 &     17 &       25 &    58 \\
V      &       4 &      24 &     65 &     38 &     15 &       21 &   167 \\
other  &       0 &       2 &      7 &      6 &      4 &       13 &    32 \\
\hline
Total  &      12 &      47 &    130 &     76 &     68 &      115 &   448 \\
\hline
\end{tabular}
\end{table*}	

Table~\ref{sptydist} shows the distribution by spectral subtypes and luminosity classes in GOSSS-DR1.1. Note that luminosity classes Ia, Iab, Ib, II, and IV are not defined for O2-O5.5. Also, in that range all supergiants are
simply called I. The O stars in the sample are heavily concentrated towards luminosity classes III-V even though selection biases favor supergiants. The concentration of O dwarfs towards the middle spectral subtypes is a 
likely selection effect, as late-type dwarfs are dimmer and tend to be less conspicuous in their spectral characteristics and their ionizing effect (hence, they have been more likely missed by the previous studies that were
used to build the sample). The proportion between middle and late subtypes is reversed for brighter luminosity classes, as expected by the elimination of that selection effect and the decrease in initial stellar mass as a
function of spectral subtype for constant luminosity class. It will be interesting to compare Table~\ref{sptydist} with future GOSSS data releases, where a larger fraction of dimmer stars will be included.

Another interesting statistical issue regarding the GOSSS sample was presented by \citet{Maizetal13b}. The false positive rate in a sample of 1014 stars previously classified as being of O type is 24.9\% (i.e. one quarter of
those stars appear as O in the literature but GOSSS reveals them not to be so). That number was calculated using a sample of 1014 objects, larger than that in GOSSS-DR1.1 and shows a clear tendency with magnitude: the quality
of the literature spectral classifications decreases as the stars become dimmer. Therefore, we expect that when GOSSS is completed, the false positive rate will be even higher. On the other hand, the false negative rate (stars
without previous classifications as O that turned out to be of that type) is 6.4\%. Those objects were observed in most cases because they had previous classifications as B0 or because they could be placed within the same slit
as an O star that was being observed. This indicates that there are still relatively bright O stars waiting to be discovered. Doing so will likely require large-scale photometric surveys of the Galactic Plane (that do not 
saturate stars around magnitude 12 and brighter) followed up by spectroscopic surveys with wide-angle multi-fiber spectrographs.

\subsection{Multiplicity}
\label{sec:Mult}

\subsubsection{The multiplicity sample}

\begin{figure}
\centerline{\includegraphics*[width=\linewidth]{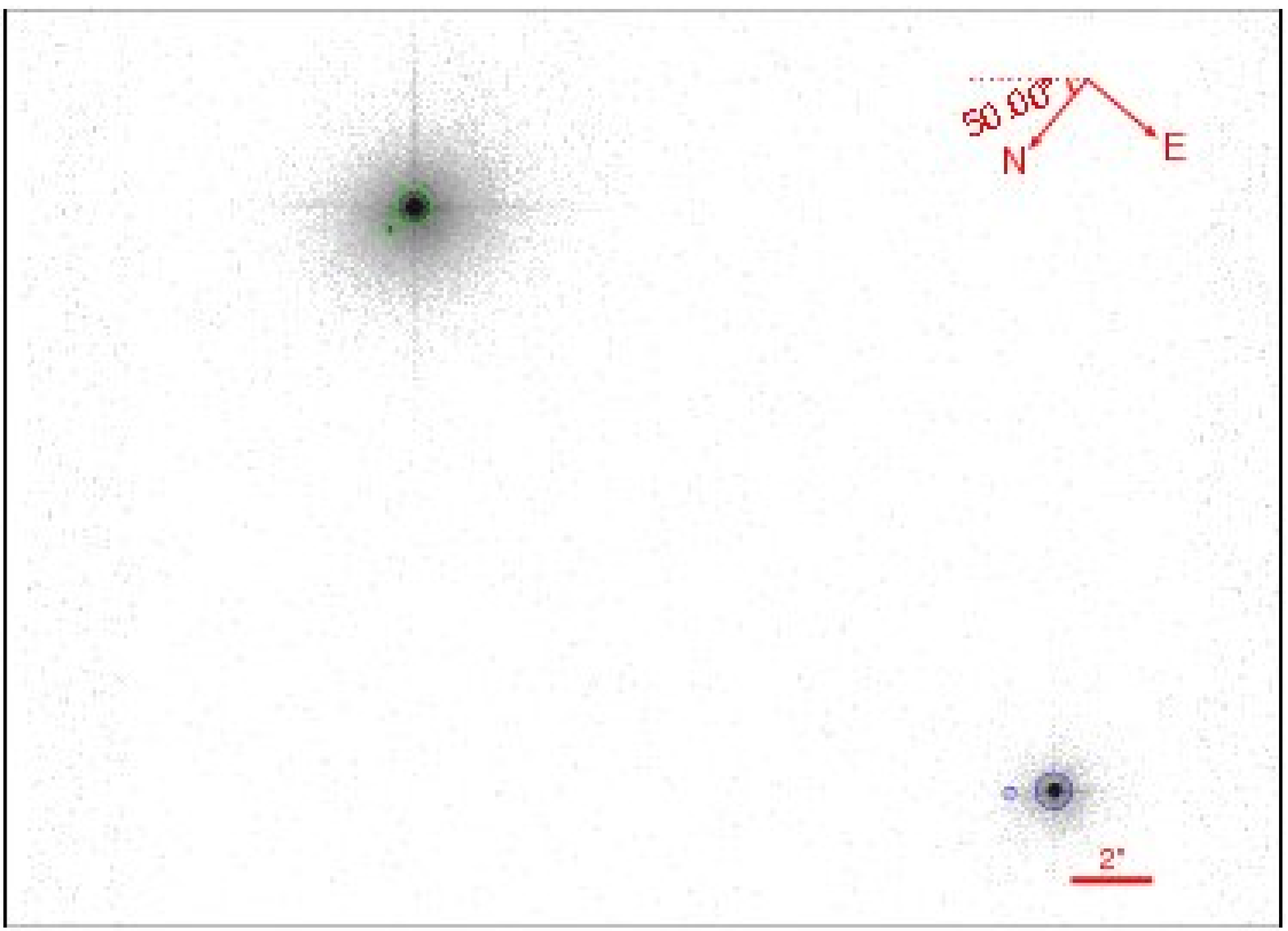}}
\centerline{\includegraphics*[width=\linewidth]{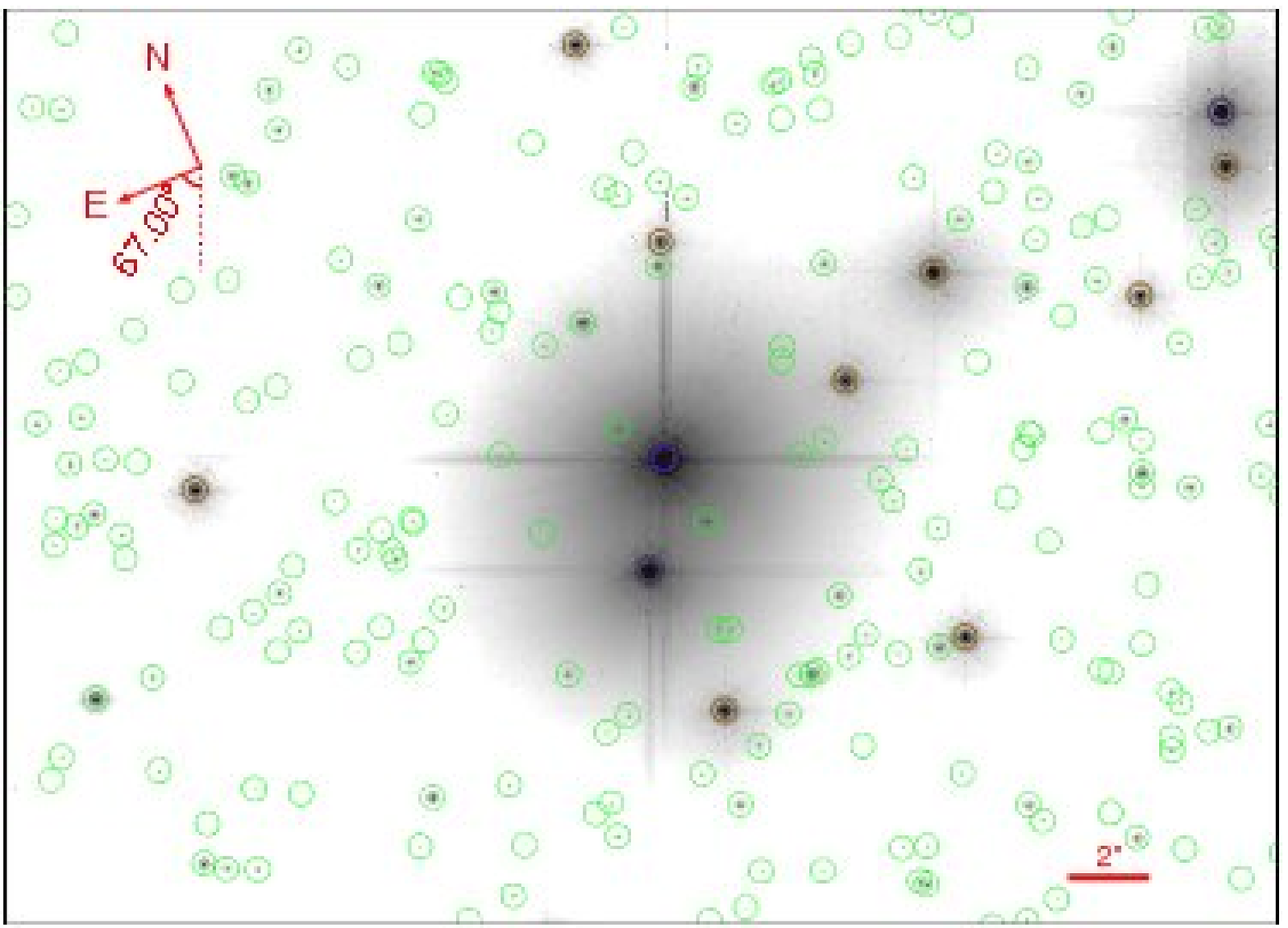}}
\caption{[top] Short exposure time (2.4 s) image of a field in the Carina Nebula obtained with ACS/HRC using the F850LP filter. Green circles mark the position of 
HD~93\,162 and its companion while blue circles mark the position of ALS~15\,210 and its companion. [bottom] Combination of short- and long-exposure (total~=~256~s) images
of the core of Trumpler~14 obtained with ACS/HRC using the F850LP filter. Blue circles mark the position of the three GOSSS systems: HD~93\,129~AaAb (center of the field,
a double system in the same image when observed at the proper contrast and magnification), HD~93\,129~B (SE of AaAb), and Trumpler 14$-$9 (top right corner). Red circles 
mark the position of stars with magnitudes similar to or brighter than those of the two companions in the top field. Green circles mark the positions of the stars 
detected with dimmer magnitudes. The low quality of the figure is caused by arXiv file size limitations.}
\label{HRC}
\end{figure}	

To study O-star multiplicity, we restrict ourselves to the sample defined in section~\ref{sec:SB}: 194 O stars with $\delta < -20\arcdeg$ with existing multi-epoch
high-resolution studies (mostly OWN). Defining spectroscopic binarity is relatively straightforward though in some cases the current data are inconclusive and the star is 
labeled as SB1?, SB2? or SB3? in Table~\ref{spectralclasS}. For that reason, SB1? and SB2? systems are considered {\it possible SBs}, while SB3? systems (which are SB2 
systems with a possible third component) and the rest of the SB-labeled objects are considered as {\it certain SBs}. This gives us two estimates (low and high) for
spectroscopic binarity according to the current data, depending on whether we include the possible SBs or not\footnote{Of course, there should be many systems with periods
comparable to those analyzed by OWN that remain undetected because the companion is too dim or the orientation is unfavorable. See \citet{Mayeetal13} for the borderline 
case of HD~165\,246. Also note that the companion mass distribution appears to be relatively flat (\citealt{Maiz08a} and references therein).}
With respect to the number of components in the system, those labeled as SB1, SB1E, SB2, SB2E, and SBE\footnote{The E stands for eclipsing.}
are considered to have two; those labeled as SB3 or SB3E, three; and those labeled as SB1+SBE, 
SB1E+SB1, SB2+SB1, SB2+SB1E, or SB2+SB2, four. SB1? and SB2? systems have one certain and two possible components while SB3? systems have two certain and three possible 
components.

Visual binarity is more difficult to ascertain, since in most cases the time baseline is not long enough to establish an orbit with certainty or even that the system is
bound and not a chance alignment with a foreground or cluster member (indeed, in many cases only a single epoch is available). Given the need to analyze a large sample, we
estimate the probability that a given candidate is a real companion. That is what \citet{Maiz10} did for his Lucky Imaging study of northern massive (mostly O) stars, 
in which he compared the number of components detected with those expected from the density of 2MASS sources in the surrounding area. That paper found that
for separations below 5\arcsec\ likely all the detected objects are real companions while in the 8\arcsec-14\arcsec\ range $78\pm 6\%$ are also so. Note that for a
distance of 1 kpc, 10\arcsec\ corresponds to 10\,000 AU. That is a large value but not an unreasonable one: Proxima Centauri is 15\,000 AU away from $\alpha$~Cen~AB and
possibly bound to it \citep{WertLaug06}, and is a low-mass, old system. Similar intermediate- and low-mass systems with separations of up to several parsecs are known 
(\citealt{DuchKrau13} and references therein). One would expect high-mass, short-lived massive stars to remain bound at large distances in time scales of a few Ma even more 
easily\footnote{Note, however, that for distances in the range 50\,000-100\,000 AU, the orbital timescale becomes larger than the life time of an O star and defining such a 
system as bound becomes meaningless, since it should become unbound by the first supernova explosion before a full orbit takes place.}. 

Our sample differs from that of \citet{Maiz10} in that it consists of southern stars and that the companions were detected in a heterogeneous manner. The distances and
spectral types, however, are similar so the southern character should not make a difference. Also, the nearest companions should also be bound in most (if not all, but
see below for an exception) cases and most of the objects between 5\arcsec\ and 10\arcsec\ are detected by 2MASS, as in \citet{Maiz10}. In any case, in analogy with SBs, 
we produce two estimates of {\it certain VBs} and {\it possible VBs}: the first corresponds to the detections of companions with separations inferior to 5\arcsec\ and the 
second to those between 5\arcsec\ and 10\arcsec\ (hence we also derive low and high estimates for visual binarity). This is a conservative approach, since we expect 
most of the {\it possible VBs} (and even some with larger separations not included here) to be bound objects except when we are observing a star in a dense cluster.

Some of those issues are illustrated by the two ACS/HRC images shown in Figure~\ref{HRC}. On the top panel, we see that the two bright stars (both in GOSSS-DR1.1 and
outside a well defined cluster) have
apparent companions with separations close to 1\arcsec. Given the absence of any other sources in the field and the proximity to the two sources, it is quite likely that
these are bound systems\footnote{Furthermore, in each case the mass in the system is of the order of 50~M$_\odot$ or higher and the physical separations in the plane of the sky
are relatively low, 2000-3000 AU.}. However, either companion would likely be undetected in a Lucky Imaging survey such as the one by \citet{Maiz10} because they are in an 
unaccessible part of the $\Delta m$-separation plane. Indeed, there is plenty of parameter space to explore for VBs and our current estimates are likely to err on the
cautious side. The bottom panel illustrates another issue: what happens when the data are extremely good (in terms of spatial resolution and exposure time) and the star is 
located at the core of a rich compact cluster such as Trumpler 14. In that case, we start detecting cluster members (even within 5\arcsec) that are likely not bound to the 
star. This contamination may turn out to be a problem for future data if one is trying to measure multiplicity but it should not worry us for this work: 
Trumpler 14 is the densest cluster in our sample and is 
also the one with the best quality ACS/HRC data. Therefore, the bottom panel in Figure~\ref{HRC} is an extreme case. Also, note that it is quite possible that
HD~93\,129~B is bound to Aa+Ab. The total mass of the system is above 100~M$_\odot$ and the projected separation is only 6000 AU. 

\subsubsection{Multiplicity statistics}

\begin{table}
\caption{Multiplicity numbers and frequencies (sample size: 194).}
\centerline{
\begin{tabular}{lrc}
 \\
\hline
\multicolumn{1}{c}{Type} & \multicolumn{1}{c}{\#} & \multicolumn{1}{c}{\%} \\
\hline
Total binaries (low):  & 126 & 64.9$\pm$3.4 \\
Total binaries (high): & 176 & 90.7$\pm$2.1 \\
SBs (low):             &  97 & 50.0$\pm$3.6 \\
VBs (low):             &  58 & 29.9$\pm$3.3 \\
SBs (high):            & 117 & 60.3$\pm$3.5 \\
VBs (high):            & 148 & 76.3$\pm$3.1 \\
S+VBs (low):           &  29 & 14.9$\pm$2.6 \\
\hline
\end{tabular}
}
\label{binstat}
\end{table}

Our multiplicity statistics for the studied systems (not for the individual components within a system) are shown in Table~\ref{binstat}. 
The difference between the low and high estimates is smaller for SBs than for VBs. This is expected, given that
the properties of spectroscopic binaries are easier to establish (if one has the required data, as it is the case with OWN) than those of visual binaries, given their
much shorter orbital periods. The values shown are not corrected for completeness.

We detect certain (spectroscopic or visual) companions in 64.9\% of the sample and the value increases to
90.7\% when the possible detections are included. To our knowledge, the second value {\it is the highest binary fraction ever measured for a large sample of massive 
stars}\footnote{Note that the $\sim$90\% values of the multiplicity fraction provided by some studies such as \citet{KimiKobu12} are extrapolated using an expected period 
distribution, not directly measured as in our data. Other homogeneous studies of multiplicity in O stars give minimum values of 43\% (\citealt{Chinetal12}, field), 69-73\% 
(\citealt{Chinetal12}, non-field), and 44-67\% \citet{Sanaetal08b,Sanaetal09,Sanaetal11a} for spectroscopic binaries, which are consistent with our results.}
Compared to the results reviewed by \citet{DuchKrau13}, the SB multiplicity frequency for high-mass stars is on the low side of their value ($70\pm 9\%$) but note that our 
results are not corrected for completeness and that our high value is consistent with theirs. On the other hand, our high value for VBs is significantly larger than theirs 
($45\pm 5\%$) and that explains why our fraction of total binaries is so high. The difference for the VB multiplicity frequency is likely caused by a better sensitivity to
low-mass systems at large separations, a problem that has long plagued this type of studies \citep{DuchKrau13}.

Indeed, the fraction of possible binaries is so large that there are only 18 stars for which no sign of binarity is observed. Of those, only four (22.2\%) are of 
luminosity class IV or V, a value that should be compared with the 99 stars (50.8\%) that belong to those classes overall. One of those fours stars is $\mu$~Col, a known
runaway, so its singularity needs not be primordial. The remaining three (HD~96\,715, HD~99\,897, and HD~101\,298) are all earlier than O7 i.e. they are more luminous than
most O stars of luminosity classes IV and V. What can cause the overabundance of apparently single systems among O giants and supergiants? One explanation is that 
this is an observational effect. All binary detection methods benefit from a small $\Delta m$, so for the earliest dwarfs and for giants and supergiants it is easier for a
companion to remain undetected. An alternative explanation for the giants and supergiants is that the difference is real and that those stars are single more frequently.
This could be because they have a higher chance of being runaways (because they are older on average than dwarfs) or, perhaps more interestingly, because some of them
are the result of mergers \citep{deMietal13}. If any of those explanations is true and we also consider the fact that we know we are not detecting some binary systems,
then it is possible that {\it all stars above 15-20 M$_\odot$ are born in multiple systems.}

Finally, a relatively large fraction (14.9\%) are confirmed S+VB systems i.e. they include at least three objects and two of the orbits have significantly different
periods (hierarchical multiple systems). The higher-order multiplicity of O-type stars also shows up when we calculate the average number of components per stellar system.
If only certain companions are included, the number is 2.0. If possible companions are added, the value grows to 3.4. Such a large value is not caused by the possible
contamination of a few objects with many possible companions, as the median of the distribution is 3 i.e. more than half of the 194 stars are in potentially triple or
higher-order multiple systems.

\begin{acknowledgements}

Support for this work was provided by: [a] the Spanish Government Ministerio de Ciencia e Innovaci\'on through 
grants AYA2007-64052, AYA2007-64712, AYA 2010-17631, AYA 2010-15081, the Ram\'on y Cajal Fellowship program, 
and FEDER funds; [b] the Junta de Andaluc\'{\i}a
grant P08-TIC-4075; [c] NASA through grants GO-10205, GO-10602, GO-10898, and GO-11981 from the Space Telescope 
Science Institute, which is operated by the Association of Universities for Research in Astronomy Inc., under 
NASA contract NAS~5-26555; [d] the Direcci\'on de Investigaci\'on de la Universidad de La Serena (DIULS 
PR09101); [e] the ESO-Government of Chile Joint Committee Postdoctoral Grant; and [f] the Chilean Government 
grants FONDECYT Regular 1120668 and FONDECYT Iniciaci\'on 11121550.
This research has made extensive use of: [a] Aladin \citep{Bonnetal00};
[b] the SIMBAD database, operated at CDS, Strasbourg, France; [c] the Washington Double Star Catalog,
maintained at the U.S. Naval Observatory \citep{Masoetal01}; and [d] the 2MASS Point Source Catalog \citep{Skruetal06}.
We would like to thank Miguel Penad\'es Ordaz for his help in the compilation of data for this paper and the anonymous referee for his/her useful comments.

\end{acknowledgements}

\bibliographystyle{apj}
\bibliography{general}

\eject

$\,\!$ 

\eject

$\,\!$ 

\addtolength{\textheight}{-5mm}
\addtolength{\topmargin}{35mm}
\addtolength{\textwidth}{50mm}
\addtolength{\oddsidemargin}{-15mm}
\addtolength{\evensidemargin}{-15mm}
\begin{deluxetable}{llllllllllllll}
\rotate
\tablecaption{Spectral classifications for stars with $\delta > -20\arcdeg$}
\tablewidth{0pt}
\tabletypesize{\scriptsize}
\tablehead{\colhead{Name} & \colhead{GOSSS ID} & \colhead{RA (J2000)} & \colhead{dec (J2000)} & \colhead{SC} & \colhead{LC} & \colhead{Qual.} & \colhead{Second.} & \colhead{Altern. classification} & \colhead{Ref.}  & \colhead{SB} & \colhead{VB}& \colhead{Sect.} & \colhead{Flag}}
\startdata 
AO~Cas                    & GOS 117.59$-$11.09\_01 & 00:17:43.059 & $+$51:25:59.12 & O9.2     & II      & \nodata      & O8 V((f))      & O9.5 III + O8 V                           & B91     & SB2E    & 1    & \ref{sec:PI}          & ch      \\ 
HD~5005~D                 & GOS 123.12$-$06.25\_01 & 00:52:48.954 & $+$56:37:30.83 & O9.2     & V       & \nodata      & \nodata        & \nodata                                   & \nodata & \nodata & 0-3  & \ref{sec:O9.2}        & ch      \\ 
HD~12\,323                & GOS 132.91$-$05.87\_01 & 02:02:30.126 & $+$55:37:26.38 & ON9.2    & V       & \nodata      & \nodata        & \nodata                                   & \nodata & SB1     & 0    & \ref{sec:O9.2}        & ch      \\ 
HD~12\,993                & GOS 133.11$-$03.40\_01 & 02:09:02.473 & $+$57:55:55.93 & O6.5     & V       & ((f)) Nstr   & \nodata        & \nodata                                   & \nodata & \nodata & 1-2  & \ref{sec:PI}          & ch      \\ 
HD~15\,558~A              & GOS 134.72$+$00.92\_01 & 02:32:42.536 & $+$61:27:21.56 & O4.5     & III     & (f)          & \nodata        & O5.5 III(f) + O7 V                        & D06     & SB2     & 2-5  & \ref{sec:PI}          & ch      \\ 
BD~$+$60~513              & GOS 134.90$+$00.92\_01 & 02:34:02.530 & $+$61:23:10.87 & O7       & V       & nz           & \nodata        & \nodata                                   & \nodata & \nodata & 0    & \ref{sec:PI}          & ch      \\ 
HD~16\,429~A              & GOS 135.68$+$01.15\_01 & 02:40:44.951 & $+$61:16:56.04 & O9       & II-III  & (n) Nwk      & \nodata        & O9.5	II	+	O8 III-IV + B0 V?               & M03     & SB3     & 2-3  & \ref{sec:PI}          & ch      \\
\enddata
\tablecomments{{\it GOSSS ID} is the identification for each star with ``GOS'' standing for ``Galactic O Star''. {\it Ref.} is the reference for the alternative classification. {\it SB} and {\it VB} are the spectroscopic and visual binary status, respectively. {\it Sect.} is the section where the star is discussed. {\it Flag} can be either ``ch'' (GOSSS O-type classification changes from \citealt{Sotaetal11a}), ``\ldots'' (O-type classification does not change from \citealt{Sotaetal11a}), ``new'' (star is not present in \citealt{Sotaetal11a}).}
\tablerefs{B91: \citet{BagnGies91}, B97: \citet{Burketal97}, B07: \citet{Boyaetal07a}, D04: \citet{DeBeetal04}, D06: \citet{DeBeetal06a}, G93: \citet{Giesetal93}, H06: \citet{Hilletal06}, M03: \citet{McSw03}, M10: \citet{Mahyetal10}, M13: \citet{Mahyetal13}, N12: \citet{Nazeetal12c}, S11: \citet{SimDetal11b}, T11: \citet{tenBetal11}} 
\label{spectralclasN}
\end{deluxetable}

\begin{deluxetable}{lllllllllllllll} 
\rotate 
\tablecaption{Spectral classifications for stars with $\delta < -20\arcdeg$} \tablewidth{0pt} \tabletypesize{\scriptsize}
\tablehead{\colhead{Name} & \colhead{GOSSS ID} & \colhead{RA (J2000)} & \colhead{dec (J2000)} & \colhead{SC} & \colhead{LC} & \colhead{Qual.} & \colhead{Second.} & \colhead{Altern. classification} & \colhead{Ref.}  & \colhead{Sam.} & \colhead{SB} & \colhead{VB}& \colhead{Sect.} & \colhead{Flag}}
\startdata                                                                      
$\mu$~Col                 & GOS 237.29$-$27.10\_01 & 05:45:59.895 & $-$32:18:23.18 & O9.5     & V       & \nodata     & \nodata    & \nodata                                        & \nodata & yes     & \nodata  & 0    & \ref{sec:Nor}         & \nodata \\ 
29~CMa                    & GOS 237.82$-$05.37\_01 & 07:18:40.378 & $-$24:33:31.32 & O7       & Ia      & fp var      & \nodata    & \nodata                                        & \nodata & \nodata & SB2E     & 0    & \ref{sec:SB2}         & ch      \\ 
$\tau$~CMa~AaAb           & GOS 238.18$-$05.54\_01 & 07:18:42.487 & $-$24:57:15.78 & O9       & II      & \nodata     & \nodata    & \nodata                                        & \nodata & yes     & SB1+SBE  & 2-3  & \ref{sec:Nor}         & \nodata \\ 
HD~57\,236                & GOS 235.64$-$04.02\_01 & 07:19:30.102 & $-$22:00:17.29 & O8.5     & V       & \nodata     & \nodata    & \nodata                                        & \nodata & yes     & \nodata  & 0-1  & \ref{sec:Nor}         & ch      \\ 
HD~64\,315~AB             & GOS 243.16$+$00.36\_01 & 07:52:20.284 & $-$26:25:46.69 & O5.5     & V       & z           & O7 V       & \nodata                                        & \nodata & yes     & SB2+SB2  & 2    & \ref{sec:SB2}         & ch      \\ 
CPD~$-$26~2716            & GOS 243.82$+$00.14\_01 & 07:53:01.007 & $-$27:06:57.75 & O6.5     & Iab     & f           & \nodata    & \nodata                                        & \nodata & yes     & \nodata  & 0-1  & \ref{sec:Nor}         & ch      \\ 
HD~64\,568                & GOS 243.14$+$00.71\_01 & 07:53:38.206 & $-$26:14:02.62 & O3       & V       & ((f*))z     & \nodata    & \nodata                                        & \nodata & yes     & SB1?     & 1-3  & \ref{sec:Nor}         & ch      \\
\enddata
\tablecomments{{\it GOSSS ID} is the identification for each star with ``GOS'' standing for ``Galactic O Star''. {\it Ref.} is the reference for the alternative classification. {\it Sam.} marks the stars belonging to the binarity sample (see section~\ref{sec:SB}). {\it SB} and {\it VB} are the spectroscopic and visual binary status, respectively. {\it Sect.} is the section where the star is discussed. {\it Flag} can be either ``ch'' (O-type classification changes from \citealt{Maizetal04b}), ``\ldots'' (O-type classification does not change from \citealt{Maizetal04b}), ``new'' (star is not present or is not O type in \citealt{Maizetal04b}).}
\tablerefs{A01: \citet{Albaetal01}, A02: \citet{Albaetal02},  A10: \citet{Ariaetal10}, C07: \citet{Campetal07},  F01: \citet{Freyetal01},  F13:  \citet{Ferretal13},   G02: \citet{Giesetal02}, 
           L07: \citet{Lindetal07}, M01: \citet{Morretal01},  M07: \citet{Maizetal07}, M12: \citet{Mahyetal12},  M13: \citet{Mayeetal13},  N88:  \citet{NiemMorr88},   N05: \citet{Nazeetal05}, 
           N06: \citet{Niemetal06}, P01: \citet{Pennetal01},  P02: \citet{Pennetal02}, P11: \citet{Parketal11},  R00: \citet{Rauwetal00}, R01a:  \citet{Rauwetal01a}, R01b: \citet{Rauwetal01b}, 
           R09: \citet{Rauwetal09} ammended in text,          R12: \citet{Rauwetal12}, S08: \citet{Sanaetal08b}, S11: \citet{Sanaetal11a}} 
\label{spectralclasS} 
\end{deluxetable}

\end{document}